\documentclass[10pt,twocolumn,superscriptaddress,amssymb,amsmath,aps,prd]{revtex4-2}
\usepackage{graphicx}

\usepackage[compat=1.1.0]{tikz-feynman}
\usepackage{adjustbox}
\tikzset{graviton/.style={decorate, decoration={snake, amplitude=.4mm, segment length=2.0mm, pre length=.25mm, post length=.25mm}, double}}
\tikzset{graviton_ends/.style={decorate, decoration={snake, amplitude=.4mm, segment length=2.0mm, pre length=.5mm, post length=.5mm}, double}}
\usepackage{slashed}

\usepackage{multirow}

\usepackage{bbm}

\usepackage[normalem]{ulem}

\PassOptionsToPackage{linktocpage}{hyperref}

\newcommand{\cev}[1]{\reflectbox{\ensuremath{\vec{\reflectbox{\ensuremath{#1}}}}}}

\newcommand\lowerdag{{\mkern3mu\raise-1.3ex\hbox{$\scriptstyle\dag$}}}

\newcommand\lowerbar[1]{\mathstrut\mkern1.5mu#1\mkern-13mu\raise2.2ex
  \hbox{$\scriptscriptstyle-$}}

\newcommand\setexp[2]{{\mkern-1mu\raise+#1ex\hbox{$\scriptstyle#2$}}}

\makeatletter
\newcommand{\oset}[3][0ex]{%
  \mathbin{\mathop{#3}\limits^{
    \vbox to#1{\kern-2\ex@
    \hbox{$\scriptstyle#2$}\vss}}}}
\makeatother

\makeatletter
\newcommand{\osetsmall}[3][0ex]{%
  \mathbin{\mathop{#3}\limits^{
    \vbox to#1{\kern-2\ex@
    \hbox{$\scriptscriptstyle#2$}\vss}}}}
\makeatother

\makeatletter

\renewcommand*{\p@subsection}{}

\renewcommand*{\p@subsubsection}{}
\makeatother

\usepackage{xr}

\makeatletter
\def\l@f@section{}%
\makeatother

\renewcommand{\figurename}{\textbf{Figure}}

\renewcommand{\tablename}{\textbf{Table}}

\renewcommand{\thetable}{\arabic{table}}
\makeatletter
\def\@caption@fignum@sep{\textbf{.}}
\def\fnum@figure{\figurename\nobreakspace\textbf{\thefigure}}
\def\fnum@table{\tablename\nobreakspace\textbf{\thetable}}
\makeatother

\begin{document}

\title{Gravity generated by four one-dimensional unitary gauge symmetries and the Standard Model}
\date{August 6, 2025}
\author{Mikko Partanen}
\email{E-mail: mikko.p.partanen@aalto.fi (corresponding author)}
\affiliation{Photonics Group, Department of Electronics and Nanoengineering, 
Aalto University, P.O. Box 13500, 00076 Aalto, Finland}
\author{Jukka Tulkki}
\affiliation{Engineered Nanosystems Group, School of Science, Aalto University, 
P.O. Box 12200, 00076 Aalto, Finland}

\begin{abstract}
The Standard Model of particle physics describes electromagnetic, weak, and strong interactions, which are three of the four known fundamental forces of nature. The unification of the fourth interaction, gravity, with the Standard Model has been challenging due to incompatibilities of the underlying theories -- general relativity and quantum field theory. While quantum field theory utilizes compact, finite-dimensional symmetries associated with the internal degrees of freedom of quantum fields, general relativity is based on noncompact, infinite-dimensional external space-time symmetries. The present work aims at deriving the gauge theory of gravity using compact, finite-dimensional symmetries in a way that resembles the formulation of the fundamental interactions of the Standard Model. For our eight-spinor representation of the Lagrangian, we define a quantity, called the space-time dimension field, which enables extracting four-dimensional space-time quantities from the eight-dimensional spinors. Four U(1) symmetries of the components of the space-time dimension field are used to derive a gauge theory, called unified gravity. The stress-energy-momentum tensor source term of gravity follows directly from these symmetries. The metric tensor enters in unified gravity through geometric conditions. We show how the teleparallel equivalent of general relativity in the Weitzenböck gauge is obtained from unified gravity by a gravity-gauge-field-dependent geometric condition. Unified gravity also enables a gravity-gauge-field-independent geometric condition that leads to an exact description of gravity in the Minkowski metric. This differs from the use of metric in general relativity, where the metric depends on the gravitational field by definition. Based on the Minkowski metric, unified gravity allows us to describe gravity within a single coherent mathematical framework together with the quantum fields of all fundamental interactions of the Standard Model. We present the Feynman rules for unified gravity and study the renormalizability and radiative corrections of the theory at 1-loop order. The equivalence principle is formulated by requiring that the renormalized values of the inertial and gravitational masses are equal. In contrast to previous gauge theories of gravity, all infinities that are encountered in the calculations of loop diagrams can be absorbed by the redefinition of the small number of parameters of the theory in the same way as in the gauge theories of the Standard Model. This result and our observation that unified gravity fulfills the Becchi--Rouet--Stora--Tyutin (BRST) symmetry and its coupling constant is dimensionless suggest that unified gravity can provide the basis for a complete, renormalizable theory of quantum gravity.
\end{abstract}

\maketitle


\tableofcontents

\section{Introduction}

Quantum field theory is a theoretical framework, which synthesizes classical field theory, quantum mechanics, and special relativity \cite{Peskin2018,Weinberg1996,Maggiore2005}. The Standard Model of particle physics arises from this framework through unitary symmetries of fermionic and Higgs fields related to invariances of a physical system \cite{Schwartz2014}. The gauge invariance of quantum electrodynamics (QED), related to the Abelian phase rotation transformations of fermions, is the most trivial example of such a symmetry [U(1)]. The Yang–Mills theory extends the gauge theory to non-Abelian special unitary symmetries \cite{Hooft2005,Peskin2018,Weinberg1996,Schwartz2014,Maggiore2005}, which enable mutually interacting force carriers. It describes the behavior of the other fundamental interactions of the Standard Model being at the core of the unification of electrodynamics to weak [SU(2)] and strong [SU(3)] interactions \cite{Schwartz2014}. The theories of these interactions are called the electroweak theory and quantum chromodynamics (QCD). The Yang--Mills gauge symmetries operate as matrix transformations of the Higgs field and doublets and triplets of fermionic fields. The symmetry groups of the Standard Model are all compact and finite dimensional. A similar compact and finite-dimensional unitary-symmetry-based approach to the description of gravity as a gauge field has remained unknown \cite{Livio2012,Maldacena2014,Misner1973,Landau1989,Dirac1975}. Therefore, alternative approaches, such as string theory \cite{Becker2007,Dine2007,Green1987}, loop quantum gravity \cite{Ashtekar1986,Jacobson1988,Rovelli1990,Rovelli2008}, asymptotic safety \cite{Addazi2022,Niedermaier2006}, noncommutative geometry \cite{Connes1994}, and causal dynamical triangulation \cite{Loll2020}, are being developed. There are also discussions on whether gravitational interaction should be quantized at all \cite{Oppenheim2023a,Oppenheim2023b}.

The current standard understanding of gravity is based on general relativity, which describes gravitational interaction through the curvature of space-time that is similarly experienced by all objects \cite{Einstein1916,Misner1973,Landau1989}. This universality is dictated by Einstein's equivalence principle \cite{Misner1973,Zych2018,Das2023}. It follows that the space-time symmetries of gravity appear fundamentally different from the symmetries of the Standard Model. In modern understanding of gravity, put forward by Einstein \cite{Einstein1928} and Cartan \cite{Cartan1925}, it is recognized that, in addition to the curvature of space-time, the gravitational interaction can also be equivalently described by different metric-affine geometries using torsion or nonmetricity \cite{Krasnov2020,Bahamonde2023a,Jimenez2019,Cabral2020,Heisenberg2019}. Thus, curvature, torsion, and non-metricity provide three seemingly different representations of the same underlying theory of general relativity as special cases of a wider class of gauge theories of gravity \cite{Krasnov2020,Bahamonde2023a,Aldrovandi2012,Heisenberg2019,Maluf2013,Blagojevic2013,Jimenez2019,Golovnev2017,Krssak2019,Jimenez2020a,Cabral2020,Jimenez2018,Barker2024a,Adak2023}. The theory, where only torsion is nonzero, is called the teleparallel equivalent of general relativity (TEGR) \cite{Bahamonde2023a,Aldrovandi2012,Heisenberg2019,Jimenez2019,Maluf2013,Cabral2020,Blagojevic2013}. TEGR is considered to be a natural way to understand the gauge field aspects of gravity since it can be formulated as the gauge theory of the translation group, and it enables the Lorentz-covariant definition of the stress-energy-momentum (SEM) tensor of gravity \cite{Andrade2000,Golovnev2017}. Modifications of TEGR have also been widely studied \cite{Hayashi1979,Ferraro2007,Linder2010,Sotiriou2010,Capozziello2010,DeFelice2010,Capozziello2021,Krssak2016,Jimenez2020b,Chakrabortty2023,Gomes2024}. The present work investigates the possibility of deriving the gauge theory of gravity by using compact, finite-dimensional gauge symmetry groups instead of the noncompact, infinite-dimensional translation gauge group of TEGR. We use the eight-spinor formalism \cite{Partanen2024a} and the associated unitary symmetries in a way that closely resembles the formulation of the interactions of the Standard Model.
Thus, the goal of the present work is to bring the gauge theory of gravity as close as possible to the gauge theory formulation of the Standard Model, and thereby to contribute to improved understanding of the relations of all four fundamental interactions of nature.

Many authors have approached the problem of unifying the Standard Model and gravity by attempting to reformulate space-time symmetries in a way compatible with the gauge symmetries of the Standard Model \cite{Utiyama1956,Kibble1961,MacDowell1977,Kaku1977,Plebanski1977,Stelle1980,Ivanov1982,Fradkin1985,Witten1988,Lasenby1998,Vasiliev2001,Fabi2014,Ho2016,Krasnov2011,Fine2014,Krasnov2017}. The difference between \emph{external space-time symmetries} and \emph{symmetries related to internal degrees of freedom}, which govern the dynamics of quantum fields via creations and annihilations of field quanta, however, represents a challenge for this gauge theory approach of gravity especially at high energies \cite{Blagojevic2013,Clifton2012,Karananas2021,Westman2013,Sciama1964,Neville1980}. In previous literature, there are at least two different ways to interpret whether a symmetry is internal or external. The first interpretation is based on observing how the given symmetry transformation operates on objects in the Lagrangian density. Internal symmetries operate via scalar and matrix multiplications, which do not depend on how the fields vary around the given space-time point. Examples are the multiplication of the Dirac field by a complex phase factor in QED and the color and weak isospin rotations of fermion field triplets and doublets and the Higgs field doublet in the strong and weak interactions of the Standard Model. This is in contrast with external symmetries, such as space-time translations, which are generated by differential operators \cite{Maggiore2005}. The second interpretation of determining the internal or external nature of a symmetry is based on the well-known Coleman--Mandula theorem \cite{Coleman1967}. This theorem states that the symmetry group of a theory that can be described by an S-matrix is locally isomorphic to the direct product of the Poincar\'e group and internal symmetry groups. Therefore, any symmetries associated with the Poincar\'e symmetry structure of the space-time are clearly not internal symmetries from the point of view of the Coleman--Mandula theorem. Consequently, symmetries associated with gravity can be interpreted internal only according to the first interpretation discussed above. The four U(1) symmetries of gravity, to be revealed in the present work, are based on the eight-spinor representation of the Lagrangian density, and they are internal according to the first interpretation above. To avoid misunderstanding, we, however, call these symmetries the U(1) symmetries of gravity instead of internal symmetries.

The main challenge of the conventional gauge theory approach of gravity, which emerges from the nature of the space-time symmetries, is the nonrenormalizability of the resulting theory without an infinite number of counterterms \cite{Bambi2023,Maggiore2005,Zee2010,Schwartz2014,Hooft1974,Deser1974a,Deser1974b,Deser1974c,Goroff1985}. In contrast, all gauge theories of the Standard Model are renormalizable, which means that their ultraviolet divergences can be reabsorbed into the redefinition of a finite number of parameters \cite{Schwartz2014,Hooft1971a,Hooft1971b}. The renormalization procedure then leads to the running of the coupling constants as a function of the energy scale \cite{Maggiore2005,Schwartz2014,Peskin2018}. The nonrenormalizability of conventional theories of gravity makes it impossible to use the quantized gauge theory of gravity to make predictions at high energies. However, the quantum field theory treatment of general relativity can be argued to be successful as a low-energy effective field theory \cite{Bambi2023,Casadio2022,Donoghue1994b,Donoghue1999,Donoghue2002,Stelle1977,Schwartz2014,Rocci2024}. The main idea of the effective field theory is that the low-energy degrees of freedom organize themselves as quantum fields in such a way that one can make predictions without knowledge of the full high-energy theory \cite{Bambi2023}. This also indicates that fundamental breakthroughs are needed to formulate a predictive quantum theory of gravity applicable to all energy scales. Such a theory can finally answer ultimate questions on the structure of the Universe in circumstances of extremely high energy densities, such as those inside black holes and at the possible beginning of time \cite{Livio2012}.

On the experimental side, general relativity has so far passed all tests planned to probe gravitational interaction. The well-known classical tests involve the precession of the perihelion of Mercury \cite{Einstein1916}, the bending of light by the Sun \cite{Watson1920}, and the gravitational redshift of light \cite{Pound1959}. New experiments are continuously developed \cite{Schiff1960,Shapiro1964,Nordtvedt1968,Bertotti2003,Milani2002,Schettino2016,Kapner2007,Fomalont2009,Everitt2011,Yunes2013}. The waveforms recorded in recent measurements of gravitational waves \cite{Abbott2016a,Abbott2016b,Abbott2017} are in good agreement with general relativity. The first image of a black hole is consistent with the shadow of a Kerr black hole predicted by the theory \cite{Akiyama2019}. Other recent measurements involve the study of the effect of gravity on the motion of antimatter \cite{Anderson2023} and the measurement of gravity with milligram-scale masses \cite{Fuchs2024,Westphal2021} and with bending beam resonators \cite{Brack2022,Brack2023}.

In a recent work \cite{Partanen2024a}, we have reformulated QED based on an eight-component spinorial representation of the electromagnetic field. The eight-spinor formulation reveals a profound connection between the unitary symmetries of the Lagrangian density in the eight-dimensional spinor space and the symmetric SEM tensor of the Dirac and electromagnetic fields. Since the SEM tensor is the well-known source of the gravitational field in general relativity, it becomes obvious that the gauge theory obtained by studying the appropriate Lagrangian density under unitary symmetry transformations describes gravitational interaction. In such a theory, the SEM tensor appears analogously to the pertinent source terms of the quantum fields of the Standard Model. This is the basis for the development of the gauge theory of gravity, called \emph{unified gravity}, in the present work. The present work also essentially generalizes some of the key mathematical concepts of our preliminary study \cite{Partanen2024a}.

In this work, we extend the eight-spinor formulation to cover the full Standard Model and derive the gauge theory of unified gravity. Our theory is based on introducing the concept of \emph{the space-time dimension field}, a geometric object which, by definition, enables extracting four-dimensional space-time quantities from the eight-dimensional spinor space. Introducing the space-time dimension field enables identifying compact, finite-dimensional unitary symmetries in the Lagrangian density, which can be utilized to form a gauge theory in analogy to how gauge symmetries are used in the Standard Model. While the interaction symmetries of the Standard Model are based on symmetry transformations of fermionic and Higgs fields as discussed above, our extension of the Standard Model to cover gravity is based on four U(1) symmetry transformations, which act on the components of the space-time dimension field. Therefore, the symmetries of the space-time dimension field form a hierarchy separate from the symmetries of the Standard Model. The U(1) symmetry transformations of the space-time dimension field components allow us to couple the quantum fields of the Standard Model to a tensor gauge field in a way that is formally analogous to the gauge couplings of the fields in the electromagnetic, weak, and strong interactions. Once the gauge field and its Lagrangian density are introduced, we obtain unified gravity and its dynamical equations, which are shown to describe the behavior of the gravitational field.

The space-time metric tensor enters unified gravity through geometric conditions. We are allowed to use geometric conditions, in which \emph{the space-time metric tensor is independent of the gravity gauge field}. This leads us to study unified gravity in the Minkowski metric (UGM) in an exact way. This differs from the use of metric in general relativity, where the metric depends on the gravitational field by definition \cite{Einstein1916,Misner1973,Landau1989,Dirac1975}, and whose effective quantization requires expansion of the metric about the flat or smooth background with an assumption that the deviation is small  \cite{Einstein1918,Gupta1952,Voronov1973,vanNieuwenhuizen1973,Donoghue1994b,Choi1995,Balbus2016,Yunes2013,Jaranowski2009,Olyaei2018}. In this respect, the conventional translation gauge formulation of TEGR is not significantly different from general relativity since a similar expansion of the tetrad is needed \cite{Obukhov2003b,Ulhoa2017}. Alternatively, in unified gravity, we are allowed to use geometric conditions, in which \emph{the space-time metric tensor depends on the gravity gauge field}, as in general relativity and in TEGR \cite{Bahamonde2023a,Aldrovandi2012,Jimenez2019,Maluf2013,Cabral2020,Blagojevic2013}. We show that, within a particular Weitzenböck gauge fixing approach, the representation of unified gravity becomes equivalent to the known representation of TEGR in the Weitzenböck gauge (TEGRW), where the teleparallel spin connection vanishes. This shows that unified gravity is in perfect agreement with the known nonlinear field equations of general relativity.

The harmonic gauge fixing of UGM is analogous to the Feynman gauge fixing of QED. It enables us to determine the Feynman rules for unified gravity. These rules are also written in a more general form for an arbitrary gauge fixing parameter. As a gauge theory similar to those of the Standard Model \cite{Peskin2018,Weinberg1996}, unified gravity is subject to quantization. The quanta of the gauge field, the gravitons, are spin-2 tensor bosons. These quanta are to be added in the spectrum of the known elementary particles extending the Standard Model to describe gravity. However, this can be done only after the nonrenormalizability problem of quantum gravity has been fully resolved making the quantum theory predictive at all energies. In this work, we study the renormalizability and the radiative corrections of UGM at 1-loop order. We show that, in contrast to previous gauge theories of gravity, all infinities that are encountered in the calculations of loop diagrams can be absorbed by the redefinition of the small number of parameters of the theory in the same way as in the gauge theories of the Standard Model. Furthermore, Einstein's equivalence principle is formulated by requiring that the renormalized values of the gravitational and inertial masses are equal. The direct relation between the equivalence principle and renormalization is obviously absent in previous studies of the equivalence principle in the quantum regime \cite{Zych2018,Das2023}. Based on the dimensionless coupling constant and the fulfillment of the Becchi--Rouet--Stora--Tyutin (BRST) symmetry \cite{Becchi1976,Iofa1976,Peskin2018,Schwartz2014,Fuster2005}, we expect that unified gravity is renormalizable to all loop orders. The complete proof extending our 1-loop results to all loop orders is, however, left as a topic of further works.

Gravity couples to all fields and matter. Therefore, one cannot exclude any field or matter from the complete dynamical description of gravity. However, to make our theory of unified gravity more transparent and easier to understand for nonexpert readers, we limit, in the first part of this work, our study to the coupling between gravity and electrodynamics. The system of the electromagnetic field, Dirac electron--positron field, and the gravitational field provides all the insight needed for obtaining a unified description of gravity in a coherent framework with the other known fundamental forces of nature. The extension of unified gravity to cover the full Standard Model is presented in the later part of the present work.

This work is organized as follows: Section \ref{sec:foundations} describes the theoretical concepts and conventions, including the eight-spinor formulation, originally introduced in \cite{Partanen2024a}. The kernel matrices, the space-time dimension field, and the generating Lagrangian density of gravity are presented in section~\ref{sec:L0}. We also formulate the equivalence principle in unified gravity. Furthermore, the generating Lagrangian density of gravity is shown to be equal to the Lagrangian density of QED in flat space-time. Section \ref{sec:symmetry} discusses the symmetries of the generating Lagrangian density of gravity and derives the conservation law of the SEM tensor of the Dirac and electromagnetic fields. The gravity gauge field is introduced through the gauge-covariant derivative in section~\ref{sec:gravityfield}. The gravity gauge field strength tensor is also discussed. In section~\ref{sec:L}, we present the locally gauge-invariant Lagrangian density of unified gravity. This section also provides the scaled representation of unified gravity to allow easier comparison with the gauge theories of the Standard Model. Section \ref{sec:UGM} discusses the representation of unified gravity in the Minkowski metric in UGM and derives the corresponding dynamical equations. Section \ref{sec:Feynman} presents the Feynman rules and their selected applications in UGM. Section \ref{sec:renormalization} proves the renormalizability of unified gravity at 1-loop order and determines the values of the related renormalization factors using the conventional on-shell renormalization scheme and dimensional regularization \cite{Schwartz2014}. In section~\ref{sec:radiative}, radiative corrections to the Coulomb and Newtonian potentials and to the anomalous magnetic moment of the electron are calculated to exemplify the calculation of radiative corrections using unified gravity. Section \ref{sec:teleparallel} derives TEGRW from unified gravity by applying the Weitzenböck gauge fixing approach. The dynamical equations for the Dirac, electromagnetic, and gravitational fields in TEGRW are also presented. The extension of the theory to cover all quantum fields of the Standard Model is developed in section~\ref{sec:StandardModel}. The results are discussed and compared with previous theories in section~\ref{sec:discussion}. Finally, conclusions are drawn in section~\ref{sec:conclusion}.

\section{\label{sec:foundations}Theoretical concepts and conventions}

This section presents the theoretical concepts and conventions used in the present work. These include the index conventions, tetrads, metric tensors, connection coefficients, torsions, contortions, derivative operators, and the eight-spinor formalism of \cite{Partanen2024a}.

\subsection{\label{sec:indices}Index conventions, tetrads, and metric tensors}

The Latin indices $a,b,c,d\in\{0,x,y,z\}$ in this work range over four Cartesian Minkowski space-time coordinates. The Latin indices starting from $i$ range over the three spatial dimensions, i.e., $i,j,k\in\{x,y,z\}$, or over other values separately specified. The Greek indices denote the general space-time indices, which range over the four general space-time dimensions $\mu,\nu,\rho,\sigma\in\{x^0,x^1,x^2,x^3\}$. The Latin Cartesian Minkowski space-time indices are lowered and raised by the Cartesian Minkowski metric tensor $\eta_{ab}$ and its inverse $\eta^{ab}$. We use the sign convention $\eta^{00}=1$ and $\eta^{xx}=\eta^{yy}=\eta^{zz}=-1$. The Greek general space-time indices are raised and lowered by the general space-time metric $g_{\mu\nu}$ and its inverse $g^{\mu\nu}$. The determinant of the general space-time metric tensor is denoted by $g=\det(g_{\mu\nu})$.

The Latin Cartesian Minkowski space-time indices and any Greek general space-time indices can be converted into each other by the tetrad $e_a^{\;\,\mu}$ and the inverse tetrad $e_{\;\,\mu}^a$ \cite{Aldrovandi2012,Maluf2013,Krasnov2020,Blagojevic2013,Cartan1925,Einstein1928}. In \emph{flat} space-time, i.e., in the absence of the gravitational field, and more generally, in a Minkowski manifold with the torsion and the spin connection equal to zero, the tetrad and the inverse tetrad are given by $\oset{\circ}{e}_a^{\;\,\mu}=\partial^\mu x_a$ and $\oset{\circ}{e}_{\;\,\mu}^a=\partial_\mu x^a$, where $x^a$ is a four-vector of Minkowski space-time coordinates, e.g., $x^a=(ct,x,y,z)$ in Cartesian coordinates and $x^a=(ct,r\sin\theta\cos\phi,r\sin\theta\sin\phi,r\cos\theta)$ in spherical coordinates. Respectively, we use the symbol $\oset{\bullet}{e}_a^{\;\,\mu}$ to highlight the tetrad of TEGRW, discussed in section~\ref{sec:teleparallel}. The generic tetrad symbol $e_a^{\;\,\mu}$ is used to indicate that an equation is independent of the definition of the tetrad and one is not restricted to using the tetrad of the Minkowski manifold or that of TEGRW. In any definition of the tetrad, the general space-time metric tensor is given in terms of the Cartesian Minkowski metric tensor and inverse tetrads as $g_{\mu\nu}=\eta_{ab}e_{\;\,\mu}^ae_{\;\,\nu}^b$. In the special case of the Minkowski space-time in Cartesian coordinates, used in UGM, the Latin and Greek indices are identical and the tetrad and the metric tensor are trivial as $\oset{\circ}{e}_\mu^{\;\,\nu}=\delta_\mu^\nu$ and $g_{\mu\nu}=\eta_{\mu\nu}$, where $\delta_\mu^\nu$ is the Kronecker delta.

Throughout this work, with a few exceptions, we use the Einstein convention for the summation over repeated indices. Exceptions to the summation convention are separately discussed and indicated by parentheses around the indices.

\subsection{Connection coefficients, torsions, and contortions}

The Levi-Civita connection coefficients of standard general relativity, i.e., the Christoffel symbols $\oset{\circ}{\Gamma}^{\setexp{-1}{\mu}}_{\;\,\sigma\nu}$, are associated with all Greek space-time indices. They are used to define the Levi-Civita coordinate-covariant derivative as presented in section~\ref{sec:derivatives}. The Christoffel symbols can always be written in terms of the metric tensor as \cite{Misner1973,Kobayashi1963}
\begin{equation}
 \oset{\circ}{\Gamma}^{\setexp{-1}{\mu}}_{\;\,\sigma\nu}=\frac{1}{2}g^{\mu\rho}(\partial_\sigma g_{\rho\nu}+\partial_\nu g_{\rho\sigma}-\partial_\rho g_{\sigma\nu}).
 \label{eq:Christoffel_general}
\end{equation}
The Christoffel symbol contraction $\oset{\circ}{\Gamma}^{\setexp{-1}{\sigma}}_{\;\,\rho\sigma}$, used in many equations of this work, can be given in terms of the determinant of the metric tensor as
\begin{equation}
 \oset{\circ}{\Gamma}^{\setexp{-1}{\sigma}}_{\;\,\rho\sigma}=\frac{1}{\sqrt{-g}}\partial_\rho(\sqrt{-g}).
 \label{eq:Christoffel0}
\end{equation}

In a Minkowski manifold with the torsion and the spin connection equal to zero, the Christoffel symbols can be written in terms of tetrads as \cite{Aldrovandi2012,Bahamonde2023a}
\begin{equation}
 \oset{\circ}{\Gamma}^{\setexp{-1}{\mu}}_{\;\,\sigma\nu}=\oset{\circ}{e}_a^{\;\,\mu}\partial_\nu \oset{\circ}{e}_{\;\,\sigma}^a=-\oset{\circ}{e}_{\;\,\sigma}^a\partial_\nu \oset{\circ}{e}_a^{\;\,\mu}.
 \label{eq:Christoffel1}
\end{equation}
The condition of zero torsion tensor, $\oset{\circ}{T}_{\;\,\mu\nu}^{\setexp{-1}{\rho}}$, is written as
\begin{equation}
 \oset{\circ}{T}_{\;\,\mu\nu}^{\setexp{-1}{\rho}}=\oset{\circ}{\Gamma}^{\setexp{-1}{\rho}}_{\;\,\mu\nu}-\oset{\circ}{\Gamma}^{\setexp{-1}{\rho}}_{\;\,\nu\mu}
 =0.
\end{equation}
The related contortion tensor, $\oset{\circ}{K}^{\setexp{-1}{\mu\nu\rho}}$, is trivially zero as
\begin{equation}
 \oset{\circ}{K}^{\setexp{-1}{\mu\nu\rho}}=\frac{1}{2}(\oset{\circ}{T}^{\setexp{-1}{\nu\mu\rho}}+\oset{\circ}{T}^{\setexp{-1}{\mu\rho\nu}}-\oset{\circ}{T}^{\setexp{-1}{\rho\nu\mu}})=0.
 \label{eq:contortionM}
\end{equation}

In TEGRW, where the spin connection is zero but the torsion is generally nonzero, the relation of the Christoffel symbols is given by \cite{Aldrovandi2012,Andrade2000}
\begin{equation}
 \oset{\circ}{\Gamma}^{\setexp{-1}{\mu}}_{\;\,\sigma\nu}=\oset{\bullet}{\Gamma}^{\setexp{-1}{\mu}}_{\;\,\sigma\nu}-\oset{\bullet}{K}_{\;\,\sigma\nu}^{\setexp{-1}{\mu}}.
 \label{eq:Christoffel2}
\end{equation}
Here $\oset{\bullet}{\Gamma}^{\setexp{-1}{\mu}}_{\;\,\sigma\nu}$ is the is the so-called Weitzenböck connection and $\oset{\bullet}{K}_{\;\,\sigma\nu}^{\setexp{-1}{\mu}}$ is the corresponding contortion tensor. The Weitzenböck connection is given by \cite{Bahamonde2023a,Krasnov2020}
\begin{equation}
 \oset{\bullet}{\Gamma}^{\setexp{-1}{\mu}}_{\;\,\sigma\nu}=\oset{\bullet}{e}_a^{\;\,\mu}\partial_\nu \oset{\bullet}{e}_{\;\,\sigma}^a=-\oset{\bullet}{e}_{\;\,\sigma}^a\partial_\nu \oset{\bullet}{e}_a^{\;\,\mu},
 \label{eq:torsionW}
\end{equation}
The torsion tensor of the Weitzenböck connection is given by \cite{Aldrovandi2012,Bahamonde2023a,Andrade2000}
\begin{equation}
 \oset{\bullet}{T}_{\;\,\mu\nu}^{\setexp{-1}{\rho}}=\oset{\bullet}{\Gamma}^{\setexp{-1}{\rho}}_{\;\,\mu\nu}-\oset{\bullet}{\Gamma}^{\setexp{-1}{\rho}}_{\;\,\nu\mu}=\oset{\bullet}{e}_a^{\;\,\rho}\partial_\mu \oset{\bullet}{e}_{\;\,\nu}^a
 -\oset{\bullet}{e}_a^{\;\,\rho}\partial_\nu \oset{\bullet}{e}_{\;\,\mu}^a.
\end{equation}
From this definition, it follows that the torsion tensor is antisymmetric in its last two indices as $\oset{\bullet}{T}_{\;\,\mu\nu}^{\setexp{-1}{\rho}}=-\oset{\bullet}{T}_{\;\,\nu\mu}^{\setexp{-1}{\rho}}$. In terms of the torsion tensor, the contortion tensor is written as
\begin{equation}
 \oset{\bullet}{K}^{\setexp{-1}{\mu\nu\rho}}=\frac{1}{2}(\oset{\bullet}{T}^{\setexp{-1}{\nu\mu\rho}}+\oset{\bullet}{T}^{\setexp{-1}{\mu\rho\nu}}-\oset{\bullet}{T}^{\setexp{-1}{\rho\nu\mu}}).
 \label{eq:contortion0}
\end{equation}
The contortion tensor is antisymmetric in its first two indices as $\oset{\bullet}{K}^{\setexp{-1}{\mu\nu\rho}} =-\oset{\bullet}{K}^{\setexp{-1}{\nu\mu\rho}}$.

The definition of $\oset{\bullet}{e}_{\;\,\sigma}^a$ in the gauge theory formulation of TEGRW obtained from unified gravity is discussed in section~\ref{sec:teleparalleltetrad}. If the so-called spin connection is added to the Weitzenböck connection of TEGRW, the connection is called the teleparallel connection \cite{Bahamonde2023a}. However, there is varying terminology in the literature \cite{Aldrovandi2012}. The spin connection is not used in the present work. For the related discussion on the local Lorentz invariance of the Lagrangian density, see section~\ref{sec:local_Lorentz_invariance}.

\subsection{\label{sec:derivatives}Derivative operators}

The partial derivative operator is denoted by $\partial_\rho$ as conventional. In Cartesian Minkowski coordinates, we, thus, have $\partial_a=(\frac{1}{c}\partial_t,\nabla)$, where the three-dimensional vector-differential operator $\nabla$ is defined as $\nabla=(\partial_x,\partial_y,\partial_z)$. For the contraction of $\partial_\rho$ with itself, we use the shorthand notation $\partial_\rho\partial^\rho=\partial^2$.

As another short notation, we define the space-time derivative operator $\widetilde{\nabla}_\rho$. In distinction to the conventional Levi-Civita coordinate-covariant derivative, to be defined below, the operator $\widetilde{\nabla}_\rho$ operates through a contraction with a single space-time index of a tensor or pseudotensor and does not care about other indices as
\begin{align}
 &\widetilde{\nabla}_\rho{V^{\mu_1\ldots\rho\ldots\mu_n}}_{\nu_1\ldots\nu_m}\nonumber\\
 &=\frac{1}{\sqrt{-g}}\partial_\rho(\sqrt{-g}\,{V^{\mu_1\ldots\rho\ldots\mu_n}}_{\nu_1\ldots\nu_m})\nonumber\\
 &=\partial_\rho{V^{\mu_1\ldots\rho\ldots\mu_n}}_{\nu_1\ldots\nu_m}+\oset{\circ}{\Gamma}^{\setexp{-1}{\sigma}}_{\;\,\rho\sigma}{V^{\mu_1\ldots\rho\ldots\mu_n}}_{\nu_1\ldots\nu_m}.
 \label{eq:nabla}
\end{align}
Here ${V^{\mu_1\ldots\rho\ldots\mu_n}}_{\nu_1\ldots\nu_m}$ is a generic tensor or pseudotensor with an arbitrary number of upper and lower space-time indices.

Another derivative to be used is the well-known Levi-Civita coordinate-covariant derivative, denoted by $\oset{\circ}{\nabla}_\nu$ and defined as \cite{Misner1973,Kobayashi1963}
\begin{align}
 &\oset{\circ}{\nabla}_\rho{V^{\mu_1\ldots\mu_n}}_{\nu_1\ldots\nu_m}=\partial_\rho{V^{\mu_1\ldots\mu_n}}_{\nu_1\ldots\nu_m}\nonumber\\
 &+{\oset{\circ}{\Gamma}^{\setexp{-1}{\mu_1}}}_{\sigma\rho}{V^{\sigma\mu_2\ldots\mu_n}}_{\nu_1\ldots\nu_m}
 +\ldots+{\oset{\circ}{\Gamma}^{\setexp{-1}{\mu_n}}}_{\sigma\rho}{V^{\mu_1\ldots\mu_{n-1}\sigma}}_{\nu_1\ldots\nu_m}\nonumber\\
 &-{\oset{\circ}{\Gamma}^{\setexp{-1}{\sigma}}}_{\nu_1\rho}{V^{\mu_1\ldots\mu_n}}_{\sigma\nu_2\ldots\nu_m}
 -\ldots-{\oset{\circ}{\Gamma}^{\setexp{-1}{\sigma}}}_{\nu_m\rho}{V^{\mu_1\ldots\mu_n}}_{\nu_1\ldots\nu_{m-1}\sigma}.
 \label{eq:nablaLeviCivita}
\end{align}

In the present work, the Levi-Civita coordinate-covariant derivative is the only coordinate-covariant derivative used. For completeness and for comparison, we present the teleparallel coordinate-covariant derivative, which is in analogy to equation~\eqref{eq:nablaLeviCivita} given by
\begin{align}
 &\oset{\bullet}{\nabla}_\rho{V^{\mu_1\ldots\mu_n}}_{\nu_1\ldots\nu_m}=\partial_\rho{V^{\mu_1\ldots\mu_n}}_{\nu_1\ldots\nu_m}\nonumber\\
 &+{\oset{\bullet}{\Gamma}^{\setexp{-1}{\mu_1}}}_{\sigma\rho}{V^{\sigma\mu_2\ldots\mu_n}}_{\nu_1\ldots\nu_m}
 +\ldots+{\oset{\bullet}{\Gamma}^{\setexp{-1}{\mu_n}}}_{\sigma\rho}{V^{\mu_1\ldots\mu_{n-1}\sigma}}_{\nu_1\ldots\nu_m}\nonumber\\
 &-{\oset{\bullet}{\Gamma}^{\setexp{-1}{\sigma}}}_{\nu_1\rho}{V^{\mu_1\ldots\mu_n}}_{\sigma\nu_2\ldots\nu_m}
 -\ldots-{\oset{\bullet}{\Gamma}^{\setexp{-1}{\sigma}}}_{\nu_m\rho}{V^{\mu_1\ldots\mu_n}}_{\nu_1\ldots\nu_{m-1}\sigma}.
 \label{eq:nablateleparallel}
\end{align}
Using the definition of $\oset{\bullet}{\nabla}_\rho$ in equation~\eqref{eq:nablateleparallel}, the tetrad $\oset{\bullet}{e}_{\;\,\nu}^a$ satisfies the so-called distant parallelism condition, given by $\oset{\bullet}{\nabla}_\rho \oset{\bullet}{e}_{\;\,\nu}^a=0$ \cite{Aldrovandi2012}.
The name of teleparallel gravity originates from this condition \cite{Aldrovandi2012}.

For the Dirac field, we also define the conventional right and left electromagnetic-gauge-covariant derivative\vspace{-1pt} operators $\vec{D}_\nu$ and $\cev{D}_\nu$, given by \cite{Peskin2018,Feynman1961}
\begin{equation}
 \vec{D}_\nu=\vec{\partial}_\nu+i\frac{q_\mathrm{e}}{\hbar}A_\nu,
 \hspace{0.5cm}\cev{D}_\nu=\cev{\partial}_\nu-i\frac{q_\mathrm{e}}{\hbar}A_\nu.
 \label{eq:Da}
\end{equation}
Here $\hbar$ is the reduced Planck constant, $A_\nu$ is the electromagnetic four-potential gauge field, and $q_\mathrm{e}=Qe$ is the electric charge of the Dirac field, where $e$ is the elementary charge and $Q=\pm 1$ for positrons and electrons. The vector arrows in equation~\eqref{eq:Da} indicate the direction in which the differential operators operate. In the case of other fundamental interactions of the Standard Model, the derivative operators in equation~\eqref{eq:Da} are complemented by the pertinent gauge field terms as discussed in section~\ref{sec:StandardModel}. The eight-spinor forms of the partial differential and electromagnetic-gauge-covariant derivative operators are presented in the next section.

\subsection{\label{sec:eightspinors}Eight-spinor formalism}

The eight-spinor formulation of QED, presented in \cite{Partanen2024a}, is not the first spinorial representation of the electromagnetic field  \cite{Sachs1961,Sachs1964,Perkins1978,Kulyabov2017,Kiessling2018,Hong2019}. However, the eight-spinor formalism has certain advantages over other known formulations. For example, it enables the representation of all Maxwell's equations by a single spinorial equation. As discussed below and in \cite{Partanen2024a}, the term eight-spinor is used in a meaning that \emph{covers different types of eight-component physical quantities} whose certain components can be defined to be zero and whose Lorentz transformation properties depend on the type of the eight-spinor. Here we briefly review the key quantities of the eight-spinor formulation.

The eight-spinor theory is formulated in terms of four $8\times8$ bosonic gamma matrices $\boldsymbol{\gamma}_\mathrm{B}^a$ and $\boldsymbol{\gamma}_\mathrm{B}^5=i\boldsymbol{\gamma}_\mathrm{B}^0\boldsymbol{\gamma}_\mathrm{B}^x\boldsymbol{\gamma}_\mathrm{B}^y\boldsymbol{\gamma}_\mathrm{B}^z$ \cite{Partanen2024a}. These matrices and their equivalence transformations are explicitly presented in section~1 of the supplementary material. The matrices $\boldsymbol{\gamma}_\mathrm{B}^a$ satisfy the Dirac algebra, i.e., the Clifford algebra $\mathcal{C}\ell_{1,3}(\mathbb{C})$. The defining property of the Dirac algebra of $\boldsymbol{\gamma}_\mathrm{B}^a$ is the anticommutation relation $\{\boldsymbol{\gamma}_\mathrm{B}^a,\boldsymbol{\gamma}_\mathrm{B}^b\}=2\eta^{ab}\mathbf{I}_8$, where $\mathbf{I}_8$ is the $8\times8$ identity matrix.

The conventional $4\times 4$ Dirac gamma matrices are denoted by $\boldsymbol{\gamma}_\mathrm{F}^a$. Here we use the subscript $\mathrm{F}$ in distinction to the subscript $\mathrm{B}$ used for bosonic gamma matrices discussed above. The Clifford algebra relation for the Dirac gamma matrices is given by $\{\boldsymbol{\gamma}_\mathrm{F}^a,\boldsymbol{\gamma}_\mathrm{F}^b\}=2\eta^{ab}\mathbf{I}_4$, where $\mathbf{I}_4$ is the $4\times4$ identity matrix. In some places, we use the Feynman slash notation for the Dirac gamma matrices to write equations more compactly. For example, one can write $\boldsymbol{\gamma}_\mathrm{F}^\rho\partial_\rho=\slashed{\partial}$ and $\boldsymbol{\gamma}_\mathrm{F}^\rho p_\rho=\slashed{p}$.

In analogy with the conventional Dirac adjoint $\bar{\psi}=\psi^\dag\boldsymbol{\gamma}_\mathrm{F}^0$ \cite{Peskin2018}, where $\psi$ is the four-component Dirac spinor and the superscript $\dag$ denotes the Hermitian conjugate, the zeroth bosonic gamma matrix $\boldsymbol{\gamma}_\mathrm{B}^0$ is used to define eight-spinor adjoints. For a generic eight-spinor $Y$, the adjoint operation is defined as $\bar{Y}=Y^\dag\boldsymbol{\gamma}_\mathrm{B}^0$. For a generic $8\times8$ matrix $\mathbf{M}$, the corresponding adjoint operation is defined as $\bar{\mathbf{M}}=\boldsymbol{\gamma}_\mathrm{B}^0\mathbf{M}^\dag\boldsymbol{\gamma}_\mathrm{B}^0$.

To express four-vector quantities in the eight-spinor notation as discussed below, we define four eight-spinor unit vectors $\boldsymbol{\mathfrak{e}}_a$ as
\begin{align}
 \boldsymbol{\mathfrak{e}}_0 &=[0,0,0,0,1,0,0,0]^T,\nonumber\\
 \boldsymbol{\mathfrak{e}}_x &=[0,1,0,0,0,0,0,0]^T,\nonumber\\
 \boldsymbol{\mathfrak{e}}_y &=[0,0,1,0,0,0,0,0]^T,\nonumber\\
 \boldsymbol{\mathfrak{e}}_z &=[0,0,0,1,0,0,0,0]^T.
\end{align}
Here the superscript $T$ denotes the transpose. The eight-spinor unit vectors $\boldsymbol{\mathfrak{e}}_a$ and their adjoints $\bar{\boldsymbol{\mathfrak{e}}}_a=\boldsymbol{\mathfrak{e}}_a^\dag\boldsymbol{\gamma}_\mathrm{B}^0$ satisfy the following index-raising identities:
\begin{align}
 \boldsymbol{\mathfrak{e}}^a &=\eta^{ab}\boldsymbol{\mathfrak{e}}_b=-\boldsymbol{\gamma}_\mathrm{B}^a\boldsymbol{\mathfrak{e}}_0,\nonumber\\
 \bar{\boldsymbol{\mathfrak{e}}}^a &=\eta^{ab}\bar{\boldsymbol{\mathfrak{e}}}_b=-\bar{\boldsymbol{\mathfrak{e}}}_0\boldsymbol{\gamma}_\mathrm{B}^a.
\end{align}

The electromagnetic field is described by an eight-component electromagnetic potential spinor $\Theta$, which is a four-vector-type eight-spinor formed from the components of the electromagnetic four-potential $A^a$ \cite{Partanen2024a}. In the present work, we always use real-valued gauge fields and potentials and omit the subscript $\Re$ used in eight-spinors in \cite{Partanen2024a}. In terms of the electromagnetic scalar and vector potential components of $A^a=(\phi_\mathrm{e}/c,\mathbf{A})$, where $\phi_\mathrm{e}$ is the scalar potential and $\mathbf{A}=(A^x,A^y,A^z)$ is the vector potential, the electromagnetic potential spinor is given by \cite{Partanen2024a} 
\begin{equation}
 \Theta=\sqrt{\frac{\varepsilon_0}{2}}\,cA^a\boldsymbol{\mathfrak{e}}_a=\sqrt{\frac{\varepsilon_0}{2}}\,[0,cA^x,cA^y,cA^z,\phi_\mathrm{e},0,0,0]^T.
 \label{eq:potentialspinor}
\end{equation}
Here $\varepsilon_0$ is the vacuum permittivity, and $c$ is the speed of light in vacuum. The adjoint spinor $\bar{\Theta}$ is given by $\bar{\Theta}=\Theta^\dag\boldsymbol{\gamma}_\mathrm{B}^0=\sqrt{\varepsilon_0/2}\,[0,cA^x,cA^y,cA^z,-\phi_\mathrm{e},0,0,0]$. As seen from equation~\eqref{eq:potentialspinor}, there are only four nonzero components in $\Theta$. More nonzero components are needed in the eight-spinor representation of the electric and magnetic fields as described below.

The gauge-invariant electromagnetic spinor $\Psi$, associated with the electric and magnetic fields, is a spin-1-field-type eight-spinor, given in terms of the electromagnetic potential spinor $\Theta$ in equation~\eqref{eq:potentialspinor} by \cite{Partanen2024a}
\begin{align}
 \Psi &=-(\mathbf{I}_8+\boldsymbol{\mathfrak{e}}_0\bar{\boldsymbol{\mathfrak{e}}}_0)\boldsymbol{\gamma}_\mathrm{B}^a\partial_a\Theta\nonumber\\
 &=\sqrt{\frac{\varepsilon_0}{2}}\,[0,E^x,E^y,E^z,0,icB^x,icB^y,icB^z]^T.
 \label{eq:electromagneticspinor}
\end{align}
To obtain the last form of equation~\eqref{eq:electromagneticspinor}, we have used the well-known expressions of the electric and magnetic fields in terms of the scalar and vector potentials, given by $\mathbf{E}=(E^x,E^y,E^z)=-\nabla\phi_\mathrm{e}-\frac{\partial}{\partial t}\mathbf{A}$ and $\mathbf{B}=(B^x,B^y,B^z)=\nabla\times\mathbf{A}$ \cite{Jackson1999}. In the special case of the Lorenz gauge, defined by the gauge condition $\partial_a A^a=0$, one can set the term $-\boldsymbol{\mathfrak{e}}_0\bar{\boldsymbol{\mathfrak{e}}}_0\boldsymbol{\gamma}_\mathrm{B}^a\partial_a\Theta$ in equation~\eqref{eq:electromagneticspinor} to zero since $\bar{\boldsymbol{\mathfrak{e}}}_0\boldsymbol{\gamma}_\mathrm{B}^a\partial_a\Theta=\sqrt{\varepsilon_0/2}\,c\partial_a A^a$. This choice also corresponds to the Feynman gauge as discussed in section~\ref{sec:QEDfixing}. The electromagnetic adjoint spinor $\bar{\Psi}$ is given by $\bar{\Psi}=\Psi^\dag\boldsymbol{\gamma}_\mathrm{B}^0=\sqrt{\varepsilon_0/2}\,[0,E^x,E^y,E^z,0,icB^x,icB^y,icB^z]$.

As a convenient brief notation, the conventional four-component Dirac spinor field $\psi$ is used to form a Dirac eight-spinor $\psi_8$, given by
\begin{equation}
 \psi_8=\psi\boldsymbol{\mathfrak{e}}_0=[\mathbf{0},\mathbf{0},\mathbf{0},\mathbf{0},\psi,\mathbf{0},\mathbf{0},\mathbf{0}]^T.
 \label{eq:Diraceightspinor}
\end{equation}
Here the transpose operates on the eight-spinor degree of freedom and not on the Dirac spinor $\psi$. As is evident from equation~\eqref{eq:Diraceightspinor}, the components of $\psi_8$ are not scalars but four-component spinors. The use of a four-component Dirac spinor as a component of the eight-spinor in equation~\eqref{eq:Diraceightspinor} may seem unusual at first sight but, apart from zero components, it is similar to how the Dirac spinors of quarks and leptons are used as triplets and doublets in the description of strong and electroweak interactions of the Standard Model. The adjoint Dirac eight-spinor is given by $\bar{\psi}_8=\psi_8^\dag\boldsymbol{\gamma}_\mathrm{B}^0\boldsymbol{\gamma}_\mathrm{F}^0=\bar{\psi}\bar{\boldsymbol{\mathfrak{e}}}_0=[\mathbf{0},\mathbf{0},\mathbf{0},\mathbf{0},-\bar{\psi},\mathbf{0},\mathbf{0},\mathbf{0}]$. Here $\bar{\psi}=\psi^\dag\boldsymbol{\gamma}_\mathrm{F}^0$ is the conventional Dirac adjoint for four-component spinors. All quantum operators of conventional four-component Dirac spinors extend to Dirac eight-spinors trivially so that they operate in the usual way on the nonzero component of $\psi_8$, which is the conventional four-component Dirac spinor.

As discussed in \cite{Partanen2024a}, under the Lorentz transformation, the electromagnetic spinor $\Psi$ in equation~\eqref{eq:electromagneticspinor} transforms as $\Psi\rightarrow\boldsymbol{\Lambda}_\mathrm{S}\Psi$, where the Lorentz transformation matrix $\boldsymbol{\Lambda}_\mathrm{S}$ is given in equation (29) of \cite{Partanen2024a}. This transformation preserves $\Psi$ invariant in $2\pi$ rotations, which is the characteristic property of spin-1 fields \cite{Landau1982}. In contrast, the Dirac eight-spinor $\psi_8$ is invariant in arbitrary transformations of the form $\psi_8\rightarrow\boldsymbol{\Lambda}_\mathrm{S}\psi_8$. The reader can verify this by experimenting with the Lorentz transformation matrices of \cite{Partanen2024a}. Therefore, the eight-spinor degrees of freedom do not participate in the Lorentz transformation of $\psi_8$. Thus, the Lorentz transformation of $\psi_8$ is given by $\psi_8\rightarrow\boldsymbol{\Lambda}_\mathrm{F}\psi_8$, where $\boldsymbol{\Lambda}_\mathrm{F}$ is the conventional Lorentz transformation matrix of four-component Dirac spinors, which operates on the components of $\psi_8$ in equation~\eqref{eq:Diraceightspinor}, i.e., on the conventional four-component Dirac spinor $\psi$ as $\psi\rightarrow\boldsymbol{\Lambda}_\mathrm{F}\psi$. Therefore, $\psi_8$ is not invariant in $2\pi$ rotations, but it is invariant in $4\pi$ rotations, which is the characteristic property of spin-$\frac{1}{2}$ fields \cite{Penrose1987,Cartan1981,Landau1982}.

In the conventional QED \cite{Landau1982}, reviewed in section~2 of the supplementary material, the electric four-current density $J_\mathrm{e}^a=(c\rho_\mathrm{e},\mathbf{J}_\mathrm{e})$, where $\rho_\mathrm{e}$ is the electric charge density and $\mathbf{J}_\mathrm{e}$ is the electric current density, is given in terms of the Dirac spinors as $J_\mathrm{e}^a=q_\mathrm{e}c\bar{\psi}\boldsymbol{\gamma}_\mathrm{F}^a\psi=-q_\mathrm{e}c\bar{\psi}_8\boldsymbol{\gamma}_\mathrm{F}^a\psi_8$. In the eight-spinor formalism, the electric four-current density is described by the charge-current spinor $\Phi$, which is a four-vector-type eight-spinor, given by \cite{Partanen2024a}
\begin{align}
 \Phi &=\frac{q_\mathrm{e}}{\sqrt{2\epsilon_0}}\bar{\psi}(\boldsymbol{\gamma}_\mathrm{F})\psi
 =\sqrt{\frac{\varepsilon_0}{2}}\,\mu_0cJ_\mathrm{e}^a\boldsymbol{\mathfrak{e}}_a\nonumber\\
 &=\sqrt{\frac{\varepsilon_0}{2}}\,[0,\mu_0cJ_\mathrm{e}^x,\mu_0cJ_\mathrm{e}^y,\mu_0cJ_\mathrm{e}^z,\rho_\mathrm{e}/\varepsilon_0,0,0,0]^T.
 \label{eq:chargecurrentspinor}
\end{align}
Here $\boldsymbol{\gamma}_\mathrm{F}$ is an eight-spinor made of the Dirac gamma matrices, and it is presented below. The eight-spinor adjoint of $\Phi$ is given by $\bar{\Phi}=\Phi^\dag\boldsymbol{\gamma}_\mathrm{B}^0=q_\mathrm{e}(2\varepsilon_0)^{-1/2}\bar{\psi}(\bar{\boldsymbol{\gamma}}_\mathrm{F})\psi$.

The Cartesian Minkowski coordinate forms of the conventional $4\times 4$ Dirac gamma matrices $\boldsymbol{\gamma}_\mathrm{F}^a$, partial derivatives $\partial_a$, and the electromagnetic-gauge-covariant derivatives $\vec{D}_a$ in equation~\eqref{eq:Da} form eight-component spinors $\boldsymbol{\gamma}_\mathrm{F}$, $\vec{\partial}$ and $\vec{D}$, given by \cite{Partanen2024a}
\begin{equation}
 \boldsymbol{\gamma}_\mathrm{F}=\boldsymbol{\gamma}_\mathrm{F}^a\boldsymbol{\mathfrak{e}}_a=[\mathbf{0},\boldsymbol{\gamma}_\mathrm{F}^x,\boldsymbol{\gamma}_\mathrm{F}^y,\boldsymbol{\gamma}_\mathrm{F}^z,\boldsymbol{\gamma}_\mathrm{F}^0,\mathbf{0},\mathbf{0},\mathbf{0}]^T,
 \label{eq:gamma8}
\end{equation}
\begin{equation}
 \vec{\partial}=-\boldsymbol{\mathfrak{e}}^a\vec{\partial}_a=[0,\vec{\partial}_x,\vec{\partial}_y,\vec{\partial}_z,-\vec{\partial}_0,0,0,0]^T,
 \label{eq:partialderivativespinor}
\end{equation}
\begin{equation}
 \vec{D}=-\boldsymbol{\mathfrak{e}}^a\vec{D}_a=[0,\vec{D}_x,\vec{D}_y,\vec{D}_z,-\vec{D}_0,0,0,0]^T.
 \label{eq:covariantderivativespinor}
\end{equation}
Here, again, the transpose operates on the eight-spinor degree of freedom and not on the Dirac gamma matrices $\boldsymbol{\gamma}_\mathrm{F}^a$ in equation~\eqref{eq:gamma8}. The adjoint spinors $\bar{\boldsymbol{\gamma}}_\mathrm{F}$, $\cev{\partial}$, and $\cev{D}$, are given by $\bar{\boldsymbol{\gamma}}_\mathrm{F}=[\mathbf{0},\boldsymbol{\gamma}_\mathrm{F}^x,\boldsymbol{\gamma}_\mathrm{F}^y,\boldsymbol{\gamma}_\mathrm{F}^z,-\boldsymbol{\gamma}_\mathrm{F}^0,\mathbf{0},\mathbf{0},\mathbf{0}]$, $\cev{\partial}=[0,\cev{\partial}_x,\cev{\partial}_y,\cev{\partial}_z,\cev{\partial}_0,0,0,0]$, and $\cev{D}=[0,\cev{D}_x,\cev{D}_y,\cev{D}_z,\cev{D}_0,0,0,0]$. Using the partial derivative spinor in equation~\eqref{eq:partialderivativespinor} and the electromagnetic potential spinor in equation~\eqref{eq:potentialspinor}, the electromagnetic gauge-covariant-derivative spinor in equation~\eqref{eq:covariantderivativespinor} and its adjoint can be rewritten as \cite{Partanen2024a}
\begin{equation}
 \vec{D}=\vec{\partial}-i\frac{q_\mathrm{e}\sqrt{2/\varepsilon_0}}{\hbar c}\Theta,\hspace{0.5cm}\cev{D}=\cev{\partial}+i\frac{q_\mathrm{e}\sqrt{2/\varepsilon_0}}{\hbar c}\bar{\Theta}.
 \label{eq:eightspinorderivatives}
\end{equation}

The eight-spinor formulation enables writing field quantities in a compact way. For example, the electromagnetic scalar and pseudoscalar are given by $\bar{\Psi}\Psi$ and $i\bar{\Psi}\boldsymbol{\gamma}_\mathrm{B}^5\Psi$ \cite{Partanen2024a}. In terms of the electric and magnetic fields, we have
\begin{equation}
 \bar{\Psi}\Psi=-\frac{1}{4\mu_0}F_{ab}F^{ab}
 =\frac{1}{2}\Big(\varepsilon_0\mathbf{E}^2-\frac{1}{\mu_0}\mathbf{B}^2\Big),
 \label{eq:scalar}
\end{equation}
\begin{equation}
 i\bar{\Psi}\boldsymbol{\gamma}_\mathrm{B}^5\Psi
 =\frac{1}{4\mu_0}F_{ab}\widetilde{F}^{ab}
 =-\varepsilon_0c\mathbf{E}\cdot\mathbf{B}.
 \label{eq:pseudoscalar}
\end{equation}
Here $F^{\mu\nu}$ is the electromagnetic field strength tensor. The Cartesian Minkowski coordinate form of $F^{ab}$ is given in terms of the electromagnetic four-potential $A^a$ and the corresponding electric and magnetic fields as \cite{Jackson1999,Landau1989}
\begin{align}
 F^{ab} &=\partial^a A^b-\partial^b A^a\nonumber\\
 &=\left[\begin{array}{cccc}
0 & -E^x/c & -E^y/c & -E^z/c\\
E^x/c & 0 & -B^z & B^y\\
E^y/c & B^z & 0 & -B^x\\
E^z/c & -B^y & B^x & 0
\end{array}\right].
\label{eq:Ftensor}
\end{align}
The dual electromagnetic field tensor $\widetilde{F}^{\mu\nu}$ in equation~\eqref{eq:pseudoscalar} is defined as \cite{Jackson1999}
\begin{equation}
 \widetilde{F}^{ab}=\frac{1}{2}\varepsilon^{abcd}F_{cd}=\left[\begin{array}{cccc}
0 & -B^x & -B^y & -B^z\\
B^x & 0 & E^z/c & -E^y/c\\
B^y & -E^z/c & 0 & E^x/c\\
B^z & E^y/c & -E^x/c & 0
\end{array}\right].
\label{eq:chiralFtensor}
\end{equation}
Here $\varepsilon^{abcd}$ is the four-dimensional Levi-Civita symbol.

\section{\label{sec:L0}Space-time dimension field and the generating Lagrangian density of gravity}

In this section, we present the concept of the space-time dimension field and use it to formulate the generating Lagrangian density of gravity. Together with the symmetry transformation in section~\ref{sec:symmetry}, the space-time dimension field forms the foundations of unified gravity. While the Standard Model symmetries are associated with fermionic fields, their doublets and triplets, and the Higgs field doublet, the symmetries of unified gravity are associated with the components of the space-time dimension field as discussed in section~\ref{sec:symmetry}. It follows from these foundations that unified gravity does not explicitly include internal degrees of freedom of quantum fields and their symmetry properties in the description of gravity. This limitation is common to all space-time-based formulations of general relativity.

In flat space-time, the generating Lagrangian density of gravity is equivalent to the known Lagrangian density of QED as shown in section~\ref{sec:equivalence}. Due to this equivalence, the space-time dimension field is a mathematical, precisely defined tool, which is associated with space-time but does not assume any specific definition of the space-time metric tensor. The relation to the space-time metric tensor is obtained only after applying separate geometric conditions, to be discussed in this section and in Secs.~\ref{sec:geomUGM} and \ref{sec:teleparalleltetrad}. The gravitational interaction, which determines the structure of space-time in general relativity and TEGR, is shown to arise from symmetries of the space-time dimension field as discussed in section~\ref{sec:symmetry} and in sections thereafter.

We have ended up to the concept of the space-time dimension field in a process of trial and error. The main goal has been to enable the gauge-theory description of gravity using a compact, finite-dimensional gauge group, similar to those of the fundamental interactions of the Standard Model. Without an explicit, separate quantity, like the space-time dimension field, the Lagrangian density of QED satisfies only the well-known U(1) phase rotation symmetry of QED and external space-time symmetries, which notoriously have a noncompact, infinite-dimensional gauge group. Therefore, the space-time dimension field is added in the Lagrangian density to enable additional symmetries. The SEM tensor is the well-known source of the gravitational field in general relativity. Thus, our goal is to construct the generating Lagrangian density, for which the variation with respect to the symmetry transformation parameters gives the SEM tensor. Through this procedure, the SEM tensor appears analogously to the pertinent source terms of the quantum fields of the Standard Model. The eight-spinor representations of the SEM tensors of the Dirac and electromagnetic fields in terms of the kernel matrices, found in \cite{Partanen2024a}, provided us the hint to use the kernel matrices, discussed in section~\ref{sec:kernel}, in the construction of the space-time dimension field. After investigating several alternatives, we have found a single meaningful definition of the space-time dimension field, to be given below.

We finally point out that the fundamental novelty of unified gravity is the addition of the space-time dimension field to the Lagrangian density of the Standard Model without introducing any free physical parameters. Thus, the fundamental hypothesis of unified gravity should sooner be considered to be the symmetry properties of the space-time dimension field and the way how it appears in the generating Lagrangian density of gravity.

\subsection{\label{sec:kernel}Kernel matrices of the eight-spinor theory}

In the eight-spinor theory, we define four kernel matrices $\mathbf{t}^a$ as a mathematical tool to extract four-dimensional space-time quantities from the eight-dimensional spinors. These matrices are needed in the definition of the space-time dimension field in section~\ref{sec:Ig}. The kernel matrices also play the role of the symmetry transformation generators in the gauge theory of unified gravity as shown in section~\ref{sec:fundamentalsymmetry}. The space-time dimension field further enables an eight-spinor representation of the Lagrangian density, whose unitary symmetries lead to the gauge theory of unifed gravity. The kernel matrices $\mathbf{t}^a$ can be expressed in terms of the four bosonic gamma matrices as
\begin{equation}
 \mathbf{t}^a=(\boldsymbol{\gamma}_\mathrm{B}^0\boldsymbol{\gamma}_\mathrm{B}^5\boldsymbol{\gamma}_\mathrm{B}^a)^*.
 \label{eq:kernel}
\end{equation}
Here the superscript $*$ denotes the complex conjugate. The matrices $\mathbf{t}^a$ appear to be constant, traceless Hermitian, unitary, and involutory. The explicit expressions of $\mathbf{t}^a$ in terms of the Lorentz group generators of the four-vector representation are given in section~1.1 of the supplementary material. The definition of $\mathbf{t}^a$ through equation~\eqref{eq:kernel} is unique up to the given representation of the bosonic gamma matrices. For the equivalence transformations of the bosonic gamma matrices, see section~1.2 of the supplementary material. The kernel of $\mathbf{t}^a$, with the restriction that the operation of $\mathbf{t}^a$ is projected to the space of four-vector-type eight-spinors, is the full space of four-vector-type eight-spinors \cite{Partanen2024a}. Therefore, we call the matrices $\mathbf{t}^a$ kernel matrices.

\begin{table}
\setlength{\tabcolsep}{4.5pt}
\renewcommand{\arraystretch}{1.5}
\caption{\label{tbl:kernel}
Restricted kernel, commutation, Hermiticity, unitarity, and trace properties of the kernel matrices $\mathbf{t}^a$ of the eight-spinor theory. Here $\Phi_\mathrm{L1}$ and $\Phi_\mathrm{L2}$ are arbitrary four-vector-type eight-spinors \cite{Partanen2024a}, $V_\mathrm{L}$ denotes the group of such eight-spinors, $\varepsilon^{ijk}$ is the three-dimensional Levi-Civita symbol, and $\delta^{ab}$ is the Kronecker delta. The commutation relation of the spatial kernel matrices $\mathbf{t}^i$ is the well-known commutation relation of $\mathrm{SU(2)}$. Therefore, the matrices $\mathbf{t}^a$ form the representation of $\mathrm{U(1)\otimes SU(2)}$ in the eight-spinor theory.}
\begin{tabular}{c}
   \hline\hline
   \hspace{-2mm}Restricted kernel property\\[0pt]
$\bar{\Phi}_\mathrm{L2}\mathbf{t}^a\Phi_\mathrm{L1}=0,\hspace{0.5cm}\text{for all }\Phi_\mathrm{L1},\Phi_\mathrm{L2}\in V_\mathrm{L}$\\[5pt]
   \hline
   \hspace{-2mm}Mutual commutation relations\\[0pt]
$[\mathbf{t}^0,\mathbf{t}^i]=\mathbf{0},\hspace{0.5cm}[\mathbf{t}^i,\mathbf{t}^j]=2i\varepsilon^{ijk}\mathbf{t}^k$\\[5pt]
   \hline
   \hspace{-2mm}Commutativity with bosonic gamma matrices\\[0pt]
$[\mathbf{t}^a,\boldsymbol{\gamma}_\mathrm{B}^5]=\mathbf{0},\hspace{0.5cm}
 [\mathbf{t}^a,\boldsymbol{\gamma}_\mathrm{B}^b\boldsymbol{\gamma}_\mathrm{B}^c]=\mathbf{0}$\\[5pt]
   \hline
   \hspace{-2mm}Hermiticity and unitarity\\[0pt]
$\mathbf{t}^{a\dag}=\mathbf{t}^a,\hspace{0.5cm}(\mathbf{t}^{a})^{-1}=\mathbf{t}^{a\dag}$\\[5pt]
   \hline
   \hspace{-2mm}Trace properties\\[0pt]
$\mathrm{Tr}(\mathbf{t}^a)=0,\hspace{0.5cm}\mathrm{Tr}(\mathbf{t}^a\mathbf{t}^b)=8\delta^{ab}$\\[5pt]
   \hline\hline
 \end{tabular}
\end{table}

The restricted kernel, commutation, Hermiticity, unitarity, and trace properties of $\mathbf{t}^a$ are summarized in table~\ref{tbl:kernel}. The commutation relation of the spatial kernel matrices $\mathbf{t}^i$ in table~\ref{tbl:kernel} is the well-known commutation relation of $\mathrm{SU(2)}$. The relation to the $\mathrm{SU(2)}$ structure is a feature that is common to our theory and to the theory of Ashtekar variables for classical and quantum gravity \cite{Ashtekar1986}. Detailed study of the relationship of the theories is a topic of further work. In our case, we, furthermore, observe that the four kernel matrices $\mathbf{t}^a$ form the representation of $\mathrm{U(1)\otimes SU(2)}$ in the eight-spinor theory. 

The main novelty of the kernel matrices $\mathbf{t}^a$ in equation~\eqref{eq:kernel} is that their properties in table~\ref{tbl:kernel} reveal the connection between the Cartesian Minkowski four-vectors and the unitary and special unitary groups on which the Standard Model is based on. The present work is founded on this insight.

\subsection{Using kernel matrices to obtain four-dimensional space-time quantities}

The kernel matrices can be used to extract four-dimensional space-time quantities from the eight-dimensional spinors. For example, the Minkowski inner product correspondence is given by $x_ax^a=\bar{\Phi}_X\boldsymbol{\gamma}_\mathrm{B}^5\mathbf{t}^0\Phi_X=c^2t^2-x^2-y^2-z^2$, where the eight-spinor corresponding to the Cartesian Minkowski position four-vector $x^a=(ct,x,y,z)$ is given by $\Phi_X=x^a\boldsymbol{\mathfrak{e}}_a=[0,x,y,z,ct,0,0,0]^T$. In the relation above, only the zeroth kernel matrix $\mathbf{t}^0$ is needed together with the fifth bosonic gamma matrix. Since $\bar{\Phi}_X\boldsymbol{\gamma}_\mathrm{B}^5\mathbf{t}^i\Phi_X=0$, we can write the Minkowski inner product correspondence also as a sum $x_ax^a=\sum_a\bar{\Phi}_X\boldsymbol{\gamma}_\mathrm{B}^5\mathbf{t}^a\Phi_X$, where the right-hand side must contain an explicit summation sign since there are no repeated indices. This relation is implicitly utilized in the definition of the space-time dimension field as the fundamental geometric object of space-time in section~\ref{sec:Ig} below. Examples of relations, where all four kernel matrices are relevant, are the expressions of the SEM tensors of the Dirac and electromagnetic fields, studied in section~\ref{sec:variation}.

\subsection{\label{sec:Ig}Space-time dimension field}
Here we define the fundamental object associated with space-time, which we call the space-time dimension field. Postulating this quantity, denoted by $\mathbf{I}_\mathrm{g}$, enables us to rewrite the conventional Lagrangian density of QED in an eight-spinor form, which we call the generating Lagrangian density of gravity, discussed in section~\ref{sec:L00}. The space-time dimension field $\mathbf{I}_\mathrm{g}=[\mathbf{I}_\mathrm{g}^0,\mathbf{I}_\mathrm{g}^x,\mathbf{I}_\mathrm{g}^y,\mathbf{I}_\mathrm{g}^z]^T$ is described by four space-time-dependent $8\times8$ matrix-valued fields $\mathbf{I}_\mathrm{g}^a$, defined in terms of the kernell matrices $\mathbf{t}^a$ of equation~\eqref{eq:kernel} by the relations
\begin{gather}
 \vec{\partial}_\nu\mathbf{I}_\mathrm{g}^{(a)}=-ig_\mathrm{g}(\partial_\nu X_{(a)})\mathbf{t}^{(a)}\mathbf{I}_\mathrm{g}^{(a)},\nonumber\\
 [\mathbf{t}^{(a)},\mathbf{I}_\mathrm{g}^{(a)}]=\mathbf{0},\hspace{0.5cm} \mathbf{I}_\mathrm{g}^{(a)\dag}\mathbf{I}_\mathrm{g}^{(a)}=\mathbf{I}_8/g_\mathrm{g}.
 \label{eq:Ig_equation}
\end{gather}
Here $X_a$, for the four values of the index $a$, are called \emph{the space-time-dependent phase factors} of $\mathbf{I}_\mathrm{g}$, and $g_\mathrm{g}$ is \emph{the scale constant of unified gravity}, to be discussed in more detail below. The parentheses around the Latin indices in equation~\eqref{eq:Ig_equation} indicate that these indices are not summed over. From equation~\eqref{eq:Ig_equation}, it follows that $\mathbf{I}_\mathrm{g}^\dag\vec{\partial}_\nu\mathbf{I}_\mathrm{g}=-i(\partial_\nu X_a)\mathbf{t}^a$, where the summation over the index $a$ is carried out on the right as conventional.

Solving equation~\eqref{eq:Ig_equation} for the fields $\mathbf{I}_\mathrm{g}^a$ and arranging a four-component vector from the four $\mathbf{I}_\mathrm{g}^a$ fields, we then have
\begin{equation}
 \mathbf{I}_\mathrm{g}=\left[\begin{array}{c}
 \mathbf{I}_\mathrm{g}^0\\[2pt]
 \mathbf{I}_\mathrm{g}^x\\[2pt]
 \mathbf{I}_\mathrm{g}^y\\[2pt]
 \mathbf{I}_\mathrm{g}^z\end{array}\right]
 =\left[\begin{array}{c}
 \frac{1}{\sqrt{g_\mathrm{g}}}e^{-ig_\mathrm{g}\mathbf{t}^0X_0}\\
 \frac{1}{\sqrt{g_\mathrm{g}}}e^{-ig_\mathrm{g}\mathbf{t}^xX_x}\\
 \frac{1}{\sqrt{g_\mathrm{g}}}e^{-ig_\mathrm{g}\mathbf{t}^yX_y}\\
 \frac{1}{\sqrt{g_\mathrm{g}}}e^{-ig_\mathrm{g}\mathbf{t}^zX_z}\end{array}\right].
 \label{eq:Ig}
\end{equation}
The form of $\mathbf{I}_\mathrm{g}$ in this equation explains why $X_a$ are called the phase factors of $\mathbf{I}_\mathrm{g}$.

The space-time dimension field $\mathbf{I}_\mathrm{g}$ is related to geometry and it is not a dynamical field itself. Therefore, we must fix the phase factors, $X_a$, so that they are not free functions of space-time. Consequently, we fix $X_a$ in terms of the Cartesian Minkowski space-time coordinates so that
\begin{equation}
 \partial_\nu X_a=\partial_\nu x_a.
 \label{eq:geom}
\end{equation}
This is the geometric condition that is applied whenever $\mathbf{I}_\mathrm{g}$ is used in calculations. We return to the geometric condition of equation~\eqref{eq:geom} in section~\ref{sec:equivalence}, where we prove the equivalence of the generating Lagrangian density of gravity and the Lagrangian density of QED. In the presence of gravity, we return to the geometric condition in section~\ref{sec:geomUGM} for UGM and in section~\ref{sec:teleparalleltetrad} for TEGRW.

The main novelty of the definition of $\mathbf{I}_\mathrm{g}$ above is that it makes the generating Lagrangian density of gravity, to be defined in section~\ref{sec:L00}, consistently equal to the known Lagrangian density of QED. It also enables a hierarchy of symmetries separate from the fermionic and Higgs field internal symmetries of the Standard Model. The symmetry transformations of $\mathbf{I}_\mathrm{g}$, studied in section~\ref{sec:symmetry}, lead to the gauge theory of unified gravity. The conservation law of the SEM tensor is shown to be obtained in accordance with Noether's theorem, which states that each generator of a continuous symmetry is associated with a conserved current \cite{Noether1918,Zee2010}.

In the introduction of the generating Lagrangian density of gravity in \cite{Partanen2024a}, the matrix $\mathbf{I}_\mathrm{g}$ was assumed to be equal to $\mathbf{I}_8/\sqrt{g_\mathrm{g}}$, where we have accounted for the different scaling of $\mathbf{I}_\mathrm{g}$ by the constant $\sqrt{g_\mathrm{g}}$ in the present work. This assumption led to the special-unitary-symmetry-based derivation of the SEM tensors of the Dirac and electromagnetic fields starting from the form of the generating Lagrangian density of gravity used in \cite{Partanen2024a}. With $\mathbf{I}_\mathrm{g}=\mathbf{I}_8/\sqrt{g_\mathrm{g}}$ and $\vec{\partial}_\nu\mathbf{I}_\mathrm{g}=\mathbf{0}$, the form of the generating Lagrangian density of gravity presented in \cite{Partanen2024a} is, however, not equal to the well-known Lagrangian density of QED, which includes all other fields of the theory except the gravitational field. Such an equality is desired on the basis of how the Lagrangian density of the free Dirac field, which excludes the electromagnetic field, generates the Lagrangian density of QED in the conventional electrodynamic gauge theory, see section~2 of the supplementary material.

\subsection{Scale constant of unified gravity}

The scale constant of unified gravity, $g_\mathrm{g}$, appearing in the definition of the space-time dimension field in equation~\eqref{eq:Ig_equation}, is given in terms of the fundamental physical constants $\hbar$ and $c$, and the constant $E_\mathrm{g}$, called the \emph{energy scale constant of unified gravity}, as
\begin{equation}
 g_\mathrm{g}=\frac{E_\mathrm{g}}{\hbar c}.
 \label{eq:gg0}
\end{equation}
The value of the energy scale constant $E_\mathrm{g}$ depends on the energy scale. The scale invariance of the dynamical equations of unified gravity means that these equations are independent of the values of $g_\mathrm{g}$ and $E_\mathrm{g}$. For a particle with a four-momentum $p$, the energy scale constant is defined as
\begin{equation}
 E_\mathrm{g}=c\sqrt{p^2}.
 \label{eq:Eg_general}
\end{equation}
By this definition, $E_\mathrm{g}$ is Lorentz invariant. Thus, $g_\mathrm{g}$ is also Lorentz invariant.
The energy scale constant $E_\mathrm{g}$ is used to define the energy-scale-dependent gravity fine-structure constant in section~\ref{sec:scaled}. As a scale constant, $E_\mathrm{g}$ does not need to be fixed to any specific energy scale. For example, the natural energy scale in the case of the Dirac electron--positron field is provided by the electron rest energy. In this special case, we have $p^2=m_\mathrm{e}^2c^2$, and thus, $E_\mathrm{g}=m_\mathrm{e}c^2$. In this work, we use $m_\mathrm{e}$ to denote the inertial mass of the electron, and $m'_\mathrm{e}$ denotes the gravitational mass. According to Einstein's equivalence principle, these masses are equal as discussed in section~\ref{sec:equivalenceprinciple}. In natural units with $\hbar=c=m_\mathrm{e}=1$, the values of $g_\mathrm{g}$ and $E_\mathrm{g}$ for the electron become simply equal to 1. However, in the present work, we choose to use the SI units so that the appearance of different constants in equations is easier to follow.

\subsection{\label{sec:equivalenceprinciple}Equivalence principle in unified gravity}
In the sections below, the Lagrangian density of unified gravity is formulated by using $m_\mathrm{e}$ to denote the \emph{inertial} mass of the electron and $m'_\mathrm{e}$ to denote the \emph{gravitational} mass.
According to Einstein's equivalence principle, these masses are equal. Accordingly, in the classical theory, these masses could be set equal in the first place. However, in the renormalization of unified gravity in Se.~\ref{sec:renormalization}, we have found that $m_\mathrm{e}$ and $m'_\mathrm{e}$ satisfy different renormalization relations and only \emph{the renormalized values of these quantities are required to be equal}. This aspect of the quantum theory of gravity is obviously absent in earlier works on the equivalence principle in the quantum regime \cite{Zych2018,Das2023}. Accordingly, we make distinction with $m_\mathrm{e}$ and $m'_\mathrm{e}$ in the formulation of the theory starting from the very beginning.

Unified gravity also has another equivalence relation that is satisfied in the classical and quantum theories. It is related to the scale invariance. When introducing the gauge-covariant derivative of unified gravity in section~\ref{sec:derivative}, we define the coupling constant of unified gravity, denoted by $g'_\mathrm{g}$. In the classical theory, the value of $g'_\mathrm{g}$ is equal to the scale constant $g_\mathrm{g}$ in equation~\eqref{eq:gg0}. In the quantum theory, the renormalized value of $g'_\mathrm{g}$ is equal to $g_\mathrm{g}$.

Therefore, the equivalence principle in unified gravity contains two equalities. These equalities are written as
\begin{equation}
 m'_\mathrm{e}=m_\mathrm{e},\hspace{0.5cm}g'_\mathrm{g}=g_\mathrm{g}.
 \label{eq:equivalenceprinciple}
\end{equation}
The first equality in equation~\eqref{eq:equivalenceprinciple} is called \emph{the equivalence principle of mass}, and the second equality is called \emph{the equivalence principle of scale}. The equivalence relations in equation~\eqref{eq:equivalenceprinciple} and the independence of the dynamical equations of unified gravity on the exact value of $g_\mathrm{g}$ fundamentally mean that unified gravity does not introduce any free parameters beyond the physical constants determined in previous experiments. This is one of the most prominent properties of the theory.

\subsection{\label{sec:L00}Generating Lagrangian density of gravity}

The generating Lagrangian density of gravity is a Lagrangian density that contains all other fields of the theory except the gravitational field. Its requirements, such as diffeomorphism invariance and the relations to symmetry and the SEM tensor, are discussed in \cite{Partanen2024a}. The original representation of the generating Lagrangian density of gravity for QED in section VIII of \cite{Partanen2024a} contained a matrix-valued quantity $\mathbf{I}_\mathrm{g}$, which was assumed a constant identity matrix in \cite{Partanen2024a}. In the present work, we \emph{generalize} $\mathbf{I}_\mathrm{g}$ into a four-component matrix-valued \emph{space-time-dependent field} as described in section~\ref{sec:Ig} and as elaborated in a trial and error process discussed at the beginning of this section. Using this generalized $\mathbf{I}_\mathrm{g}$, given in equation~\eqref{eq:Ig}, we write the generating Lagrangian density of gravity for QED as
\begin{align}
 \mathcal{L}_0 &\!=\!\Big[\frac{\hbar c}{4}\bar{\psi}_8(\bar{\boldsymbol{\gamma}}_\mathrm{F}\bar{\mathbf{I}}_\mathrm{g}\boldsymbol{\gamma}_\mathrm{B}^5\boldsymbol{\gamma}_\mathrm{B}^\nu\vec{\partial}_\nu\mathbf{I}_\mathrm{g}\vec{D}-\cev{D}\bar{\mathbf{I}}_\mathrm{g}\boldsymbol{\gamma}_\mathrm{B}^5\boldsymbol{\gamma}_\mathrm{B}^\nu\vec{\partial}_\nu\mathbf{I}_\mathrm{g}\boldsymbol{\gamma}_\mathrm{F})\psi_8\nonumber\\
 &\hspace{0.4cm}+\frac{im'_\mathrm{e}c^2}{2}\bar{\psi}_8\mathbf{I}^\dag_\mathrm{g}\boldsymbol{\gamma}_\mathrm{B}^5\boldsymbol{\gamma}_\mathrm{B}^\nu\vec{\partial}_\nu\bar{\mathbf{I}}^\dag_\mathrm{g}\psi_8
 -(2m'_\mathrm{e}-m_\mathrm{e})c^2\bar{\psi}_8\psi_8\nonumber\\
 &\hspace{0.4cm}+i\bar{\Psi}\mathbf{I}^\dag_\mathrm{g}\boldsymbol{\gamma}_\mathrm{B}^5\boldsymbol{\gamma}_\mathrm{B}^\nu\vec{\partial}_\nu\bar{\mathbf{I}}^\dag_\mathrm{g}\Psi
 +\bar{\Psi}\Psi\Big]\sqrt{-g}.
 \label{eq:L0}
\end{align}
Here $m'_\mathrm{e}$ is the gravitational mass of the electron and $m_\mathrm{e}$ is the inertial mass. The metric tensor $g^{\mu\nu}$ corresponding to equation~\eqref{eq:L0} is in any coordinates in flat Minkowski space-time. The space-time can be called \emph{flat} since the gravitational field has not been introduced to define a space-time-dependent metric. The explicitly shown partial derivatives $\vec{\partial}_\nu$ in equation~\eqref{eq:L0} act on $\mathbf{I}_\mathrm{g}$ and do not extend to the spinors $\Psi$ and $\psi_8$. These partial derivatives couple the Dirac and electromagnetic fields to gravity when generalized to gauge-covariant derivatives in section~\ref{sec:derivative}. The constant $g_\mathrm{g}$, which appears in the generating Lagrangian density of gravity in \cite{Partanen2024a}, has been here absorbed in the definition of $\mathbf{I}_\mathrm{g}$, given through equations \eqref{eq:Ig_equation} and \eqref{eq:Ig}.

The tensor gauge field of gravity, to be introduced in section~\ref{sec:gravityfield}, arises from unitary symmetries of the generating Lagrangian density of gravity in equation~\eqref{eq:L0} in the same way as the gauge fields of the other fundamental interactions of the Standard Model arise from unitary symmetries. However, the tensor gauge field is associated with a hierarchy of symmetries associated with the transformations of the space-time dimension field, to be discussed in section~\ref{sec:symmetry}. This hierarchy of symmetries is separate from the symmetries of the Standard Model related to the transformations of the fermionic and Higgs fields.

\subsection{\label{sec:equivalence}Equivalence of the generating Lagrangian density of gravity and the Lagrangian density of QED in flat space-time}
Here we show that the generating Lagrangian density of gravity in equation~\eqref{eq:L0} is equivalent to the known Lagrangian density of QED in flat space-time. The calculations of this section justify the definition of the space-time dimension field in equation~\eqref{eq:Ig_equation}. Using equations \eqref{eq:Ig_equation} and \eqref{eq:Ig}, the geometric condition in equation~\eqref{eq:geom}, and the commutation relations of the matrices $\mathbf{t}^a$, given in table~\ref{tbl:kernel}, we obtain $\bar{\mathbf{I}}_\mathrm{g}\boldsymbol{\gamma}_\mathrm{B}^5\boldsymbol{\gamma}_\mathrm{B}^\nu\vec{\partial}_\nu\mathbf{I}_\mathrm{g}=-i(\partial_\nu x_a)\boldsymbol{\gamma}_\mathrm{B}^5\boldsymbol{\gamma}_\mathrm{B}^\nu\mathbf{t}^a=-i\eta_{ab}\boldsymbol{\gamma}_\mathrm{B}^5\boldsymbol{\gamma}_\mathrm{B}^b\mathbf{t}^a$, where we have used the tetrad relation of flat space-time, given by $\partial_\nu x_a=\eta_{ab}\oset{\circ}{e}_\nu^{\;\,b}$, and $\oset{\circ}{e}_\nu^{\;\,b}\boldsymbol{\gamma}_\mathrm{B}^\nu=\boldsymbol{\gamma}_\mathrm{B}^b$. Correspondingly, we obtain $\mathbf{I}_\mathrm{g}^\dag\boldsymbol{\gamma}_\mathrm{B}^5\boldsymbol{\gamma}_\mathrm{B}^\nu\vec{\partial}_\nu\bar{\mathbf{I}}_\mathrm{g}^\dag=-i\eta_{ab}\boldsymbol{\gamma}_\mathrm{B}^5\boldsymbol{\gamma}_\mathrm{B}^b\bar{\mathbf{t}}^a$. Using these identities, the generating Lagrangian density of gravity in equation~\eqref{eq:L0} becomes
\begin{align}
 \mathcal{L}_0 &=\Big[\frac{i\hbar c}{4}\eta_{ab}\bar{\psi}_8(\cev{D}\boldsymbol{\gamma}_\mathrm{B}^5\boldsymbol{\gamma}_\mathrm{B}^b\mathbf{t}^a\boldsymbol{\gamma}_\mathrm{F}-\bar{\boldsymbol{\gamma}}_\mathrm{F}\boldsymbol{\gamma}_\mathrm{B}^5\boldsymbol{\gamma}_\mathrm{B}^b\mathbf{t}^a\vec{D})\psi_8\nonumber\\
 &\hspace{0.4cm}+\frac{m'_\mathrm{e}c^2}{2}\eta_{ab}\bar{\psi}_8\boldsymbol{\gamma}_\mathrm{B}^5\boldsymbol{\gamma}_\mathrm{B}^b\bar{\mathbf{t}}^a\psi_8
 -(2m'_\mathrm{e}-m_\mathrm{e})c^2\bar{\psi}_8\psi_8\nonumber\\
 &\hspace{0.4cm}+\eta_{ab}\bar{\Psi}\boldsymbol{\gamma}_\mathrm{B}^5\boldsymbol{\gamma}_\mathrm{B}^b\bar{\mathbf{t}}^a\Psi
 +\bar{\Psi}\Psi\Big]\sqrt{-g}.
 \label{eq:L01}
\end{align}

Next, we express equation~\eqref{eq:L01} in terms of the electromagnetic field strength tensor and the conventional four-component Dirac spinors using the expressions of the eight-spinors in equations \eqref{eq:electromagneticspinor} and \eqref{eq:Diraceightspinor}. Using the identities $\eta_{ab}\boldsymbol{\gamma}_\mathrm{B}^5\boldsymbol{\gamma}_\mathrm{B}^b\mathbf{t}^a\vec{D}=2\vec{D}$,
$\eta_{ab}\boldsymbol{\gamma}_\mathrm{B}^5\boldsymbol{\gamma}_\mathrm{B}^b\mathbf{t}^a\boldsymbol{\gamma}_\mathrm{F}=2\boldsymbol{\gamma}_\mathrm{F}$, $\eta_{ab}\bar{\psi}_8\boldsymbol{\gamma}_\mathrm{B}^5\boldsymbol{\gamma}_\mathrm{B}^b\bar{\mathbf{t}}^a\psi_8=-4\bar{\psi}\psi$, $\bar{\psi}_8\psi_8=-\bar{\psi}\psi$, $\eta_{ab}\boldsymbol{\gamma}_\mathrm{B}^5\boldsymbol{\gamma}_\mathrm{B}^b\bar{\mathbf{t}}^a\Psi=\mathbf{0}$, $\bar{\boldsymbol{\gamma}}_\mathrm{F}\vec{D}=\boldsymbol{\gamma}_\mathrm{F}^\nu\vec{D}_\nu$, and $\cev{D}\boldsymbol{\gamma}_\mathrm{F}=\cev{D}_\nu\boldsymbol{\gamma}_\mathrm{F}^\nu$, and rewriting the electromagnetic Lagrangian density using equation~\eqref{eq:scalar}, the Lagrangian density in equation~\eqref{eq:L01} becomes
\begin{align}
 \mathcal{L}_0\! &=\!\!\Big[\frac{i\hbar c}{2}\bar{\psi}(\boldsymbol{\gamma}_\mathrm{F}^\nu\vec{D}_\nu\!-\!\cev{D}_\nu\boldsymbol{\gamma}_\mathrm{F}^\nu)\psi
 \!-\!m_\mathrm{e}c^2\bar{\psi}\psi\!-\!\frac{1}{4\mu_0}\!F_{\mu\nu}F^{\mu\nu}\Big]\!\sqrt{-g}\nonumber\\
 &=\Big[\frac{i\hbar c}{2}\bar{\psi}(\boldsymbol{\gamma}_\mathrm{F}^\nu\vec{\partial}_\nu-\cev{\partial}_\nu\boldsymbol{\gamma}_\mathrm{F}^\nu)\psi
 -m_\mathrm{e}c^2\bar{\psi}\psi
 -J_\mathrm{e}^\nu A_\nu\nonumber\\
 &\hspace{0.4cm}-\frac{1}{4\mu_0}\!F_{\mu\nu}F^{\mu\nu}\Big]\!\sqrt{-g}.
 \label{eq:L02}
\end{align}
Here we observe that the terms proportional to the gravitational mass $m'_\mathrm{e}$ have cancelled out and the inertial mass $m_\mathrm{e}$ is the only mass that remains. In the last form of equation~\eqref{eq:L02}, $J_\mathrm{e}^\nu$ is the well-known electric four-current density, which is also the source of the electromagnetic field, given in terms of the Dirac field by
\begin{equation}
 J_\mathrm{e}^\nu=q_\mathrm{e}c\bar{\psi}\boldsymbol{\gamma}_\mathrm{F}^\nu\psi.
 \label{eq:Je0}
\end{equation}
Equation \eqref{eq:L02} is the well-known Lagrangian density of QED \cite{Landau1982,Peskin2018}. This calculation then justifies the definitions of the generating Lagrangian density of gravity in equation~\eqref{eq:L0} and the space-time dimension field $\mathbf{I}_\mathrm{g}$, defined through equations \eqref{eq:Ig_equation} and \eqref{eq:Ig}. Further support is provided by the unitary symmetry transformations associated with $\mathbf{I}_\mathrm{g}$, discussed in section~\ref{sec:symmetry}, which, by following the gauge theory approach, lead to unified gravity, to the conservation law of the SEM tensor, and to agreement with TEGRW as discussed in Secs.~\ref{sec:gravityfield}--\ref{sec:teleparallel}.

\section{\label{sec:symmetry}Symmetries of the generating Lagrangian density of gravity}

In this section, we investigate the symmetries of the generating Lagrangian density of gravity. Symmetries are termed to be global when they are independent of the space-time coordinates. Correspondingly, symmetries are local when they depend on the space-time coordinates.

As fundamental interaction symmetries, we consider unitary and special unitary symmetries associated with compact, finite-dimensional symmetry groups. We follow the gauge theory procedure to seek for global symmetries with respect to which the generating Lagrangian density of gravity in equation~\eqref{eq:L0} is invariant. Then, we introduce gauge fields to make global symmetries local. The generating Lagrangian density in equation~\eqref{eq:L0} trivially satisfies the U(1) symmetry of QED. This symmetry is satisfied locally since the electromagnetic gauge field is included and the electromagnetic-gauge-covariant derivative $\vec{D}$ is used. Unified gravity is shown to be associated with four U(1) symmetries of the components of the space-time dimension field. The associated gauge field is introduced in section~\ref{sec:gravityfield}.

We also introduce chiral symmetries but leave their detailed study as a topic of further work. Furthermore, we study the Lorentz invariance. These investigations are followed by the discussion on the Coleman--Mandula theorem, the variation of the generating Lagrangian density of gravity with respect to the parameters of the fundamental interaction symmetries, and the derivation of the conservation law of the SEM tensor of the Dirac and electromagnetic fields.

\subsection{\label{sec:fundamentalsymmetry}U(1) symmetries of unified gravity}

Next, we investigate the invariance of the generating Lagrangian density of gravity in equation~\eqref{eq:L0} under the unitary symmetry transformation of $\mathbf{I}_\mathrm{g}$. Our goal is that the gauge field generated by the symmetries of $\mathbf{I}_\mathrm{g}$, discussed in section~\ref{sec:gravityfield}, describes gravitational interaction. The symmetries of $\mathbf{I}_\mathrm{g}$ are \emph{independent} of the internal symmetries of the Dirac and electromagnetic fields. The symmetry transformation of $\mathbf{I}_\mathrm{g}$ is formed from four symmetries of its components, which we call \emph{the U(1) symmetries of unified gravity}, given by
\begin{equation}
 \mathbf{I}_\mathrm{g} \rightarrow \mathbf{U}\mathbf{I}_\mathrm{g},\hspace{0.5cm}
 \mathbf{U} =\bigotimes_a\mathbf{U}_a,\hspace{0.5cm}
 \mathbf{U}_a =e^{i\phi_{(a)}\mathbf{t}^{(a)}}.
 \label{eq:psitransformation}
\end{equation}
Here $\mathbf{U}_a$ are the four U(1) symmetries with kernel matrix generators $\mathbf{t}^a$ and symmetry transformation parameters $\phi_a$, which are real-valued constants for a global symmetry. Each of the four U(1) symmetries operates on one of the four components of $\mathbf{I}_\mathrm{g}$ in equation~\eqref{eq:Ig} so that the symmetry transformation $\mathbf{U}_a$ is associated with the component $\mathbf{I}_\mathrm{g}^a$. After observation of the symmetries above, all that follows below is a direct consequence of the gauge theory approach analogous to that in the Standard Model.

From the expressions of the space-time dimension field in equation~\eqref{eq:Ig} and the symmetry transformation in equation~\eqref{eq:psitransformation}, it follows that the transformed space-time dimension field is given by
\begin{equation}
 \mathbf{I}_\mathrm{g}\rightarrow\mathbf{U}\mathbf{I}_\mathrm{g}=\left[\begin{array}{c}
 \frac{1}{\sqrt{g_\mathrm{g}}}e^{-ig_\mathrm{g}\mathbf{t}^0(X_0-\phi_0/g_\mathrm{g})}\\
 \frac{1}{\sqrt{g_\mathrm{g}}}e^{-ig_\mathrm{g}\mathbf{t}^x(X_x-\phi_x/g_\mathrm{g})}\\
 \frac{1}{\sqrt{g_\mathrm{g}}}e^{-ig_\mathrm{g}\mathbf{t}^y(X_y-\phi_y/g_\mathrm{g})}\\
 \frac{1}{\sqrt{g_\mathrm{g}}}e^{-ig_\mathrm{g}\mathbf{t}^z(X_z-\phi_z/g_\mathrm{g})}\end{array}\right].
 \label{eq:Ig_gauged}
\end{equation}
Thus, the effect of $\mathbf{U}$ on $\mathbf{I}_\mathrm{g}$ is equivalent to the translation of the space-time-dependent phase factors of $\mathbf{I}_\mathrm{g}$ as
\begin{equation}
 X_a\rightarrow X_a-\phi_a/g_\mathrm{g}.
 \label{eq:xatransformation}
\end{equation}
Therefore, the symmetry transformation of $\mathbf{I}_\mathrm{g}$ in equation~\eqref{eq:psitransformation} can be replaced by the translation of the phase factors of $\mathbf{I}_\mathrm{g}$ as given in equation~\eqref{eq:xatransformation}. This is analogous to how the U(1) gauge symmetry of QED is equivalent to the phase translation of the Dirac field. In comparison with the conventional translation gauge theory of TEGRW, briefly reviewed in section~3 of the supplementary material, unified gravity is fundamentally different since the translation symmetry in unified gravity only applies to the phase factors of $\mathbf{I}_\mathrm{g}$ and not to the tangent-space coordinates of all fields in the Lagangian density. However, in section~\ref{sec:teleparallel}, we show how TEGRW is derived from unified gravity by applying the so-called Weitzenböck gauge fixing approach.

The U(1) symmetries of equation~\eqref{eq:psitransformation} are different from the U(1) symmetry of QED. While the U(1) symmetry transformation of QED operates on all components of the Dirac field by a common phase factor, the four U(1) symmetries of unified gravity in equation~\eqref{eq:psitransformation} operate on different components of $\mathbf{I}_\mathrm{g}$. Another specialty of the U(1) symmetries of unified gravity is that the generators $\mathbf{t}^a$ are not identity matrices in contrast to the generator of the U(1) symmetry of QED. This is related to the definition of $\mathbf{I}_\mathrm{g}$ through equations \eqref{eq:Ig_equation} and \eqref{eq:Ig}. That the symmetry transformation matrices $\mathbf{U}_a$ in equation~\eqref{eq:psitransformation} are noncommuting for different values of $a$ reminds Yang--Mills theories, but it has no influence on the operation of these symmetries since they operate on different components of $\mathbf{I}_\mathrm{g}$. Thus, these operations trivially commute, and our gauge theory of unified gravity is Abelian.

\vspace{-0.1cm}
\subsection{Chiral symmetries}
We observe that the generating Lagrangian density of gravity in equation~\eqref{eq:L0} is invariant under the chiral symmetry transformations, given by
\begin{equation}
 \mathbf{I}_\mathrm{g} \rightarrow\widetilde{\mathbf{U}}\mathbf{I}_\mathrm{g},\hspace{0.5cm}
 \widetilde{\mathbf{U}} =\bigotimes_{a}\widetilde{\mathbf{U}}_{a},\hspace{0.5cm}
 \widetilde{\mathbf{U}}_a =e^{i\theta_{(a)}\boldsymbol{\gamma}_\mathrm{B}^5\mathbf{t}^{(a)}}.
 \label{eq:chiraltransformation}
\end{equation}
\vspace{-0.4cm}

\noindent Here $\widetilde{\mathbf{U}}_a$ are the four chiral symmetries with chiral generators $\boldsymbol{\gamma}_\mathrm{B}^5\mathbf{t}^a$ and symmetry transformation parameters $\theta_a$, which are real-valued constants for a global symmetry. Each of the four chiral symmetries operates on one of the four components of $\mathbf{I}_\mathrm{g}$ in equation~\eqref{eq:Ig} so that the symmetry transformation $\widetilde{\mathbf{U}}_a$ is associated with the component $\mathbf{I}_\mathrm{g}^a$. There are also mixed symmetry transformations in which $\mathbf{U}_a$ or $\widetilde{\mathbf{U}}_a$ is applied to an arbitrary component of $\mathbf{I}_\mathrm{g}$ other than $\mathbf{I}_\mathrm{g}^a$. In contrast to the U(1) symmetries of unified gravity in equation~\eqref{eq:psitransformation}, the chiral symmetries cannot be replaced by translations of the space-time-dependent phase factors of $\mathbf{I}_\mathrm{g}$. The study of chiral and mixed symmetry transformations is left as a topic of further work.

\subsection{\label{sec:local_Lorentz_invariance}Lorentz invariance}

Using the Lorentz transformations of the eight-spinor formalism, given in \cite{Partanen2024a}, it is relatively straightforward to show that the generating Lagrangian density of gravity satisfies the Lorentz invariance globally. For a related discussion on the equivalence relation of the bosonic gamma matrices, see section~1.2 of the supplementary material.

The local form of the Lorentz invariance is typically interpreted throught the change of the tangent-space coordinates \cite{Aldrovandi2012}. In the present theory, this means the Lorentz transformation of the Cartesian Minkowski coordinates $x^a$ is given by
\begin{equation}
 x^a\rightarrow\Lambda_{\;\,b}^a(\mathrm{x})x^b.
 \label{eq:Lorentz}
\end{equation}
Here $\Lambda_{\;\,b}^a(\mathrm{x})=(e^{\frac{1}{2}\Omega_{cd}(\mathrm{x})K^{cd}})_{\;\,b}^a$ are components of the position-dependent Lorentz transformation matrix $\boldsymbol{\Lambda}(\mathrm{x})=e^{\frac{1}{2}\Omega_{cd}(\mathrm{x})K^{cd}}$, in which $\Omega_{cd}(\mathrm{x})$ parametrizes the transformation, and $(K^{cd})^a_{\;\,b}=\eta^{ca}\delta^d_{\;\,b}-\eta^{da}\delta^c_{\;\,b}$ combines the representations of the Lorentz generators of the four-vector representation, given in section~1.1 of the supplementary material \cite{Partanen2024a}. In the present theory, the Lorentz transformation of $x^a$ in equation~\eqref{eq:Lorentz} is accompanied by the invariance of the kernel matrices, which also appear in the definition of $\mathbf{I}_\mathrm{g}$ in equation~\eqref{eq:Ig}, as
\begin{equation}
 \mathbf{t}^a\rightarrow\boldsymbol{\Lambda}_\mathrm{J}(\mathrm{x})\mathbf{t}^a\boldsymbol{\Lambda}_\mathrm{J}^{-1}(x)=\mathbf{t}^a.
 \label{eq:ta_transformation}
\end{equation}
Here $\boldsymbol{\Lambda}_\mathrm{J}(\mathrm{x})=e^{\frac{1}{8}\Omega_{cd}(\mathrm{x})[\boldsymbol{\gamma}_\mathrm{B}^c,\boldsymbol{\gamma}_\mathrm{B}^d]}$ is the position-dependent Lorentz transformation for $8\times8$ matrix forms of the eight-spinor theory \cite{Partanen2024a}. Since $[\mathbf{t}^a,\boldsymbol{\Lambda}_\mathrm{J}(\mathrm{x})]=\mathbf{0}$, the transformation in equation~\eqref{eq:ta_transformation} preserves the kernel matrices unchanged.

The generating Lagrangian density of gravity in equation~\eqref{eq:L0} is manifestly invariant in the transformation of equation~\eqref{eq:ta_transformation}, but it satisfies the transformation of equation~\eqref{eq:Lorentz} only if the Lorentz transformation matrix $\Lambda_{\;\,b}^a(\mathrm{x})$ is independent of the position, i.e., a global transformation. Therefore, the generating Lagrangian density of gravity in equation~\eqref{eq:L0} does not satisfy local Lorentz invariance as interpreted through equation~\eqref{eq:Lorentz}. As is known from previous studies of local Lorentz transformations in TEGR \cite{Hohmann2022}, theories that do not satisfy local Lorentz invariance do not, however, directly imply any experimentally observable Lorentz violation. The closely related topic of spontaneous breaking of local Lorentz symmetry has also attracted attention in recent literature \cite{Zlosnik2018,Gallagher2024,Nikjoo2024,Jacobson2001,ArkaniHamed2004}.

In the known Poincar\'e gauge theories of gravity \cite{Kibble1961,Sciama1964,Blagojevic2013,Bahamonde2023a}, the local Lorentz invariance is obtained by introducing the spin connection, which is effectively a gauge field of local Lorentz transformations.  The spin connection is conventionally used to describe coupling of gravity to fermionic fields. It is typically used in TEGR and its modifications \cite{Ferraro2022,Bahamonde2023a,Aldrovandi2012,Sotiriou2011,Golovnev2021,Jimenez2019,Cabral2020}, and the case of zero spin connection is called the Weitzenböck gauge and abbreviated in this work by TEGRW. In the present work, we do not choose to introduce the spin connection, but use only the gauge fields associated with the U(1) symmetries of unified gravity in equation~\eqref{eq:psitransformation}. The introduction of the spin connection in the present theoretical framework as a separate field that enables local Lorentz invariance is, nevertheless, possible. As we show in section~\ref{sec:teleparallel}, applying a particular Weitzenböck gauge fixing approach to unified gravity, we can derive the field equations equivalent to those of TEGRW.

\subsection{Coleman--Mandula theorem}
The Coleman--Mandula theorem is known to require that all generators of internal symmetries must commute with the Poincar\'e generators \cite{Coleman1967}. Therefore, the overall symmetry group can only be a direct product of the Poincar\'e group and the internal symmetry groups. From the point of view of this theorem, the four U(1) symmetries of the space-time dimension field in equation~\eqref{eq:psitransformation} are not internal symmetries separate from the Poincar\'e group since they are associated with the translational degrees of freedom of the Cartesian Minkowski coordinates $x_a$. Therefore, instead of introducing new internal symmetries considered by the Coleman--Mandula theorem, we are dealing with degrees of freedom associated with Poincar\'e symmetries. However, the way how the U(1) symmetries are introduced in equation~\eqref{eq:psitransformation} is analogous to how the fermionic and Higgs field internal symmetries of the Standard Model are introduced. Therefore, one can call our U(1) symmetries as internal symmetries of the space-time dimension field, but one must keep in mind that these symmetries are not separate from the Poincar\'e symmetries. For clarity, in this work, we call these symmetries the U(1) symmetries of unified gravity instead of internal symmetries.

\subsection{\label{sec:variation}Variation of the generating Lagrangian of gravity in the symmetry transformation of the space-time dimension field}
Here we calculate the variation of the generating Lagrangian density of gravity in equation~\eqref{eq:L0} with respect to the symmetry transformation parameter $\phi_a$ of the space-time dimension field in equation~\eqref{eq:psitransformation}. We note that only the component $\mathbf{I}_\mathrm{g}^a$ of $\mathbf{I}_\mathrm{g}$ contributes to the variation with respect to $\phi_a$.
The infinitesimal variation of $\mathbf{I}_\mathrm{g}^a$ in the U(1) symmetry transformation of equation~\eqref{eq:psitransformation} with respect to the transformation parameter $\phi_a$ is given by
\begin{equation}
 \delta\mathbf{I}_\mathrm{g}^a=i\mathbf{t}^{(a)}\mathbf{I}_\mathrm{g}^{a}\delta\phi_{(a)}.
 \label{eq:infinitesimalvariation}
\end{equation}
Using the infinitesimal variation in equation~\eqref{eq:infinitesimalvariation}, the variation of the generating Lagrangian density of gravity in equation~\eqref{eq:L0} with respect to $\phi_a$ is then given by
\begin{align}
 &\delta\mathcal{L}_0\nonumber\\[-2pt]
 &\!=\!\sum_a\!\Big[\frac{\hbar c}{4}\bar{\psi}_8(\bar{\boldsymbol{\gamma}}_\mathrm{F}\overline{\delta\mathbf{I}_\mathrm{g}^{(a)}}\boldsymbol{\gamma}_\mathrm{B}^5\boldsymbol{\gamma}_\mathrm{B}^\nu\vec{\partial}_\nu\mathbf{I}_\mathrm{g}^{(a)}\!\vec{D}
 \!+\!\bar{\boldsymbol{\gamma}}_\mathrm{F}\overline{\mathbf{I}_\mathrm{g}^{(a)}}\boldsymbol{\gamma}_\mathrm{B}^5\boldsymbol{\gamma}_\mathrm{B}^\nu\vec{\partial}_\nu\delta\mathbf{I}_\mathrm{g}^{(a)}\!\vec{D}\nonumber\\[-5pt]
 &\hspace{0.4cm}\!-\cev{D}\overline{\delta\mathbf{I}_\mathrm{g}^{(a)}}\boldsymbol{\gamma}_\mathrm{B}^5\boldsymbol{\gamma}_\mathrm{B}^\nu\vec{\partial}_\nu\mathbf{I}_\mathrm{g}^{(a)}\boldsymbol{\gamma}_\mathrm{F}
 \!-\!\cev{D}\overline{\mathbf{I}_\mathrm{g}^{(a)}}\boldsymbol{\gamma}_\mathrm{B}^5\boldsymbol{\gamma}_\mathrm{B}^\nu\vec{\partial}_\nu\delta\mathbf{I}_\mathrm{g}^{(a)}\boldsymbol{\gamma}_\mathrm{F})\psi_8\nonumber\\[-2pt]
 &\hspace{0.4cm}\!+\frac{im'_\mathrm{e}c^2}{2}\bar{\psi}_8(\delta\mathbf{I}_\mathrm{g}^{(a)\dag}\boldsymbol{\gamma}_\mathrm{B}^5\boldsymbol{\gamma}_\mathrm{B}^\nu\vec{\partial}_\nu\overline{\mathbf{I}_\mathrm{g}^{(a)}}^\dag
 \!\!\!+\!\mathbf{I}_\mathrm{g}^{(a)\dag}\boldsymbol{\gamma}_\mathrm{B}^5\boldsymbol{\gamma}_\mathrm{B}^\nu\vec{\partial}_\nu\overline{\delta\mathbf{I}_\mathrm{g}^{(a)}}^\dag)\psi_8\nonumber\\[-2pt]
 &\hspace{0.4cm}\!+i\bar{\Psi}(\delta\mathbf{I}_\mathrm{g}^{(a)\dag}\boldsymbol{\gamma}_\mathrm{B}^5\boldsymbol{\gamma}_\mathrm{B}^\nu\vec{\partial}_\nu\overline{\mathbf{I}_\mathrm{g}^{(a)}}^\dag
 \!\!+\!\mathbf{I}_\mathrm{g}^{(a)\dag}\boldsymbol{\gamma}_\mathrm{B}^5\boldsymbol{\gamma}_\mathrm{B}^\nu\vec{\partial}_\nu\overline{\delta\mathbf{I}_\mathrm{g}^{(a)}}^\dag)\Psi\Big]\sqrt{-g}\nonumber\\[-2pt]
 &\!=\frac{\sqrt{-g}}{g_\mathrm{g}}\Big[\frac{i\hbar c}{4}\bar{\psi}_8(\bar{\boldsymbol{\gamma}}_\mathrm{F}\boldsymbol{\gamma}_\mathrm{B}^5\boldsymbol{\gamma}_\mathrm{B}^\nu\mathbf{t}^{a}\vec{D}-\cev{D}\boldsymbol{\gamma}_\mathrm{B}^5\boldsymbol{\gamma}_\mathrm{B}^\nu\mathbf{t}^{a}\boldsymbol{\gamma}_\mathrm{F})\psi_8\nonumber\\[-2pt]
 &\hspace{0.4cm}-\frac{m'_\mathrm{e}c^2}{2}\bar{\psi}_8\boldsymbol{\gamma}_\mathrm{B}^5\boldsymbol{\gamma}_\mathrm{B}^\nu\bar{\mathbf{t}}^a\psi_8
 -\bar{\Psi}\boldsymbol{\gamma}_\mathrm{B}^5\boldsymbol{\gamma}_\mathrm{B}^\nu\bar{\mathbf{t}}^a\Psi\Big]\vec{\partial}_\nu\delta\phi_{a}\nonumber\\[-2pt]
 &\!=\frac{\sqrt{-g}}{g_\mathrm{g}}\Big[\frac{i\hbar c}{4}\bar{\psi}_8(\bar{\boldsymbol{\gamma}}_\mathrm{F}\boldsymbol{\gamma}_\mathrm{B}^5\boldsymbol{\gamma}_\mathrm{B}^\nu\mathbf{t}^{a}\vec{D}-\cev{D}\boldsymbol{\gamma}_\mathrm{B}^5\boldsymbol{\gamma}_\mathrm{B}^\nu\mathbf{t}^{a}\boldsymbol{\gamma}_\mathrm{F})\psi_8\nonumber\\[-2pt]
 &\hspace{0.4cm}+\frac{m'_\mathrm{e}c^2}{2}\bar{\psi}_8\mathbf{t}^{a}\boldsymbol{\gamma}_\mathrm{B}^\nu\boldsymbol{\gamma}_\mathrm{B}^5\psi_8
 +\bar{\Psi}\mathbf{t}^{a}\boldsymbol{\gamma}_\mathrm{B}^\nu\boldsymbol{\gamma}_\mathrm{B}^5\Psi\Big]\vec{\partial}_\nu\delta\phi_{a}\nonumber\\[-2pt]
 &\!=\frac{\sqrt{-g}}{g_\mathrm{g}}T_\mathrm{m}^{a\nu}\vec{\partial}_\nu\delta\phi_{a}.
 \label{eq:Lagrangiandensityvariation}
\end{align}
In the second equality, we have used equation~\eqref{eq:infinitesimalvariation}, the commutativity of $\mathbf{I}_\mathrm{g}^{(a)}$ with $\boldsymbol{\gamma}_\mathrm{B}^\mu\boldsymbol{\gamma}_\mathrm{B}^\nu$, $\boldsymbol{\gamma}_\mathrm{B}^5$, and $\mathbf{t}^a$, and the normalization condition $\mathbf{I}_\mathrm{g}^{(a)\dag}\mathbf{I}_\mathrm{g}^{(a)}=\mathbf{I}_8/g_\mathrm{g}$. The only terms that are not cancelled by other terms are those where the partial derivative operates on $\delta\phi_a$. In the third equality of equation~\eqref{eq:Lagrangiandensityvariation}, we have used the identity
$\boldsymbol{\gamma}_\mathrm{B}^5\boldsymbol{\gamma}_\mathrm{B}^\nu\bar{\mathbf{t}}^a=-\mathbf{t}^{a}\boldsymbol{\gamma}_\mathrm{B}^\nu\boldsymbol{\gamma}_\mathrm{B}^5$. The last equality of equation~\eqref{eq:Lagrangiandensityvariation} defines the quantity $T_\mathrm{m}^{a\nu}=e_{\;\,\mu}^{a}T_\mathrm{m}^{\mu\nu}$, which is the symmetric SEM tensor of the Dirac and electromagnetic fields. The SEM tensor $T_\mathrm{m}^{\mu\nu}$ is written in terms of the Dirac field SEM tensor $T_\mathrm{D}^{\mu\nu}$ and the electromagnetic field SEM tensor $T_\mathrm{em}^{\mu\nu}$ as \cite{Partanen2024a}
\begin{align}
 T_\mathrm{m}^{\mu\nu} &=T_\mathrm{D}^{\mu\nu}+T_\mathrm{em}^{\mu\nu},\nonumber\\
 T_\mathrm{D}^{\mu\nu} &=\frac{i\hbar c}{4}\bar{\psi}_8(\bar{\boldsymbol{\gamma}}_\mathrm{F}\boldsymbol{\gamma}_\mathrm{B}^5\boldsymbol{\gamma}_\mathrm{B}^\nu\mathbf{t}^{\mu}\vec{D}-\cev{D}\boldsymbol{\gamma}_\mathrm{B}^5\boldsymbol{\gamma}_\mathrm{B}^\nu\mathbf{t}^{\mu}\boldsymbol{\gamma}_\mathrm{F})\psi_8\nonumber\\
 &\hspace{0.4cm}+\frac{m'_\mathrm{e}c^2}{2}\bar{\psi}_8\mathbf{t}^{\mu}\boldsymbol{\gamma}_\mathrm{B}^\nu\boldsymbol{\gamma}_\mathrm{B}^5\psi_8\nonumber\\
 &=\frac{c}{2}P^{\mu\nu,\rho\sigma}[i\hbar\bar{\psi}(\boldsymbol{\gamma}_\mathrm{F\rho}\vec{D}_\sigma-\cev{D}_\rho\boldsymbol{\gamma}_\mathrm{F\sigma})\psi-m'_\mathrm{e}cg_{\rho\sigma}\bar{\psi}\psi]\nonumber\\
 &=\frac{i\hbar c}{4}\bar{\psi}(\boldsymbol{\gamma}_\mathrm{F}^\mu\vec{D}^\nu
 +\boldsymbol{\gamma}_\mathrm{F}^\nu\vec{D}^\mu
 -\cev{D}^\nu\boldsymbol{\gamma}_\mathrm{F}^\mu
 -\cev{D}^\mu\boldsymbol{\gamma}_\mathrm{F}^\nu)\psi\nonumber\\
 &\hspace{0.4cm}-\frac{1}{2}g^{\mu\nu}\Big[\frac{i\hbar c}{2}\bar{\psi}(\boldsymbol{\gamma}_\mathrm{F}^\rho\vec{D}_\rho-\cev{D}_\rho\boldsymbol{\gamma}_\mathrm{F}^\rho)\psi-m'_\mathrm{e}c^2\bar{\psi}\psi\Big],\nonumber\\
 T_\mathrm{em}^{\mu\nu} &=\bar{\Psi}\mathbf{t}^{\mu}\boldsymbol{\gamma}_\mathrm{B}^\nu\boldsymbol{\gamma}_\mathrm{B}^5\Psi
 =\frac{1}{\mu_0}\Big(F_{\;\;\rho}^{\mu}F^{\rho\nu}+\frac{1}{4}g^{\mu\nu}F_{\rho\sigma}F^{\rho\sigma}\Big)\nonumber\\
 &=\frac{1}{2\mu_0}P^{\mu\nu,\rho\sigma,\eta\lambda}\partial_\rho A_\sigma\partial_\eta A_\lambda.
 \label{eq:semtensors0}
\end{align}
In the second form of $T_\mathrm{em}^{\mu\nu}$ in equation~\eqref{eq:semtensors0}, $F^{\mu\nu}$ is the electromagnetic field strength tensor, whose Cartesian Minkowski coordinate form $F^{ab}$ is given in equation~\eqref{eq:Ftensor}. In the last two forms of $T_\mathrm{D}^{\mu\nu}$ in equation~\eqref{eq:semtensors0}, we have used the identities $\bar{\boldsymbol{\gamma}}_\mathrm{F}\boldsymbol{\gamma}_\mathrm{B}^5\boldsymbol{\gamma}_\mathrm{B}^\nu\mathbf{t}^\mu\vec{D}=-2P^{\mu\nu,\rho\sigma}\boldsymbol{\gamma}_\mathrm{F\rho}\vec{D}_\sigma=-\boldsymbol{\gamma}_\mathrm{F}^\mu\vec{D}^\nu-\boldsymbol{\gamma}_\mathrm{F}^\nu\vec{D}^\mu+g^{\mu\nu}\boldsymbol{\gamma}_\mathrm{F}^\rho\vec{D}_\rho$, $\cev{D}\boldsymbol{\gamma}_\mathrm{B}^5\boldsymbol{\gamma}_\mathrm{B}^\nu\mathbf{t}^\mu\boldsymbol{\gamma}_\mathrm{F}=-\cev{D}_\rho\boldsymbol{\gamma}_\mathrm{F\sigma}2P^{\mu\nu,\rho\sigma}=-\cev{D}^\mu\boldsymbol{\gamma}_\mathrm{F}^\nu-\cev{D}^\nu\boldsymbol{\gamma}_\mathrm{F}^\mu+\cev{D}_\rho\boldsymbol{\gamma}_\mathrm{F}^\rho g^{\mu\nu}$, $\bar{\psi}_8\mathbf{t}^\mu\boldsymbol{\gamma}_\mathrm{B}^\nu\boldsymbol{\gamma}_\mathrm{B}^5\psi_8
 =-g^{\mu\nu}\bar{\psi}_8\psi_8$, and $P^{\mu\nu,\rho\sigma}g_{\rho\sigma}=-g^{\mu\nu}$ and the definition of the Dirac eight-spinor $\psi_8$ in terms of $\psi$ in equation~\eqref{eq:Diraceightspinor}. The quantity $P^{\mu\nu,\rho\sigma}$ is defined as
\begin{equation}
P^{\mu\nu,\rho\sigma}=\frac{1}{2}(g^{\mu\sigma}g^{\rho\nu}+g^{\mu\rho}g^{\nu\sigma}-g^{\mu\nu}g^{\rho\sigma}).
\label{eq:P}
\end{equation}
This quantity satisfies the symmetry relations $P^{\mu\nu,\rho\sigma}=P^{\nu\mu,\rho\sigma}=P^{\mu\nu,\sigma\rho}=P^{\rho\sigma,\mu\nu}$ and the multiplicative identity $P^{\mu\nu,\alpha\beta}P_{\alpha\beta,\rho\sigma}=I_{\rho\sigma}^{\mu\nu}$, where $I_{\rho\sigma}^{\mu\nu}$ is the identity tensor, defined as
\begin{equation}
 I_{\rho\sigma}^{\mu\nu}=\frac{1}{2}(\delta^\mu_\rho\delta^\nu_\sigma+\delta^\mu_\sigma\delta^\nu_\rho).
 \label{eq:II}
\end{equation}

The last two forms of $T_\mathrm{em}^{\mu\nu}$ in equation~\eqref{eq:semtensors0} follow from the definition of the bosonic gamma matrices, the electromagnetic spinor in equation~\eqref{eq:electromagneticspinor}, and the electromagnetic field strength tensor in equation~\eqref{eq:Ftensor}. We have defined the quantity $P^{\mu\nu,\rho\sigma,\eta\lambda}$ in equation~\eqref{eq:semtensors0} as
\begin{align}
 &P^{\mu\nu,\rho\sigma,\eta\lambda}\nonumber\\
 &=g^{\eta\sigma}g^{\lambda\mu}g^{\nu\rho}-g^{\eta\mu}g^{\lambda\sigma}g^{\nu\rho}-g^{\eta\rho}g^{\lambda\mu}g^{\nu\sigma}+g^{\eta\mu}g^{\lambda\rho}g^{\nu\sigma}\nonumber\\
 &\hspace{0.3cm}-g^{\mu\sigma}g^{\nu\lambda}g^{\rho\eta}+g^{\mu\sigma}g^{\nu\eta}g^{\rho\lambda}+g^{\mu\rho}g^{\nu\lambda}g^{\sigma\eta}-g^{\mu\rho}g^{\nu\eta}g^{\sigma\lambda}\nonumber\\
 &\hspace{0.3cm}+\frac{1}{2}g^{\mu\nu}(g^{\eta\rho}g^{\lambda\sigma}-g^{\eta\sigma}g^{\lambda\rho}-g^{\rho\lambda}g^{\sigma\eta}+g^{\rho\eta}g^{\sigma\lambda}).
 \label{eq:P3}
\end{align}
This quantity satisfies the symmetry relations $P^{\mu\nu,\rho\sigma,\eta\lambda}=P^{\nu\mu,\rho\sigma,\eta\lambda}=P^{\mu\nu,\eta\lambda,\rho\sigma}$ and the antisymmetry relations $P^{\mu\nu,\rho\sigma,\eta\lambda}=-P^{\mu\nu,\sigma\rho,\eta\lambda}=-P^{\mu\nu,\rho\sigma,\lambda\eta}$.

In the Minkowski space-time, where the last term of the last form of $T_\mathrm{D}^{\mu\nu}$ in equation~\eqref{eq:semtensors0} is zero when the Dirac equation is satisfied and the equivalence principle applies with $m'_\mathrm{e}=m_\mathrm{e}$, the SEM tensor terms $T_\mathrm{D}^{\mu\nu}$ and $T_\mathrm{em}^{\mu\nu}$ of equation~\eqref{eq:semtensors0} agree, respectively, with the well-known symmetric SEM tensors of the Dirac and electromagnetic fields \cite{Peskin2018,Landau1989}. However, in comparison with the Dirac field SEM tensor in \cite{Peskin2018}, our $T_\mathrm{D}^{\mu\nu}$ has both left and right derivatives, which makes it always symmetric, which is not the case with the SEM tensor in \cite{Peskin2018}, e.g., for spherical electron states. We also note that the SEM tensor of the electromagnetic field is traceless since photons are massless. See section~\ref{sec:TEGRgravity} for further discussion on the SEM tensors.

Our approach of deriving the SEM tensor in equation~\eqref{eq:Lagrangiandensityvariation} is fundamentally different from the conventional derivation of the canonical SEM tensor of the field and matter. The conventional derivation is based on varying the Lagrangian density with respect to the space-time coordinates \cite{Landau1989,Partanen2019b}, and the resulting canonical SEM tensor is typically asymmetric if no additional symmetrization procedures are introduced, such as the Belinfante--Rosenfeld symmetrization \cite{Schroder1968,Belinfante1940,Rosenfeld1940,Ramos2015,Partanen2021b}. In contrast, in our equation~\eqref{eq:Lagrangiandensityvariation}, the generating Lagrangian density of gravity is varied with respect to the symmetry transformation parameters of equation~\eqref{eq:psitransformation}. Being associated with the transformation parameters of a continuous symmetry of the generating Lagrangian density, our derivation of the SEM tensor is analogous to the derivations of the conserved currents in the gauge theories of the Standard Model \cite{Peskin2018}. The comparison with the case of QED is discussed in more detail below.

The variation of the generating Lagrangian density of gravity in equation~\eqref{eq:Lagrangiandensityvariation} is analogous to the variation of the generating Lagrangian density of QED, i.e., the Lagrangian density of the free Dirac field, with respect to the electromagnetic U(1) symmetry transformation parameter $\theta$. This variation is discussed in section~2.3 of the supplementary material, and it is given by $\delta\mathcal{L}_\mathrm{QED,0}=-\frac{\sqrt{-g}\,\hbar}{e}J_\mathrm{e}^\nu\partial_\nu\delta\theta$. In the case of QED, the conserved current is the electric four-current density $J_\mathrm{e}^\nu$, appearing in this variation. In the case of the U(1) symmetries of unified gravity in the present work, the corresponding conserved current is the SEM tensor of the Dirac and electromagnetic fields, appearing in the variation of equation~\eqref{eq:Lagrangiandensityvariation}, as derived in section~\ref{sec:conservation} below. However, note that gravity can also contribute to the total SEM tensor if the gravitational field is associated with the definition of the space-time metric as is the case in the Weitzenböck gauge fixing approach discussed in section~\ref{sec:TEGRgravity}.

When the tensor gauge field is added in the present theory through the gauge-covariant derivative, to be introduced in section~\ref{sec:derivative} below, and through the kinetic term of the gauge field, we obtain the gauge-invariant Lagrangian density $\mathcal{L}$, to be given in section~\ref{sec:L}, for which the variation with respect to $\phi_a$ gives
\begin{equation}
 \delta\mathcal{L}=0.
 \label{eq:Lagrangiandensityvariation2}
\end{equation}
Thus, the addition of the gauge field cancels the variation of the generating Lagrangian density of gravity in equation~\eqref{eq:Lagrangiandensityvariation} so that the total variation of the gauge-invariant Lagrangian density is zero.

\subsection{\label{sec:conservation}Conservation law of the SEM tensor}
Here we show that the variation of the generating Lagrangian density of gravity with respect to the U(1) symmetry transformation parameters $\phi_a$, given in equation~\eqref{eq:Lagrangiandensityvariation}, leads to the conservation law of the SEM tensor in accordance with Noether's theorem \cite{Noether1918,Zee2010}. The variation of the action integral, corresponding to $\mathcal{L}_0$, with respect to $\phi_a$ is given by
\begin{align}
 &\delta S_0\nonumber\\
 &\!=\int\delta\mathcal{L}_0d^4x=\int\frac{\sqrt{-g}}{g_\mathrm{g}}T_\mathrm{m}^{a\nu}\partial_\nu\delta\phi_a d^4x\nonumber\\
 &\!=\int\partial_\nu\Big(\frac{\sqrt{-g}}{g_\mathrm{g}}T_\mathrm{m}^{a\nu}\delta\phi_a\Big)d^4x
 -\int\partial_\nu\Big(\frac{\sqrt{-g}}{g_\mathrm{g}}T_\mathrm{m}^{a\nu}\Big)\delta\phi_a d^4x\nonumber\\
 &\!=-\int\oset{\circ}{e}_a^{\;\,\mu}\partial_\nu\Big(\frac{\sqrt{-g}}{g_\mathrm{g}}T_\mathrm{m}^{a\nu}\Big)\delta\phi_\mu d^4x\nonumber\\
 &\!=-\int\Big[\partial_\nu\Big(\oset{\circ}{e}_a^{\;\,\mu}\frac{\sqrt{-g}}{g_\mathrm{g}}T_\mathrm{m}^{a\nu}\Big)
 -\frac{\sqrt{-g}}{g_\mathrm{g}}T_\mathrm{m}^{a\nu}(\partial_\nu\oset{\circ}{e}_a^{\;\,\mu})\Big]\delta\phi_\mu d^4x\nonumber\\
 &\!=-\int\Big[\partial_\nu\Big(\frac{\sqrt{-g}}{g_\mathrm{g}}T_\mathrm{m}^{\mu\nu}\Big)
 -\frac{\sqrt{-g}}{g_\mathrm{g}}T_\mathrm{m}^{\sigma\nu}(\oset{\circ}{e}_{\;\,\sigma}^a\partial_\nu\oset{\circ}{e}_a^{\;\,\mu})\Big]\delta\phi_\mu d^4x\nonumber\\
 &\!=-\int\Big[\partial_\nu\Big(\frac{\sqrt{-g}}{g_\mathrm{g}}T_\mathrm{m}^{\mu\nu}\Big)
 +\frac{\sqrt{-g}}{g_\mathrm{g}}\oset{\circ}{\Gamma}^{\setexp{-1}{\mu}}_{\;\,\sigma\nu}T_\mathrm{m}^{\sigma\nu}\Big]\delta\phi_\mu d^4x\nonumber\\
 &\!=-\int\frac{\sqrt{-g}}{g_\mathrm{g}}(\partial_\nu T_\mathrm{m}^{\mu\nu}+\oset{\circ}{\Gamma}^{\setexp{-1}{\mu}}_{\;\,\sigma\nu}T_\mathrm{m}^{\sigma\nu}+\oset{\circ}{\Gamma}^{\setexp{-1}{\nu}}_{\;\,\sigma\nu}T_\mathrm{m}^{\mu\sigma})\delta\phi_\mu d^4x\nonumber\\
 &\!=-\int\frac{\sqrt{-g}}{g_\mathrm{g}}(\oset{\circ}{\nabla}_\nu T_\mathrm{m}^{\mu\nu})\delta\phi_\mu d^4x.
 \label{eq:actionvariation}
\end{align}
In the second equality, we have applied equation~\eqref{eq:Lagrangiandensityvariation}. In the third equality, we have applied partial integration. In the fourth equality, we have set the first term to zero since it is a total divergence, which can be transformed to a boundary integral, and the fields are assumed to vanish at the boundary. We have also used the identity $\delta\phi_a=\oset{\circ}{e}_a^{\;\,\mu}\delta\phi_\mu$. In the fifth equality, we have applied the product rule of differentiation. In the sixth equality, we have applied $T_\mathrm{m}^{\mu\nu}=\oset{\circ}{e}_a^{\;\,\mu}T_\mathrm{m}^{a\nu}$ and $T_\mathrm{m}^{a\nu}=\oset{\circ}{e}_{\;\,\sigma}^aT_\mathrm{m}^{\sigma\nu}$. In the seventh equality, we have applied $\oset{\circ}{\Gamma}^{\setexp{-1}{\mu}}_{\;\,\sigma\nu}=-\oset{\circ}{e}_{\;\,\sigma}^a\partial_\nu\oset{\circ}{e}_a^{\;\,\mu}$ following from equation~\eqref{eq:Christoffel1}. In the eighth equality of equation~\eqref{eq:actionvariation}, we have used the definition of the operator $\widetilde{\nabla}_\rho$ in equation~\eqref{eq:nabla}. Finally, in the ninth equality of equation~\eqref{eq:actionvariation}, we have applied the well-known definition of the Levi-Civita coordinate-covariant derivative $\oset{\circ}{\nabla}_\nu$, given in equation~\eqref{eq:nablaLeviCivita}.

The final result of equation~\eqref{eq:actionvariation} shows that the variation of the action integral vanishes for arbitrary $\delta\phi_\mu$ when
\begin{equation}
 \oset{\circ}{\nabla}_\nu T_\mathrm{m}^{\mu\nu}=0.
 \label{eq:Tmconservation}
\end{equation}
This is the well-known conservation law of the SEM tensor of all fields and matter except the gravitational field. Since $T_\mathrm{m}^{\mu\nu}$ is symmetric, angular momentum is also conserved. With the pertinent Levi-Civita coordinate-covariant derivative, the conservation law in equation~\eqref{eq:Tmconservation} also holds in a space-time curved by the gravitational field as is well known from general relativity \cite{Misner1973}. The derivation of equation~\eqref{eq:Tmconservation} above is fully analogous to the well-known derivation of the conservation law of the electric four-current density reviewed in section~2.4 of the supplementary material. For general reference, we also finally give the local Lorentz scalar, which is the trace of the SEM tensor of QED in equation~\eqref{eq:semtensors0}, given by
\begin{equation}
 {T_\mathrm{m}}_{\;\,\nu}^\nu=(2m'_\mathrm{e}-m_\mathrm{e})c^2\bar{\psi}\psi.
 \label{eq:contraction}
\end{equation}
Since the SEM tensor of the electromagnetic field is traceless, it does not contribute to equation~\eqref{eq:contraction}.

\section{\label{sec:gravityfield}Gravity gauge field}

Until this point, we have assumed flat space-time, where the gravitational field is nonexistent. In this section, we introduce the tensor gauge field of gravity following the standard gauge theory approach of quantum field theory. The tensor gauge field is introduced without reference to the local Poincar\'e invariance of the theory. This highlights the gauge theory nature of unified gravity based on the invariance of the Lagrangian density in the gauge symmetry transformations of the space-time dimension field, discussed in section~\ref{sec:symmetry}. This also makes unified gravity fundamentally different from previous gauge theory approaches to gravity. Later, in section~\ref{sec:teleparallel}, we show that it is also possible to define the space-time dependent tetrad and metric tensor in the framework of the present theory, in which case unified gravity leads to TEGRW.

\subsection{\label{sec:derivative}Gauge-covariant derivative}
The generating Lagrangian density of gravity in equation~\eqref{eq:L0} is \emph{globally gauge-invariant} in the symmetry transformation of equation~\eqref{eq:psitransformation} for constant values of $\phi_a$. To promote this global symmetry to a local symmetry, we allow $\phi_a$ to depend on the space-time coordinates $x^\mu$. As mentioned in section~\ref{sec:fundamentalsymmetry}, the symmetry transformation matrices $\mathbf{U}_a$ are noncommuting for different values of $a$, but this noncommutativity has no influence on the applicability of the U(1) symmetries, which operate on different components of the space-time dimension field $\mathbf{I}_\mathrm{g}$ in equation~\eqref{eq:Ig}. Following the gauge theory approach \cite{Peskin2018,Weinberg1996}, the generating Lagrangian density of gravity in equation~\eqref{eq:L0} can be made \emph{locally gauge-invariant} in the symmetry transformation of equation~\eqref{eq:psitransformation} when we generalize the partial derivative that acts on $\mathbf{I}_\mathrm{g}$ into a gauge-covariant derivative $\vec{\boldsymbol{\mathcal{D}}}_\nu$, given by
\begin{equation}
 \vec{\boldsymbol{\mathcal{D}}}_\nu =\vec{\partial}_\nu-ig'_\mathrm{g}H_{a\nu}\mathbf{t}^a.
 \label{eq:covariantderivative}
\end{equation}
Here $g'_\mathrm{g}$ is the coupling constant of unified gravity, whose renormalized value is defined to be equal to the scale constant $g_\mathrm{g}$ in equation~\eqref{eq:gg0}. The Hermitian gauge field $H_{a\nu}\mathbf{t}^{a}$ combines the four U(1) gauge fields and it is given in terms of real-valued components $H_{a\nu}$, which describe gravitational interaction as shown below. The gauge theory allows us to obtain a locally gauge-invariant Lagrangian density independent of the choice of the constant in the gauge field term of equation~\eqref{eq:covariantderivative}. The use of the coupling constant $g'_\mathrm{g}$ in equation~\eqref{eq:covariantderivative} allows us to define $H_{a\nu}$ as a dimensionless quantity, which enables easy comparison of the present theory with TEGR as discussed in section~\ref{sec:discussion}.

In the symmetry transformation of equation~\eqref{eq:psitransformation},
the gauge-covariant derivative transforms as $\vec{\boldsymbol{\mathcal{D}}}_\nu\mathbf{I}_\mathrm{g} \rightarrow\mathbf{U}\vec{\boldsymbol{\mathcal{D}}}_\nu\mathbf{I}_\mathrm{g}$. It follows that the transformation of $H_{a\nu}\mathbf{t}^{a}$ is given by $H_{a\nu}\mathbf{t}^{a}\rightarrow(\mathbf{U}H_{a\nu}\mathbf{t}^{a}-\frac{i}{g'_\mathrm{g}}\partial_\nu\mathbf{U})\mathbf{U}^\dag=(H_{a\nu}+\frac{1}{g'_\mathrm{g}}\partial_\nu\phi_a)\mathbf{t}^a$. Thus, the transformation of $H_{a\nu}$ is written as
\begin{equation}
 H_{a\nu}\rightarrow H_{a\nu}+\frac{1}{g'_\mathrm{g}}\partial_\nu\phi_a.
 \label{eq:Hanutransformation}
\end{equation}
Using the gauge-covariant derivative operator $\vec{\boldsymbol{\mathcal{D}}}_\nu$ in place of the partial derivatives $\vec{\partial}_\nu$ makes the generating Lagrangian density of gravity in equation~\eqref{eq:L0} gauge-invariant with respect to the local form of the symmetry transformation in equation~\eqref{eq:psitransformation}.

\subsection{Gravity gauge field strength tensor}
To write the complete gauge-invariant Lagrangian density, we must include a gauge-invariant term that depends only on the gauge field $H_{a\nu}$. However, in the case of gravity, we must apply the soldered character of the gauge theory as discussed below. Utilizing the gauge theory, one obtains an unambiguous form for the gauge field strength tensor from the commutator of the gauge-covariant derivatives \cite{Peskin2018,Weinberg1996}. The commutation relation ${[\vec{\boldsymbol{\mathcal{D}}}_\mu,\vec{\boldsymbol{\mathcal{D}}}_\nu]}=-ig'_\mathrm{g}\mathbf{H}_{\mu\nu}$ defines the antisymmetric gravity field strength tensor $\mathbf{H}_{\mu\nu}$ as
\begin{equation}
 \mathbf{H}_{\mu\nu}=H_{a\mu\nu}\mathbf{t}^a,\hspace{0.5cm}H_{a\mu\nu}=\partial_\mu H_{a\nu}-\partial_\nu H_{a\mu}.
 \label{eq:Hdefinition}
\end{equation}
This equation does not contain commutator terms of $\mathbf{t}^a$ since the symmetry generators operate on different degrees of freedom of the space-time dimension field and, thus, these operations trivially commute as discussed in section~\ref{sec:fundamentalsymmetry}. Therefore, the gauge theory is Abelian, and the gravity field strength tensor in equation~\eqref{eq:Hdefinition} is a gauge-invariant quantity as in elecromagnetism. That $H_{a\mu\nu}$ in equation~\eqref{eq:Hdefinition} is invariant in the transformation of $H_{a\nu}$ in equation~\eqref{eq:Hanutransformation} follows directly from the commutativity of partial derivatives. The redundant degrees of freedom in $H_{a\nu}$ are analogous to those of the electromagnetic four-potential. To treat redundant degrees of freedom, one must fix the gauge as in conventional gauge theories \cite{Peskin2018}. The gauge fixing in unified gravity is discussed in section~\ref{sec:gaugefixing}. In comparison with TEGR \cite{Bahamonde2023a,Aldrovandi2012,Krssak2019}, the form of the gravity gauge field strength tensor $H_{a\mu\nu}$ in equation~\eqref{eq:Hdefinition} is seen to be equivalent to the form of the torsion in the Weitzenböck gauge, i.e., in the absence of the spin connection.

The generic inverse tetrad $e_{\;\,\rho}^a$ allows us to obtain a three-space-time-index form of the gravity gauge field strength tensor in equation~\eqref{eq:Hdefinition} as
\begin{equation}
 H_{\rho\mu\nu}=e_{\;\,\rho}^a H_{a\mu\nu}.
 \label{eq:Hthree}
\end{equation}
This is a necessary property of unified gravity and any gauge theories describing a tensor gauge field. In previous literature, this property is known as the soldered character \cite{Aldrovandi2012,Krasnov2020}. It makes unified gravity fundamentally different from the gauge theories of the Standard Model, which all describe vector fields. We use the generic inverse tetrad symbol $e_{\;\,\rho}^a$ in equation~\eqref{eq:Hthree} and below to indicate that this tetrad can be the tetrad of the Minkowski manifold or it can have a different definition, such as that of TEGRW, discussed in section~\ref{sec:teleparallel}.

\subsection{\label{sec:dual}Dual gravity gauge field strength tensor}

Next, we present the dual gravity gauge field strength tensor, which is a quantity needed to form the Lagrangian density of gravity field strength in section~\ref{sec:L}. In analogy with the gauge theories of the Standard Model, the dual gravity gauge field strength tensor $\widetilde{H}^a_{\;\,\sigma\lambda}$ is given by
\begin{equation}
 \widetilde{H}^a_{\;\,\sigma\lambda}=\frac{1}{2}\varepsilon_{\sigma\lambda\mu\nu}S^{a\mu\nu}.
 \label{eq:dualHdefinition}
\end{equation}
Here $S^{a\mu\nu}$ is called the superpotential, and it is antisymmetric in its last two indices as $S^{a\mu\nu}=-S^{a\nu\mu}$. In vector-field gauge theories, i.e., in the absence of soldered character of the gauge theory, the superpotential is trivially given by the gauge-field strength tensor itself, e.g., see equation~\eqref{eq:chiralFtensor} for QED. This is not the case in soldered gauge theories. Therefore, following the definition of the superpotential for soldered gauge theories, e.g., in \cite{Aldrovandi2012}, we obtain
\begin{equation}
 S^{a\mu\nu}\!=\!e_{\;\,\rho}^a\Big[\frac{1}{2}(H^{\nu\mu\rho}+H^{\mu\rho\nu}-H^{\rho\nu\mu})+g^{\rho\mu}H_{\;\;\;\;\,\sigma}^{\sigma\nu}-g^{\rho\nu}H_{\;\;\;\;\,\sigma}^{\sigma\mu}\Big].
 \label{eq:superpotential}
\end{equation}
This superpotential is shown to appear under the derivative of the dynamical equation of unified gravity, in the case of UGM in section~\ref{sec:UGM}, and in the case of TEGRW in section~\ref{sec:teleparallel}.

\section{\label{sec:L}Locally gauge-invariant Lagrangian density of unified gravity}

In this section, we present the locally gauge-invariant Lagrangian density of unified gravity. In section~\ref{sec:Lg}, we write the kinetic Lagrangian density term, which depends on the gravity gauge field only, and is necessary for the description of the dynamics of gravity. Then, we write the complete locally gauge-invariant Lagrangian density of unified gravity in section~\ref{sec:LL}, and elaborate the reduced form of the gauge-invariant Lagrangian density in section~\ref{sec:reduced}.

\subsection{\label{sec:Lg}Lagrangian density of gravity gauge field strength}
In analogy with the gauge theories of the Standard Model \cite{Peskin2018}, the Lagrangian density of the gravity gauge field strength does not follow from the gauge invariance alone without further assumptions. In the gauge theories of the Standard Model, restrictions are set by parity and time-reversal symmetries and renormalizability of the interactions \cite{Peskin2018}. However, the gauge field Lagrangian densities in all gauge theories of the Standard Model are of the same well-established form. Therefore, following the gauge theories of the Standard Model, the Lagrangian density of the gravity gauge field strength is written in terms of the field strength tensor and the dual field strength tensor, and the constant prefactor is determined by experiments. Thus, the Lagrangian density of the gravity gauge field strength is written as
\begin{equation}
 \mathcal{L}_\mathrm{g,kin}=\frac{1}{8\kappa}H_{a\mu\nu}\widetilde{H}^a_{\;\,\sigma\lambda}\varepsilon^{\mu\nu\sigma\lambda}\sqrt{-g}=\frac{1}{4\kappa}H_{a\mu\nu}S^{a\mu\nu}\sqrt{-g}.
 \label{eq:Lg}
\end{equation}
Here $\kappa=8\pi G/c^4$ is Einstein's constant, where $G$ is the gravitational constant. The prefactor of equation~\eqref{eq:Lg} has been determined by comparison of the theoretical prediction of the strength of the gravitational interaction to experiments on gravitation. The Lagrangian density of the gravity gauge field in equation~\eqref{eq:Lg} is also seen to be of the same mathematical form as the Lagrangian density of gravity in previous works on TEGR \cite{Bahamonde2023a,Aldrovandi2012,Maluf2013,Krssak2019}. Using the Weitzenböck gauge fixing approach, TEGRW is derived from the present theory in section~\ref{sec:teleparallel}. In comparison with QED, the determination of the constant prefactor in equation~\eqref{eq:Lg} is analogous to the determination of the permeability of vacuum in the Lagrangian density of the electromagnetic gauge field based on related experiments or known Maxwell's equations as discussed in section~2.7 of the supplementary material. Comparison of $\mathcal{L}_\mathrm{g,kin}$ in equation~\eqref{eq:Lg} with the Einstein-Hilbert Lagrangian density of general relativity is discussed in the case of the Weitzenböck gauge fixing approach leading to TEGRW in section~6 of the supplementary material.

The second form of the Lagrangian density of the gravity field strength in equation~\eqref{eq:Lg} is obtained from the first form as follows:
\begin{align}
 &H_{a\alpha\beta}\widetilde{H}^a_{\;\,\sigma\lambda}\varepsilon^{\alpha\beta\sigma\lambda}
 =\frac{1}{2}H_{a\alpha\beta}S^{a\mu\nu}\varepsilon_{\sigma\lambda\mu\nu}\varepsilon^{\alpha\beta\sigma\lambda}\nonumber\\
 &=H_{a\mu\nu}S^{a\mu\nu}-H_{a\nu\mu}S^{a\mu\nu}
 =2H_{a\mu\nu}S^{a\mu\nu}.
 \label{eq:Lg_2nd}
\end{align}
In the first equality of equation~\eqref{eq:Lg_2nd}, we have used the definition of the dual field strength tensor in equation~\eqref{eq:dualHdefinition}.
In the second equality, we have used the identity
$\varepsilon_{\sigma\lambda\mu\nu}\varepsilon^{\alpha\beta\sigma\lambda}=2\delta_\mu^\alpha\delta_\nu^\beta-2\delta_\mu^\beta\delta_\nu^\alpha$. In the third equality, we have used the antisymmetry of $H_{a\mu\nu}$ in its space-time indices.

As discussed in section~\ref{sec:dual}, the superpotentials of vector-field gauge theories, such as all gauge theories of the Standard Model, are trivially given by the field strength tensors themselves. Therefore, in formulating vector-field gauge theories, one typically jumps over the definition of the dual field strength tensor and writes the Lagrangian density of the gauge field directly as the field strength tensor contracted with itself. Thus, despite being more general, the forms of Lagrangian densities corresponding to the first form of equation~\eqref{eq:Lg} are not conventionally used in representations of vector-field gauge theories.

\subsection{\label{sec:LL}Locally gauge-invariant Lagrangian density of unified gravity}

Using the gauge-covariant derivative in equation~\eqref{eq:covariantderivative} and adding the gauge field term of gravity in equation~\eqref{eq:Lg} to the generating Lagrangian density of gravity in equation~\eqref{eq:L0}, we obtain the locally gauge-invariant Lagrangian density of unified gravity. This Lagrangian density satisfies locally the electromagnetic [U(1)] gauge-invariance in the same way as the conventional Lagrangian density of QED discussed in Appendix 2. Furthermore, this Lagrangian density satisfies locally the gravity [4$\times$U(1)] gauge-invariance with respect to the symmetry transformation of equation~\eqref{eq:psitransformation}. We recall that, in the locally gauge-invariant Lagrangian density, to be defined below, we do not assume any particular definition of the space-time metric or tetrad. The locally gauge-invariant Lagrangian density of unified gravity is, thus, given by
\begin{align}
 \mathcal{L} &\!=\!\Big[\frac{\hbar c}{4}\bar{\psi}_8(\bar{\boldsymbol{\gamma}}_\mathrm{F}\bar{\mathbf{I}}_\mathrm{g}\boldsymbol{\gamma}_\mathrm{B}^5\boldsymbol{\gamma}_\mathrm{B}^\nu\vec{\boldsymbol{\mathcal{D}}}_\nu\mathbf{I}_\mathrm{g}\vec{D}-\cev{D}\bar{\mathbf{I}}_\mathrm{g}\boldsymbol{\gamma}_\mathrm{B}^5\boldsymbol{\gamma}_\mathrm{B}^\nu\vec{\boldsymbol{\mathcal{D}}}_\nu\mathbf{I}_\mathrm{g}\boldsymbol{\gamma}_\mathrm{F})\psi_8\nonumber\\
 &\hspace{0.4cm}\!\!+\!\frac{im'_\mathrm{e}c^2}{2}\bar{\psi}_8\mathbf{I}^\dag_\mathrm{g}\boldsymbol{\gamma}_\mathrm{B}^5\boldsymbol{\gamma}_\mathrm{B}^\nu\lowerbar{\vec{\boldsymbol{\mathcal{D}}}}_\nu^\lowerdag\bar{\mathbf{I}}^\dag_\mathrm{g}\psi_8
 \!-\!(2m'_\mathrm{e}-m_\mathrm{e})c^2\bar{\psi}_8\psi_8\nonumber\\
 &\hspace{0.4cm}\!\!+\!i\bar{\Psi}\mathbf{I}^\dag_\mathrm{g}\boldsymbol{\gamma}_\mathrm{B}^5\boldsymbol{\gamma}_\mathrm{B}^\nu\lowerbar{\vec{\boldsymbol{\mathcal{D}}}}_\nu^\lowerdag\bar{\mathbf{I}}^\dag_\mathrm{g}\Psi
 +\bar{\Psi}\Psi
 +\frac{1}{4\kappa}H_{a\mu\nu}S^{a\mu\nu}\Big]\sqrt{-g}.
 \label{eq:L}
\end{align}
If we set $H_{a\nu}$ to zero in equations \eqref{eq:covariantderivative}, \eqref{eq:tetrad}, and \eqref{eq:Hdefinition}, the locally gauge-invariant Lagrangian density in equation~\eqref{eq:L} becomes equivalent to the generating Lagrangian density of gravity in equation~\eqref{eq:L0}, which is shown to be equivalent to the Lagrangian density of QED in section~\ref{sec:equivalence}. In the general case, the gauge field $H_{a\nu}$ is obtained as a solution of the Euler-Lagrange equations. The pertinent Euler-Lagrange equations of gravity are discussed in section~\ref{sec:UGMgravity} for UGM and in section~\ref{sec:TEGRgravity} for TEGRW.

\subsection{\label{sec:reduced}Reduced form of the locally gauge-invariant Lagrangian density of unified gravity}

Here we apply the definition of the space-time dimension field in equations \eqref{eq:Ig_equation} and \eqref{eq:Ig} to obtain a reduced form of the locally gauge-invariant Lagrangian density of unified gravity in equation~\eqref{eq:L}. In the reduced Lagrangian density, to be derived below, the space-time dimension field $\mathbf{I}_\mathrm{g}$ does not appear explicitly but there is an explicit dependence on the derivative $\partial_\nu X_a$ of the phase factors of $\mathbf{I}_\mathrm{g}$ instead. Accordingly, the symmetry transformation of $\mathbf{I}_\mathrm{g}$ in equation~\eqref{eq:psitransformation} is replaced by the translation of $X_a$ in equation~\eqref{eq:xatransformation}.

To write out the terms of the gauge-invariant Lagrangian density in equation~\eqref{eq:L} in reduced forms,\vspace{-1pt} we first derive expressions for $\bar{\mathbf{I}}_\mathrm{g}\boldsymbol{\gamma}_\mathrm{B}^5\boldsymbol{\gamma}_\mathrm{B}^\nu\vec{\boldsymbol{\mathcal{D}}}_\nu\mathbf{I}_\mathrm{g}$\vspace{-3pt} and $\mathbf{I}^\dag_\mathrm{g}\boldsymbol{\gamma}_\mathrm{B}^5\boldsymbol{\gamma}_\mathrm{B}^\nu\lowerbar{\vec{\boldsymbol{\mathcal{D}}}}_\nu^\lowerdag\bar{\mathbf{I}}^\dag_\mathrm{g}$. The calculation for $\bar{\mathbf{I}}_\mathrm{g}\boldsymbol{\gamma}_\mathrm{B}^5\boldsymbol{\gamma}_\mathrm{B}^\nu\vec{\boldsymbol{\mathcal{D}}}_\nu\mathbf{I}_\mathrm{g}$ is given by
\begin{align}
 &\bar{\mathbf{I}}_\mathrm{g}\boldsymbol{\gamma}_\mathrm{B}^5\boldsymbol{\gamma}_\mathrm{B}^\nu\vec{\boldsymbol{\mathcal{D}}}_\nu\mathbf{I}_\mathrm{g}
 =\sum_a\bar{\mathbf{I}}_\mathrm{g}^{(a)}\boldsymbol{\gamma}_\mathrm{B}^5\boldsymbol{\gamma}_\mathrm{B}^\nu(\vec{\partial}_\nu-ig'_\mathrm{g}H_{(a)\nu}\mathbf{t}^{(a)})\mathbf{I}_\mathrm{g}^{(a)}\nonumber\\
 &=-ig_\mathrm{g}\sum_a\Big(\partial_\nu X_{(a)}+\frac{g'_\mathrm{g}}{g_\mathrm{g}}H_{(a)\nu}\Big)\bar{\mathbf{I}}_\mathrm{g}^{(a)}\boldsymbol{\gamma}_\mathrm{B}^5\boldsymbol{\gamma}_\mathrm{B}^\nu\mathbf{t}^{(a)}\mathbf{I}_\mathrm{g}^{(a)}\nonumber\\
 &=-i\Big(\partial_\nu X_a+\frac{g'_\mathrm{g}}{g_\mathrm{g}}H_{a\nu}\Big)\boldsymbol{\gamma}_\mathrm{B}^5\boldsymbol{\gamma}_\mathrm{B}^\nu\mathbf{t}^a.
 \label{eq:simplify1}
\end{align}
Correspondingly, for $\mathbf{I}^\dag_\mathrm{g}\boldsymbol{\gamma}_\mathrm{B}^5\boldsymbol{\gamma}_\mathrm{B}^\nu\lowerbar{\vec{\boldsymbol{\mathcal{D}}}}_\nu^\lowerdag\bar{\mathbf{I}}^\dag_\mathrm{g}$, we obtain
\begin{align}
 &\mathbf{I}^\dag_\mathrm{g}\boldsymbol{\gamma}_\mathrm{B}^5\boldsymbol{\gamma}_\mathrm{B}^\nu\lowerbar{\vec{\boldsymbol{\mathcal{D}}}}_\nu^\lowerdag\bar{\mathbf{I}}^\dag_\mathrm{g}
 =\sum_a\mathbf{I}_\mathrm{g}^{(a)\dag}\boldsymbol{\gamma}_\mathrm{B}^5\boldsymbol{\gamma}_\mathrm{B}^\nu(\vec{\partial}_\nu-ig'_\mathrm{g}H_{(a)\nu}\bar{\mathbf{t}}^{(a)})\bar{\mathbf{I}}_\mathrm{g}^{(a)\dag}\nonumber\\
 &=-ig_\mathrm{g}\sum_a\Big(\partial_\nu X_{(a)}+\frac{g'_\mathrm{g}}{g_\mathrm{g}}H_{(a)\nu}\Big)\mathbf{I}_\mathrm{g}^{(a)\dag}\boldsymbol{\gamma}_\mathrm{B}^5\boldsymbol{\gamma}_\mathrm{B}^\nu\bar{\mathbf{t}}^{(a)}\bar{\mathbf{I}}_\mathrm{g}^{(a)\dag}\nonumber\\
 &=-i\Big(\partial_\nu X_a+\frac{g'_\mathrm{g}}{g_\mathrm{g}}H_{a\nu}\Big)\boldsymbol{\gamma}_\mathrm{B}^5\boldsymbol{\gamma}_\mathrm{B}^\nu\bar{\mathbf{t}}^a\nonumber\\
 &=i\Big(\partial_\nu X_a+\frac{g'_\mathrm{g}}{g_\mathrm{g}}H_{a\nu}\Big)\mathbf{t}^{a}\boldsymbol{\gamma}_\mathrm{B}^\nu\boldsymbol{\gamma}_\mathrm{B}^5.
 \label{eq:simplify2}
\end{align}
In the first equalities of equations \eqref{eq:simplify1} and \eqref{eq:simplify2}, we have used the definition of the gauge-covariant derivative in equation~\eqref{eq:covariantderivative} and the representation of $\mathbf{I}_\mathrm{g}$ in terms of $\mathbf{I}_\mathrm{g}^a$ in equation~\eqref{eq:Ig}. In the second equalities, we have applied equation~\eqref{eq:Ig_equation}. In the third equalities, we have applied the commutativity of $\mathbf{I}_\mathrm{g}^a$ with $\boldsymbol{\gamma}_\mathrm{B}^\mu\boldsymbol{\gamma}_\mathrm{B}^\nu$, $\boldsymbol{\gamma}_\mathrm{B}^5$, and $\mathbf{t}^a$ and the normalization condition $\mathbf{I}_\mathrm{g}^{(a)\dag}\mathbf{I}_\mathrm{g}^{(a)}=\mathbf{I}_8/g_\mathrm{g}$. 
In the last equality of equation~\eqref{eq:simplify2}, we have used the identity
$\boldsymbol{\gamma}_\mathrm{B}^5\boldsymbol{\gamma}_\mathrm{B}^\nu\bar{\mathbf{t}}^a=-\mathbf{t}^{a}\boldsymbol{\gamma}_\mathrm{B}^\nu\boldsymbol{\gamma}_\mathrm{B}^5$.

Substituting the quantities of equations \eqref{eq:simplify1} and \eqref{eq:simplify2} to the locally gauge-invariant Lagrangian density in equation~\eqref{eq:L}, we obtain
\begin{align}
 \mathcal{L} &=\Big\{\Big(\partial_\nu X_a+\frac{g'_\mathrm{g}}{g_\mathrm{g}}H_{a\nu}\Big)\nonumber\\
 &\hspace{0.4cm}\times\Big[\frac{i\hbar c}{4}\bar{\psi}_8(\cev{D}\boldsymbol{\gamma}_\mathrm{B}^5\boldsymbol{\gamma}_\mathrm{B}^\nu\mathbf{t}^a\boldsymbol{\gamma}_\mathrm{F}-\bar{\boldsymbol{\gamma}}_\mathrm{F}\boldsymbol{\gamma}_\mathrm{B}^5\boldsymbol{\gamma}_\mathrm{B}^\nu\mathbf{t}^a\vec{D})\psi_8\nonumber\\
 &\hspace{0.4cm}-\frac{m'_\mathrm{e}c^2}{2}\bar{\psi}_8\mathbf{t}^{a}\boldsymbol{\gamma}_\mathrm{B}^\nu\boldsymbol{\gamma}_\mathrm{B}^5\psi_8
 -\bar{\Psi}\mathbf{t}^{a}\boldsymbol{\gamma}_\mathrm{B}^\nu\boldsymbol{\gamma}_\mathrm{B}^5\Psi\Big]\nonumber\\
 &\hspace{0.4cm}-(2m'_\mathrm{e}-m_\mathrm{e})c^2\bar{\psi}_8\psi_8
 +\bar{\Psi}\Psi
 +\frac{1}{4\kappa}H_{a\mu\nu}S^{a\mu\nu}\Big\}\sqrt{-g}\nonumber\\
 &=\Big[\!-\!\Big(\partial_\nu X_a+\frac{g'_\mathrm{g}}{g_\mathrm{g}}H_{a\nu}\Big)T_\mathrm{m}^{a\nu}
 -(2m'_\mathrm{e}\!-\!m_\mathrm{e})c^2\bar{\psi}_8\psi_8
 +\bar{\Psi}\Psi\nonumber\\
 &\hspace{0.4cm}
 +\frac{1}{4\kappa}H_{a\mu\nu}S^{a\mu\nu}\Big]\sqrt{-g}.
 \label{eq:LL}
\end{align}
We call this Lagrangian density the reduced form of the locally gauge-invariant Lagrangian density of unified gravity. In the last equality of equation~\eqref{eq:LL}, we have used the definition of the SEM tensor of the Dirac and electromagnetic fields in equation~\eqref{eq:semtensors0}. The gauge-invariance of the Lagrangian density in equation~\eqref{eq:LL} means that it remains invariant when $X_a$ is transformed by the transformation law in equation~\eqref{eq:xatransformation} and $H_{a\nu}$ is transformed by the transformation law in equation~\eqref{eq:Hanutransformation}.

The locally gauge-invariant Lagrangian density of unified gravity in equation~\eqref{eq:LL} allows the use of different definitions of the tetrad and gauge fixing.  The use of the tetrad of the Cartesian Minkowski manifold and the harmonic gauge fixing, which lead to UGM, are studied in section~\ref{sec:UGM}. The Weitzenböck gauge fixing, which leads to TEGRW, is discussed in section~\ref{sec:teleparallel}.

\subsection{\label{sec:scaled}Scaled representation of unified gravity and the dimensionless coupling constant}
Next, we discuss the scaled representation of unified gravity, which is analogous to the scaled representation of QED discussed in section~2.9 of the supplementary material. The scaled representation of unified gravity enables transparent comparison with the gauge theories of the Standard Model and the conventional translation gauge theory of TEGRW as presented here and in section~\ref{sec:discussion}.

In the scaled representation of unified gravity, the gravity gauge field is scaled so that the Lagrangian density of the gravity gauge field in equation~\eqref{eq:Lg} no longer contains Einstein's constant in the prefactor. The scaling of the gravity gauge field is given by
\begin{equation}
 H_{a\nu}\rightarrow \sqrt{\kappa}H_{a\nu}'.
 \label{eq:Hscaling}
\end{equation}
The gravity gauge field strength tensor and the dual field strength tensor become then replaced by scaled quantities as
\begin{equation}
 H_{a\mu\nu}\rightarrow\sqrt{\kappa}H_{a\mu\nu}',
\end{equation}
\begin{equation}
 \widetilde{H}^a_{\;\,\sigma\lambda}\rightarrow\sqrt{\kappa}\widetilde{H}'^a_{\;\;\,\sigma\lambda}.
\end{equation}
This scaling makes the Lagrangian density of the gravity gauge field in equation~\eqref{eq:Lg} to become
\begin{equation}
 \mathcal{L}_\mathrm{g,kin}\rightarrow\frac{1}{8}H_{a\mu\nu}'\widetilde{H}'^a_{\;\;\,\sigma\lambda}\varepsilon^{\mu\nu\sigma\lambda}\sqrt{-g}.
\end{equation}
Here, in comparison with equation~\eqref{eq:Lg}, Einstein's constant is no longer present in the prefactor.

In analogy with QED, discussed in section~2.9 of the supplementary material, we next define \emph{the dimensionless coupling constant of unified gravity}, $E_\mathrm{g}'$, by scaling the dimensionful coupling constant $g'_\mathrm{g}=E_\mathrm{g}/(\hbar c)$. Here we set the coupling constant of unified gravity equal to the scale constant in equation~\eqref{eq:gg0}. With further scaling, we then write
\begin{equation}
 E'_\mathrm{g}=E_\mathrm{g}\sqrt{\frac{\kappa}{\hbar c}}
 =\sqrt{8\pi\alpha_\mathrm{g}}.
 \label{eq:gg}
\end{equation}
In analogy with the electric fine-structure constant $\alpha_\mathrm{e}$ in section~2.9 of the supplementary material, equation~\eqref{eq:gg} defines the gravity fine-structure constant $\alpha_\mathrm{g}$. Solving equation~\eqref{eq:gg} for $\alpha_\mathrm{g}$ and using equation~\eqref{eq:Eg_general}, we obtain
\begin{equation}
 \alpha_\mathrm{g}=\frac{\kappa cp^2}{8\pi\hbar}=\frac{Gp^2}{\hbar c^3}.
 \label{eq:alphag_general}
\end{equation}
The gravity fine-structure constant depends on the energy scale through $p^2$. In the special case of the electron, we have $p^2=m_\mathrm{e}^2c^2$. Using equation~\eqref{eq:alphag_general}, this gives the gravity fine-structure constant of the electron as
\begin{equation}
 \alpha_\mathrm{g}=\frac{\kappa m_\mathrm{e}^2c^3}{8\pi\hbar}=\frac{Gm_\mathrm{e}^2}{\hbar c}.
 \label{eq:alphag}
\end{equation}
Using the experimental value of the gravitational constant, the numerical value of $\alpha_\mathrm{g}$ is given by $\alpha_\mathrm{g}\approx 1.75181\times 10^{-45}$. This small value characterizes the weakness of the gravitational interaction in comparison with the fundamental interactions of the Standard Model. The ratio of $\alpha_\mathrm{g}$ in equation~\eqref{eq:alphag} and the well-known electric fine-structure constant $\alpha_\mathrm{e}=e^2/(4\pi\varepsilon_0\hbar c)$ is equal to the known ratio of the strengths of the gravitational and electromagnetic forces or potential energies between two electrons, i.e., $|\mathbf{F}_\mathrm{g}|/|\mathbf{F}_\mathrm{e}|=|V_\mathrm{g}|/|V_\mathrm{e}|=4\pi\varepsilon_0Gm_\mathrm{e}^2/e^2$. This result justifies the physical meaningfulness of the gravity fine structure constant $\alpha_\mathrm{g}$ in characterizing the strength of the gravitational interaction. The term gravity fine-structure constant has also appeared in previous literature. In \cite{Weisskopf1970,Ashtekar2014}, its formula, similar to equation~\eqref{eq:alphag}, has been expressed in terms of the proton mass. The analogy with QED in section~2.9 of the supplementary material suggests that the dimensionful and dimensionless coupling constants $E_\mathrm{g}$ and $E_\mathrm{g}'$ are subject to the phenomenon known as the running of the coupling constant \cite{Deur2016,Abel2023,Leuchs2020,Aad2023}. Detailed study of this effect in unified gravity is left as a topic of further work.

The scaling of the gravity gauge field in equation~\eqref{eq:Hscaling} and the definition of the dimensionless coupling constant of unified gravity in equation~\eqref{eq:gg} make the gauge-covariant derivative in equation~\eqref{eq:covariantderivative} to become
\begin{equation}
 \vec{\boldsymbol{\mathcal{D}}}_\nu\rightarrow\vec{\partial}_\nu-i\frac{E_\mathrm{g}'}{\sqrt{\hbar c}}H'_{a\nu}\mathbf{t}^a.
 \label{eq:covariantderivativescaled}
\end{equation}
This form of the gauge-covariant derivative of unified gravity allows easy comparison with the gauge-covariant derivative of the scaled representation of QED, discussed in section~2.9 of the supplementary material, and the gauge-covariant derivative of the Standard Model, discussed in section~\ref{sec:vectorbosons}. In natural units with $\hbar=c=1$ complemented with $\kappa=1$, the scaled representation above is trivial since these units imply $H_{a\nu}=H_{a\nu}'$ and $E_\mathrm{g}=E_\mathrm{g}'$.

It is well known from previous literature that the coupling constant of a theory must be either dimensionless or have a positive mass dimension for the theory to be renormalizable \cite{Zee2010,Schwartz2014,Maggiore2005}. In the latter case, the theory is called super-renormalizable. If the coupling constant has negative mass dimension, the theory is nonrenormalizable. Therefore, the dimensionless coupling constant of unified gravity strongly suggests that the theory is renormalizable. The renormalizability of unified gravity is studied at 1-loop order in section~\ref{sec:renormalization}.

\section{\label{sec:UGM}Unified gravity in the Minkowski metric}

It is especially interesting to study unified gravity in the Cartesian Minkowski metric, i.e., UGM, since the Minkowski metric is the obvious starting point for the quantization of gravitational interaction \cite{Paston2011}. When the space-time metric is equal to the Cartesian Minkowski metric, there is no difference between the Latin and Greek indices. Therefore, since it is conventional to use Greek letters for space-time indices, in UGM, we choose to use Greek indices exclusively.

\subsection{\label{sec:geomUGM}Geometric conditions of UGM: tetrad and metric tensor}
Unified gravity allows freedom in the definition of the tetrad and the space-time metric tensor. Thus, in unified gravity, we can use a gauge-field-independent tetrad and a metric tensor as done for UGM in this section or a gauge-field-dependent tetrad and a metric tensor as done for TEGRW in section~\ref{sec:teleparallel}. In UGM, we use the geometric condition in equation~\eqref{eq:geom} together with the tetrad of the Cartesian Minkowski manifold, given by
\begin{equation}
 \oset{\circ}{e}^{\;\,\nu}_\mu=\partial^\nu x_\mu=\partial^\nu X_\mu=\delta_\mu^\nu.
 \label{eq:tetradMinkowskimanifold}
\end{equation}
Since the two indices of $\oset{\circ}{e}^{\;\,\nu}_\mu$ in equation~\eqref{eq:tetradMinkowskimanifold} are of the same type, $\oset{\circ}{e}^{\;\,\nu}_\mu$ becomes equal to $\delta_\mu^\nu$. Therefore, in the Cartesian Minkowski manifold, $\partial_\nu x_\mu=\partial_\nu X_\mu$ is equal to the Cartesian Minkowski metric tensor as
\begin{equation}
 \eta_{\mu\nu}=\partial_\nu x_\mu=\partial_\nu X_\mu.
 \label{eq:metricMinkowskimanifold}
\end{equation}
Accordingly, in UGM, we set the metric tensor determinant associated factor $\sqrt{-g}$ to unity in the Lagrangian densities since $g=\det(\eta_{\mu\nu})=-1$. Equations \eqref{eq:tetradMinkowskimanifold} and \eqref{eq:metricMinkowskimanifold} together with equation~\eqref{eq:geom} are called \emph{the geometric conditions of UGM}.

If one applies the geometric condition in equation~\eqref{eq:geom} to the reduced form of the gauge-invariant Lagrangian density of unified gravity in equation~\eqref{eq:LL}, the Lagrangian density loses its explicit dependence on $X_\mu$ and one can no longer perform local gauge transformations on $X_\mu$ using equation~\eqref{eq:xatransformation}. Therefore, the gauge invariance of the Lagrangian density of unified gravity in equation~\eqref{eq:L} or \eqref{eq:LL} with respect to the four U(1) symmetry transformations of gravity in equation~\eqref{eq:psitransformation} or \eqref{eq:xatransformation} must be interpreted \emph{before} the application of equation~\eqref{eq:geom} to the Lagrangian density. The gauge fixing of UGM in section~\ref{sec:gaugefixing} below is also performed before the geometric conditions of UGM are applied to the Lagrangian density.

In the conventional TEGRW, the translation gauge field has been particularly introduced to make the tetrad independent of translations of the tangent-space coordinates $x_a$. Hence, the translation gauge field is, by definition, included in the expression of the tetrad in TEGRW, which is thus different from the tetrad of the Minkowski manifold. This highlights the fundamental difference in the foundations of unified gravity and the conventional formulation of TEGRW \cite{Aldrovandi2012,Bahamonde2023a}.

\subsection{\label{sec:gaugefixing}Gauge fixing in UGM}

As is well known, the path integral formulation of the field theory and the gauge field propagators cannot be consistently formulated without fixing the gauge \cite{Peskin2018,Schwartz2014}. This is because, by definition, gauge theories represent each distinct field configuration of the physical system as an equivalence class of field configurations. The equivalence classes of field configurations are defined by gauge transformations. Therefore, they are associated with redundant degrees of freedom in gauge field variables. Gauge fixing must be applied to effectively remove the redundant degrees of freedom \cite{Peskin2018,Schwartz2014}. The relation of gauge fixing to the determination of gauge field propagators of field theories is discussed in more detail in section~\ref{sec:Feynman} below and in section~4 of the supplementary material.

In analogy with the gauge theories of the Standard Model \cite{Peskin2018,Schwartz2014}, the starting point for gauge fixing in UGM is the non-gauge-fixed functional integral, given by
\begin{equation}
 \int e^{iS[\bar{\psi},\psi,A,H]/(\hbar c)}\mathcal{D}\bar{\psi}\mathcal{D}\psi\mathcal{D}A\mathcal{D}H.
 \label{eq:pathintegralREV}
\end{equation}
Here $S[\bar{\psi},\psi,A,H]$ is the action integral, i.e., the integral of the locally gauge-invariant Lagrangian density of unified gravity in equations \eqref{eq:L} and \eqref{eq:LL} over the space-time coordinates, written as
\begin{equation}
 S[\bar{\psi},\psi,A,H]=\int\mathcal{L}\,d^4x.
 \label{eq:action}
\end{equation}
The functional integral in equation~\eqref{eq:pathintegralREV} is over the components of the Dirac spinor $\psi$ and adjoint spinor $\bar{\psi}$, over the four components of the electromagnetic four-potential $A_\mu$ as $\mathcal{D}A=\prod_\mu\mathcal{D}A_\mu$, and over all components of the gravity gauge field $H_{\mu\nu}$ as $\mathcal{D}H=\prod_{\mu,\nu}\mathcal{D}H_{\mu\nu}$. The gauge fixing is performed by the Faddeev--Popov method as explained below.

\subsubsection{\label{sec:QEDfixing}QED gauge fixing}
Gauge fixing must be carried out for all gauge fields of the theory. We start with the well-known gauge fixing of QED. A particularly convenient gauge is the Feynman gauge, whose gauge condition is in QED equivalent to the Lorenz gauge condition, given by $\partial_{\nu}A^{\nu}=0$ \cite{Feynman1949}. The Feynman gauge condition is written as
\begin{equation}
 C_\mathrm{em}(A) \equiv\partial_{\nu}A^{\nu}
 =\sqrt{2\mu_0}\,\bar{\boldsymbol{\mathfrak{e}}}_0\boldsymbol{\gamma}_\mathrm{B}^\nu\partial_\nu\Theta=0.
 \label{eq:Lorenz}
\end{equation}
Equation \eqref{eq:Lorenz} defines the gauge-fixing function $C_\mathrm{em}(A)$. The Feynman gauge condition in equation~\eqref{eq:Lorenz} is satisfied in the electromagnetic gauge transformation of the form $A_\nu\rightarrow A_\nu-\frac{\hbar}{e}\partial_\nu\theta$ when the gauge function $\theta$ is a solution of the wave equation $\partial^2\theta=0$. Thus, the Feynman gauge does not determine the electromagnetic four-potential uniquely. The remaining residual degrees of freedom are called the residual gauge symmetry \cite{Swanson2022,Partanen2024a}. Consequently, it is not sufficient to use the Feynman gauge condition alone, but also an integral over the electromagnetic gauge function is necessary. The complete gauge fixing is formally obtained by the Faddeev--Popov method \cite{Peskin2018,Schwartz2014,Faddeev1967}, to be applied below.

In the Faddeev--Popov method for gauge fixing in QED, one inserts 1 under the non-gauge-fixed functional integral of equation~\eqref{eq:pathintegralREV} in the following form \cite{Peskin2018,Schwartz2014}:
\begin{equation}
 1=\int\delta[C_\mathrm{em}(A^{(\theta)})]\det\!\Big[\frac{\delta C_\mathrm{em}(A^{(\theta)})}{\delta\theta}\Big]\mathcal{D}\theta.
 \label{eq:FP}
\end{equation}
Here the functional integral is over the electromagnetic gauge function $\theta$ and $A^{(\theta)}$ is the gauge-transformed field, given by
\begin{equation}
 A_\nu^{(\theta)}=A_\nu-\frac{\hbar}{e}\partial_\nu\theta.
\end{equation}
The term $\delta[C_\mathrm{em}(A^{(\theta)})]$ in equation~\eqref{eq:FP} is the functional delta function and $\det[\delta C_\mathrm{em}(A^{(\theta)})/\delta\theta]=\det(-\frac{\hbar}{e}\partial^2)$ is called the Faddeev--Popov determinant.

Due to the local gauge invariance of the Lagrangian density in equation~\eqref{eq:L} or \eqref{eq:LL}, the action integral in equation~\eqref{eq:action} satisfies $S[\bar{\psi},\psi,A,H]=S[\bar{\psi},\psi,A^{(\theta)},H]$. Therefore, after inserting the Faddeev--Popov unity from equation~\eqref{eq:FP} under the functional integral of equation~\eqref{eq:pathintegralREV} and performing the shift $A_\nu\rightarrow A_\nu+\frac{\hbar}{e}\partial_\nu\theta$, $\mathcal{D}A\rightarrow\mathcal{D}A$, the functional integral in equation~\eqref{eq:pathintegralREV} becomes
\begin{align}
 &\int \delta[C_\mathrm{em}(A)]\det\!\Big(-\frac{\hbar}{e}\partial^2\Big)\nonumber\\
 &\times e^{iS[\bar{\psi},\psi,A,H]/(\hbar c)}\mathcal{D}\theta\mathcal{D}\bar{\psi}\mathcal{D}\psi\mathcal{D}A\mathcal{D}H.
 \label{eq:pathintegralREV2}
\end{align}
The functional delta function in equation~\eqref{eq:pathintegralREV2} indicates that only such field configurations, which satisfy the Feynman gauge condition in equation~\eqref{eq:Lorenz}, give nonzero contributions.

The remaining task is to express the functional delta function and the Faddeev--Popov determinant in equation~\eqref{eq:pathintegralREV2} as Lagrangian density functionals. The functional delta function in equation~\eqref{eq:pathintegralREV2} is rewritten as \cite{Peskin2018}
\begin{align}
 &\delta[C_\mathrm{em}(A)]\nonumber\\
 &=N_\mathrm{em}\int\exp\Big(\frac{-i}{2\mu_0\hbar c{\xi_\mathrm{e}}}\int w^2d^4x\Big)\delta[C_\mathrm{em}(A)-w]\mathcal{D}w\nonumber\\
 &=N_\mathrm{em}\exp\!\Big(\frac{-i}{2\mu_0\hbar c{\xi_\mathrm{e}}}\int[C_\mathrm{em}(A)]^2d^4x\Big)\nonumber\\
 &=N_\mathrm{em}\exp\!\Big(\frac{i}{\hbar c}\!\int\!\mathcal{L}_\mathrm{em,gf}d^4x\Big).
 \label{eq:deltaCem}
\end{align}
Here $\xi_\mathrm{e}$ is the electromagnetic gauge-fixing parameter, $N_\mathrm{em}$ is an unimportant normalization constant, and $w$ is an arbitrary scalar function. The gauge-fixing parameter $\xi_\mathrm{e}$ is generally any finite real-valued constant. In this picture, the choice of $\xi_\mathrm{e}=1$ corresponds to the Feynman gauge condition in equation~\eqref{eq:Lorenz} \cite{Peskin2018,Schwartz2014}. For a discussion of this correspondence, see section~4.1 of the supplementary material. The choice $\xi_\mathrm{e}=0$ is known as the Landau gauge \cite{Peskin2018} but there is varying terminology in the literature as this choice is in some works called the Lorenz gauge \cite{Schwartz2014}. In the second equality of equation~\eqref{eq:deltaCem}, we have used the functional delta function to integrate over $w$. The last equality of equation~\eqref{eq:deltaCem} defines the gauge-fixing Lagrangian density of QED, given by
\begin{align}
 \mathcal{L}_\mathrm{em,gf} &
 =-\frac{1}{2\mu_0\xi_\mathrm{e}}[C_\mathrm{em}(A)]^2=-\frac{1}{2\mu_0\xi_\mathrm{e}}(\partial_{\nu}A^{\nu})^2\nonumber\\
 &=-\frac{1}{\xi_\mathrm{e}}\bar{\Theta}\cev{\partial}_\rho\boldsymbol{\gamma}_\mathrm{B}^\rho\boldsymbol{\mathfrak{e}}_0\bar{\boldsymbol{\mathfrak{e}}}_0\boldsymbol{\gamma}_\mathrm{B}^\sigma\vec{\partial}_\sigma\Theta.
 \label{eq:QEDfixing}
\end{align}
The last form of equation~\eqref{eq:QEDfixing} is written using the eight-spinor notation. For a discussion on the relation between the gauge-fixing Lagrangian density of QED in equation~\eqref{eq:QEDfixing} and the definition of the photon propagator, see section~\ref{sec:photonpropagator}.

Next, we turn on to the Faddeev--Popov determinant part of the functional integral in equation~\eqref{eq:pathintegralREV2}. The Faddeev--Popov determinant is rewritten as \cite{Peskin2018}
\begin{align}
 &\det\!\Big(-\frac{\hbar}{e}\partial^2\Big)\nonumber\\
 &=\int\!\exp\Big(i\!\int\!\bar{c}_\mathrm{em}\partial^2 c_\mathrm{em}d^4x\Big)\mathcal{D}c_\mathrm{em}\mathcal{D}\bar{c}_\mathrm{em}\nonumber\\
 &=\int\!\exp\Big(\frac{i}{\hbar c}\int\!\mathcal{L}_\mathrm{em,ghost}d^4x\Big)\mathcal{D}c_\mathrm{em}\mathcal{D}\bar{c}_\mathrm{em}.
 \label{eq:FP2}
\end{align}
In the first equality of equation~\eqref{eq:FP2}, we have followed the conventional Faddeev--Popov method to represent the determinant as a functional integral over anticommuting fields $c_\mathrm{em}$ and $\bar{c}_\mathrm{em}$ belonging to the adjoint representation and called the Faddeev--Popov ghosts \cite{Faddeev1967,Peskin2018,Schwartz2014}. The coefficient $\hbar/e$ has been absorbed in the normalization of $c_\mathrm{em}$ and $\bar{c}_\mathrm{em}$. The last equality of equation~\eqref{eq:FP2} defines the Faddeev--Popov ghost Lagrangian density of QED, given by
\begin{equation}
 \mathcal{L}_\mathrm{em,ghost} =\hbar c\bar{c}_\mathrm{em}\partial^2 c_\mathrm{em}
 =-\hbar c\bar{c}_\mathrm{em8}\partial^2 c_\mathrm{em8}.
 \label{eq:Lemghost}
\end{equation}
In the last form of equation~\eqref{eq:Lemghost}, we have used $c_\mathrm{em8}=c_\mathrm{em}\boldsymbol{\mathfrak{e}}_0$ and $\bar{c}_\mathrm{em8}=\bar{c}_\mathrm{em}\bar{\boldsymbol{\mathfrak{e}}}_0$. Since $\mathcal{L}_\mathrm{em,ghost}$ in equation~\eqref{eq:Lemghost} does not contain the electromagnetic gauge field, $\mathcal{L}_\mathrm{em,ghost}$ is typically absorbed in the unimportant normalization constant of the functional integral of QED \cite{Peskin2018}. Consequently, in the Standard Model, the Faddeev--Popov ghosts are important only in the Yang--Mills gauge theories.

\subsubsection{Gravity gauge fixing}
Next, we discuss gravity gauge fixing following the analogy with the QED gauge fixing above. In the case of general relativity, the corresponding gauge fixing analog has been applied in previous literature \cite{Hooft1974,Donoghue1994b}. The gravity gauge fixing to be introduced below fundamentally differs from the Weitzenböck gauge fixing of TEGRW, discussed in section~\ref{sec:teleparallel}, and the gauge fixing of general relativity \cite{Hooft1974,DeWitt1967c,DeWitt1967b,DeWitt1967c,Donoghue1994a} since here we do not relate the gauge field to the definition of the tetrad or metric. Consequently, UGM treats the gravity gauge field on the same formal footing with the gauge fields of the Standard Model.

We fix the gravity gauge field by the gravitational analog of the Feynman gauge fixing of QED, discussed in section~\ref{sec:QEDfixing}. In general relativity, a particularly convenient gauge is the harmonic gauge, also known as the de Donder gauge \cite{Donoghue1994b,Donoghue2002,Balbus2016,Kibble1965,DeWitt1967c,Gupta1952,Thirring1961}. In unified gravity, the gauge field $H_{\mu\nu}$ is not assumed symmetric by definition, which leads to modification of the harmonic gauge condition. The harmonic gauge condition is generalized to unified gravity as
\begin{align}
 C_\mathrm{g}^\mu(H) &\equiv\partial_\rho H^{\mu\rho}+\partial_\rho H^{\rho\mu}-\partial^\mu H_{\;\,\rho}^\rho\nonumber\\
 &=2P^{\alpha\beta,\rho\mu}\partial_\rho H_{\alpha\beta}
 =0.
 \label{eq:harmonicgaugeREV}
\end{align}
Equation \eqref{eq:harmonicgaugeREV} defines the gauge-fixing four-vector $C_\mathrm{g}^\mu(H)$. The gauge condition in equation~\eqref{eq:harmonicgaugeREV} is satisfied in the gauge transformation of the form in equation~\eqref{eq:Hanutransformation} when the gravity gauge function $\phi_\mu$ is a solution of the wave equation $\partial^2\phi_\mu=0$. Thus, the harmonic gauge does not determine the gravitational gauge field $H_{\mu\nu}$ uniquely. The remaining residual degrees of freedom are analogous to the residual gauge symmetry of QED, discussed in section~\ref{sec:QEDfixing}.

In analogy with the case of QED in section~\ref{sec:QEDfixing}, we apply the Faddeev--Popov method to gravity gauge fixing. Therefore, we insert 1 under the non-gauge-fixed functional integral of equation~\eqref{eq:pathintegralREV} in the following form:
\begin{equation}
 1=\prod_\mu\int\delta[C_\mathrm{g}^{(\mu)}(H^{(\phi)})]\det\!\Big[\frac{\delta C_\mathrm{g}^{(\mu)}(H^{(\phi)})}{\delta\phi^{(\mu)}}\Big]\mathcal{D}\phi_{(\mu)}.
 \label{eq:FPg}
\end{equation}
Here the functional integrals are over the gravity gauge functions $\phi_\mu$, and $H^{(\phi)}$ is the gauge-transformed field, given by
\begin{equation}
 H_{\mu\nu}^{(\phi)}=H_{\mu\nu}+\frac{1}{g'_\mathrm{g}}\partial_\nu\phi_\mu.
\end{equation}
The term $\delta[C_\mathrm{g}^{(\mu)}(H^{(\phi)})]$ in equation~\eqref{eq:FPg} is the functional delta function and $\det[\delta C_\mathrm{g}^{(\mu)}(H^{(\phi)})/\delta\phi^{(\mu)}]=\det(\frac{1}{g'_\mathrm{g}}\partial^2)$ is the Faddeev--Popov determinant of unified gravity.

Due to the local gauge invariance of the Lagrangian density in equation~\eqref{eq:L} or \eqref{eq:LL}, the action integral in equation~\eqref{eq:action} satisfies $S[\bar{\psi},\psi,A,H]=S[\bar{\psi},\psi,A,H^{(\phi)}]$. Since $\mathcal{L}_\mathrm{em,gf}$ and $\mathcal{L}_\mathrm{em,ghost}$ in equations \eqref{eq:QEDfixing} and \eqref{eq:Lemghost} are locally gauge invariant with respect to the U(1) symmetries of unified gravity, the equality $S[\bar{\psi},\psi,A,H]=S[\bar{\psi},\psi,A,H^{(\phi)}]$ remains satisfied after the QED gauge fixing discussed in section~\ref{sec:QEDfixing} above. Therefore, after inserting the Faddeev--Popov unity from equation~\eqref{eq:FPg} under the functional integral of equation~\eqref{eq:pathintegralREV} and performing the shift $H_{\mu\nu}\rightarrow H_{\mu\nu}-\frac{1}{g'_\mathrm{g}}\partial_\nu\phi_\mu$, $\mathcal{D}H\rightarrow\mathcal{D}H$, the functional integral in equation~\eqref{eq:pathintegralREV} becomes
\begin{align}
 &\int\Big\{\prod_\mu\delta[C_\mathrm{g}^{\mu}(H)]\det\!\Big(-\frac{1}{g'_\mathrm{g}}\partial^2\Big)\Big\}\nonumber\\
 &\times e^{iS[\bar{\psi},\psi,A,H]/(\hbar c)}\mathcal{D}\phi\mathcal{D}\bar{\psi}\mathcal{D}\psi\mathcal{D}A\mathcal{D}H.
 \label{eq:pathintegralg}
\end{align}
Here we have defined the functional measure of the gravity gauge functions as $\mathcal{D}\phi=\prod_\mu\mathcal{D}\phi_\mu$.

Next, we express the functional delta functions and the Faddeev--Popov determinants in equation~\eqref{eq:pathintegralg} as Lagrangian densities. The functional delta functions in equation~\eqref{eq:pathintegralg} are rewritten as
\begin{align}
 \prod_\mu\delta[C_\mathrm{g}^\mu(H)]&=N_\mathrm{g}\prod_\mu\int\exp\Big(\frac{i}{4\kappa\hbar c\xi_\mathrm{g}}\int w^{(\mu)}w_{(\mu)}d^4x\Big)\nonumber\\
 &\hspace{0.4cm}\times\delta[C_\mathrm{g}^{(\mu)}(H)-w^{(\mu)}]\mathcal{D}w^{(\mu)}\nonumber\\
 &=N_\mathrm{g}\exp\!\Big(\frac{i}{4\kappa\hbar c\xi_\mathrm{g}}\int\!C_\mathrm{g}^\mu(H) C_{\mathrm{g}\mu}(H)d^4x\Big)\nonumber\\
 &=N_\mathrm{g}\exp\!\Big(\frac{i}{\hbar c}\!\int\!\mathcal{L}_\mathrm{g,gf}d^4x\Big).
 \label{eq:deltaCg}
\end{align}
Here $\xi_\mathrm{g}$ is the gravity gauge-fixing parameter, $N_\mathrm{g}$ is an unimportant normalization constant, and $w^\mu$ are arbitrary scalar functions. In analogy with the QED gauge fixing in section~\ref{sec:QEDfixing}, the gravity gauge-fixing parameter $\xi_\mathrm{g}$ is any finite real-valued constant. The choice of $\xi_\mathrm{g}=1$ corresponds to the harmonic gauge condition in equation~\eqref{eq:harmonicgaugeREV}. For a discussion of this correspondence, see section~4.2 of the supplementary material. In the second equality of equation~\eqref{eq:deltaCg}, we have used the functional delta function to integrate over $w^\mu$. The last equality of equation~\eqref{eq:deltaCg} defines the gauge-fixing Lagrangian density of unified gravity, given by
\begin{align}
 \mathcal{L}_\mathrm{g,gf}
 &=\frac{1}{4\kappa\xi_\mathrm{g}}C_\mathrm{g}^\mu(H) C_{\mathrm{g}\mu}(H)\nonumber\\
 &=\frac{1}{4\kappa\xi_\mathrm{g}}(\partial_\rho H^{\mu\rho}+\partial_\rho H^{\rho\mu}-\partial^\mu H_{\;\,\rho}^\rho)\nonumber\\
 &\hspace{0.4cm}\times(\partial_\eta H_\mu^{\;\,\eta}+\partial_\eta H_{\;\,\mu}^{\eta}-\partial_\mu H_{\;\,\eta}^\eta)\nonumber\\
 &=\frac{1}{\kappa\xi_\mathrm{g}}\eta_{\gamma\delta}P^{\alpha\beta,\lambda\gamma}P^{\rho\sigma,\eta\delta}\partial_\lambda H_{\alpha\beta}\partial_\eta H_{\rho\sigma}.
 \label{eq:LggfREV}
\end{align}
For a discussion on the relation between the gauge-fixing Lagrangian density of unified gravity in equation~\eqref{eq:LggfREV} and the definition of the graviton propagator, see section~\ref{sec:gravitonpropagator}.

Next, we turn on to the Faddeev--Popov determinant part of the functional integral in equation~\eqref{eq:pathintegralg}. In analogy with the case of QED in section~\ref{sec:QEDfixing} above, the Faddeev--Popov determinants are rewritten as
\begin{align}
 &\prod_\mu\det\Big(\frac{1}{g'_\mathrm{g}}\partial^2\Big)\nonumber\\
 &=\int\!\exp\Big(-i\!\int\!\bar{c}_\mathrm{g}\partial^2c_\mathrm{g}d^4x\Big)\mathcal{D}c_\mathrm{g}\mathcal{D}\bar{c}_\mathrm{g}\nonumber\\
 &=\int\!\exp\Big(\frac{i}{\hbar c}\int\!\mathcal{L}_\mathrm{g,ghost}d^4x\Big)\mathcal{D}c_\mathrm{g}\mathcal{D}\bar{c}_\mathrm{g}.
 \label{eq:FPg2}
\end{align}
In the first equality of equation~\eqref{eq:FPg2}, we have represented the determinant as a functional integral over anticommuting fields $c_\mathrm{g}=[c_\mathrm{g}^0,c_\mathrm{g}^x,c_\mathrm{g}^y,c_\mathrm{g}^z]^T$ and $\bar{c}_\mathrm{g}=[\bar{c}_\mathrm{g}^0,\bar{c}_\mathrm{g}^x,\bar{c}_\mathrm{g}^y,\bar{c}_\mathrm{g}^z]$ belonging to the adjoint representation and called the Faddeev--Popov ghost fields of unified gravity. We have also defined $\mathcal{D}c_\mathrm{g}=\prod_\mu\mathcal{D}c_\mathrm{g}^\mu$ and $\mathcal{D}\bar{c}_\mathrm{g}=\prod_\mu\mathcal{D}\bar{c}_\mathrm{g}^\mu$. The coefficient $1/g'_\mathrm{g}$ has been absorbed in the normalization of $c_\mathrm{g}$ and $\bar{c}_\mathrm{g}$. The last equality of equation~\eqref{eq:FPg2} defines the Faddeev--Popov ghost Lagrangian density of unified gravity, given by
\begin{equation}
 \mathcal{L}_\mathrm{g,ghost}=-\hbar c\bar{c}_\mathrm{g}\partial^2 c_\mathrm{g}=\hbar c\bar{c}_\mathrm{g8}\partial^2 c_\mathrm{g8}.
 \label{eq:Lgghost}
\end{equation}
In the last form of equation~\eqref{eq:Lgghost}, we have used $c_\mathrm{g8}=c_\mathrm{g}\boldsymbol{\mathfrak{e}}_0$ and $\bar{c}_\mathrm{g8}=\bar{c}_\mathrm{g}\bar{\boldsymbol{\mathfrak{e}}}_0$. Since $\mathcal{L}_\mathrm{g,ghost}$ in equation~\eqref{eq:Lgghost} does not contain the gravity gauge field, $\mathcal{L}_\mathrm{g,ghost}$ can be absorbed in the unimportant normalization constant of the functional integral in analogy with the case of QED \cite{Peskin2018}.

\subsection{\label{sec:BRST}Faddeev--Popov Lagrangian density of UGM and the BRST invariance}
Here we conclude the Faddeev--Popov gauge fixing approach of UGM by presenting the complete locally gauge-fixed Lagrangian density of UGM. This Lagrangian density is a sum of the locally gauge-invariant Lagrangian density in equation~\eqref{eq:L} or \eqref{eq:LL} and the gauge-fixing and ghost Lagrangian densities in equations \eqref{eq:QEDfixing}, \eqref{eq:Lemghost}, \eqref{eq:LggfREV}, and \eqref{eq:Lgghost}. Thus, the locally gauge-fixed Faddeev--Popov Lagrangian density of UGM is given by
\begin{equation}
 \mathcal{L}_\mathrm{FP}=\mathcal{L}+\mathcal{L}_\mathrm{em,gf}+\mathcal{L}_\mathrm{em,ghost}+\mathcal{L}_\mathrm{g,gf}+\mathcal{L}_\mathrm{g,ghost}.
 \label{eq:LFP}
\end{equation}

The Faddeev--Popov Lagrangian density of UGM in equation~\eqref{eq:LFP} is found to satisfy an exact global symmetry, which is known in the case of the gauge theories of the Standard Model as BRST invariance, named after Becchi, Rouet, Stara, and Tyutin \cite{Becchi1976,Iofa1976,Peskin2018,Schwartz2014,Fuster2005}. In the BRST method, the local electromagnetic U(1) gauge transformation parameter is defined to be proportional to the electromagnetic ghost field as $\theta=\theta' c_\mathrm{em}$, where $\theta'$ is a space-time-independent anticommuting Grassmann number satisfying $\theta'^2=0$. Correspondingly, the four U(1) transformation parameters of unified gravity are defined to be proportional to the four components of the gravitational ghost field as $\phi_\mu=\phi'c_\mathrm{g\mu}$, where $\phi'$ is a space-time-independent anticommuting Grassmann number satisfying $\phi'^2=0$. Then, the Faddeev--Popov Lagrangian density of UGM in equation~\eqref{eq:LFP} is invariant under the following transformations associated with electromagnetism \cite{Schwartz2014}:
\begin{align}
 \psi &\rightarrow e^{i\theta' c_\mathrm{em}Q}\psi,\label{eq:BRSTstart}\\
 A_\nu &\rightarrow A_\nu-\frac{\hbar}{e}\theta'\vec{\partial}_\nu c_\mathrm{em},\\
 \bar{c}_\mathrm{em} &\rightarrow\bar{c}_\mathrm{em}-\frac{1}{\mu_0ce\xi_\mathrm{e}}\theta'C_\mathrm{em}(A),\\
 c_\mathrm{em} &\rightarrow c_\mathrm{em},
\end{align}
and under the following transformations associated with gravity:
\begin{align}
 \mathbf{I}_\mathrm{g}&\rightarrow\Big(\bigotimes_{\mu}e^{i\phi'c_{\mathrm{g}(\mu)}\mathbf{t}^{(\mu)}}\Big)\mathbf{I}_\mathrm{g},\\
 X_\mu &\rightarrow X_\mu-\frac{1}{g_\mathrm{g}}\phi'c_{\mathrm{g}\mu},\\
 H_{\mu\nu} &\rightarrow H_{\mu\nu}+\frac{1}{g'_\mathrm{g}}\phi'\vec{\partial}_\nu c_{\mathrm{g}\mu},\\
 \bar{c}_\mathrm{g}^\mu &\rightarrow\bar{c}_\mathrm{g}^\mu-\frac{1}{\kappa\hbar cg'_\mathrm{g}\xi_\mathrm{g}}\phi'C_\mathrm{g}^\mu(H),\\
 c_\mathrm{g}^\mu &\rightarrow c_\mathrm{g}^\mu\label{eq:BRSTend}.
\end{align}
The BRST symmetry transformations in equations \eqref{eq:BRSTstart}--\eqref{eq:BRSTend} are global since the Grassmann numbers $\theta'$ and $\phi'$ are independent of the space-time coordinates.

In previous literature \cite{Peskin2018,Schwartz2014}, the BRST symmetry is understood to elucidate the introduction of the Faddeev--Popov ghosts. In Yang-Mills gauge theories, where ghost fields appear in virtual states, it also explains the exclusion of the ghosts from physical asymptotic states when performing quantum field theory calculations. The BRST symmetry of the path integral is known to be preserved at each loop order \cite{Schwartz2014}. Therefore, the BRST symmetry of unified gravity, observed above, strongly suggests that unified gravity is a renormalizable gauge theory like the gauge theories of the Standard Model. This conclusion is further supported by the successful renormalization of unified gravity at 1-loop order in section~\ref{sec:renormalization}. The BRST symmetry of unified gravity differs from symmetries of conventional theories of gravity, where the gauge transformation generators are not constant matrices, and thus, the BRST symmetry must be replaced by a more general Batalin--Vilkovisky formalism \cite{Batalin1981,Batalin1983,Weinberg1996,Costello2011,Henneaux1992,Gomis1995}.

\subsection{Dynamical equations of the ghost fields}

Here we investigate the dynamical equations of the ghost fields. The starting point is the Lagrangian density of the electromagnetic ghost field in equation~\eqref{eq:Lemghost} and that of the gravitational ghost field in equation~\eqref{eq:Lgghost}. The Euler--Lagrange equations for $c_\mathrm{em}$ and $\bar{c}_\mathrm{em}$ are given by
\begin{align}
 &\frac{\partial\mathcal{L}_\mathrm{em,ghost}}{\partial c_\mathrm{em}}-\partial_\rho\Big[\frac{\partial\mathcal{L}_\mathrm{em,ghost}}{\partial(\partial_\rho c_\mathrm{em})}\Big]+\partial_\rho\partial_\sigma\Big[\frac{\partial\mathcal{L}_\mathrm{em,ghost}}{\partial(\partial_\rho\partial_\sigma c_\mathrm{em})}\Big]=0,\nonumber\\
 &\frac{\partial\mathcal{L}_\mathrm{em,ghost}}{\partial \bar{c}_\mathrm{em}}-\partial_\rho\Big[\frac{\partial\mathcal{L}_\mathrm{em,ghost}}{\partial(\partial_\rho \bar{c}_\mathrm{em})}\Big]+\partial_\rho\partial_\sigma\Big[\frac{\partial\mathcal{L}_\mathrm{em,ghost}}{\partial(\partial_\rho\partial_\sigma \bar{c}_\mathrm{em})}\Big]=0.
 \label{eq:EulerLagrangeGravitycem}
\end{align}
Using the Lagrangian density in equation~\eqref{eq:Lemghost}, for the derivatives with respect to $c_\mathrm{em}$, we obtain $\partial\mathcal{L}_\mathrm{em,ghost}/\partial c_\mathrm{em}=0$, $\partial\mathcal{L}_\mathrm{em,ghost}/\partial(\partial_\rho c_\mathrm{em})=0$, and $\partial\mathcal{L}_\mathrm{em,ghost}/\partial(\partial_\rho\partial_\sigma c_\mathrm{em})=\hbar c\bar{c}_\mathrm{em}\eta^{\rho\sigma}$. For the derivatives with respect to $\bar{c}_\mathrm{em}$, we obtain $\partial\mathcal{L}_\mathrm{em,ghost}/\partial \bar{c}_\mathrm{em}=\hbar c\partial^2 c_\mathrm{em}$, $\partial\mathcal{L}_\mathrm{em,ghost}/\partial(\partial_\rho \bar{c}_\mathrm{em})=0$, $\partial\mathcal{L}_\mathrm{em,ghost}/\partial(\partial_\rho\partial_\sigma \bar{c}_\mathrm{em})=0$. Substituting these terms into equation~\eqref{eq:EulerLagrangeGravitycem} and dividing the equations by $\hbar c$, we obtain
\begin{equation}
 \partial^2\bar{c}_\mathrm{em}=0,
 \hspace{0.5cm}\partial^2 c_\mathrm{em}=0.
 \label{eq:cem_dynamics}
\end{equation}
Similar equations are obtained for the gravitational ghost fields, written as
\begin{equation}
 \partial^2\bar{c}_\mathrm{g}=0,
 \hspace{0.5cm}\partial^2 c_\mathrm{g}=0.
 \label{eq:cg_dynamics}
\end{equation}
Equations \eqref{eq:cem_dynamics} and \eqref{eq:cg_dynamics} are the dynamical equations of the electromagnetic and gravitational ghost fields in UGM. The solutions for $c_\mathrm{em}$, $\bar{c}_\mathrm{em}$, $c_\mathrm{g}$, and $\bar{c}_\mathrm{g}$ are arbitrary harmonic functions.

From now on, in UGM, we assume that the degrees of freedom of the electromagnetic and gravitational ghost fields have been integrated out of the path integral in equation~\eqref{eq:pathintegralREV}. Accordingly, we drop out the ghost field Lagrangian densities from the considerations below. In analogy with the conventional QED, we do not need to study the dynamics of the ghost fields to investigate the dynamics of the physical fields. Since the BRST invariance is known to be preserved at each loop order \cite{Peskin2018}, one does not either need to study the ghost fields when performing higher-order quantum field theory calculations in Abelian gauge theories, such as QED. This must extend to the theory of gravity in UGM.

\subsection{\label{sec:LUGM}Geometric Lagrangian density of UGM}
Next, we present the \emph{geometric} Lagrangian density of UGM. We call the Lagrangian density geometric after the tetrad and metric relations in equations \eqref{eq:tetradMinkowskimanifold} and \eqref{eq:metricMinkowskimanifold} have been applied. The geometric Lagrangian density of UGM is obtained from the locally gauge-fixed Faddeev--Popov Lagrangian density of UGM in equation~\eqref{eq:LFP} by dropping out the ghost field terms and by applying the Minkowski metric relation in equation~\eqref{eq:metricMinkowskimanifold}. The geometric Lagrangian density of UGM is then given by
\begin{align}
 \mathcal{L}_\mathrm{UGM} &=\mathcal{L}_\mathrm{D,kin}+\mathcal{L}_\mathrm{em,kin}+\mathcal{L}_\mathrm{em,int}+\mathcal{L}_\mathrm{g,kin}+\mathcal{L}_\mathrm{g,int}\nonumber\\
 &\hspace{0.4cm}+\mathcal{L}_\mathrm{em,gf}+\mathcal{L}_\mathrm{g,gf}.
 \label{eq:LUGM}
\end{align}
This Lagrangian density is the starting point for the derivation of the Feynman rules of UGM in section~\ref{sec:Feynman}. The terms of the geometric Lagrangian density of UGM in equation~\eqref{eq:LUGM} are given by
\begin{align}
 \mathcal{L}_\mathrm{D,kin} &=\frac{i\hbar c}{2}\bar{\psi}(\boldsymbol{\gamma}_\mathrm{F}^\nu\vec{\partial}_\nu-\cev{\partial}_\nu\boldsymbol{\gamma}_\mathrm{F}^\nu)\psi-m_\mathrm{e}c^2\bar{\psi}\psi,\nonumber\\
 \mathcal{L}_\mathrm{em,kin} &=-\frac{1}{4\mu_0}F_{\mu\nu}F^{\mu\nu}=\bar{\Psi}\Psi\nonumber\\
 &=\bar{\Theta}\cev{\partial}_\rho\boldsymbol{\gamma}_\mathrm{B}^\rho(\mathbf{I}_8+\boldsymbol{\mathfrak{e}}_0\bar{\boldsymbol{\mathfrak{e}}}_0)^2\boldsymbol{\gamma}_\mathrm{B}^\sigma\vec{\partial}_\sigma\Theta,\nonumber\\
 \mathcal{L}_\mathrm{em,int} &=-J_\mathrm{e}^\nu A_\nu=-q_\mathrm{e}c\bar{\psi}\boldsymbol{\gamma}_\mathrm{F}^\nu\psi A_\nu=\bar{\Phi}\Theta+\bar{\Theta}\Phi,\nonumber\\
 \mathcal{L}_\mathrm{g,kin} &=\frac{1}{4\kappa}H_{\rho\mu\nu}S^{\rho\mu\nu},\nonumber\\
 \mathcal{L}_\mathrm{g,int} &=-\frac{g'_\mathrm{g}}{g_\mathrm{g}}T_\mathrm{m}^{\mu\nu}H_{\mu\nu}\nonumber\\
 &=-\frac{g'_\mathrm{g}}{g_\mathrm{g}}\Big\{\frac{c}{2}P^{\mu\nu,\rho\sigma}[i\hbar\bar{\psi}(\boldsymbol{\gamma}_\mathrm{F\rho}\vec{\partial}_\sigma-\cev{\partial}_\rho\boldsymbol{\gamma}_\mathrm{F\sigma})\psi\nonumber\\
 &\hspace{0.4cm}-q_\mathrm{e}\bar{\psi}(\boldsymbol{\gamma}_\mathrm{F\rho}A_\sigma+A_\rho\boldsymbol{\gamma}_\mathrm{F\sigma})\psi
 -m'_\mathrm{e}c\eta_{\rho\sigma}\bar{\psi}\psi]\nonumber\\
 &\hspace{0.4cm}+\frac{1}{2\mu_0}P^{\mu\nu,\rho\sigma,\eta\lambda}\partial_\rho A_\sigma\partial_\eta A_\lambda\Big\} H_{\mu\nu},\nonumber\\
 \mathcal{L}_\mathrm{em,gf} &=-\frac{1}{2\mu_0\xi_\mathrm{e}}(\partial_{\nu}A^{\nu})^2
 =-\frac{1}{\xi_\mathrm{e}}\bar{\Theta}\cev{\partial}_\rho\boldsymbol{\gamma}_\mathrm{B}^\rho\boldsymbol{\mathfrak{e}}_0\bar{\boldsymbol{\mathfrak{e}}}_0\boldsymbol{\gamma}_\mathrm{B}^\sigma\vec{\partial}_\sigma\Theta,\nonumber\\
 \mathcal{L}_\mathrm{g,gf}
 &=\frac{1}{\kappa\xi_\mathrm{g}}\eta_{\gamma\delta}P^{\alpha\beta,\lambda\gamma}P^{\rho\sigma,\eta\delta}\partial_\lambda H_{\alpha\beta}\partial_\eta H_{\rho\sigma}.
 \label{eq:LUGMterms}
\end{align}
In equation~\eqref{eq:LUGMterms}, $J_\mathrm{e}^\nu$ is the electric four-current density, given in equation~\eqref{eq:Je0}, and $T_\mathrm{m}^{\mu\nu}$ is the symmetric SEM tensor of the Dirac and electromagnetic fields, given in equation~\eqref{eq:semtensors0}. Note that, in equation~\eqref{eq:LUGMterms}, the SEM tensor $T_\mathrm{m}^{\mu\nu}$ is the source of gravity in analogy to how the electromagnetic four-current density $J_\mathrm{e}^\nu$ is the source of electromagnetism in QED. Correspondingly, $-(g'_\mathrm{g}/g_\mathrm{g})T_\mathrm{m}^{\mu\nu}H_{\mu\nu}$ of equation~\eqref{eq:LUGMterms} is the interaction term of gravity analogous to the electromagnetic interaction term $-J_\mathrm{e}^\nu A_\nu$ of QED. With a different prefactor, the interaction term of gravity of the form $-(g'_\mathrm{g}/g_\mathrm{g})T_\mathrm{m}^{\mu\nu}H_{\mu\nu}$ can also be found from previous literature \cite{Berends1976}. In the sections below, we use the kinetic and gauge-fixing Lagragian density terms of the electromagnetic and gravitational fields in equation~\eqref{eq:LUGMterms} to define the gauge-fixed kinetic Lagrangian densities of the electromagnetic and gravitational fields.

\subsubsection{Gauge-fixed kinetic Lagrangian density of the electromagnetic field}

We find that the Feynman gauge fixing in the eight-spinor formulation of UGM with $\xi_\mathrm{e}=1$ corresponds to dropping out the term $\boldsymbol{\mathfrak{e}}_0\bar{\boldsymbol{\mathfrak{e}}}^a\partial_a\Theta$ of the definition of the electromagnetic spinor in equation~\eqref{eq:electromagneticspinor}. When this dropping is performed, $\Psi$ is replaced by $\Psi'$, given by
\begin{equation}
 \Psi'=-\boldsymbol{\gamma}_\mathrm{B}^\rho\partial_\rho\Theta.
 \label{eq:dropping}
\end{equation}
Consequently, instead of the electromagnetic Lagrangian density $\mathcal{L}_\mathrm{em,kin}=\bar{\Psi}\Psi$ in equation~\eqref{eq:LUGMterms}, we obtain $\mathcal{L}_\mathrm{em,kin}^{{\prime(\xi_\mathrm{e}=1)}}$, given by
\begin{align}
 \mathcal{L}_\mathrm{em,kin}^{{\prime(\xi_\mathrm{e}=1)}} &=\bar{\Psi}'\Psi'=\mathcal{L}_\mathrm{em,kin}+\mathcal{L}_\mathrm{em,gf}^{{(\xi_\mathrm{e}=1)}}\nonumber\\
 &=-\frac{1}{4\mu_0}F_{\mu\nu}F^{\mu\nu}-\frac{1}{2\mu_0}(\partial_{\nu}A^{\nu})^2.
 \label{eq:Lem_fixed}
\end{align}
Equation equation~\eqref{eq:Lem_fixed} is the standard form of the gauge-fixed electromagnetic Lagrangian density of QED in the Feynman gauge \cite{Peskin2018}.
The second equality of equation~\eqref{eq:Lem_fixed} shows that using $\Psi'$ provides a simple expression for the sum of the non-gauge-fixed electromagnetic Lagrangian density $\mathcal{L}_\mathrm{em,kin}$ and the gauge-fixing Lagrangian density $\mathcal{L}_\mathrm{em,gf}^{(\xi_\mathrm{e}=1)}$, given by equation~\eqref{eq:LUGMterms} for $\xi_\mathrm{e}=1$.

The generalization of $\mathcal{L}_\mathrm{em,kin}^{{\prime(\xi_\mathrm{e}=1)}}$ in equation~\eqref{eq:Lem_fixed} is given for arbitrary $\xi_\mathrm{e}$ by
\begin{align}
 \mathcal{L}'_\mathrm{em,kin}
 &=\mathcal{L}_\mathrm{em,kin}+\mathcal{L}_\mathrm{em,gf}\nonumber\\
 &=-\frac{1}{4\mu_0}F_{\mu\nu}F^{\mu\nu}-\frac{1}{2\mu_0\xi_\mathrm{e}}(\partial_{\nu}A^{\nu})^2.
 \label{eq:Lem_general}
\end{align}
The Lagrangian density $\mathcal{L}_\mathrm{em,kin}^{{\prime(\xi_\mathrm{e}=1)}}$ in equation~\eqref{eq:Lem_fixed} and its generalization $\mathcal{L}'_\mathrm{em,kin}$ in equation~\eqref{eq:Lem_general} enable the determination of the photon propagator as discussed in section~\ref{sec:photonpropagator} below and in section~4 of the supplementary material.

\subsubsection{Gauge-fixed kinetic Lagrangian density of the gravitational field}

The gauge-fixed Lagrangian density of the gravity gauge field in UGM in the harmonic gauge with $\xi_\mathrm{g}=1$ is given by the sum of the non-gauge-fixed gravity Lagrangian density $\mathcal{L}_\mathrm{g,kin}$ and the harmonic gauge-fixing Lagrangian density $\mathcal{L}_\mathrm{g,gf}^{(\xi_\mathrm{g}=1)}$, given by equation~\eqref{eq:LUGMterms} for $\xi_\mathrm{g}=1$, as
\begin{align}
 &\mathcal{L}_\mathrm{g,kin}^{{\prime(\xi_\mathrm{g}=1)}}
 =\mathcal{L}_\mathrm{g,kin}+\mathcal{L}_\mathrm{g,gf}^{(\xi_\mathrm{g}=1)}\nonumber\\
 &=\frac{1}{4\kappa}H_{\rho\mu\nu}S^{\rho\mu\nu}
 +\frac{1}{\kappa}\eta_{\gamma\delta}P^{\alpha\beta,\lambda\gamma}P^{\rho\sigma,\eta\delta}\partial_\lambda H_{\alpha\beta}\partial_\eta H_{\rho\sigma}.
 \label{eq:Lg_fixed}
\end{align}
This equation is analogous to the corresponding equation of the electromagnetic field in UGM in equation~\eqref{eq:Lem_fixed}.

The generalization of $\mathcal{L}_\mathrm{g,kin}^{\prime(\xi_\mathrm{g}=1)}$ in equation~\eqref{eq:Lg_fixed} is given for arbitrary $\xi_\mathrm{g}$ by
\begin{align}
 &\mathcal{L}'_\mathrm{g,kin}
 =\mathcal{L}_\mathrm{g,kin}+\mathcal{L}_\mathrm{g,gf}\nonumber\\
 &=\frac{1}{4\kappa}H_{\rho\mu\nu}S^{\rho\mu\nu}
 +\frac{1}{\kappa\xi_\mathrm{g}}\eta_{\gamma\delta}P^{\alpha\beta,\lambda\gamma}P^{\rho\sigma,\eta\delta}\partial_\lambda H_{\alpha\beta}\partial_\eta H_{\rho\sigma}.
 \label{eq:Lg_general}
\end{align}
The Lagrangian density $\mathcal{L}_\mathrm{g,kin}^{\prime(\xi_\mathrm{g}=1)}$ in equation~\eqref{eq:Lg_fixed} and its generalization $\mathcal{L}'_\mathrm{g,kin}$ in equation~\eqref{eq:Lg_general} enable the determination of the graviton propagator as discussed in section~\ref{sec:gravitonpropagator} below and in section~4 of the supplementary material.

\subsection{\label{sec:UGM_dynamics}Dynamical equations of UGM}

The dynamics of all fields in the Lagrangian density of equation~\eqref{eq:LUGM} are described by the Euler--Lagrange equations. In the sections below, we derive the dynamical equations for the gravity gauge field, the electromagnetic gauge field, and the Dirac field in UGM.

\subsubsection{\label{sec:UGMgravity}Field equation of gravity in UGM}

Here we derive the dynamical equation of gravity in UGM. The starting point is the Lagrangian density of UGM in equation~\eqref{eq:LUGM} in the harmonic gauge with $\xi_\mathrm{g}=1$. The Euler--Lagrange equation for $H_{\mu\nu}$ in UGM is given by
\begin{equation}
 \frac{\partial\mathcal{L}_\mathrm{UGM}}{\partial H_{\mu\nu}}-\partial_\rho\Big[\frac{\partial\mathcal{L}_\mathrm{UGM}}{\partial(\partial_\rho H_{\mu\nu})}\Big]=0.
 \label{eq:EulerLagrangeGravityUGM}
\end{equation}
Using the Lagrangian density in equation~\eqref{eq:LUGM} with $g'_\mathrm{g}=g_\mathrm{g}$ according to the equivalence principle in equation~\eqref{eq:equivalenceprinciple}, we obtain $\partial\mathcal{L}_\mathrm{UGM}/\partial H_{\mu\nu}=\partial\mathcal{L}_\mathrm{g,int}/\partial H_{\mu\nu}=T_\mathrm{m}^{\mu\nu}$ and $\partial\mathcal{L}_\mathrm{UGM}/\partial(\partial_\rho H_{\mu\nu})=-\frac{1}{\kappa}S^{\mu\nu\rho}+\frac{1}{\kappa}P^{\mu\nu,\rho\sigma}(\partial_\lambda H_\sigma^{\;\,\lambda}+\partial_\lambda H_{\;\,\sigma}^\lambda-\partial_\sigma H_{\;\,\lambda}^\lambda)$. Substituting these terms into equation~\eqref{eq:EulerLagrangeGravityUGM}, multiplying the equation by $\kappa$, and rearranging the terms, we obtain
\begin{equation}
 -P^{\mu\nu,\rho\sigma}\partial^2 H_{\rho\sigma}
 =\kappa T_\mathrm{m}^{\mu\nu}.
 \label{eq:UGMgravity}
\end{equation}
This equation is the dynamical equation of the gravitational field in UGM. The source term of the gravitational field is the SEM tensor of the Dirac and electromagnetic fields in equation~\eqref{eq:semtensors0}.

\subsubsection{Maxwell's equations in UGM}
Here we derive the dynamical equation of the electromagnetic field in UGM in the eight-spinor notation. As discussed in \cite{Partanen2024a}, the electromagnetic potential spinors $\Theta$ and $\bar{\Theta}$ can be treated as independent dynamical variables. The Euler--Lagrange equation for $\bar{\Theta}$ is given by~\cite{Partanen2024a}
\begin{equation}
 \frac{\partial\mathcal{L}_\mathrm{UGM}}{\partial\bar{\Theta}}-\partial_\rho\Big[\frac{\partial\mathcal{L}_\mathrm{UGM}}{\partial(\partial_\rho\bar{\Theta})}\Big]=0.
 \label{eq:EulerLagrangeThetaUGM}
\end{equation}
We use the Lagrangian density of UGM in equation~\eqref{eq:LUGM}, set $g'_\mathrm{g}=g_\mathrm{g}$ according to the equivalence principle in equation~\eqref{eq:equivalenceprinciple}, and assume the Feynman gauge with $\xi_\mathrm{e}=1$ to obtain $\partial\mathcal{L}_\mathrm{UGM}/\partial\bar{\Theta}=\Phi+\frac{1}{2}H_{\mu\nu}\boldsymbol{\gamma}_\mathrm{B}^5\boldsymbol{\gamma}_\mathrm{B}^\nu\mathbf{t}^\mu\Phi$ and $\partial\mathcal{L}_\mathrm{UGM}/\partial(\partial_\rho\bar{\Theta})=-\boldsymbol{\gamma}_\mathrm{B}^\rho(\mathbf{I}_8+\boldsymbol{\mathfrak{e}}_0\bar{\boldsymbol{\mathfrak{e}}}_0)(\Psi-H_{\mu\nu}\mathbf{t}^\mu\boldsymbol{\gamma}_\mathrm{B}^\nu\boldsymbol{\gamma}_\mathrm{B}^5\Psi)-\boldsymbol{\gamma}_\mathrm{B}^\rho\boldsymbol{\mathfrak{e}}_0\bar{\boldsymbol{\mathfrak{e}}}_0\boldsymbol{\gamma}_\mathrm{B}^\sigma\partial_\sigma\Theta$. Substituting these derivatives into equation~\eqref{eq:EulerLagrangeThetaUGM}, using $\bar{\boldsymbol{\mathfrak{e}}}_0\Psi=0$, equation~\eqref{eq:electromagneticspinor}, and the identity $\boldsymbol{\gamma}_\mathrm{B}^\rho\boldsymbol{\gamma}_\mathrm{B}^\sigma\partial_\rho\partial_\sigma=\mathbf{I}_8\partial^2$, and rearranging the terms, we obtain
\begin{align}
 \partial^2\Theta
 &=\Phi+\boldsymbol{\gamma}_\mathrm{B}^\rho(\mathbf{I}_8+\boldsymbol{\mathfrak{e}}_0\bar{\boldsymbol{\mathfrak{e}}}_0)\mathbf{t}^\mu\boldsymbol{\gamma}_\mathrm{B}^\nu\boldsymbol{\gamma}_\mathrm{B}^5(\mathbf{I}_8+\boldsymbol{\mathfrak{e}}_0\bar{\boldsymbol{\mathfrak{e}}}_0)\nonumber\\
 &\hspace{0.4cm}
 \times\partial_\rho(H_{\mu\nu}\boldsymbol{\gamma}_\mathrm{B}^\sigma\partial_\sigma\Theta)
 +\frac{1}{2}\boldsymbol{\gamma}_\mathrm{B}^5\boldsymbol{\gamma}_\mathrm{B}^\nu\mathbf{t}^\mu\Phi H_{\mu\nu}.
 \label{eq:MaxwellUGM1}
\end{align}
This equation is the eight-spinor representation of the wave equation of the electromagnetic four-potential in UGM in the presence of sources. As a consistency check, if we had started from the Euler--Lagrange equation for $\Theta$, we would have ended up to a dynamical equation that is equivalent to equation~\eqref{eq:MaxwellUGM1}.

Next, we derive the dynamical equations of the electromagnetic field in UGM using the electromagnetic four-potential as the dynamical variable. These equations are equivalent to the eight-spinor equations above. The Euler--Lagrange equation for the electromagnetic four-potential is given by \cite{Landau1989,Partanen2019b}
\begin{equation}
 \frac{\partial\mathcal{L}_\mathrm{UGM}}{\partial A_\sigma}-\partial_\rho\Big[\frac{\partial\mathcal{L}_\mathrm{UGM}}{\partial(\partial_\rho A_\sigma)}\Big]=0.
 \label{eq:EulerLagrangeUGMA}
\end{equation}
Using the Lagrangian density of UGM in equation~\eqref{eq:LUGM}, setting $g'_\mathrm{g}=g_\mathrm{g}$ according to the equivalence principle in equation~\eqref{eq:equivalenceprinciple}, and assuming the Feynman gauge with $\xi_\mathrm{e}=1$, we obtain $\partial\mathcal{L}_\mathrm{UGM}/\partial A_{\sigma}=-q_\mathrm{e}c\bar{\psi}\boldsymbol{\gamma}_\mathrm{F}^\sigma\psi+q_\mathrm{e}cP^{\mu\nu,\rho\sigma}\bar{\psi}\boldsymbol{\gamma}_\mathrm{F\rho}\psi H_{\mu\nu}=-J_\mathrm{e}^\sigma+P^{\mu\nu,\rho\sigma}J_\mathrm{e\rho}H_{\mu\nu}$ and $\partial\mathcal{L}_\mathrm{UGM}/\partial(\partial_\rho A_{\sigma})=-\frac{1}{\mu_0}F^{\rho\sigma}-\frac{1}{\mu_0}\eta^{\rho\sigma}\partial_\lambda A^\lambda-\frac{1}{\mu_0}P^{\mu\nu,\rho\sigma,\eta\lambda}H_{\mu\nu}\partial_\eta A_\lambda$. Substituting these terms into equation~\eqref{eq:EulerLagrangeUGMA}, multiplying the equation by $\mu_0$, and rearranging the terms, we obtain
\begin{align}
 \partial^2A^\sigma &=\mu_0J_\mathrm{e}^\sigma
 -P^{\mu\nu,\rho\sigma,\eta\lambda}\partial_\rho(H_{\mu\nu}\partial_\eta A_\lambda)\nonumber\\
 &\hspace{0.4cm}-\mu_0P^{\mu\nu,\rho\sigma}J_\mathrm{e\rho}H_{\mu\nu}.
 \label{eq:MaxwellUGM2}
\end{align}
This equation is the wave equation of the electromagnetic four-potential in UGM in the presence of sources. To the best knowledge of the authors, equation~\eqref{eq:MaxwellUGM2} has not been presented in previous literature. The same naturally applies to the eight-spinor form in equation~\eqref{eq:MaxwellUGM1}.

The right-hand sides of equations \eqref{eq:MaxwellUGM1} and \eqref{eq:MaxwellUGM2} represent the sources of the electromagnetic field. The first terms on the right in equations \eqref{eq:MaxwellUGM1} and \eqref{eq:MaxwellUGM2} are associated with the well-known electron--photon vertex, the second terms on the right are associated with the photon--graviton vertex, and the third terms on the right are associated with the electron--photon--graviton vertex. The interaction vertices are discussed in more detail in section~\ref{sec:Feynman}.

\subsubsection{Dirac equation in UGM}
Next, we derive the dynamical equation of the Dirac field in UGM. As conventional, we treat $\psi$ and $\bar{\psi}$ as independent dynamical variables. The Euler--Lagrange equation for $\bar{\psi}$ is given by \cite{Peskin2018}
\begin{equation}
 \frac{\partial\mathcal{L}_\mathrm{UGM}}{\partial\bar{\psi}}-\partial_\rho\Big[\frac{\partial\mathcal{L}_\mathrm{UGM}}{\partial(\partial_\rho\bar{\psi})}\Big]=0.
 \label{eq:EulerLagrangeDiracUGM}
\end{equation}
Using the Lagrangian density of UGM in equation~\eqref{eq:LUGM} and setting $g'_\mathrm{g}=g_\mathrm{g}$ and $m'_\mathrm{e}=m_\mathrm{e}$ according to the equivalence principle in equation~\eqref{eq:equivalenceprinciple}, we obtain $\partial\mathcal{L}_\mathrm{UGM}/\partial\bar{\psi}=\frac{i}{2}\hbar c\boldsymbol{\gamma}_\mathrm{F}^\rho\vec{\partial}_\rho\psi-q_\mathrm{e}c\boldsymbol{\gamma}_\mathrm{F}^\rho A_\rho-m_\mathrm{e}c^2\psi-H_{\mu\nu}[\frac{i\hbar c}{4}(\eta^{\nu\rho}\boldsymbol{\gamma}_\mathrm{F}^\mu+\eta^{\mu\rho}\boldsymbol{\gamma}_\mathrm{F}^\nu-\eta^{\mu\nu}\boldsymbol{\gamma}_\mathrm{F}^\rho)\vec{\partial}_\rho\psi+\frac{m_\mathrm{e}c^2}{2}\eta^{\mu\nu}\psi-\frac{q_\mathrm{e}c}{2}(\eta^{\nu\rho}\boldsymbol{\gamma}_\mathrm{F}^\mu
+\eta^{\mu\rho}\boldsymbol{\gamma}_\mathrm{F}^\nu-\eta^{\mu\nu}\boldsymbol{\gamma}_\mathrm{F}^\rho)A_\rho\psi]$ and $\partial\mathcal{L}_\mathrm{UGM}/\partial(\partial_\rho\bar{\psi})=-\frac{i}{2}\hbar c\boldsymbol{\gamma}_\mathrm{F}^\rho\psi+\frac{i\hbar c}{4}H_{\mu\nu}(\eta^{\nu\rho}\boldsymbol{\gamma}_\mathrm{F}^\mu+\eta^{\mu\rho}\boldsymbol{\gamma}_\mathrm{F}^\nu-\eta^{\mu\nu}\boldsymbol{\gamma}_\mathrm{F}^\rho)\psi$. Substituting these derivatives into equation~\eqref{eq:EulerLagrangeDiracUGM}, we obtain
\begin{align}
 &i\hbar c\boldsymbol{\gamma}_\mathrm{F}^\rho\vec{\partial}_\rho\psi\!-\!m_\mathrm{e}c^2\psi
 \!=q_\mathrm{e}c\boldsymbol{\gamma}_\mathrm{F}^\rho\psi A_\rho
 +P^{\mu\nu,\rho\sigma}\Big(i\hbar c\boldsymbol{\gamma}_\mathrm{F\sigma}\vec{\partial}_\rho\psi\nonumber\\
 &-\frac{m_\mathrm{e}c^2}{2}\eta_{\rho\sigma}\psi+\frac{i\hbar c}{2}\boldsymbol{\gamma}_\mathrm{F\sigma}\psi\vec{\partial}_\rho
 -q_\mathrm{e}c\boldsymbol{\gamma}_\mathrm{F\sigma}\psi A_\rho\Big)H_{\mu\nu}.
 \label{eq:DiracUGM}
\end{align}
Here we have used the definition of $P^{\mu\nu,\rho\sigma}$ in equation~\eqref{eq:P} and the identity $\eta^{\mu\nu}=-P^{\mu\nu,\rho\sigma}\eta_{\rho\sigma}$. Equation \eqref{eq:DiracUGM} is the Dirac equation in UGM. To the best knowledge of the authors, this form of the Dirac equation has not been presented in previous literature. As a consistency check, if we had started from the Euler--Lagrange equation for $\psi$, we would have ended up to a dynamical equation that is equivalent to equation~\eqref{eq:DiracUGM}.

The left-hand side of equation~\eqref{eq:DiracUGM} is equivalent to all terms of the Dirac equation in free space. The right-hand side of equation~\eqref{eq:DiracUGM} represents the source terms. The first term on the right in equation~\eqref{eq:DiracUGM} corresponds to the electron--photon vertex, and the second term on the right in equation~\eqref{eq:DiracUGM} corresponds to the electron--graviton and electron--photon--graviton vertices, which are discussed in section~\ref{sec:Feynman}.

\section{\label{sec:Feynman}Feynman rules and their application in UGM}

Above, we have unambiguously fixed the electromagnetic and gravitational gauges. Therefore, the resulting form of unified gravity allows us to perform quantum field theory calculations using the Feynman diagrams. In this section, we formulate the Feynman rules for UGM. Instead of eight-spinors, we use the conventional Dirac spinors and the electromagnetic four-potential as dynamical variables to enable direct comparison with the Feynman rules of QED and gravity in previous literature \cite{Schwartz2014,Peskin2018,Donoghue1994b,Lawrence1971,Butt2006,Ulhoa2017,Olyaei2018,Ramos2010}. For completeness, we also briefly review the derivations of the electron and photon propagators and the electron--photon vertex \cite{Peskin2018,Schwartz2014}. These derivations are presented to highlight the complete analogy with the derivations of the graviton propagator and the vertices associated with gravity, which are given below. The Feynman diagrams and rules for unified gravity in the SI units are summarized in table~\ref{tbl:Feynman}.

\begin{table*}
\setlength{\tabcolsep}{7.5pt}
\renewcommand{\arraystretch}{2.0}
\caption{\label{tbl:Feynman}
Feynman diagrams and the corresponding momentum-space rules for unified gravity in the SI units. The photon and graviton propagators are given in gauges defined by arbitrary gauge-fixing parameters $\xi_\mathrm{e}$ and $\xi_\mathrm{g}$. Each propagator and vertex has a corresponding counterterm presented below the pertinent propagator or vertex. The red crosses at the end of external lines of vertices and counterterms indicate that these lines are not included in the expressions. In the application of the Feynman rules, the inertial and gravitational masses are set equal according to Einstein's equivalence principle as $m'_\mathrm{e}=m_\mathrm{e}$, and we furthermore set the coupling constant of unified gravity equal to the scale constant as $g'_\mathrm{g}=g_\mathrm{g}$.}
\begin{tabular}{cccccccccccc}
   \hline\hline
   \multicolumn{4}{c}{Electron propagator} & \multicolumn{4}{c}{\hspace{1cm}Photon propagator} & \multicolumn{4}{c}{Graviton propagator} \\[0pt]
\multicolumn{4}{c}{$\adjustbox{valign=b, raise=-0.09cm}{\includegraphics{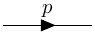}}
=\dfrac{i(\slashed{p}+m_\mathrm{e}c\mathbf{I}_4)}{p^2-m_\mathrm{e}^2c^2+i\epsilon}$} &
\multicolumn{4}{c}{\hspace{1cm}$\begin{array}{l}\adjustbox{valign=b, raise=-0.1cm}{\includegraphics{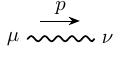}}=\dfrac{-i}{p^2+i\epsilon}\\
\hspace{0.2cm}\times\Big[\eta_{\mu\nu}\!-\!(1\!-\!\xi_\mathrm{e})\dfrac{p_\mu p_\nu}{p^2+i\epsilon}\Big]
\end{array}$} &
\multicolumn{4}{c}{$\begin{array}{l}\adjustbox{valign=b, raise=-0.05cm}{\includegraphics{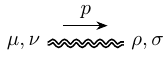}}=\dfrac{i}{p^2\!+\!i\epsilon}\\
\times\Big[P_{\mu\nu,\rho\sigma}^{(D)}\!-\!\dfrac{1-\xi_\mathrm{g}}{p^2+i\epsilon}(\eta_{\alpha\rho}p_\beta p_\sigma\\
+\eta_{\alpha\sigma}p_\beta p_\rho+\eta_{\beta\rho}p_\alpha p_\sigma+\eta_{\beta\sigma}p_\alpha p_\rho)\Big]\hspace{-0.5cm}
\end{array}$}\\[7pt]
\multicolumn{4}{c}{$\begin{array}{l}\adjustbox{valign=c}{\includegraphics{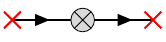}}\\
=i[(Z_\psi-1)\slashed{p}-(Z_\psi Z_\mathrm{m}-1)m_\mathrm{e}c\mathbf{I}_4]
\end{array}\hspace{-0.8cm}$} &
\multicolumn{4}{c}{\hspace{1cm}$\begin{array}{l}\adjustbox{valign=c}{\includegraphics{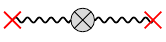}}\\
=-i(Z_\mathrm{A}-1)(p^2\eta^{\mu\nu}-p^\mu p^\nu)
\end{array}$} &
\multicolumn{4}{c}{$\begin{array}{l}\adjustbox{valign=c}{\includegraphics{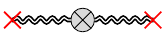}}\\
=i(Z_H-1)p^2\hat{P}_{1,2,1}^{\alpha\beta\eta\lambda}
\end{array}$}\\[12pt]
   \hline
   \multicolumn{6}{c}{Electron--photon vertex} & \multicolumn{6}{c}{Electron--graviton vertex}\\[0pt]
\multicolumn{6}{c}{$\adjustbox{valign=c}{\includegraphics{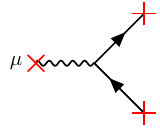}}
=\dfrac{-iq_\mathrm{e}\boldsymbol{\gamma}_\mathrm{F}^\mu}{\sqrt{\varepsilon_0\hbar c}}$} &
\multicolumn{6}{c}{$\adjustbox{valign=c}{\includegraphics{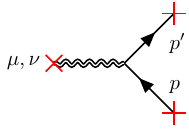}}
 =\!\begin{array}{l}
 \\
 -\dfrac{i}{2}\sqrt{\dfrac{\kappa c}{\hbar}}\dfrac{g'_\mathrm{g}}{g_\mathrm{g}}P^{\mu\nu,\rho\sigma}\\
\times[\boldsymbol{\gamma}_{\mathrm{F}\rho}(p+p')_\sigma-m'_\mathrm{e}c\eta_{\rho\sigma}\mathbf{I}_4]
 \end{array}$}\\[22pt]
\multicolumn{6}{c}{$\adjustbox{valign=c}{\includegraphics{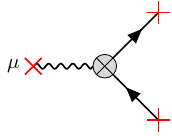}}
=(Z_\psi-1)\dfrac{-iq_\mathrm{e}\boldsymbol{\gamma}_\mathrm{F}^\mu}{\sqrt{\varepsilon_0\hbar c}}$} &
\multicolumn{6}{c}{$\hspace{-0.5cm}\adjustbox{valign=c}{\includegraphics{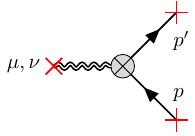}}
 =\!\!\begin{array}{l}
 \\
 -\dfrac{i}{2}\sqrt{\dfrac{\kappa c}{\hbar}}\dfrac{g'_\mathrm{g}}{g_\mathrm{g}}P^{\mu\nu,\rho\sigma}[(Z_\mathrm{g\psi}-1)\boldsymbol{\gamma}_{\mathrm{F}\rho}(p+p')_\sigma\hspace{-0.3cm}\nonumber\\
 -(Z_\mathrm{g\psi} Z_{\mathrm{g}m}-1)m'_\mathrm{e}c\eta_{\rho\sigma}\mathbf{I}_4]
 \end{array}$}\\[27pt]
   \hline
   \multicolumn{6}{c}{Photon--graviton vertex} & \multicolumn{6}{c}{Electron--photon--graviton vertex}\\[0pt]
\multicolumn{6}{c}{$\adjustbox{valign=c}{\includegraphics{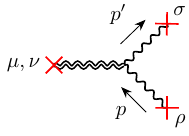}}
\!\! =-i\sqrt{\dfrac{\kappa c}{\hbar}}\dfrac{g'_\mathrm{g}}{g_\mathrm{g}}P^{\mu\nu,\sigma\eta,\rho\lambda}p'_\eta p_\lambda$} &
\multicolumn{6}{c}{$\adjustbox{valign=c}{\includegraphics{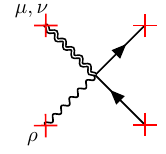}}
\!\! =\dfrac{iq_\mathrm{e}}{\hbar}\sqrt{\dfrac{\kappa}{\varepsilon_0}}\dfrac{g'_\mathrm{g}}{g_\mathrm{g}}P^{\mu\nu,\rho\sigma}\boldsymbol{\gamma}_{\mathrm{F}\sigma}$}\\[22pt]
\multicolumn{6}{c}{$\hspace{-0.3cm}\adjustbox{valign=c, raise=-0.04cm}{\includegraphics{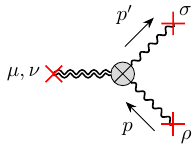}}
\!\! =-i(Z_{\mathrm{g}A}-1)\sqrt{\dfrac{\kappa c}{\hbar}}\dfrac{g'_\mathrm{g}}{g_\mathrm{g}}P^{\mu\nu,\sigma\eta,\rho\lambda}p'_\eta p_\lambda$} &
\multicolumn{6}{c}{$\adjustbox{valign=c}{\includegraphics{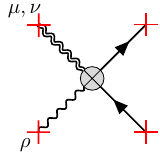}}
\!\! =(Z_{\mathrm{g}\psi}-1)\dfrac{iq_\mathrm{e}}{\hbar}\sqrt{\dfrac{\kappa}{\varepsilon_0}}\dfrac{g'_\mathrm{g}}{g_\mathrm{g}}P^{\mu\nu,\rho\sigma}\boldsymbol{\gamma}_{\mathrm{F}\sigma}$}\\[22pt]
   \hline
   \multicolumn{3}{c}{External electron lines} & \multicolumn{3}{c}{External positron lines} & \multicolumn{3}{c}{External photon lines} & \multicolumn{3}{c}{External graviton lines}\\[0pt]
\multicolumn{3}{c}{$\adjustbox{valign=c}{\includegraphics{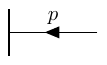}}
=u^s(p)$} &
\multicolumn{3}{c}{$\adjustbox{valign=c, raise=0.15cm}{\includegraphics{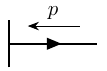}}
=\bar{v}^s(p)$} &
\multicolumn{3}{c}{$\adjustbox{valign=c, raise=0.15cm}{\includegraphics{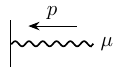}}
\!\!\!=\epsilon_\mu(p)$} &
\multicolumn{3}{c}{$\adjustbox{valign=c, raise=0.15cm}{\includegraphics{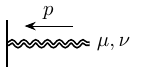}}
\!\!=\epsilon_{\mu\nu}(p)$}\\[7pt]
\multicolumn{3}{c}{$\adjustbox{valign=c}{\includegraphics{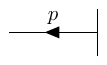}}
=\bar{u}^s(p)$} &
\multicolumn{3}{c}{$\adjustbox{valign=c, raise=0.15cm}{\includegraphics{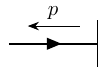}}
=v^s(p)$} &
\multicolumn{3}{c}{$\adjustbox{valign=c, raise=0.15cm}{\includegraphics{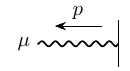}}
\!\!=\epsilon_\mu^*(p)$} &
\multicolumn{3}{c}{$\adjustbox{valign=c, raise=0.15cm}{\includegraphics{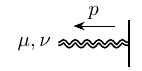}}
\!\!=\epsilon_{\mu\nu}^*(p)$}\\[7pt]
   \hline\hline
 \end{tabular}
\end{table*}

\subsection{\label{sec:electronpropagator}Electron propagator}
To derive the electron propagator, we consider the functional integral \cite{Peskin2018}
\begin{equation}
 \int e^{iS_\mathrm{D,kin}[\bar{\psi},\psi]/(\hbar c)}\mathcal{D}\bar{\psi}\mathcal{D}\psi.
 \label{eq:pathintegralD}
\end{equation}
Here $S_\mathrm{D,kin}[\bar{\psi},\psi]$ is the action integral of the Lagrangian density of the Dirac field, $\mathcal{L}_\mathrm{D,kin}$ in equation~\eqref{eq:LUGMterms}. Thus, we write \cite{Peskin2018}
\begin{equation}
 S_\mathrm{D,kin}[\bar{\psi},\psi]=\!\int\!\mathcal{L}_\mathrm{D,kin}d^4x
 =c\!\int\! \bar{\psi}(x)(i\hbar\slashed{\partial}-m_\mathrm{e}c)\psi(x)d^4x.
 \label{eq:action_D}
\end{equation}
Here we have applied partial integration to transform the integral in a form, where all derivatives apply to the Dirac spinor on the right. The electron propagator, denoted by $D_\mathrm{D}(x_1-x_2)$, is defined to be the Green's function of the operator $i\hbar\slashed{\partial}-m_\mathrm{e}c$, obtained from equation~\eqref{eq:action_D}. Thus, it is the solution of the eigenvalue equation of this operator for a delta function source term, written as \cite{Peskin2018}
\begin{equation}
 (i\hbar\slashed{\partial}-m_\mathrm{e}c)D_\mathrm{D}(x_1-x_2)=i\mathbf{I}_4\delta^{4}(x_1-x_2).
 \label{eq:electronpropagator_equation1}
\end{equation}
Here the derivative in $\slashed{\partial}$ operates on $x_1$. Applying the Fourier transform $\tilde f(p)\!=\!\int\! f(x)e^{ip\cdot x/\hbar}d^4x$ with $x=x_1-x_2$ to equation~\eqref{eq:electronpropagator_equation1}, we obtain
\begin{equation}
 (\slashed{p}-m_\mathrm{e}c\mathbf{I}_4)D_\mathrm{D}(p)=i\mathbf{I}_4.
\end{equation}
The solution of this equation, the momentum-space electron propagator $\tilde D_\mathrm{D}(p)$, is given by \cite{Peskin2018}
\begin{equation}
 \tilde D_\mathrm{D}(p)=\dfrac{i(\slashed{p}+m_\mathrm{e}c\mathbf{I}_4)}{p^2-m_\mathrm{e}^2c^2+i\epsilon}.
\end{equation}
Here the infinitesimal constant $\epsilon$ displaces the poles above and below the real axis so that $D_\mathrm{D}(x_1-x_2)$ is consistently obtained as the inverse Fourier transform of $\tilde D_\mathrm{D}(p)$.

\subsection{\label{sec:photonpropagator}Photon propagator}
To derive the photon propagator, the Faddeev--Popov method leads us to use the gauge-fixed Lagrangian density of the electromagnetic field in the functional integral of the path integral formulation of QED \cite{Peskin2018}. Thus, we consider the functional integral, given by
\begin{equation}
 \int e^{iS_\mathrm{em,kin}^{\prime(\xi_\mathrm{e}=1)}[A]/(\hbar c)}\mathcal{D}A.
 \label{eq:pathintegralem}
\end{equation}
The Feynman gauge-fixed electromagnetic action integral $S_\mathrm{em,kin}^{\prime(\xi_\mathrm{e}=1)}[A]$ in equation~\eqref{eq:pathintegralem} is an integral of $\mathcal{L}_\mathrm{em,kin}^{\prime(\xi_\mathrm{e}=1)}$ in equation~\eqref{eq:Lem_fixed} over the Cartesian Minkowski coordinates, rewritten as \cite{Peskin2018}
\begin{align}
 S_\mathrm{em,kin}^{\prime(\xi_\mathrm{e}=1)}[A] &=\!\int\!\mathcal{L}_\mathrm{em,kin}^{\prime(\xi_\mathrm{e}=1)}d^4x\nonumber\\
 &=\frac{1}{2\mu_0\hbar^2}\!\int\! A_\mu(x)(\eta^{\mu\nu}\hbar^2\partial^2)A_\nu(x)d^4x.
 \label{eq:action_em}
\end{align}
Here we have applied partial integration to transform the integral in a form, where all derivatives apply to the electromagnetic four-potential on the right. The photon propagator corresponding to the applied Feynman gauge, denoted by $D^\mathrm{(em,\xi_\mathrm{e}=1)}_{\nu\rho}(x_1-x_2)$, is defined to be the Green's function of the operator $\eta^{\mu\nu}\hbar^2\partial^2$, obtained from equation~\eqref{eq:action_em}. Thus, it is the solution of the eigenvalue equation of this operator for a delta function source term, written as \cite{Peskin2018}
\begin{equation}
 \eta^{\mu\nu}\hbar^2\partial^2 D^\mathrm{(em,\xi_\mathrm{e}=1)}_{\nu\rho}(x_1-x_2)=i\delta^\mu_\rho\delta^{4}(x_1-x_2).
 \label{eq:photonpropagator_equation1}
\end{equation}
Here the derivative $\partial^2$ operates on $x_1$. Applying the Fourier transform, $\tilde f(p)\!=\!\int\! f(x)e^{ip\cdot x/\hbar}d^4x$ with $x=x_1-x_2$, to equation~\eqref{eq:photonpropagator_equation1}, we obtain
\begin{equation}
 -\eta^{\mu\nu}p^2\tilde D^\mathrm{(em,\xi_\mathrm{e}=1)}_{\nu\rho}(p)=i\delta^\mu_\rho.
\end{equation}
The solution of this equation is the well-known momentum-space photon propagator $\tilde D^\mathrm{(em,\xi_\mathrm{e}=1)}_{\nu\rho}(p)$ in the Feynman gauge, given by \cite{Peskin2018}
\begin{equation}
 \tilde D^\mathrm{(em,\xi_\mathrm{e}=1)}_{\nu\rho}(p)=\frac{-i\eta_{\nu\rho}}{p^2+i\epsilon}.
 \label{eq:photonpropagator1}
\end{equation}
As in the case of the electron propagator in section~\ref{sec:electronpropagator}, the infinitesimal constant $\epsilon$ in equation~\eqref{eq:photonpropagator1} displaces the poles above and below the real axis so that $D^\mathrm{(em,\xi_\mathrm{e}=1)}_{\nu\rho}(x_1-x_2)$ is consistently obtained as the inverse Fourier transform of $\tilde D^\mathrm{(em,\xi_\mathrm{e}=1)}_{\nu\rho}(p)$.

Finally, we note that the general expression of the photon propagator for an arbitrary gauge-fixing parameter $\xi_\mathrm{e}$ is given by \cite{Schwartz2014}
\begin{equation}
 \tilde D^\mathrm{(em)}_{\nu\rho}(p)=\frac{-i}{p^2+i\epsilon}\Big[\eta_{\nu\rho}-(1-\xi_\mathrm{e})\frac{p_\nu p_\rho}{p^2+i\epsilon}\Big].
\end{equation}
This form is obtained by using $\mathcal{L}'_\mathrm{em,kin}$, given in equation~\eqref{eq:Lem_general}, in place of $\mathcal{L}_\mathrm{em,kin}^{\prime(\xi_\mathrm{e}=1)}$ in equation~\eqref{eq:action_em}.

\subsection{\label{sec:gravitonpropagator}Graviton propagator}
To derive the graviton propagator, we follow the analogy with the derivation of the photon propagator in QED, discussed in section~\ref{sec:photonpropagator}. Correspondingly, the Faddeev--Popov method leads us to use the gauge-fixed Lagrangian density of the gravity gauge field in the functional integral of the path integral formulation of UGM. Therefore, we consider the functional integral, given by
\begin{equation}
 \int e^{iS_\mathrm{g,kin}^{\prime(\xi_\mathrm{e}=1)}[H]/(\hbar c)}\mathcal{D}H.
 \label{eq:pathintegralH}
\end{equation}
The harmonic gauge-fixed gravity action integral $S_\mathrm{g,kin}^{\prime(\xi_\mathrm{e}=1)}[H]$ in equation~\eqref{eq:pathintegralH} is an integral of $\mathcal{L}_\mathrm{g,kin}^{\prime(\xi_\mathrm{e}=1)}$ in equation~\eqref{eq:Lg_fixed}, written as
\begin{align}
 S_\mathrm{g,kin}^{\prime(\xi_\mathrm{e}=1)}[H] &=\!\int\!\mathcal{L}_\mathrm{g,kin}^{\prime(\xi_\mathrm{e}=1)}d^4x\nonumber\\
 &=\frac{1}{\kappa\hbar^2}\!\int\! H_{\mu\nu}(x)(-P^{\mu\nu,\alpha\beta}\hbar^2\partial^2)H_{\alpha\beta}(x)d^4x.
 \label{eq:action_g}
\end{align}
The graviton propagator corresponding to the applied harmonic gauge, denoted by $D^\mathrm{(g,\xi_\mathrm{g}=1)}_{\alpha\beta,\rho\sigma}(x_1-x_2)$, is defined to be the Green's function of the operator $-P^{\mu\nu,\alpha\beta}\hbar^2\partial^2$, obtained from equation~\eqref{eq:action_g}. Thus, it is the solution of the eigenvalue equation of this operator for a delta function source term, written as
\begin{equation}
 -P^{\mu\nu,\alpha\beta}\hbar^2\partial^2 D^\mathrm{(g,\xi_\mathrm{g}=1)}_{\alpha\beta,\rho\sigma}(x_1-x_2)=iI_{\rho\sigma}^{\mu\nu}\delta^{4}(x_1-x_2).
 \label{eq:gravitonpropagator_equation1}
\end{equation}
Here the derivative $\partial^2$ operates on $x_1$ and the identity tensor $I_{\rho\sigma}^{\mu\nu}$ is defined in equation~\eqref{eq:II}. Applying the Fourier transform, $\tilde f(p)\!=\!\int\! f(x)e^{ip\cdot x/\hbar}d^4x$ with $x=x_1-x_2$, to equation~\eqref{eq:photonpropagator_equation1}, we obtain
\begin{equation}
 P^{\mu\nu,\alpha\beta}p^2\tilde D^\mathrm{(g,\xi_\mathrm{g}=1)}_{\alpha\beta,\rho\sigma}(p)=iI_{\rho\sigma}^{\mu\nu}.
\end{equation}
The solution of this equation, the momentum-space graviton propagator $\tilde D^\mathrm{(g,\xi_\mathrm{g}=1)}_{\alpha\beta,\rho\sigma}(p)$, is given by
\begin{equation}
 \tilde D^\mathrm{(g,\xi_\mathrm{g}=1)}_{\alpha\beta,\rho\sigma}(p)=\frac{iP_{\alpha\beta,\rho\sigma}^{(D)}}{p^2+i\epsilon}.
 \label{eq:gravitonpropagator}
\end{equation}
Here the infinitesimal constant $\epsilon$ has the same role as discussed in the case of the electron and photon propagators above. The graviton propagator in equation~\eqref{eq:gravitonpropagator} is equal to the well-known graviton propagator of linearized general relativity in the harmonic gauge \cite{Schwartz2014,Donoghue1994b,Kibble1965,DeWitt1967c}.

The quantity $P_{\mu\nu,\rho\sigma}^{(D)}$, which we have used in writing the graviton propagator in equation~\eqref{eq:gravitonpropagator}, is defined as the $D$-dimensional space-time generalization of the quantity $P_{\mu\nu,\rho\sigma}$, given in equation~\eqref{eq:P}. This generalization is known to be given by \cite{Veltman1976}
\begin{equation}
P_{\mu\nu,\rho\sigma}^{(D)}=\frac{1}{2}\Big(\eta_{\mu\sigma}\eta_{\rho\nu}+\eta_{\mu\rho}\eta_{\nu\sigma}-\frac{2}{D-2}\eta_{\mu\nu}\eta_{\rho\sigma}\Big).
\label{eq:PD}
\end{equation}
If we set the space-time dimension to $D=4$, $P_{\mu\nu,\rho\sigma}^{(D)}$ in equation~\eqref{eq:PD} becomes equivalent to $P_{\mu\nu,\rho\sigma}$ in equation~\eqref{eq:P}. Defining the $D$-dimensional form of the graviton propagator is essential for dimensional regularization, used in the renormalization of unified gravity in section~\ref{sec:renormalization}.

Finally, we note that the general expression of the graviton propagator for an arbitrary gauge-fixing parameter $\xi_\mathrm{g}$ is given by
\begin{align}
 \tilde D^\mathrm{(g)}_{\alpha\beta,\rho\sigma}(p) &=\frac{i}{p^2\!+\!i\epsilon}\Big[P_{\alpha\beta,\rho\sigma}^{(D)}-\frac{1-\xi_\mathrm{g}}{p^2+i\epsilon}(\eta_{\alpha\rho}p_\beta p_\sigma\nonumber\\
 &\hspace{0.4cm}+\eta_{\alpha\sigma}p_\beta p_\rho+\eta_{\beta\rho}p_\alpha p_\sigma+\eta_{\beta\sigma}p_\alpha p_\rho)\Big].
 \label{eq:gravitonpropagator_general}
\end{align}
This form is obtained by using $\mathcal{L}'_\mathrm{g,kin}$, given in equation~\eqref{eq:Lg_general}, in place of $\mathcal{L}_\mathrm{g,kin}^{\prime(\xi_\mathrm{g}=1)}$ in equation~\eqref{eq:action_g}. The general graviton propagator in equation~\eqref{eq:gravitonpropagator_general} can also be derived using the conventional effective field theory of gravity as shown in recent previous literature \cite{Brandt2022}.

\subsection{\label{sec:electromagnetic_interaction}Electron--photon vertex}

Next, we determine the electron--photon vertex. The electromagnetic interaction term of the Lagrangian density in equation~\eqref{eq:LUGMterms} is rewritten as
\begin{align}
 \mathcal{L}_\mathrm{em,int} &=-J_\mathrm{e}^{\mu}A_{\mu}
 =-i\sqrt{\frac{\hbar c}{\mu_0}}\bar{\psi}\Big(\frac{-iq_\mathrm{e}\boldsymbol{\gamma}_\mathrm{F}^\mu}{\sqrt{\varepsilon_0\hbar c}}\Big)\psi A_\mu\nonumber\\
 &=-i\tilde u_0^2\tilde A_0\sqrt{\frac{\hbar c}{\mu_0}}\bar{u}(p')\Big(\frac{-iq_\mathrm{e}\boldsymbol{\gamma}_\mathrm{F}^\mu}{\sqrt{\varepsilon_0\hbar c}}\Big)u(p)\epsilon_\mu(k).
 \label{eq:Linte}
\end{align}
The last form of equation~\eqref{eq:Linte} is written assuming momentum eigenstates $\psi=\tilde u_0u(p)$, $\bar{\psi}=\tilde u_0\bar{u}(p')$, and $A_\mu=\tilde A_0\epsilon_\mu(k)$, where $\tilde u_0$ and $\tilde A_0$ are normalization constants and $\epsilon_\mu(k)$ is the photon polarization vector. Then, we identify the electron--photon vertex from equation~\eqref{eq:Linte} as the term between the Dirac spinors.

\subsection{Electron--graviton, photon--graviton, and electron--photon--graviton vertices}

Here we use the gravitational interaction terms of the Lagrangian density in equation~\eqref{eq:LUGM} to obtain vertex functions associated with the coupling of gravitons to electromagnetic and Dirac fields. The approach is analogous to how the standard electron--photon vertex is obtained from the interaction Lagrangian density of QED in section~\ref{sec:electromagnetic_interaction}. The gravitational interaction term $\mathcal{L}_\mathrm{g,int}$ of the Lagrangian density in equation~\eqref{eq:LUGMterms} is rewritten as
\begin{align}
 &\mathcal{L}_\mathrm{g,int}=-\frac{g'_\mathrm{g}}{g_\mathrm{g}}T_\mathrm{m}^{\mu\nu}H_{\mu\nu}\nonumber\\
 &\!=\!-i\sqrt{\frac{\hbar c}{\kappa}}\Big\{\!-\!\frac{i}{2}\sqrt{\frac{\kappa c}{\hbar}}\frac{g'_\mathrm{g}}{g_\mathrm{g}}P^{\mu\nu,\rho\sigma}\nonumber\\
&\times[i\hbar\bar{\psi}\boldsymbol{\gamma}_{\mathrm{F}\rho}(\vec{\partial}_\sigma-\cev{\partial}_\sigma)\psi-m'_\mathrm{e}c\eta_{\rho\sigma}\bar{\psi}\psi]\Big\}H_{\mu\nu}\nonumber\\
&-\!\frac{i\hbar}{\sqrt{\mu_0\kappa}}\bar{\psi}\Big\{\!\frac{iq_\mathrm{e}}{\hbar}\sqrt{\!\frac{\kappa}{\varepsilon_0}}\frac{g'_\mathrm{g}}{g_\mathrm{g}}P^{\mu\nu,\rho\sigma}\boldsymbol{\gamma}_{\mathrm{F}\sigma}\Big\}\psi A_\rho H_{\mu\nu}\nonumber\\
&-\!i\epsilon_0\hbar c\sqrt{\!\frac{c}{\kappa\hbar}}\Big\{\!\!-\!\frac{i}{2}\sqrt{\frac{\kappa c}{\hbar}}\frac{g'_\mathrm{g}}{g_\mathrm{g}}P^{\mu\nu,\sigma\eta,\rho\lambda}\partial_\eta A_\sigma\partial_\lambda A_\rho\!\Big\}H_{\mu\nu}\nonumber\\
 &\!=\!-i\tilde u_0^2\tilde H_0\sqrt{\frac{\hbar c}{\kappa}}\bar{u}(p')\Big\{\!-\!\frac{i}{2}\sqrt{\frac{\kappa c}{\hbar}}\frac{g'_\mathrm{g}}{g_\mathrm{g}}P^{\mu\nu,\rho\sigma}\nonumber\\
&\times[\boldsymbol{\gamma}_{\mathrm{F}\rho}(p+p')_\sigma-m'_\mathrm{e}c\eta_{\rho\sigma}\mathbf{I}_4]\Big\}u(p)\epsilon_{\mu\nu}(q)\nonumber\\
&-\!\frac{i\hbar\tilde u_0^2\tilde A_0\tilde H_0}{\sqrt{\mu_0\kappa}}\bar{u}(p')\Big\{\!\frac{iq_\mathrm{e}}{\hbar}\sqrt{\!\frac{\kappa}{\varepsilon_0}}\frac{g'_\mathrm{g}}{g_\mathrm{g}}P^{\mu\nu,\rho\sigma}\boldsymbol{\gamma}_{\mathrm{F}\sigma}\!\Big\}u(p)\epsilon_\rho(k)\epsilon_{\mu\nu}(q)\nonumber\\
&-\!\frac{i\tilde A_0^2\tilde H_0}{2}\sqrt{\!\frac{\epsilon_0^2c^3}{\kappa\hbar^3}}\epsilon_\sigma^*(k')\Big\{\!\!-\!i\sqrt{\!\frac{\kappa c}{\hbar}}\frac{g'_\mathrm{g}}{g_\mathrm{g}}P^{\mu\nu,\sigma\eta,\rho\lambda}k'_\eta k_\lambda\!\Big\}\nonumber\\
&\times\epsilon_\rho(k)\epsilon_{\mu\nu}(q).
 \label{eq:Lintg}
\end{align}
In analogy with equation~\eqref{eq:Linte}, the last form of equation~\eqref{eq:Lintg} is written assuming momentum eigenstates $\psi=\tilde u_0u(p)$, $\bar{\psi}=\tilde u_0\bar{u}(p')$, $A_\rho=\tilde A_0\epsilon_\rho(k)$, $A_\sigma=\tilde A_0\epsilon_\sigma^*(k')$, and $H_{\mu\nu}=\tilde H_0\epsilon_{\mu\nu}(q)$. Here $\tilde H_0$ is the normalization constant and $\epsilon_{\mu\nu}$ is the graviton polarization tensor known from previous literature \cite{Schwartz2014,Jaranowski2009}. The quantities $P^{\mu\nu,\rho\sigma}$ and $P^{\mu\nu,\rho\sigma,\eta\lambda}$ in equation~\eqref{eq:Lintg} are given in equations \eqref{eq:P} and \eqref{eq:P3}.

In analogy with the well-known case of the electron--photon vertex in QED, discussed in section~\ref{sec:electromagnetic_interaction}, we identify the electron--graviton, electron--photon--graviton, and photon--graviton vertices from equation~\eqref{eq:Lintg} as the terms in the wave brackets. In the application of the Feynman rules, we set $m'_\mathrm{e}=m_\mathrm{e}$ and $g'_\mathrm{g}=g_\mathrm{g}$ according to the equivalence principle in equation~\eqref{eq:equivalenceprinciple}. In this work, we additionally assume the Feynman gauge of QED with $\xi_\mathrm{e}=1$ and the harmonic gauge of unified gravity with $\xi_\mathrm{g}=1$. The tree-level calculations of section~\ref{sec:gravitational_scattering} show that these vertices lead to physically meaningful scattering amplitudes consistent with previous literature. The calculations at 1-loop order in section~\ref{sec:renormalization} show how the vertices lead to consistent renormalization of the theory. More detailed comparison of the vertices of UGM to those of the effective field theory quantization of general relativity \cite{Donoghue1994b,Lawrence1971,Butt2006,Ulhoa2017,Olyaei2018,Ramos2010} is a topic of further work.

\subsection{\label{sec:gravitational_scattering}Gravitational interaction potential and scattering processes}

\subsubsection{\label{sec:Vg}Gravitational interaction potential}

Here we show how the electron--graviton vertex and the graviton propagator, given in table~\ref{tbl:Feynman}, can be used to derive the known classical limit of the gravitational interaction potential energy between two electrons. Our calculation is completely analogous to how the electron-photon vertex and the photon propagator are used to derive the classical Coulomb interaction potential energy limit of QED as summarized in section~2.10 of the supplementary material. In previous literature \cite{Ulhoa2017,Santos2019,Santos2020,Ramos2010}, the gravitational scattering of two electrons is also called the gravitational M{\o}ller scattering since it is analogous to the conventional electromagnetic M{\o}ller scattering \cite{Peskin2018,Schwartz2014,Moller1932,Ashkin1954}. The gravitational M{\o}ller scattering is studied in more detail in section~\ref{sec:gravitational_Moller}.

In analogy with QED, there are two leading-order Feynman diagrams, and the processes are called the t- and u-channels. In the classical consideration, the effect of the indistinguishability of electrons vanishes, and thus, it is sufficient to consider the t-channel scattering. The scattering amplitude of the t-channel is given by
\begin{align}
 &i\mathcal{M}
 =\adjustbox{valign=c, raise=0.4mm}{\includegraphics{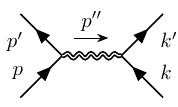}}\nonumber\\
 &=\!\dfrac{iP_{\alpha\mu,\beta\nu}}{(p\!-\!p')^2\!+\!i\epsilon}\nonumber\\
 &\hspace{0.4cm}\times\!\bar{u}(p')\Big\{\!-\!\dfrac{i}{2}\sqrt{\dfrac{\kappa c}{\hbar}}P_{\;\;\;\;\;\rho\sigma}^{\alpha\mu,}
 [\boldsymbol{\gamma}_\mathrm{F}^\rho(p\!+\!p')^\sigma\!-\!m_\mathrm{e}c\eta^{\rho\sigma}\mathbf{I}_4]\Big\}u(p)\nonumber\\
 &\hspace{0.4cm}\times\!\bar{u}(k')\Big\{\!-\!\dfrac{i}{2}\sqrt{\dfrac{\kappa c}{\hbar}}P_{\;\;\;\;\;\lambda\eta}^{\beta\nu,}
 [\boldsymbol{\gamma}_\mathrm{F}^\lambda(k\!+\!k')^\eta\!-\!m_\mathrm{e}c\eta^{\lambda\eta}\mathbf{I}_4]\Big\}u(k)\nonumber\\
 &=\!-\frac{i\kappa c}{4\hbar}\dfrac{P_{\rho\sigma,\lambda\eta}}{(p\!-\!p')^2\!+\!i\epsilon}\bar{u}(p')[\boldsymbol{\gamma}_\mathrm{F}^\rho(p\!+\!p')^\sigma\!-\!m_\mathrm{e}c\eta^{\rho\sigma}\mathbf{I}_4]u(p)\nonumber\\
 &\hspace{0.4cm}\times\!\bar{u}(k')[\boldsymbol{\gamma}_\mathrm{F}^\lambda(k+k')^\eta-m_\mathrm{e}c\eta^{\lambda\eta}\mathbf{I}_4]u(k).
 \label{eq:gravitational_scattering}
\end{align}
In the last equality of equation~\eqref{eq:gravitational_scattering}, we have applied the identity $P_{\alpha\mu,\beta\nu}P_{\;\;\;\;\;\rho\sigma}^{\alpha\mu,}P_{\;\;\;\;\;\lambda\eta}^{\beta\nu,}=P_{\rho\sigma,\lambda\eta}$.

To derive the classical expression of the gravitational interaction potential energy, we use the nonrelativistic limit of the Dirac spinors, in which case we have
\begin{align}
 &\bar{u}^{s'}(p')\boldsymbol{\gamma}_\mathrm{F}^\rho u^s(p)\rightarrow 2m_\mathrm{e}c\delta^{ss'}\delta^{\rho0}\!,
 \hspace{0.2cm}\bar{u}^{s'}(p')u^s(p)\rightarrow 2m_\mathrm{e}c\delta^{ss'}\!\!,\nonumber\\
 &\bar{u}^{r'}(k')\boldsymbol{\gamma}_\mathrm{F}^\lambda u^r(k)\rightarrow 2m_\mathrm{e}c\delta^{rr'}\delta^{\lambda0}\!,
 \hspace{0.2cm}\bar{u}^{r'}(k')u^r(k)\rightarrow 2m_\mathrm{e}c\delta^{rr'}\!\!.
 \label{eq:nonrelativisticspinors}
\end{align}
Here $s$, $s'$, $r$, and $r'$ denote spin states. As in the case of QED, discussed in section~2.10 of the supplementary material, the spin is conserved in the nonrelativistic limit. Furthermore, in the nonrelativistic limit, we have
\begin{align}
 (p+p')^\sigma &= (k+k')^\sigma= 2m_\mathrm{e}c\delta^{\sigma0}+\mathcal{O}(|\mathbf{p}''|),\nonumber\\
 (p-p')^2 &=(k'-k)^2=-|\mathbf{p}''|^2+\mathcal{O}(|\mathbf{p}''|^4).
\end{align}
Here $\mathbf{p}''$ is the three-dimensional momentum vector component of $p''=p-p'=k'-k=(p''^0,\mathbf{p}'')$. Therefore, in the nonrelativistic limit, the scattering amplitude in equation~\eqref{eq:gravitational_scattering} becomes
\begin{align}
 i\mathcal{M}\! &=\!\frac{i\kappa m_\mathrm{e}^4c^5}{\hbar}\dfrac{\delta^{ss'}\delta^{rr'}\!P_{\rho\sigma,\lambda\eta}}{|\mathbf{p}''|^2-i\epsilon}(2\delta^{\rho0}\delta^{\sigma0}\!-\!\eta^{\rho\sigma})
 (2\delta^{\lambda0}\delta^{\eta0}\!-\!\eta^{\lambda\eta})\nonumber\\
 &=\!\frac{2i\kappa m_\mathrm{e}^4c^5}{\hbar}\frac{\delta^{ss'}\delta^{rr'}}{|\mathbf{p}''|^2-i\epsilon}.
 \label{eq:M_g}
\end{align}
This result is compared with the Born approximation in nonrelativistic quantum mechanics, according to which the scattering amplitude is proportional to the Fourier transform $\widetilde{V}(\mathbf{p}'')$ of the potential function $V(\mathbf{r})$, where $\mathbf{r}$ is the position vector whose length is the distance between the particles \cite{Peskin2018}. Accounting for the same coefficient of proportionality as in the case of deriving the Coulomb interaction potential energy of QED in section~2.10 of the supplementary material, and summing over the final spin states, we then obtain
\begin{align}
 \widetilde{V}_\mathrm{g}(\mathbf{p}'') &=-\frac{\hbar^3}{4m_\mathrm{e}^2c}\sum_{s',r'}\mathcal{M}
 =-\frac{\kappa\hbar^2m_\mathrm{e}^2c^4}{2(|\mathbf{p}''|^2-i\epsilon)}.
 \label{eq:Vg_Fourier}
\end{align}

Then, performing the inverse Fourier transform from the momentum space to the position space gives
\begin{align}
 &V_\mathrm{g}(\mathbf{r})=\int\widetilde{V}_\mathrm{g}(\mathbf{p}'')e^{i\mathbf{p}''\cdot\mathbf{r}/\hbar}\frac{d^3p''}{(2\pi\hbar)^3}\nonumber\\
 &=-\frac{\kappa m_\mathrm{e}^2c^4}{2(2\pi)^3\hbar}\int\frac{e^{i\mathbf{p}''\cdot\mathbf{r}/\hbar}}{|\mathbf{p}''|^2-i\epsilon}d^3p''\nonumber\\
 &=-\frac{\kappa m_\mathrm{e}^2c^4}{8\pi|\mathbf{r}|}e^{i|\mathbf{r}|\sqrt{i\epsilon}/\hbar}.
 \label{eq:Vg_derivation}
\end{align}
The intermediate steps of the calculation of this integral are given in section~7 of the supplementary material.

In the limit of $\epsilon$ approaching zero as $\epsilon\rightarrow0$, we obtain the gravitational interaction potential energy from equation~\eqref{eq:Vg_derivation} as
\begin{equation}
 V_\mathrm{g}(\mathbf{r})=-\frac{\kappa m_\mathrm{e}^2c^4}{8\pi|\mathbf{r}|}
 =-\frac{Gm_\mathrm{e}^2}{|\mathbf{r}|}=-\frac{\hbar c\alpha_\mathrm{g}}{|\mathbf{r}|}.
 \label{eq:Vg}
\end{equation}
Here $\alpha_\mathrm{g}$ is the gravity fine structure constant, defined in equation~\eqref{eq:alphag}. The negative sign of equation~\eqref{eq:Vg} is associated with the fact that the gravitational force is attractive. The gravitational interaction potential energy formula in equation~\eqref{eq:Vg} is equivalent to Newton's law of gravitation written for two electrons. Therefore, it is the correct nonrelativistic limit of general relativity. In the case of nonzero momenta, one must account for radiative corrections. For scalar particles, such radiative corrections have been studied using the previous effective field theory quantization of gravity in \cite{Donoghue1994a,BjerrumBohr2003,Faller2008,Netto2022}. In our case of UGM, the radiative corrections to the Newtonian potential are studied in section~\ref{sec:radiativeNewton}.

\subsubsection{\label{sec:gravitational_Moller}Gravitational M{\o}ller scattering}

Next, we investigate the momentum dependence of the gravitational M{\o}ller scattering in the nonrelativistic limit. Accounting for both the t- and u-channel scattering processes, in analogy to the electromagnetic M{\o}ller scattering \cite{Peskin2018,Schwartz2014}, the total scattering amplitude of the gravitational M{\o}ller scattering is given by
\begin{align}
 &i\mathcal{M}
 =\adjustbox{valign=c, raise=0.4mm}{\includegraphics{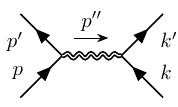}}
\;\;+\;
\adjustbox{valign=c, raise=-0.6mm}{\includegraphics{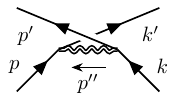}}\nonumber\\
&=\!\dfrac{iP_{\alpha\mu,\beta\nu}}{(p\!-\!p')^2}\nonumber\\
 &\hspace{0.4cm}\times\!\bar{u}(p')\Big\{\!-\!\dfrac{i}{2}\sqrt{\dfrac{\kappa c}{\hbar}}P_{\;\;\;\;\;\rho\sigma}^{\alpha\mu,}
 [\boldsymbol{\gamma}_\mathrm{F}^\rho(p\!+\!p')^\sigma\!-\!m_\mathrm{e}c\eta^{\rho\sigma}\mathbf{I}_4]\Big\}u(p)\nonumber\\
 &\hspace{0.4cm}\times\!\bar{u}(k')\Big\{\!-\!\dfrac{i}{2}\sqrt{\dfrac{\kappa c}{\hbar}}P_{\;\;\;\;\;\lambda\eta}^{\beta\nu,}
 [\boldsymbol{\gamma}_\mathrm{F}^\lambda(k\!+\!k')^\eta\!-\!m_\mathrm{e}c\eta^{\lambda\eta}\mathbf{I}_4]\Big\}u(k)\nonumber\\
 &\hspace{0.4cm}-\!\dfrac{iP_{\alpha\mu,\beta\nu}}{(p\!-\!k')^2}\nonumber\\
 &\hspace{0.4cm}\times\!\bar{u}(k')\Big\{\!-\!\dfrac{i}{2}\sqrt{\dfrac{\kappa c}{\hbar}}P_{\;\;\;\;\;\rho\sigma}^{\alpha\mu,}
 [\boldsymbol{\gamma}_\mathrm{F}^\rho(p\!+\!k')^\sigma\!-\!m_\mathrm{e}c\eta^{\rho\sigma}\mathbf{I}_4]\Big\}u(p)\nonumber\\
 &\hspace{0.4cm}\times\!\bar{u}(p')\Big\{\!-\!\dfrac{i}{2}\sqrt{\dfrac{\kappa c}{\hbar}}P_{\;\;\;\;\;\lambda\eta}^{\beta\nu,}
 [\boldsymbol{\gamma}_\mathrm{F}^\lambda(k\!+\!p')^\eta\!-\!m_\mathrm{e}c\eta^{\lambda\eta}\mathbf{I}_4]\Big\}u(k)\nonumber\\
 &=\!-\frac{i\kappa c}{4\hbar}P_{\rho\sigma,\lambda\eta}\Big\{\dfrac{1}{t}\bar{u}(p')[\boldsymbol{\gamma}_\mathrm{F}^\rho(p\!+\!p')^\sigma\!-\!m_\mathrm{e}c\eta^{\rho\sigma}\mathbf{I}_4]u(p)\nonumber\\
 &\hspace{0.4cm}\times\!\bar{u}(k')[\boldsymbol{\gamma}_\mathrm{F}^\lambda(k+k')^\eta-m_\mathrm{e}c\eta^{\lambda\eta}\mathbf{I}_4]u(k)\nonumber\\
 &\hspace{0.4cm}-\dfrac{1}{u}\bar{u}(k')[\boldsymbol{\gamma}_\mathrm{F}^\rho(p\!+\!k')^\sigma\!-\!m_\mathrm{e}c\eta^{\rho\sigma}\mathbf{I}_4]u(p)\nonumber\\
 &\hspace{0.4cm}\times\!\bar{u}(p')[\boldsymbol{\gamma}_\mathrm{F}^\lambda(k+p')^\eta-m_\mathrm{e}c\eta^{\lambda\eta}\mathbf{I}_4]u(k)\Big\}.
 \label{eq:gravitationalMoller}
\end{align}
In the last equality of equation~\eqref{eq:gravitationalMoller}, we have applied the identity $P_{\alpha\mu,\beta\nu}P_{\;\;\;\;\;\rho\sigma}^{\alpha\mu,}P_{\;\;\;\;\;\lambda\eta}^{\beta\nu,}=P_{\rho\sigma,\lambda\eta}$.
We have also used the Mandelstam variables, defined as \cite{Mandelstam1958,Peskin2018,Schwartz2014}
\begin{align}
 s &=(p+k)^2=(p'+k')^2=E_\mathrm{cm}^2/c^2,\nonumber\\
 t &=(p-p')^2=(k'-k)^2=-2p_\mathrm{r}^2(1-\cos\theta_\mathrm{r}),\nonumber\\
 u &=(p-k')^2=(p'-k)^2=-2p_\mathrm{r}^2(1+\cos\theta_\mathrm{r}).
 \label{eq:Mandelstam}
\end{align}
Here $E_\mathrm{cm}=2E_\mathrm{r}$ is the total energy of the electrons in the center of mass frame. The last forms of the Mandelstam variables in equation~\eqref{eq:Mandelstam} correspond to the parametrization of the four-momenta of the incoming and outgoing electrons, given by
\begin{align}
 p^\mu &=(E_\mathrm{r}/c,0,0,p_\mathrm{r}),\nonumber\\
 k^\mu &=(E_\mathrm{r}/c,0,0,-p_\mathrm{r}),\nonumber\\
 p'^\mu &=(E_\mathrm{r}/c,p_\mathrm{r}\sin\theta_\mathrm{r},0,p_\mathrm{r}\cos\theta_\mathrm{r}),\nonumber\\
 k'^\mu &=(E_\mathrm{r}/c,-p_\mathrm{r}\sin\theta_\mathrm{r},0,-p_\mathrm{r}\cos\theta_\mathrm{r}).
 \label{eq:pkparameters}
\end{align}

Using equation~\eqref{eq:gravitationalMoller}, we obtain the square of the scattering amplitude as\pagebreak
\vspace*{-0.4cm}
\begin{align}
 |\mathcal{M}|^2 &=\frac{\kappa^2c^2}{16\hbar^2}P_{\rho\sigma,\lambda\eta}P_{\alpha\beta,\gamma\delta}\nonumber\\
 &\hspace{0.4cm}\times\Big(\frac{1}{t^2}\mathrm{Tr}\{[\boldsymbol{\gamma}_\mathrm{F}^\rho(p+p')^\sigma-m_\mathrm{e}c\eta^{\rho\sigma}\mathbf{I}_4]\mathbf{u}_p\nonumber\\
 &\hspace{0.4cm}\times[\boldsymbol{\gamma}_\mathrm{F}^\alpha(p+p')^\beta-m_\mathrm{e}c\eta^{\alpha\beta}\mathbf{I}_4]\mathbf{u}_{p'}\}\nonumber\\
 &\hspace{0.4cm}\times\mathrm{Tr}\{[\boldsymbol{\gamma}_\mathrm{F}^\lambda(k+k')^\eta-m_\mathrm{e}c\eta^{\lambda\eta}\mathbf{I}_4]\mathbf{u}_k\nonumber\\
 &\hspace{0.4cm}\times[\boldsymbol{\gamma}_\mathrm{F}^\gamma(k+k')^\delta-m_\mathrm{e}c\eta^{\gamma\delta}\mathbf{I}_4]\mathbf{u}_{k'}\}\nonumber\\
 &\hspace{0.4cm}+\frac{1}{u^2}\mathrm{Tr}\{[\boldsymbol{\gamma}_\mathrm{F}^\rho(p+k')^\sigma-m_\mathrm{e}c\eta^{\rho\sigma}\mathbf{I}_4]\mathbf{u}_p\nonumber\\
 &\hspace{0.4cm}\times[\boldsymbol{\gamma}_\mathrm{F}^\alpha(p+k')^\beta-m_\mathrm{e}c\eta^{\alpha\beta}\mathbf{I}_4]\mathbf{u}_{k'}\}\nonumber\\
 &\hspace{0.4cm}\times\mathrm{Tr}\{[\boldsymbol{\gamma}_\mathrm{F}^\lambda(k+p')^\eta-m_\mathrm{e}c\eta^{\lambda\eta}\mathbf{I}_4]\mathbf{u}_k\nonumber\\
 &\hspace{0.4cm}\times[\boldsymbol{\gamma}_\mathrm{F}^\gamma(k+p')^\delta-m_\mathrm{e}c\eta^{\gamma\delta}\mathbf{I}_4]\mathbf{u}_{p'}\}\nonumber\\
 &\hspace{0.4cm}-\frac{1}{tu}\mathrm{Tr}\{[\boldsymbol{\gamma}_\mathrm{F}^\rho(p+p')^\sigma-m_\mathrm{e}c\eta^{\rho\sigma}\mathbf{I}_4]\mathbf{u}_p\nonumber\\
 &\hspace{0.4cm}\times[\boldsymbol{\gamma}_\mathrm{F}^\alpha(p+k')^\beta-m_\mathrm{e}c\eta^{\alpha\beta}\mathbf{I}_4]\mathbf{u}_{k'}\nonumber\\
 &\hspace{0.4cm}\times[\boldsymbol{\gamma}_\mathrm{F}^\lambda(k+k')^\eta-m_\mathrm{e}c\eta^{\lambda\eta}\mathbf{I}_4]\mathbf{u}_k\nonumber\\
 &\hspace{0.4cm}\times[\boldsymbol{\gamma}_\mathrm{F}^\gamma(k+p')^\delta-m_\mathrm{e}c\eta^{\gamma\delta}\mathbf{I}_4]\mathbf{u}_{p'}\}\nonumber\\
 &\hspace{0.4cm}-\frac{1}{tu}\mathrm{Tr}\{[\boldsymbol{\gamma}_\mathrm{F}^\rho(p+k')^\sigma-m_\mathrm{e}c\eta^{\rho\sigma}\mathbf{I}_4]\mathbf{u}_p\nonumber\\
 &\hspace{0.4cm}\times[\boldsymbol{\gamma}_\mathrm{F}^\alpha(p+p')^\beta-m_\mathrm{e}c\eta^{\alpha\beta}\mathbf{I}_4]\mathbf{u}_{p'}\nonumber\\
 &\hspace{0.4cm}\times[\boldsymbol{\gamma}_\mathrm{F}^\lambda(k+p')^\eta-m_\mathrm{e}c\eta^{\lambda\eta}\mathbf{I}_4]\mathbf{u}_k\nonumber\\[-2pt]
 &\hspace{0.4cm}\times[\boldsymbol{\gamma}_\mathrm{F}^\gamma(k+k')^\delta-m_\mathrm{e}c\eta^{\gamma\delta}\mathbf{I}_4]\mathbf{u}_{k'}\}\Big).
\label{eq:Mg2}
\end{align}
In equation~\eqref{eq:Mg2}, we have expressed the terms in terms of traces and applied the cyclicity of the trace to write all dependencies on the Dirac spinors using the shorthand notation $\mathbf{u}_p=u(p)\bar{u}(p)$.

To calculate the unpolarized scattering cross section, we average $|\mathcal{M}|^2$ in equation~\eqref{eq:Mg2} over the initial spins, sum over the final spins, apply the completeness relations of the Dirac spinors, given by $\sum_s\mathbf{u}_p^s=\sum_s u^s(p)\bar{u}^s(p)=\boldsymbol{\gamma}_\mathrm{F}^\rho p_\rho+m_\mathrm{e}c\mathbf{I}_4$ \cite{Peskin2018}, and finally use the trace relations of the Dirac gamma matrices \cite{Peskin2018,Schwartz2014} to obtain
\begin{align}
 &\langle|\mathcal{M}|^2\rangle=\frac{1}{4}\sum_{s,s',r,r'}|\mathcal{M}|^2\nonumber\\[-5pt]
 &=\frac{\kappa^2c^2}{64\hbar^2t^2u^2}\{(s-t)^2t^2(5s^2-6st+5t^2)\nonumber\\
 &\hspace{0.4cm}-tu(-9s^4+12s^3t-4st^3+t^4)\nonumber\\
 &\hspace{0.4cm}+u^2(5s^4-12s^3t+8s^2t^2+4st^3-3t^4)\nonumber\\
 &\hspace{0.4cm}+2u^3(-8s^3+2st^2+t^3)+u^4(22s^2+4st-3t^2)\nonumber\\
 &\hspace{0.4cm}-u^5(16s+t)+5u^6+4m_\mathrm{e}^2c^2[-2s^3(t^2+7tu+u^2)\nonumber\\
 &\hspace{0.4cm}-(t+u)(t^2+u^2)(2t^2-7tu+2u^2)\nonumber\\
 &\hspace{0.4cm}+s^2(t+u)(2t^2+31tu+2u^2)\nonumber\\
 &\hspace{0.4cm}+2s(t^4-12t^3u-13t^2u^2-12tu^3+u^4)]\nonumber\\
 &\hspace{0.4cm}-16m_\mathrm{e}^4c^4[-s(2t-3u)(3t-2u)(t+u)\nonumber\\
 &\hspace{0.4cm}+(t^2+tu+u^2)(3s^2+3t^2-11tu+3u^2)]\nonumber\\
 &\hspace{0.4cm}-64m_\mathrm{e}^6c^6tu[5(t+u)-8s]+256m_\mathrm{e}^8c^8(t^2-tu+u^2)\}.
 \label{eq:Mg2ave}
\end{align}

Using the last forms of the Mandelstam variables in equation~\eqref{eq:Mandelstam} and the equation for the unpolarized scattering cross section of particles with equal masses in the center-of-mass frame in terms of $\langle|\mathcal{M}|^2\rangle$ \cite{Schwartz2014}, we then obtain
\begin{align}
 &\Big(\frac{d\sigma}{d\Omega}\Big)_\mathrm{cm}=\frac{\hbar^2c^2}{64\pi^2E_\mathrm{cm}^2}\langle|\mathcal{M}|^2\rangle\nonumber\\
 &=\frac{\hbar^2\alpha_\mathrm{g}^2}{4m_\mathrm{e}^4c^2E_\mathrm{cm}^2p_\mathrm{r}^4\sin^4\theta_\mathrm{r}}\Big[16(m_\mathrm{e}^4c^4+8m_\mathrm{e}^2c^2p_\mathrm{r}^2+8p_\mathrm{r}^4)^2\nonumber\\
 &\hspace{0.4cm}-4(3m_\mathrm{e}^8c^8\!+\!53m_\mathrm{e}^6c^6p_\mathrm{r}^2\!+\!270m_\mathrm{e}^4c^4p_\mathrm{r}^4\!+\!456m_\mathrm{e}^2c^2p_\mathrm{r}^6\nonumber\\
 &\hspace{0.4cm}+240p_\mathrm{r}^8)\sin^2\theta_\mathrm{r}+p_\mathrm{r}^4(61m_\mathrm{e}^4c^4\!+\!240m_\mathrm{e}^2c^2p_\mathrm{r}^2\nonumber\\
 &\hspace{0.4cm}+\!186p_\mathrm{r}^4)\sin^4\theta_\mathrm{r}
 -p_\mathrm{r}^8\sin^6\theta_\mathrm{r}\Big].
 \label{eq:gravitationalMollerCS}
\end{align}
Equation \eqref{eq:gravitationalMollerCS} is the scattering cross section of the unpolarized gravitational M{\o}ller scattering in unified gravity. In the nonrelativistic limit, $p_\mathrm{r}\ll m_\mathrm{e}c$, the unpolarized gravitational M{\o}ller scattering cross section in equation~\eqref{eq:gravitationalMollerCS} becomes
\begin{equation}
 \Big(\frac{d\sigma}{d\Omega}\Big)_\mathrm{cm}
 =\frac{\hbar^2\alpha_\mathrm{g}^2m_\mathrm{e}^4c^6}{E_\mathrm{cm}^2p_\mathrm{r}^4\sin^4\theta_\mathrm{r}}(4-3\sin^2\theta_\mathrm{r}).
 \label{eq:gravitationalMollerCSNR}
\end{equation}
In comparison with the nonrelativistic limit of the unpolarized electromagnetic M{\o}ller scattering cross section, summarized in section~2.11 of the supplementary material, the only difference of equation~\eqref{eq:gravitationalMollerCSNR} is that the gravity fine structure constant appears in place of the electromagnetic fine structure constant. This is as expected in the nonrelativistic limit. The same result follows from the somewhat different scattering amplitudes calculated in previous works \cite{Ramos2010}. For larger momenta, in analogy with the electromagnetic M{\o}ller scattering, one must account for the radiative corrections. This is left as a topic of further work.

\section{\label{sec:renormalization}Renormalization of unified gravity}

\subsection{Introduction to renormalization of quantum gravity}

To obtain finite values for physical quantities in quantum field theories, one often needs to apply the renormalization procedure \cite{Schwartz2014,Peskin2018}. This procedure removes infinities that arise in the description of self-energies and higher-order scattering processes involving loops in the Feynman diagrams. The general complexity of this procedure is compounded by the need to renormalize different types of interactions and to account for the perturbative and nonperturbative regimes of the theories. Further subtleties follow from possible spontaneous symmetry breaking, anomalies, and nonrenormalizable interactions. Despite these challenges, renormalization is essential in making quantum field theories physically predictive and mathematically consistent.

In the case of gravity, it is known that general relativity does not fit the paradigm of describing fundamental interactions by renormalizable quantum field theories \cite{Schwartz2014,Hooft1974,Deser1974a,Deser1974b,Deser1974c,Goroff1985}. This is because the gravitational interactions in general relativity are of such a form that the induced divergences cannot be absorbed by the redefinition of the parameters of the theory. Therefore, general relativity is considered inherently nonrenormalizable \cite{Schwartz2014,Zee2010,Maggiore2005}. In the conventional effective-field-theory-based quantization of gravity, this problem is circumvented by introducing an infinite series of new coupling constants and higher-order gravitational terms in the Lagrangian density to absorb the divergences. The infinite number of free parameters, however, ruins the predictivity of the theory. Thus, only the leading low-energy, long-distance quantum corrections can be reliably calculated using the effective-field-theory-based quantum gravity \cite{Donoghue1994a,Donoghue1994b}.

The goal of renormalizing quantum gravity is to develop a consistent theory unifying the principles of general relativity with quantum mechanics. Many previous approaches, such as string theory \cite{Becker2007,Dine2007,Green1987} and loop quantum gravity \cite{Ashtekar1986,Jacobson1988,Rovelli1990,Rovelli2008}, have been developed but they have not led to the ultimate success. Thus, a complete and widely accepted renormalizable theory of quantum gravity remains unknown.

Below, we present a restricted investigation of the renormalizability of unified gravity. We aim at showing that, because of the different gauge symmetries, the renormalizability properties of unified gravity are fundamentally different from those in the conventional nonrenormalizable effective field theory of gravity. We prove that the theory is renormalizable at 1-loop order and leave the complete proof of renormalizability to all loop orders as a topic of further work. In comparison, conventional gravity is well known to be nonrenormalizable already at 1-loop order \cite{Hooft1974,Deser1974a,Deser1974b,Deser1974c}.

\subsection{Renormalization of unified gravity in the UGM formulation}

In this section, we study the renormalization of unified gravity in the Minkowski metric (UGM) at 1-loop order. As in previous sections, we limit our study to the system of the Dirac electron--positron field, the electromagnetic field, and the gravitational field. The renormalized theory is expressed in terms of the renormalization factors, to be presented and discussed in section~\ref{sec:renormalizationfactors}.

In this work, we use the on-shell renormalization scheme \cite{Schwartz2014,Peskin2018}, also known as the pole scheme or the physical scheme. The on-shell scheme is especially suitable for particles that can travel over asymptotically large distances, such as electrons, photons, and gravitons, studied in this work. The on-shell scheme also has an advantage that its mass parameters straightforwardly correspond to the physical masses and no separate corrections of the parameters are needed \cite{Schwartz2014}.

We present the relevant Feynman diagrams and calculate all 1-loop contributions to the on-shell renormalization factors of the theory. The calculations are found to be analogous to those of the renormalization of the gauge theories of the Standard Model \cite{Schwartz2014}. The radiative corrections of selected physical quantities are calculated in section~\ref{sec:radiative}. Here we use the unscaled representation of unified gravity and QED, instead of the scaled representations discussed in section~\ref{sec:scaled} above and in section~2.9 of the supplementary material. The 1-loop contributions to the on-shell renormalization factors of UGM using the conventional dimensional regularization are collected from the sections below and summarized in table~\ref{tbl:renormalization}.

\begin{table*}
\setlength{\tabcolsep}{7pt}
\renewcommand{\arraystretch}{1.5}
\caption{\label{tbl:renormalization}
One-loop contributions of the renormalization factors of unified gravity in the on-shell renormalization scheme obtained by using dimensional regularization. The electromagnetic part of the renormalization factors follows from electromagnetic interactions through the electron--photon and electron--photon--graviton vertices. The gravitational part follows from gravitational interactions through the electron--graviton, photon--graviton, and electron--photon--graviton vertices.}
\begin{tabular}{ccc}
 \hline\hline
 Renormalization factor & Electromagnetic part of $\delta Z_i^{(1)}$ & Gravity part of $\delta Z_i^{(1)}$\\
 \hline\\[-12pt]
 $Z_\psi=1+\delta Z_\psi^{(1)}+\ldots$ & $-\frac{\alpha_\mathrm{e}}{4\pi}\Big[\frac{1}{\epsilon_\mathrm{UV}}+\frac{2}{\epsilon_\mathrm{IR}}+4+3\log\!\Big(\frac{4\pi\mu^2e^{-\gamma}}{m_\mathrm{e}^2c^2}\Big)\Big]$ & $-\frac{\kappa c p^2}{64\pi^2\hbar}\Big[\frac{7}{\epsilon_\mathrm{UV}}-\frac{4}{\epsilon_\mathrm{IR}}+10+3\log\!\Big(\frac{4\pi\mu^2e^{-\gamma}}{m_\mathrm{e}^2c^2}\Big)\Big]$\\[5pt]
 $Z_m=1+\delta Z_m^{(1)}+\ldots$ & $-\frac{3\alpha_\mathrm{e}}{4\pi}\Big[\frac{1}{\epsilon_\mathrm{UV}}+\frac{4}{3}+\log\!\Big(\frac{4\pi\mu^2e^{-\gamma}}{m_\mathrm{e}^2c^2}\Big)\Big]$ & $\frac{\kappa c p^2}{16\pi^2\hbar}\Big[\frac{1}{\epsilon_\mathrm{UV}}+1+\log\!\Big(\frac{4\pi\mu^2e^{-\gamma}}{m_\mathrm{e}^2c^2}\Big)\Big]$\\[5pt]
 $Z_A=1+\delta Z_A^{(1)}+\ldots$ & $-\frac{\alpha_\mathrm{e}}{3\pi}\Big[\frac{1}{\epsilon_\mathrm{UV}}+\log\!\Big(\frac{4\pi\mu^2e^{-\gamma}}{m_\mathrm{e}^2c^2}\Big)\Big]$ & $-\frac{\kappa c p^2}{24\pi^2\hbar}\Big[\frac{1}{\epsilon_\mathrm{UV}}+\frac{1}{6}+\log(4\pi\mu^2e^{-\gamma})\Big]$\\[5pt]
 $Z_\mathrm{g\psi}=1+\delta Z_\mathrm{g\psi}^{(1)}+\ldots$ & $-\frac{\alpha_\mathrm{e}}{4\pi}\Big[\frac{1}{\epsilon_\mathrm{UV}}+\frac{2}{\epsilon_\mathrm{IR}}+4+3\log\!\Big(\frac{4\pi\mu^2e^{-\gamma}}{m_\mathrm{e}^2c^2}\Big)\Big]$ & $\frac{\kappa c p^2}{192\pi^2\hbar}\Big[\frac{11}{\epsilon_\mathrm{UV}}+\frac{12}{\epsilon_\mathrm{IR}}+\frac{172}{3}+23\log\!\Big(\frac{4\pi\mu^2e^{-\gamma}}{m_\mathrm{e}^2c^2}\Big)\Big]$\\[5pt]
 $Z_{\mathrm{g}m}=1+\delta Z_{\mathrm{g}m}^{(1)}+\ldots$ & $-\frac{3\alpha_\mathrm{e}}{4\pi}\Big[\frac{1}{\epsilon_\mathrm{UV}}+\frac{4}{3}+\log\!\Big(\frac{4\pi\mu^2e^{-\gamma}}{m_\mathrm{e}^2c^2}\Big)\Big]$ & $\frac{5\kappa c p^2}{192\pi^2\hbar}\Big[\frac{1}{\epsilon_\mathrm{UV}}-\frac{31}{30}+\log\!\Big(\frac{4\pi\mu^2e^{-\gamma}}{m_\mathrm{e}^2c^2}\Big)\Big]$\\[5pt]
 $Z_{\mathrm{g}A}=1+\delta Z_{\mathrm{g}A}^{(1)}+\ldots$ & $-\frac{\alpha_\mathrm{e}}{3\pi}\Big[\frac{1}{\epsilon_\mathrm{UV}}+\log\!\Big(\frac{4\pi\mu^2e^{-\gamma}}{m_\mathrm{e}^2c^2}\Big)\Big]$ & $\frac{\kappa cp^2}{8\pi^2\hbar}\Big[\frac{1}{\epsilon_\mathrm{UV}}+\frac{43}{12}+\log\!\Big(\frac{4\pi\mu^2e^{-\gamma}}{(m_\mathrm{e}^2c^2)^{4/3}}\Big)\Big]$\\[5pt]
 \hline
 Renormalization factor & Electron-loop part of $\delta Z_H^{(1)}$ & Photon-loop part of $\delta Z_H^{(1)}$\\
 \hline\\[-12pt]
 \multirow{2}{*}{$Z_H=1+\delta Z_H^{(1)}+\ldots$} & $-\frac{\kappa cp^2}{192\pi^2\hbar}\Big\{\frac{37}{15}-\frac{4m_\mathrm{e}^2c^2}{p^2}-\frac{24m_\mathrm{e}^4c^4}{p^4}+\log(m_\mathrm{e}^2c^2)$ & \multirow{2}{*}{$-\frac{\kappa c p^2}{96\pi^2\hbar}\Big[\frac{1}{\epsilon_\mathrm{UV}}+\frac{29}{30}+\log(4\pi\mu^2e^{-\gamma})\Big]$}\\[2pt]
 & $+\Big(1-\frac{4m_\mathrm{e}^2c^2}{p^2}-\frac{16m_\mathrm{e}^2c^2}{p^4}\Big)\Big[\frac{1}{\epsilon_\mathrm{UV}}+\log\!\Big(\frac{4\pi\mu^2e^{-\gamma}}{m_\mathrm{e}^2c^2}\Big)\Big]\Big\}$\\[5pt]
 \hline\hline
 \end{tabular}
\end{table*}

\subsection{\label{sec:renormalizationfactors}Dressed states and renormalization factors}

We start our study of the renormalization of unified gravity with the discussion of the bare and dressed states and the renormalization factors. The renormalization is based on recognizing that the fields and physical constants in the classical Lagrangian density, in the absence of quantum-field-theoretical interactions, are bare quantities. In the presence of interactions, the bare particles, like electrons, interact through the gauge fields, leading to the creation of particle-antiparticle pairs, virtual photons, and other excitations that modify the bare states to become dressed states. Thus, a dressed state refers to a quantum state that is the fundamental excitation of the theory in the presence of interactions.  Therefore, the bare quantities neglect the quantum-field-theoretical virtual particle corrections associated with the Feynman loop diagrams. The bare fields and the bare physical constants are typically infinite and written in terms of the renormalized quantities as follows:
\begin{equation}
 \begin{array}{rclrcl}
 \psi_\mathrm{bare} & \!\!=\!\! & \sqrt{Z_\psi}\psi,\hspace{0.5cm}
 & A_\mathrm{bare}^\mu & \!\!=\!\! & \sqrt{Z_A}A^\mu,\\[5pt]
 m_\mathrm{e,bare} & \!\!=\!\! & Z_m m_\mathrm{e},
 & e_\mathrm{bare} & \!\!=\!\! & Z_e e,\\[5pt]
 \xi_\mathrm{e,bare} & \!\!=\!\! & Z_\mathrm{\xi e}\xi_\mathrm{e},
 & H_\mathrm{bare}^{\mu\nu} & \!\!=\!\! & \sqrt{Z_H} H^{\mu\nu},\\[5pt]
 m'_\mathrm{e,bare} & \!\!=\!\! & Z_{\mathrm{g}m}m'_\mathrm{e},
 & g'_\mathrm{g,bare} & \!\!=\!\! & Z_\mathrm{g}g'_\mathrm{g},\\[5pt]
 \xi_\mathrm{g,bare} & \!\!=\!\! & Z_\mathrm{\xi g}\xi_\mathrm{g}.
 \label{eq:renormalization}
 \end{array}
\end{equation}
The renormalization relations for $\psi$, $A^\mu$, $m_\mathrm{e}$, $e$ and $\xi_\mathrm{e}$ in equation~\eqref{eq:renormalization} are well known in QED \cite{Schwartz2014,Peskin2018}. In unified gravity, these relations are complemented with the renormalization of the gravity gauge field $H^{\mu\nu}$, the gravitational mass $m'_\mathrm{e}$, the gravity coupling constant $g'_\mathrm{g}$, and the gravity gauge-fixing parameter $\xi_\mathrm{g}$.

That the gravitational and inertial masses are associated with different renormalization factors in equation~\eqref{eq:renormalization} does not mean that the gravitational and inertial masses are not equal. According to the equivalence principle in equation~\eqref{eq:equivalenceprinciple}, we assume that the renormalized values of these masses are equal. Together with the generally different renormalization factors, the equivalence principle thus means that the bare values of the gravitational and inertial masses are generally different. This is not in conflict with any experiments since the bare masses have anyway infinite values and they are not experimental observables.

For convenience, we define the scaled renormalization factors $Z_{e\psi}$, $Z_\mathrm{g\psi}$, and $Z_{\mathrm{g}A}$ by the following relations:
\begin{equation}
 Z_e =\frac{Z_{e\psi}}{Z_\psi\sqrt{Z_A}},\hspace{0.5cm}
 Z_\mathrm{g}=\frac{Z_\mathrm{g\psi}}{Z_\psi\sqrt{Z_H}}
 =\frac{Z_{\mathrm{g}A}}{Z_A\sqrt{Z_H}}.
 \label{eq:Zequalities1}
\end{equation}
As is well known, the electromagnetic gauge invariance of QED implies via the Ward--Takahashi identity that the renormalization factors $Z_{e\psi}$ and $Z_\psi$ are equal as \cite{Peskin2018,Schwartz2014}
\begin{equation}
 Z_{e\psi}=Z_\psi.
 \label{eq:WardTakahashi}
\end{equation}
The violation of this equality would be an indication of a gauge anomaly breaking the gauge symmetry of QED. Consistent theories, such as the Standard Model, are known to be free from gauge anomalies \cite{Schwartz2014}. In the sections below, we show at 1-loop order that equation~\eqref{eq:WardTakahashi} is satisfied in unified gravity in the presence of gravitational interactions. This derivation provides one strong argument for the consistency of unified gravity.

Furthermore, the renormalization factors of the gauge-fixing parameters are set equal to the renormalization factors of the gauge fields as
\begin{equation}
 Z_\mathrm{\xi e}=Z_A,\hspace{0.5cm}
 Z_\mathrm{\xi g}=Z_H.
 \label{eq:Zequalities2}
\end{equation}
These equations imply that the gauge-fixing Lagrangian densities do not obtain counterterms as shown in section~\ref{sec:LUGMrenorm} below.

In perturbation theory, the renormalization factors $Z_i$ are expanded in powers of the expansion parameter $\lambda$ according to
\begin{equation}
 Z_i=1+\lambda\delta Z_i^{(1)}+\lambda^2\delta Z_i^{(2)}+\ldots\;.
\end{equation}
Accordingly, at 1-loop order, we expand $Z_i$ up to the first power in $\lambda$ and set $\lambda\rightarrow 1$. The higher-order terms in $\lambda$ are dropped out. Order by order, the renormalization factor terms, $\delta Z_i^{(1)}$, $\delta Z_i^{(2)}$,..., become determined by the renormalization conditions, to be discussed in the sections below.

\subsection{\label{sec:LUGMrenorm}Geometric Lagrangian density of UGM in terms of renormalized quantities}

Next, we present the Lagrangian density of UGM in terms of the renormalized quantities. First, we recognize that the quantities of the Lagrangian density of UGM in equations \eqref{eq:LUGM} and \eqref{eq:LUGMterms} are bare quantities, even though \emph{this has not been indicated} by any subscripts in these equations. Using the expressions of the bare quantities in terms of the renormalized quantities in equation~\eqref{eq:renormalization} and applying the renormalization factor equalities in equations \eqref{eq:Zequalities1}--\eqref{eq:Zequalities2}, the terms of the Lagrangian density of UGM in equation~\eqref{eq:LUGMterms} become
\begin{align}
 \mathcal{L}_\mathrm{D,kin} &=\frac{i\hbar c}{2}\bar{\psi}(\boldsymbol{\gamma}_\mathrm{F}^\nu\vec{\partial}_\nu-\cev{\partial}_\nu\boldsymbol{\gamma}_\mathrm{F}^\nu)\psi-m_\mathrm{e}c^2\bar{\psi}\psi\nonumber\\
 &\hspace{0.4cm}+(Z_\psi-1)\frac{i\hbar c}{2}\bar{\psi}(\boldsymbol{\gamma}_\mathrm{F}^\nu\vec{\partial}_\nu-\cev{\partial}_\nu\boldsymbol{\gamma}_\mathrm{F}^\nu)\psi\nonumber\\
 &\hspace{0.4cm}-(Z_\psi Z_m-1)m_\mathrm{e}c^2\bar{\psi}\psi,\nonumber\\
 \mathcal{L}_\mathrm{em,kin} &=-\frac{1}{4\mu_0}F_{\mu\nu}F^{\mu\nu}
 -(Z_A-1)\frac{1}{4\mu_0}F_{\mu\nu}F^{\mu\nu},\nonumber\\
 \mathcal{L}_\mathrm{em,int} &=-q_\mathrm{e}c\bar{\psi}\boldsymbol{\gamma}_\mathrm{F}^\nu\psi A_\nu
 -(Z_\psi-1)q_\mathrm{e}c\bar{\psi}\boldsymbol{\gamma}_\mathrm{F}^\nu\psi A_\nu,\nonumber\\
 \mathcal{L}_\mathrm{g,kin} &=\frac{1}{4\kappa}H_{\rho\mu\nu}S^{\rho\mu\nu}
 +(Z_H-1)\frac{1}{4\kappa}H_{\rho\mu\nu}S^{\rho\mu\nu},\nonumber\\
 \mathcal{L}_\mathrm{g,int} &=-\frac{g'_\mathrm{g}}{g_\mathrm{g}}\Big\{\frac{c}{2}P^{\mu\nu,\rho\sigma}[i\hbar\bar{\psi}(\boldsymbol{\gamma}_\mathrm{F\rho}\vec{\partial}_\sigma-\cev{\partial}_\rho\boldsymbol{\gamma}_\mathrm{F\sigma})\psi\nonumber\\
 &\hspace{0.4cm}-q_\mathrm{e}\bar{\psi}(\boldsymbol{\gamma}_\mathrm{F\rho}A_\sigma+A_\rho\boldsymbol{\gamma}_\mathrm{F\sigma})\psi
 -m'_\mathrm{e}c\eta_{\rho\sigma}\bar{\psi}\psi]\nonumber\\
 &\hspace{0.4cm}+\frac{1}{2\mu_0}P^{\mu\nu,\rho\sigma,\eta\lambda}\partial_\rho A_\sigma\partial_\eta A_\lambda\Big\} H_{\mu\nu}\nonumber\\
 &\hspace{0.4cm}-\frac{g'_\mathrm{g}}{g_\mathrm{g}}\Big\{\frac{c}{2}P^{\mu\nu,\rho\sigma}[(Z_\mathrm{g\psi}\!-\!1)i\hbar\bar{\psi}(\boldsymbol{\gamma}_\mathrm{F\rho}\vec{\partial}_\sigma\!-\!\cev{\partial}_\rho\boldsymbol{\gamma}_\mathrm{F\sigma})\psi\nonumber\\
 &\hspace{0.4cm}-(Z_{\mathrm{g}\psi}-1)q_\mathrm{e}\bar{\psi}(\boldsymbol{\gamma}_\mathrm{F\rho}A_\sigma+A_\rho\boldsymbol{\gamma}_\mathrm{F\sigma})\psi\nonumber\\
 &\hspace{0.4cm}-(Z_\mathrm{g\psi}Z_{\mathrm{g}m}-1)m'_\mathrm{e}c\eta_{\rho\sigma}\bar{\psi}\psi]\nonumber\\
 &\hspace{0.4cm}+(Z_{\mathrm{g}A}-1)\frac{1}{2\mu_0}P^{\mu\nu,\rho\sigma,\eta\lambda}\partial_\rho A_\sigma\partial_\eta A_\lambda\Big\} H_{\mu\nu},\nonumber\\
 \mathcal{L}_\mathrm{em,gf} &=-\frac{1}{2\mu_0\xi_\mathrm{e}}(\partial_{\nu}A^{\nu})^2,\nonumber\\
 \mathcal{L}_\mathrm{g,gf}
 &=\frac{1}{\kappa\xi_\mathrm{g}}\eta_{\gamma\delta}P^{\alpha\beta,\lambda\gamma}P^{\rho\sigma,\eta\delta}\partial_\lambda H_{\alpha\beta}\partial_\eta H_{\rho\sigma}.
 \label{eq:LUGMrenormalized}
\end{align}
The terms of the Lagrangian density in equation~\eqref{eq:LUGMrenormalized} involving the renormalization factors are called the counterterms. When the renormalization factors are set equal to unity, the counterterms become zero and the Lagrangian density terms in equation~\eqref{eq:LUGMrenormalized} become identical in their form to the terms of equation~\eqref{eq:LUGMterms}. The gauge-fixing Lagrangian densities $\mathcal{L}_\mathrm{em,gf}$ and $\mathcal{L}_\mathrm{g,gf}$ in equation~\eqref{eq:LUGMrenormalized} do not obtain counterterms due to the renormalization relations of the gauge-fixing parameters, given in equation~\eqref{eq:Zequalities2}. The counterterm Feynman diagrams follow unambiguously from the Lagrangian density in equation~\eqref{eq:LUGMrenormalized}. These counterterm diagrams are presented together with the other Feynman diagrams in table~\ref{tbl:Feynman}.

\subsection{Computational approach}

In the study of the renormalization of unified gravity at 1-loop order, we need to calculate several 1-loop Feynman diagrams. The literature on the computation of Feynman loop diagrams is extensive \cite{Weinzierl2022,Smirnov2006,Harlander2024}. For all loop integrals of unified gravity, the so-called Passarino--Veltman reduction can be applied to transform the more complicated integrals into simpler ones, which can be solved by standard techniques known from previous literature. For both the Passarino--Veltman reduction and the calculation of the simpler loop integrals, there are highly automated tools available. In this work, after writing the loop integrals, we systematically reduce and calculate them using Wolfram Mathematica packages FeynCalc \cite{Shtabovenko2016} and Package-X \cite{Patel2015}.

To regularize the divergences that are obtained in the calculation of Feynman loop diagrams, we follow FeynCalc and Package-X to use dimensional regularization with the space-time dimension set to $D=4-2\epsilon$, where $\epsilon$ is a small positive number. In the regularization of UV divergences, $\epsilon$ is denoted by $\epsilon_\mathrm{UV}$, and in the regularization of IR divergences, $\epsilon$ is denoted by $\epsilon_\mathrm{IR}$. The use of these notations clarifies the origin of the divergences in our results for the renormalization factors in table~\ref{tbl:renormalization}.

\vspace{-0.1cm}
\subsection{Renormalization of the electron self-energy}
\vspace{-0.1cm}

We start the calculations of the renormalization of unified gravity from the electron self-energy. The electron self-energy refers to the quantum corrections to the electron propagator due to its interaction with the electromagnetic field. In QED, the electron self-energy correction arises from virtual photon exchanges between the electron and the vacuum, which leads to a modification of the electron’s mass and wavefunction. In unified gravity, the electron self-energy becomes also contributed by the virtual graviton exchanges.

The 1-photon-loop, 1-graviton-loop, and counterterm contributions to the electron propagator are, respectively, given by the following irreducible Feynman diagrams:
\vspace{-0.2cm}
\begin{align}
 &-i\Sigma_\mathrm{1L,photon}=\adjustbox{valign=c, raise=0.38cm}{\includegraphics{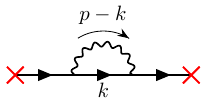}}\nonumber\\
&=\int\frac{-i\eta_{\mu\nu}}{(p-k)^2}\Big(\frac{ie\boldsymbol{\gamma}_\mathrm{F}^\mu}{\sqrt{\varepsilon_0\hbar c}}\Big)\frac{i(\slashed{k}+m_\mathrm{e}c\mathbf{I}_4)}{k^2-m_\mathrm{e}^2c^2}\Big(\frac{ie\boldsymbol{\gamma}_\mathrm{F}^\nu}{\sqrt{\varepsilon_0\hbar c}}\Big)\frac{d^Dk}{(2\pi)^D}\nonumber\\
&=\frac{i\alpha_\mathrm{e}}{4\pi}\Big\{\Big[\frac{1}{\epsilon_\mathrm{UV}}+\log\!\Big(\frac{4\pi\mu^2e^{-\gamma}}{m_\mathrm{e}^2c^2}\Big)\Big](\slashed{p}-4m_\mathrm{e}c\mathbf{I}_4)\nonumber\\
&\hspace{0.4cm}+\Big(1\!+\!\frac{m_\mathrm{e}^2c^2}{p^2}\Big)\slashed{p}\!-\!6m_\mathrm{e}c\mathbf{I}_4
\!+\!\Big(1\!-\!\frac{m_\mathrm{e}^2c^2}{p^2}\Big)\log\!\Big(\frac{m_\mathrm{e}^2c^2}{m_\mathrm{e}^2c^2-p^2}\Big)\nonumber\\
&\hspace{0.4cm}\times\Big[\Big(1+\frac{m_\mathrm{e}^2c^2}{p^2}\Big)\slashed{p}-4m_\mathrm{e}c\mathbf{I}_4\Big]\Big\},
\label{eq:Sigmaphoton}
\end{align}
\vspace{-0.2cm}
\begin{align}
 &-i\Sigma_\mathrm{1L,graviton}=\adjustbox{valign=c, raise=0.42cm}{\includegraphics{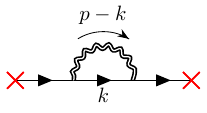}}\nonumber\\[-2pt]
&=\int\!\frac{iP_{\alpha\beta,\eta\lambda}^{(D)}}{(p-k)^2}\Big\{\!-\!\frac{i}{2}\sqrt{\frac{\kappa c}{\hbar}}P_{\;\;\;\;\;\rho\sigma}^{\alpha\beta,}[\boldsymbol{\gamma}_\mathrm{F}^\rho(p+k)^\sigma-\eta^{\rho\sigma}m_\mathrm{e}c\mathbf{I}_4]\Big\}\nonumber\\
&\hspace{0.4cm}\cdot\frac{i(\slashed{k}+m_\mathrm{e}c\mathbf{I}_4)}{k^2-m_\mathrm{e}^2c^2}
\Big\{-\frac{i}{2}\sqrt{\frac{\kappa c}{\hbar}}P_{\;\;\;\;\;\mu\nu}^{\eta\lambda,}[\boldsymbol{\gamma}_\mathrm{F}^\mu(p+k)^\nu\nonumber\\
&\hspace{0.4cm}-\eta^{\mu\nu}m_\mathrm{e}c\mathbf{I}_4]\Big\}\frac{d^Dk}{(2\pi)^D}\nonumber\\
&=-\frac{i\kappa cp^2}{64\pi^2\hbar}\Big\{\Big[\frac{1}{\epsilon_\mathrm{UV}}+\log\!\Big(\frac{4\pi\mu^2e^{-\gamma}}{m_\mathrm{e}^2c^2}\Big)\Big](\slashed{p}-5m_\mathrm{e}c\mathbf{I}_4)\nonumber\\
&\hspace{0.4cm}+\Big(1+\frac{m_\mathrm{e}^2c^2}{p^2}+\frac{2m_\mathrm{e}^4c^4}{p^4}\Big)\slashed{p}
-8m_\mathrm{e}c\mathbf{I}_4\nonumber\\
&\hspace{0.4cm}+\Big(1\!-\!\frac{m_\mathrm{e}^2c^2}{p^2}\Big)\log\!\Big(\frac{m_\mathrm{e}^2c^2}{m_\mathrm{e}^2c^2-p^2}\Big)\Big[\Big(1\!+\!\frac{m_\mathrm{e}^2c^2}{p^2}\!+\!\frac{2m_\mathrm{e}^4c^4}{p^4}\Big)\slashed{p}\nonumber\\
&\hspace{0.4cm}-\Big(5+\frac{3m_\mathrm{e}^2c^2}{p^2}\Big)m_\mathrm{e}c\mathbf{I}_4\Big]\Big\},
\label{eq:Sigmagraviton}
\end{align}
\begin{align}
 &-i\Sigma_\mathrm{1L,CT}=\adjustbox{valign=c}{\includegraphics{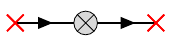}}\nonumber\\
&=i[\delta Z_\psi^{(1)}\slashed{p}-(\delta Z_\psi^{(1)}+\delta Z_\mathrm{m}^{(1)})m_\mathrm{e}c\mathbf{I}_4].
\label{eq:SigmaCT}
\end{align}
Here $\mu$ is the arbitrary scale constant of the dimensional regularization, and $\gamma$ is the Euler--Mascheroni constant. The red crosses on the external lines highlight that these lines are not included in the expressions. The renormalized electron self-energy 1-loop amplitude is given by the sum of the terms in equations \eqref{eq:Sigmaphoton}--\eqref{eq:SigmaCT} as
\begin{equation}
 \Sigma_\mathrm{1L}=\Sigma_\mathrm{1L,photon}+\Sigma_\mathrm{1L,graviton}+\Sigma_\mathrm{1L,CT}.
\end{equation}

The renormalization conditions for the electron self-energy in the on-shell renormalization scheme are well known to be given by \cite{Schwartz2014,Peskin2018}
\begin{equation}
 \Sigma_\mathrm{1L}\big|_{\slashed{p}=m_\mathrm{e}c\mathbf{I}_4}=\mathbf{0},\hspace{0.4cm}
 \frac{d\Sigma_\mathrm{1L}}{d\slashed{p}}\bigg|_{\slashed{p}=m_\mathrm{e}c\mathbf{I}_4}=\mathbf{0}.
 \label{eq:erenormcond}
\end{equation}
The zero values of these renormalization conditions mean that, at the physically meaningful on-shell renormalization point, the renormalized electron propagator has the same form as in the free-particle theory in the absence of loop corrections.

Using the renormalization conditions in equation~\eqref{eq:erenormcond}, the 1-photon-loop and 1-graviton-loop contributions to the renormalization factors $Z_\mathrm{m}$ and $Z_\psi$, in the on-shell renormalization scheme, are determined to be
\begin{equation}
 \delta Z_\mathrm{m,photon}^{(1)}=-\frac{3\alpha_\mathrm{e}}{4\pi}\Big[\frac{1}{\epsilon_\mathrm{UV}}+\frac{4}{3}+\log\!\Big(\frac{4\pi\mu^2e^{-\gamma}}{m_\mathrm{e}^2c^2}\Big)\Big],
 \label{eq:Z_m_photon}
\end{equation}
\begin{equation}
 \delta Z_\mathrm{\psi,photon}^{(1)}=-\frac{\alpha_\mathrm{e}}{4\pi}\Big[\frac{1}{\epsilon_\mathrm{UV}}+\frac{2}{\epsilon_\mathrm{IR}}+4+3\log\!\Big(\frac{4\pi\mu^2e^{-\gamma}}{m_\mathrm{e}^2c^2}\Big)\Big],
 \label{eq:Z_psi_photon}
\end{equation}
\begin{equation}
 \delta Z_\mathrm{m,graviton}^{(1)}=\frac{\kappa c p^2}{16\pi^2\hbar}\Big[\frac{1}{\epsilon_\mathrm{UV}}+1+\!\log\!\Big(\frac{4\pi\mu^2e^{-\gamma}}{m_\mathrm{e}^2c^2}\Big)\Big],
 \label{eq:Z_m_graviton}
\end{equation}
\begin{equation}
 \delta Z_\mathrm{\psi,graviton}^{(1)}=-\frac{\kappa c p^2}{64\pi^2\hbar}\Big[\frac{7}{\epsilon_\mathrm{UV}}-\frac{4}{\epsilon_\mathrm{IR}}+10+3\!\log\!\Big(\frac{4\pi\mu^2e^{-\gamma}}{m_\mathrm{e}^2c^2}\Big)\Big].
 \label{eq:Z_psi_graviton}
\end{equation}
The pole parts of equations \eqref{eq:Z_m_photon}--\eqref{eq:Z_psi_graviton} are proportional to $1/\epsilon_\mathrm{UV}$ and $1/\epsilon_\mathrm{IR}$ and they appear in some form in all renormalization schemes \cite{Schwartz2014}. The physical observables of a renormalized theory are independent of the renormalization scheme, the scale constant, the Euler--Mascheroni constant, and the regularization parameters, such as $\epsilon_\mathrm{UV}$ and $\epsilon_\mathrm{IR}$. This independence is shown for selected radiative corrections in section~\ref{sec:radiative}.

\subsection{\label{sec:photonSErenorm}Renormalization of the photon self-energy}

Next, we study the photon self-energy. In QED, the photon self-energy correction arises from the creation and annihilation of virtual electron--positron pairs. It is also called the electromagnetic vacuum polarization. In unified gravity, the photon self-energy becomes also contributed by the virtual graviton exchanges.

The 1-electron-loop, 1-graviton-loop, and counterterm contributions to the photon propagator are described by the following irreducible Feynman diagrams:
\begin{align}
 &i\Pi_\mathrm{1L,electron}^{\mu\nu}=\adjustbox{valign=c, raise=0.2mm}{\includegraphics{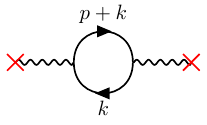}}\nonumber\\
&=-\int\mathrm{Tr}\Big[\frac{i(\slashed{k}+m_\mathrm{e}c\mathbf{I}_4)}{k^2-m_\mathrm{e}^2c^2}\Big(\frac{ie\boldsymbol{\gamma}_\mathrm{F}^\mu}{\sqrt{\varepsilon_0\hbar c}}\Big)\nonumber\\
&\hspace{0.4cm}\cdot\frac{i(\slashed{p}+\slashed{k}+m_\mathrm{e}c\mathbf{I}_4)}{(p+k)^2-m_\mathrm{e}^2c^2}\Big(\frac{ie\boldsymbol{\gamma}_\mathrm{F}^\nu}{\sqrt{\varepsilon_0\hbar c}}\Big)\Big]\frac{d^Dk}{(2\pi)^D}\nonumber\\
&=-\frac{i\alpha_\mathrm{e}}{3\pi}\Big\{\frac{1}{\epsilon_\mathrm{UV}}+\log\!\Big(\frac{4\pi\mu^2e^{-\gamma}}{m_\mathrm{e}^2c^2}\Big)+\frac{5}{3}+\frac{4m_\mathrm{e}^2c^2}{p^2}\nonumber\\
&\hspace{0.4cm}+\Big(1+\frac{2m_\mathrm{e}^2c^2}{p^2}\Big)\sqrt{1\!-\!\frac{4m_\mathrm{e}^2c^2}{p^2}}\nonumber\\
&\hspace{0.4cm}\times\log\!\Big[1\!+\!\frac{p^2}{2m_\mathrm{e}^2c^2}\Big(\sqrt{1\!-\!\frac{4m_\mathrm{e}^2c^2}{p^2}}\!-\!1\Big)\Big]\Big\}\nonumber\\
&\hspace{0.4cm}\times(p^2\eta^{\mu\nu}-p^\mu p^\nu),
\label{eq:photon1Lelectron}
\end{align}
\begin{align}
 &i\Pi_\mathrm{1L,graviton}^{\mu\nu}=\adjustbox{valign=c, raise=0.33cm}{\includegraphics{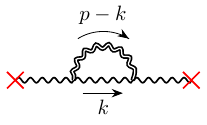}}\nonumber\\
&=\int\frac{iP_{\alpha\beta,\eta\lambda}^{(D)}}{(p-k)^2}\frac{-i\eta_{\rho\sigma}}{k^2}\Big(-i\sqrt{\frac{\kappa c}{\hbar}}P^{\alpha\beta,\nu\xi,\rho\kappa}p_\xi k_\kappa\Big)\nonumber\\
&\hspace{0.4cm}\times\Big(-i\sqrt{\frac{\kappa c}{\hbar}}P^{\eta\lambda,\sigma\zeta,\mu\delta}k_\zeta p_\delta\Big)\frac{d^Dk}{(2\pi)^D}\nonumber\\
&=-\frac{i\kappa cp^2}{24\pi^2\hbar}\Big[\frac{1}{\epsilon_\mathrm{UV}}+\frac{1}{6}+\log\!\Big(-\frac{4\pi\mu^2e^{-\gamma}}{p^2}\Big)\Big]\nonumber\\
&\hspace{0.4cm}\times(p^2\eta^{\mu\nu}-p^\mu p^\nu),
\label{eq:photon1Lgraviton}
\end{align}
\begin{align}
 i\Pi_\mathrm{1L,CT}^{\mu\nu} &=\adjustbox{valign=c}{\includegraphics{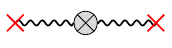}}
=-i\delta Z_\mathrm{A}^{(1)}(p^2\eta^{\mu\nu}-p^\mu p^\nu).
\label{eq:photon1LCT}
\end{align}

The 1-loop photon propagator correction is equal to $i\Pi_\mathrm{1L}^{\mu\nu}$ multiplied by the tree-level photon propagator from the left and right. Thus, the corrected photon propagator is given by
\begin{align}
&\adjustbox{valign=c}{\includegraphics{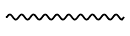}}
\;\;+\;\;
\adjustbox{valign=c, raise=-0.2mm}{\includegraphics{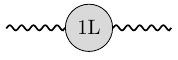}}\nonumber\\
 &=\frac{-i\eta_{\mu\nu}}{p^2}
 +\frac{-i\eta_{\mu\mu'}}{p^2}i\Pi_\mathrm{1L}^{\mu'\nu'}\frac{-i\eta_{\nu'\nu}}{p^2}\nonumber\\
 &=(1+\Pi_\mathrm{1L})\frac{-i\eta_{\mu\nu}}{p^2}+i\Pi_\mathrm{1L}\frac{p_\mu p_\nu}{p^4}.
 \label{eq:photonpropagatorrenormalized}
\end{align}
Here $\Pi_\mathrm{1L}$ is the scalar 1-loop photon self-energy renormalization factor, whose value is given below. In the S-matrix element calculations, at least one end of the renormalized photon propagator in equation~\eqref{eq:photonpropagatorrenormalized} connects to a fermion line. In the summation over all places along the line where it could connect, one finds, according to the Ward identity, that the terms proportional to $p_\mu$ or $p_\nu$ vanish \cite{Peskin2018}. Therefore, in the calculation of the S-matrix elements, the last term in the last line of equation~\eqref{eq:photonpropagatorrenormalized} can be dropped out.

The scalar 1-loop renormalization factor $\Pi_\mathrm{1L}$ of photon self-energy and its electron-loop, graviton-loop, and counterterm parts following from equations \eqref{eq:photon1Lelectron}--\eqref{eq:photonpropagatorrenormalized} are given by
\begin{equation}
 \Pi_\mathrm{1L}=\Pi_\mathrm{1L,electron}+\Pi_\mathrm{1L,graviton}+\Pi_\mathrm{1L,CT},
\end{equation}
\begin{align}
 \Pi_\mathrm{1L,electron}
 &=-\frac{\alpha_\mathrm{e}}{3\pi}\Big\{\frac{1}{\epsilon_\mathrm{UV}}+\log\!\Big(\frac{4\pi\mu^2e^{-\gamma}}{m_\mathrm{e}^2c^2}\Big)+\frac{5}{3}+\frac{4m_\mathrm{e}^2c^2}{p^2}\nonumber\\
&\hspace{0.4cm}+\Big(1+\frac{2m_\mathrm{e}^2c^2}{p^2}\Big)\sqrt{1\!-\!\frac{4m_\mathrm{e}^2c^2}{p^2}}\nonumber\\
&\hspace{0.4cm}\times\log\!\Big[1\!+\!\frac{p^2}{2m_\mathrm{e}^2c^2}\Big(\sqrt{1\!-\!\frac{4m_\mathrm{e}^2c^2}{p^2}}\!-\!1\Big)\Big]\Big\},
\end{align}
\begin{align}
 \Pi_\mathrm{1L,graviton}
 &=-\frac{\kappa cp^2}{24\pi^2\hbar}\Big[\frac{1}{\epsilon_\mathrm{UV}}+\frac{1}{6}+\log\!\Big(-\frac{4\pi\mu^2e^{-\gamma}}{p^2}\Big)\Big],
\end{align}
\begin{equation}
 \Pi_\mathrm{1L,CT}=\Pi_\mathrm{1L,CT,electron}+\Pi_\mathrm{1L,CT,graviton},
\end{equation}
\begin{equation}
 \Pi_\mathrm{1L,CT,electron}=-\delta Z_{A,\mathrm{electron}}^{(1)},
\end{equation}
\begin{equation}
 \Pi_\mathrm{1L,CT,graviton}=-\delta Z_{A,\mathrm{graviton}}^{(1)}.
\end{equation}

In the on-shell renormalization scheme, the renormalization conditions for the photon self-energy are given by
\begin{equation}
 \Pi_\mathrm{1L}\big|_{p^2=0}=0,\hspace{0.5cm}
 \frac{\Pi_\mathrm{1L}}{p^4}\bigg|_{p^2=\infty}=0.
 \label{eq:Pirenorm}
\end{equation}
The first condition in equation~\eqref{eq:Pirenorm} is the standard on-shell renormalization condition for the photon self-energy in quantum field theory \cite{Schwartz2014,Peskin2018}. It means that, at the physically meaningful on-shell renormalization point, the renormalized photon propagator has the same form as in the free-particle theory in the absence of loop corrections. The second condition in equation~\eqref{eq:Pirenorm} is imposed to guarantee that the loop corrections to the Coulomb potential are integrable functions. Through the renormalization constants, one could easily add to the quantity $\Pi_\mathrm{1L}$ arbitrary analytic terms proportional to $p^2$ or its higher powers so that the first condition in equation~\eqref{eq:Pirenorm} remains satisfied. Therefore, to avoid the addition of arbitrary terms that ruin the integrability of the Coulomb potential corrections studied in section~\ref{sec:radiativeCoulomb}, the second condition in equation~\eqref{eq:Pirenorm} is imposed.

The 1-electron-loop and 1-graviton-loop contributions to the renormalization factor $Z_\mathrm{A}$, in the on-shell renormalization scheme, are determined to be
\begin{equation}
 \delta Z_{A,\mathrm{electron}}^{(1)}=-\frac{\alpha_\mathrm{e}}{3\pi}\Big[\frac{1}{\epsilon_\mathrm{UV}}+\log\!\Big(\frac{4\pi\mu^2e^{-\gamma}}{m_\mathrm{e}^2c^2}\Big)\Big],
 \label{eq:ZAe}
\end{equation}
\begin{equation}
 \delta Z_{A,\mathrm{graviton}}^{(1)}=-\frac{\kappa c p^2}{24\pi^2\hbar}\Big[\frac{1}{\epsilon_\mathrm{UV}}+\frac{1}{6}+\log(4\pi\mu^2e^{-\gamma})\Big].
 \label{eq:ZAg}
\end{equation}

Using the values of the renormalization factors in equation~\eqref{eq:ZAe} and \eqref{eq:ZAg}, the 1-loop renormalized scalar amplitude factors of the photon self-energy become
\begin{align}
 &\Pi_\mathrm{1L,electron}^\mathrm{(r)} =\Pi_\mathrm{1L,electron}+\Pi_\mathrm{1L,CT,electron}\nonumber\\
 &=-\frac{\alpha_\mathrm{e}}{3\pi}\Big\{\frac{5}{3}+\frac{4m_\mathrm{e}^2c^2}{p^2}+\Big(1+\frac{2m_\mathrm{e}^2c^2}{p^2}\Big)\sqrt{1\!-\!\frac{4m_\mathrm{e}^2c^2}{p^2}}\nonumber\\
&\hspace{0.4cm}\times\log\!\Big[1\!+\!\frac{p^2}{2m_\mathrm{e}^2c^2}\Big(\sqrt{1\!-\!\frac{4m_\mathrm{e}^2c^2}{p^2}}\!-\!1\Big)\Big]\Big\},
\label{eq:Pi_1L_electron}
\end{align}
\begin{align}
 &\Pi_\mathrm{1L,graviton}^\mathrm{(r)} =\Pi_\mathrm{1L,graviton}+\Pi_\mathrm{1L,CT,graviton}\nonumber\\
 &=\frac{\kappa cp^2}{24\pi^2\hbar}\log(-p^2).
 \label{eq:Pi_1L_graviton}
\end{align}
In section~\ref{sec:radiativeCoulomb}, these renormalized correction factors are applied to calculate the radiative corrections to the Coulomb potential.

\subsection{Renormalization of the graviton self-energy}

Next, we study the graviton self-energy. The graviton self-energy correction arises from the creation and annihilation of virtual electron--positron pairs. It is also contributed by the virtual photon exchanges between the graviton and the vacuum. In contrast to the previous effective field theory quantization of general relativity \cite{Schwartz2014,Bambi2023,Casadio2022,Donoghue1994a,Donoghue1994b}, in unified gravity, the gravitons do not interact directly with each other in the vertex interactions. The consistent renormalization of the 1-electron-loop correction to the graviton self-energy in unified gravity is a substantial breakthrough since the conventional gravity in the presence of matter is well known to be nonrenormalizable starting from the 1-loop order \cite{Hooft1974,Deser1974a,Deser1974b,Deser1974c}.

The 1-electron-loop, 1-photon-loop, and counterterm contributions to the graviton self-energy are given by the following irreducible Feynman diagrams:
\begin{align}
 &i\Xi_\mathrm{1L,electron}^{\alpha\beta,\eta\lambda}=\adjustbox{valign=c}{\includegraphics{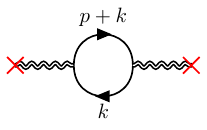}}\nonumber\\
&=-\int\mathrm{Tr}\Big[\frac{i(\slashed{k}+m_\mathrm{e}c\mathbf{I}_4)}{k^2-m_\mathrm{e}^2c^2}\Big\{\!-\!\frac{i}{2}\sqrt{\frac{\kappa c}{\hbar}}P_{\;\;\;\;\;\kappa\delta}^{\alpha\beta,}[\boldsymbol{\gamma}_\mathrm{F}^\kappa(p+2k)^\delta\nonumber\\
&\hspace{0.4cm}-\eta^{\kappa\delta}m_\mathrm{e}c\mathbf{I}_4]\Big\}\frac{i(\slashed{p}+\slashed{k}+m_\mathrm{e}c\mathbf{I}_4)}{(p+k)^2-m_\mathrm{e}^2c^2}\nonumber\\
&\hspace{0.4cm}\cdot\Big\{\!-\!\frac{i}{2}\sqrt{\frac{\kappa c}{\hbar}}P_{\;\;\;\;\;\rho\sigma}^{\eta\lambda,}[\boldsymbol{\gamma}_\mathrm{F}^\rho(p+2k)^\sigma-\eta^{\rho\sigma}m_\mathrm{e}c\mathbf{I}_4]\Big\}\Big]\frac{d^Dk}{(2\pi)^D}\nonumber\\
&=\frac{i\kappa cp^4}{480\pi^2\hbar}\Big\{\hat{P}_\mathrm{A}^{\alpha\beta,\eta\lambda}+\Big[\frac{1}{\epsilon_\mathrm{UV}}+\log\!\Big(\frac{4\pi\mu^2e^{-\gamma}}{m_\mathrm{e}^2c^2}\Big)\Big]\hat{P}_\mathrm{B}^{\alpha\beta,\eta\lambda}\nonumber\\
&\hspace{0.4cm}+\!\Big(1\!-\!\frac{4m_\mathrm{e}^2c^2}{p^2}\Big)^{3/2}\log\!\Big[1\!+\!\frac{p^2}{2m_\mathrm{e}^2c^2}\Big(\sqrt{1\!-\!\frac{4m_\mathrm{e}^2c^2}{p^2}}\!-\!1\Big)\Big]\nonumber\\
&\hspace{0.4cm}\times\hat{P}_\mathrm{C}^{\alpha\beta,\eta\lambda}\Big\},
\label{eq:graviton1Lelectron}
\end{align}
\begin{align}
 &i\Xi_\mathrm{1L,photon}^{\alpha\beta,\eta\lambda}=\adjustbox{valign=c}{\includegraphics{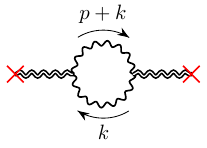}}\nonumber\\
&=\frac{1}{2}\int\frac{-i\eta^{\mu\nu}}{k^2}\frac{-i\eta^{\rho\sigma}}{(p+k)^2}\Big(-i\sqrt{\frac{\kappa c}{\hbar}}P^{\alpha\beta,\mu\kappa,\rho\xi}k_\kappa(p+k)_\xi\Big)\nonumber\\
&\hspace{0.4cm}\times\Big(-i\sqrt{\frac{\kappa c}{\hbar}}P^{\eta\lambda,\sigma\delta,\nu\zeta}(p+k)_\delta k_\zeta\Big)\frac{d^Dk}{(2\pi)^D}\nonumber\\
&=\!\frac{i\kappa cp^4}{240\pi^2\hbar}\Big\{\!\hat{P}_{81,94,30}^{\alpha\beta,\eta\lambda}
\!+\!\Big[\frac{1}{\epsilon_\mathrm{UV}}\!+\!\log\!\Big(\!\!-\!\frac{4\pi\mu^2e^{-\gamma}}{p^2}\Big)\Big]\!\hat{P}_{3,2,1}^{\alpha\beta,\eta\lambda}\Big\},
\label{eq:graviton1Lphoton}
\end{align}
\begin{align}
 i\Xi_\mathrm{1L,CT}^{\alpha\beta,\eta\lambda} &=\adjustbox{valign=c}{\includegraphics{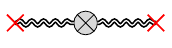}}
=i(Z_H-1)p^2\hat{P}_{1,2,1}^{\alpha\beta,\eta\lambda}.
\label{eq:graviton1LCT}
\end{align}
The factor of $\frac{1}{2}$ in front of the integral in equation~\eqref{eq:graviton1Lphoton} comes from the division by the symmetry factor of the diagram according to the standard Feynman rule. The quantities $\hat{P}_{a,b,c}^{\alpha\beta,\eta\lambda}$, $\hat{P}_\mathrm{A}^{\alpha\beta,\eta\lambda}$, $\hat{P}_\mathrm{B}^{\alpha\beta,\eta\lambda}$, and $\hat{P}_\mathrm{C}^{\alpha\beta,\eta\lambda}$ in equations \eqref{eq:graviton1Lelectron}--\eqref{eq:graviton1LCT} are defined below. First, the quantity $P_{a,b,c}^{\mu\nu,\rho\sigma}$ without a hat is a generalization of $P^{\mu\nu,\rho\sigma}$ in equation~\eqref{eq:P} for the integer parameters $a,b$ and $c$, defined as
\begin{equation}
P_{a,b,c}^{\mu\nu,\rho\sigma}=\frac{1}{2c}(a\eta^{\mu\sigma}\eta^{\rho\nu}+a\eta^{\mu\rho}\eta^{\nu\sigma}-b\eta^{\mu\nu}\eta^{\rho\sigma}).
\label{eq:Pa}
\end{equation}
Here $\eta^{\alpha\beta}$ is the Minkowski metric tensor. Analogously, the quantity $\hat{P}_{a,b,c}^{\alpha\beta,\eta\lambda}$ with a hat is defined as
\begin{equation}
 \hat{P}_{a,b,c}^{\alpha\beta,\eta\lambda}=\frac{1}{2c}(a\hat{\eta}^{\alpha\lambda}\hat{\eta}^{\beta\eta}+a\hat{\eta}^{\alpha\eta}\hat{\eta}^{\beta\lambda}-b\hat{\eta}^{\alpha\beta}\hat{\eta}^{\eta\lambda}).
\end{equation}
Here the quantity $\hat{\eta}^{\alpha\beta}$ is a projection defined as
\begin{equation}
 \hat{\eta}^{\alpha\beta}=\eta^{\alpha\beta}-\frac{p^\alpha p^\beta}{p^2}.
\end{equation}
The derived quantities $\hat{P}_\mathrm{A}^{\alpha\beta,\eta\lambda}$, $\hat{P}_\mathrm{B}^{\alpha\beta,\eta\lambda}$, and $\hat{P}_\mathrm{C}^{\alpha\beta,\eta\lambda}$ in equation~\eqref{eq:graviton1Lelectron} are defined as
\begin{align}
 \hat{P}_\mathrm{A}^{\alpha\beta,\eta\lambda} &=4\hat{P}_{27,23,15}^{\alpha\beta,\eta\lambda}-\frac{2m_\mathrm{e}^2c^2}{p^2}\hat{P}_{19,86,3}^{\alpha\beta,\eta\lambda}\nonumber\\
 &\hspace{0.4cm}-\frac{m_\mathrm{e}^4c^4}{p^4}(64\hat{P}_{1,-1,1}^{\alpha\beta,\eta\lambda}+45P_{1,0,1}^{\alpha\beta,\eta\lambda}),
\end{align}
\begin{equation}
 \hat{P}_\mathrm{B}^{\alpha\beta,\eta\lambda}=\hat{P}_{3,2,1}^{\alpha\beta,\eta\lambda}-\frac{10m_\mathrm{e}^2c^2}{p^2}\hat{P}_{1,2,1}^{\alpha\beta,\eta\lambda}-\frac{30m_\mathrm{e}^4c^4}{p^4}P_{1,0,1}^{\alpha\beta,\eta\lambda},
\end{equation}
\begin{equation}
 \hat{P}_\mathrm{C}^{\alpha\beta,\eta\lambda}=\hat{P}_{3,2,1}^{\alpha\beta,\eta\lambda}+\frac{8m_\mathrm{e}^2c^2}{p^2}\hat{P}_{1,-1,1}^{\alpha\beta,\eta\lambda}.
\end{equation}

The 1-graviton-loop propagator correction is equal to $i\Xi_\mathrm{1L}^{\alpha\beta,\eta\lambda}$ multiplied by the tree-level graviton propagator from the left and right. Thus, the corrected graviton propagator is given by
\begin{align}
 &\adjustbox{valign=c}{\includegraphics{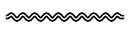}}
\;\;+\;\;
\adjustbox{valign=c}{\includegraphics{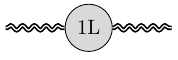}}\nonumber\\
 &=\frac{iP_{\alpha\beta,\eta\lambda}}{p^2}
 +\frac{iP_{\alpha\beta\alpha'\beta'}}{p^2}i\Xi_\mathrm{1L}^{\alpha'\beta'\eta'\lambda'}\frac{iP_{\eta'\lambda'\eta\lambda}}{p^2}\nonumber\\
 &=(1+\Xi_\mathrm{1L})\frac{iP_{\alpha\beta,\eta\lambda}}{p^2}.
\end{align}
Here the renormalized scalar factor $\Xi_\mathrm{1L}$ is formed from the 1-photon-loop, 1-electron-loop, and counterterm contributions as
\begin{equation}
 \Xi_\mathrm{1L}=\Xi_\mathrm{1L,electron}+\Xi_\mathrm{1L,photon}+\Xi_\mathrm{1L,CT},
\end{equation}
\begin{align}
 &\Xi_\mathrm{1L,electron}\nonumber\\
 &=-\frac{\kappa cp^2}{32\pi^2\hbar}\Big\{\frac{37}{15}-\frac{20m_\mathrm{e}^2c^2}{3p^2}-\frac{40m_\mathrm{e}^4c^4}{p^4}\nonumber\\
 &\hspace{0.4cm}+\Big(1-\frac{4m_\mathrm{e}^2c^2}{p^2}-\frac{16m_\mathrm{e}^4c^4}{p^4}\Big)\Big[\frac{1}{\epsilon_\mathrm{UV}}+\log\!\Big(\frac{4\pi\mu^2e^{-\gamma}}{m_\mathrm{e}^2c^2}\Big)\Big]\nonumber\\
 &\hspace{0.4cm}+\Big(1-\frac{2m_\mathrm{e}^2c^2}{p^2}-\frac{8m_\mathrm{e}^4c^4}{p^4}\Big)\sqrt{1-\frac{4m_\mathrm{e}^2c^2}{p^2}}\nonumber\\
 &\hspace{0.4cm}\times\log\Big[1+\frac{p^2}{2m_\mathrm{e}^2c^2}\Big(\sqrt{1-\frac{4m_\mathrm{e}^2c^2}{p^2}}-1\Big)\Big]\Big\},
\end{align}
\begin{equation}
 \Xi_\mathrm{1L,photon}=-\frac{\kappa cp^2}{16\pi^2\hbar}\Big[\frac{1}{\epsilon_\mathrm{UV}}+\frac{29}{30}+\log\!\Big(-\frac{4\pi\mu^2e^{-\gamma}}{p^2}\Big)\Big],
\end{equation}
\begin{equation}
 \Xi_\mathrm{1L,CT}=\Xi_\mathrm{1L,CT,electron}+\Xi_\mathrm{1L,CT,photon},
\end{equation}
\begin{equation}
 \Xi_\mathrm{1L,CT,electron}=-6\delta Z_\mathrm{H,electron}^{(1)},
\end{equation}
\begin{equation}
 \Xi_\mathrm{1L,CT,photon}=-6\delta Z_\mathrm{H,photon}^{(1)}.
\end{equation}

In the on-shell renormalization scheme, the renormalization conditions for the graviton self-energy in unified gravity are analogous to the  renormalization conditions for the photon self-energy in section~\ref{sec:photonSErenorm}. The graviton self-energy renormalization conditions are given by
\begin{equation}
 \Xi_\mathrm{1L}\big|_{p^2=0}=0,\hspace{0.5cm}
 \frac{\Xi_\mathrm{1L}}{p^4}\bigg|_{p^2=\infty}=0.
 \label{eq:Xirenorm}
\end{equation}
In analogy with the photon self-energy renormalization conditions in equation~\eqref{eq:Pirenorm}, the second condition in equation~\eqref{eq:Xirenorm} is imposed to guarantee that the loop corrections to the Newtonian potential are integrable functions. Through the renormalization factors, one could easily add to the quantity $\Xi_\mathrm{1L}$ arbitrary analytic terms proportional to $p^2$ or its higher powers so that the first condition in equation~\eqref{eq:Xirenorm} remains satisfied. Therefore, to avoid the addition of arbitrary terms that ruin the integrability of the Newtonian potential corrections studied in section~\ref{sec:radiativeNewton}, the second condition in equation~\eqref{eq:Xirenorm} is imposed.

The renormalization conditions of the graviton self-energy in equation~\eqref{eq:Xirenorm} lead to the 1-photon-loop and 1-electron-loop counterterms for the graviton self-energy, with the 1-loop contributions to the renormalization factors given by
\begin{align}
 &\delta Z_\mathrm{H,electron}^{(1)}\nonumber\\
 &=-\frac{\kappa cp^2}{192\pi^2\hbar}\Big\{\frac{37}{15}-\frac{4m_\mathrm{e}^2c^2}{p^2}-\frac{24m_\mathrm{e}^4c^4}{p^4}+\log(m_\mathrm{e}^2c^2)\nonumber\\
 &\hspace{0.4cm}+\Big(1-\frac{4m_\mathrm{e}^2c^2}{p^2}-\frac{16m_\mathrm{e}^2c^2}{p^4}\Big)\Big[\frac{1}{\epsilon_\mathrm{UV}}+\log\!\Big(\frac{4\pi\mu^2e^{-\gamma}}{m_\mathrm{e}^2c^2}\Big)\Big]\Big\},
\end{align}
\begin{equation}
 \delta Z_\mathrm{H,photon}^{(1)}=-\frac{\kappa c p^2}{96\pi^2\hbar}\Big[\frac{1}{\epsilon_\mathrm{UV}}+\frac{29}{30}+\log(4\pi\mu^2e^{-\gamma})\Big].
\end{equation}

The renormalized 1-photon and 1-electron loop correction factors to the graviton propagator are given by
\begin{align}
 &\Xi_\mathrm{1L,electron}^\mathrm{(r)}=\Xi_\mathrm{1L,electron}+\Xi_\mathrm{1L,CT,electron}\nonumber\\
 &=-\frac{\kappa cp^2}{32\pi^2\hbar}\Big\{-\frac{8m_\mathrm{e}^2c^2}{3p^2}\Big(1+\frac{6m_\mathrm{e}^2c^2}{p^2}\Big)-\log(m_\mathrm{e}^2c^2)\nonumber\\
 &\hspace{0.4cm}+\Big(1-\frac{2m_\mathrm{e}^2c^2}{p^2}-\frac{8m_\mathrm{e}^4c^4}{p^4}\Big)\sqrt{1-\frac{4m_\mathrm{e}^2c^2}{p^2}}\nonumber\\
 &\hspace{0.4cm}\times\log\Big[1+\frac{p^2}{2m_\mathrm{e}^2c^2}\Big(\sqrt{1-\frac{4m_\mathrm{e}^2c^2}{p^2}}-1\Big)\Big]\Big\},
 \label{eq:Xi_1L_electron}
\end{align}
\begin{align}
 \Xi_\mathrm{1L,photon}^\mathrm{(r)} &=\Xi_\mathrm{1L,photon}+\Xi_\mathrm{1L,CT,photon}\nonumber\\
 &=\frac{\kappa cp^2}{16\pi^2\hbar}\log(-p^2).
 \label{eq:Xi_1L_photon}
\end{align}
In section~\ref{sec:radiativeNewton}, these renormalized correction factors are used to calculate radiative corrections to the Newtonian potential.

\subsection{Photon--graviton two-point function}

Next, we study the photon--graviton two-point function, which describes possible transformation between photons and gravitons. At the tree level, such a process is nonexistent. At the 1-loop order, the photon--graviton two-point function is formed by the Feynman diagrams with a single electron loop. The explicit calculations of such loop diagrams, however, give zero as
\begin{equation}
 \adjustbox{valign=c}{\includegraphics{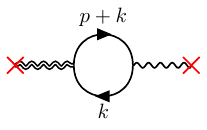}}=0,
\end{equation}
\begin{equation}
 \adjustbox{valign=c,raise=6.1mm}{\includegraphics{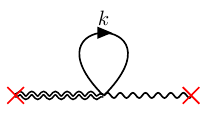}}=0.
\end{equation}
Therefore, the 1-electron-loop contribution to the photon--graviton two-point function vanishes, and we do not need to account for such a process in the renormalization of the vertices, discussed in the sections below.

\subsection{Renormalization of the electron--photon vertex}

The renormalization of the electron--photon vertex is well known to be associated with the Ward--Takahashi identity \cite{Schwartz2014,Peskin2018}. In writing the Lagrangian density in terms of the renormalized quantities in section~\ref{sec:LUGMrenorm}, we used the renormalization factor $Z_\psi$ both in the counterterm of the electron propagator and in the electron--photon vertex. This is in agreement with equation~\eqref{eq:WardTakahashi}. Therefore, in the renormalization of the electron--photon vertex in this section, we do not obtain any new relations for the renormalization factors. Instead, we use the renormalization factors obtained in previous sections to verify that the renormalization condition of the electron--photon vertex is satisfied identically.

The photon index is assumed to be $\mu$ but, for compactness, we do not show it in the diagrams below. Accordingly, the 1-photon-loop, 1-graviton-loop, and counterterm contributions to the electron--photon vertex are given by
\begin{align}
 &\frac{-iq_\mathrm{e}\Gamma_\mathrm{1L,photon}^\mathrm{\mu}}{\sqrt{\varepsilon_0\hbar c}}=\adjustbox{valign=c}{\includegraphics{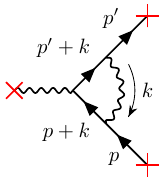}},
\label{eq:vertex1photon}
\end{align}
\begin{align}
 &\frac{-iq_\mathrm{e}\Gamma_\mathrm{1L,graviton}^\mathrm{\mu}}{\sqrt{\varepsilon_0\hbar c}}\nonumber\\
 &=\adjustbox{valign=c}{\includegraphics{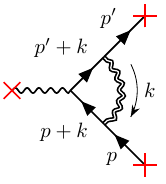}}
\;+\;
\adjustbox{valign=c}{\includegraphics{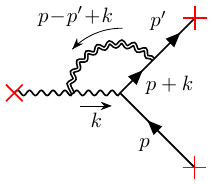}}\nonumber\\
&\hspace{0.4cm}+\;
\adjustbox{valign=c}{\includegraphics{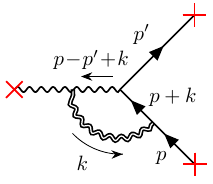}}
\;+\;
\adjustbox{valign=c}{\includegraphics{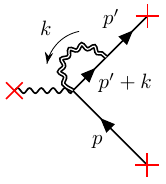}}\nonumber\\
&\hspace{0.4cm}+\;
\adjustbox{valign=c}{\includegraphics{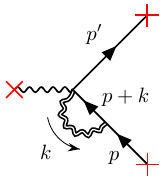}}
\;+\;
\adjustbox{valign=c}{\includegraphics{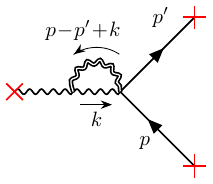}},
\label{eq:vertex1graviton}
\end{align}
\begin{align}
 \frac{-iq_\mathrm{e}\Gamma_\mathrm{1L,CT}^\mathrm{\mu}}{\sqrt{\varepsilon_0\hbar c}} &=\adjustbox{valign=c}{\includegraphics{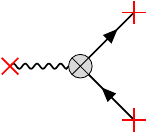}}
=\delta Z_\psi^{(1)}\frac{-iq_\mathrm{e}\boldsymbol{\gamma}_\mathrm{F}^\mu}{\sqrt{\varepsilon_0\hbar c}}.
\label{eq:vertex1CT}
\end{align}
Apart from the counterterm, we only give the Feynman diagrams since there are many diagrams and the integral expressions are straightforward to write using the Feynman rules presented in table~\ref{tbl:Feynman}.

The total 1-loop contribution to the electron--photon vertex is given by
\begin{equation}
 \Gamma_\mathrm{1L}^{\mu}=\Gamma_\mathrm{1L,photon}^{\mu}+\Gamma_\mathrm{1L,graviton}^\mathrm{\mu}+\Gamma_\mathrm{1L,CT}^{\mu}.
\end{equation}
The counterterm can also be split into photon and graviton parts $\Gamma_\mathrm{1L,CT,photon}^{\mu}$ and $\Gamma_\mathrm{1L,CT,graviton}^{\mu}$ based on substituting the renormalization factors of equations \eqref{eq:Z_psi_photon} and \eqref{eq:Z_psi_graviton} into equation~\eqref{eq:vertex1CT}.

The on-shell renormalization condition of the electron--photon vertex is imposed by requiring that $\Gamma_\mathrm{1L}^{\mu}$ and any higher-order-loop terms do not lead to corrections of the tree-level vertex for $p=p'$ \cite{Schwartz2014,Peskin2018}. For on-shell electrons, we additionally have $p^2=m_\mathrm{e}^2c^2$. Therefore, the matrix element of $\Gamma_\mathrm{1L}^{\mu}$ must be zero for $p=p'$ as
\begin{equation}
 \bar{u}(p')\Gamma_\mathrm{1L}^{\mu}u(p)\big|_{p=p'}=0.
 \label{eq:renorm_vertex_e-p}
\end{equation}
The vanishing of the quantum corrections of the electron--photon vertex for $p=p'$ effectively means that the classical form of the Coulomb law applies at asymptotically long distances \cite{Schwartz2014}.

The general structure of the 1-loop radiative correction to the electron--photon vertex matrix element is well known to be of the Lorentz-invariant form, given by \cite{Schwartz2014}
\begin{align}
 &\bar{u}(p')\Gamma_\mathrm{1L}^{\mu}u(p)\nonumber\\
 &=\bar{u}(p')\Big[F_\mathrm{1,1L}^\mathrm{(e\gamma)}(q^2)\boldsymbol{\gamma}_\mathrm{F}^\mu+F_\mathrm{2,1L}^\mathrm{(e\gamma)}(q^2)\frac{iq_\nu}{\hbar m_\mathrm{e}c}\hat{S}_\mathrm{F}^{\mu\nu}\Big]u(p).
 \label{eq:vertexstructure}
\end{align}
Here $q=p'-p$ is the photon four-momentum, $\hat{S}_\mathrm{F}^{\mu\nu}=\frac{i\hbar}{4}[\boldsymbol{\gamma}_\mathrm{F}^\mu,\boldsymbol{\gamma}_\mathrm{F}^\nu]$ is the spin operator for fermions \cite{Partanen2024a}, and the quantities $F_\mathrm{1,1L}^\mathrm{(e\gamma)}(q^2)$ and $F_\mathrm{2,1L}^\mathrm{(e\gamma)}(q^2)$ are the 1-loop radiative corrections of the total electron--photon-vertex form factors $F_1^\mathrm{(e\gamma)}(q^2)$ and $F_2^\mathrm{(e\gamma)}(q^2)$, which are generally written as series over all loop orders as
\begin{equation}
 F_1^\mathrm{(e\gamma)}(q^2)=F_\mathrm{1,0L}^\mathrm{(e\gamma)}+F_\mathrm{1,1L}^\mathrm{(e\gamma)}(q^2)+\ldots,
\end{equation}
\begin{equation}
 F_2^\mathrm{(e\gamma)}(q^2)=F_\mathrm{2,0L}^\mathrm{(e\gamma)}+F_\mathrm{2,1L}^\mathrm{(e\gamma)}(q^2)+\ldots\;.
\end{equation}
Here the leading order terms, $F_\mathrm{1,0L}^\mathrm{(e\gamma)}=1$ and $F_\mathrm{2,0L}^\mathrm{(e\gamma)}=0$, correspond to the tree-level diagram.

Since the renormalization condition in equation~\eqref{eq:renorm_vertex_e-p} is determined at $p=p'$, we have $q=p'-p=0$, and thus, the second term of equation~\eqref{eq:vertexstructure} is zero. Therefore, the renormalization condition considers the form factor $F_1^\mathrm{(e\gamma)}(q^2)$ stating that $F_1^\mathrm{(e\gamma)}(q^2)$ is determined to be $F_1^\mathrm{(e\gamma)}(q^2)=1$ for $q^2=0$ at all loop orders. Consequently, $F_\mathrm{1,1L}^\mathrm{(e\gamma)}(q^2)=0$. The second form factor $F_2^\mathrm{(e\gamma)}(q^2)$ is associated with the anomalous magnetic moment of the electron, discussed in section~\ref{sec:radiativemoment}. Since the 1-photon-loop and 1-graviton-loop radiative corrections and the associated counterterm parts are proportional to different physical constants, the quantity $F_\mathrm{1,1L}^\mathrm{(e\gamma)}(q^2)$ can be split into the 1-photon-loop and 1-graviton-loop associated parts, which are separately zero for $q^2=0$ as
\begin{equation}
 F_\mathrm{1,1L,photon}^\mathrm{(e\gamma)}=F_\mathrm{1,1L,photon}^\mathrm{(e\gamma)}+F_\mathrm{1,1L,CT,photon}^\mathrm{(e\gamma)}=0,
 \label{eq:F1e}
\end{equation}
\begin{equation}
 F_\mathrm{1,1L,graviton}^\mathrm{(e\gamma)}=F_\mathrm{1,1L,graviton}^\mathrm{(e\gamma)}+F_\mathrm{1,1L,CT,graviton}^\mathrm{(e\gamma)}=0.
 \label{eq:F1g}
\end{equation}

The explicit calculation of the 1-photon-loop, 1-graviton-loop, and counterterm contributions to the form factors following from the Feynman diagrams in equations \eqref{eq:vertex1photon}--\eqref{eq:vertex1CT} show that equations \eqref{eq:F1e} and \eqref{eq:F1g} are satisfied identically. The 1-photon-loop contribution to equation~\eqref{eq:F1e} following from equation~\eqref{eq:vertex1photon} and the associated counterterm contribution part following from equations \eqref{eq:Z_psi_photon} and \eqref{eq:vertex1CT} are given  for $q^2=0$ by
\begin{equation}
 F_\mathrm{1,1L,photon}^\mathrm{(e\gamma)}=\frac{\alpha_\mathrm{e}}{4\pi}\Big[\frac{1}{\epsilon_\mathrm{UV}}+\frac{2}{\epsilon_\mathrm{IR}}+4+3\log\!\Big(\frac{4\pi\mu^2e^{-\gamma}}{m_\mathrm{e}^2c^2}\Big)\Big],
 \label{eq:F1e1}
\end{equation}
\begin{equation}
 F_\mathrm{1,1L,CT,photon}^\mathrm{(e\gamma)}=-\frac{\alpha_\mathrm{e}}{4\pi}\Big[\frac{1}{\epsilon_\mathrm{UV}}+\frac{2}{\epsilon_\mathrm{IR}}+4+3\log\!\Big(\frac{4\pi\mu^2e^{-\gamma}}{m_\mathrm{e}^2c^2}\Big)\Big].
 \label{eq:F1eCT}
\end{equation}
The 1-graviton-loop contributions to equation~\eqref{eq:F1g} following from each Feynman diagram in equation~\eqref{eq:vertex1graviton} and the associated counterterm contribution part following from equations \eqref{eq:Z_psi_graviton} and \eqref{eq:vertex1CT} are given for $q^2=0$ by
\begin{equation}
 F_\mathrm{1,1L,graviton}^\mathrm{(e\gamma,diag1)}=-\frac{\alpha_\mathrm{g}}{8\pi}\Big[\frac{7}{\epsilon_\mathrm{UV}}+\frac{4}{\epsilon_\mathrm{IR}}+18+11\log\!\Big(\frac{4\pi\mu^2e^{-\gamma}}{m_\mathrm{e}^2c^2}\Big)\Big],
 \label{eq:F1g1}
\end{equation}
\begin{equation}
 F_\mathrm{1,1L,graviton}^\mathrm{(e\gamma,diag2)}=F_\mathrm{1,1L,graviton}^\mathrm{(e\gamma,diag3)}=F_\mathrm{1,1L,graviton}^\mathrm{(e\gamma,diag6)}=0,
\end{equation}
\begin{align}
 &F_\mathrm{1,1L,graviton}^\mathrm{(e\gamma,diag4)}
 =F_\mathrm{1,1L,graviton}^\mathrm{(e\gamma,diag5)}\nonumber\\
 &=\frac{7\alpha_\mathrm{g}}{8\pi}\Big[\frac{1}{\epsilon_\mathrm{UV}}+2+\log\!\Big(\frac{4\pi\mu^2e^{-\gamma}}{m_\mathrm{e}^2c^2}\Big)\Big],
 \label{eq:F1g4}
\end{align}
\begin{equation}
 F_\mathrm{1,1L,CT,graviton}^\mathrm{(e\gamma)}\!=\!-\frac{\alpha_\mathrm{g}}{8\pi}\!\Big[\frac{7}{\epsilon_\mathrm{UV}}-\frac{4}{\epsilon_\mathrm{IR}}+10+3\log\!\Big(\frac{4\pi\mu^2e^{-\gamma}}{m_\mathrm{e}^2c^2}\Big)\!\Big].
 \label{eq:F1gCT}
\end{equation}
The exact cancellation of the terms in equations \eqref{eq:F1e1} and \eqref{eq:F1eCT} when substituted into equation~\eqref{eq:F1e} is a consequence of the electromagnetic gauge invariance via the Ward--Takahashi identity \cite{Schwartz2014,Peskin2018}. The exact cancellation of the different nonzero terms in equations \eqref{eq:F1g1}, \eqref{eq:F1g4}, and \eqref{eq:F1gCT} when substituted into equation~\eqref{eq:F1g} is likewise a result of the electromagnetic gauge symmetry, and it shows that the Ward--Takahashi identity of QED is satisfied in unified gravity at 1-loop order. This is considered to be a necessary requirement for the consistency of the theory and it should extend to all loop orders.

\subsection{Renormalization of the electron--graviton vertex}

Next, we study the renormalization of the electron--graviton vertex, which is related to the renormalization factors $Z_\mathrm{g\psi}$ and $Z_{\mathrm{g}m}$ in the Lagrangian density terms of equation~\eqref{eq:LUGMrenormalized}. Through the on-shell renormalization conditions, we obtain unambiguous values for the 1-loop contributions to these renormalization factors.

The graviton indices are assumed to be $\mu$ and $\nu$ but, for compactness, we do not show them in the diagrams below. Accordingly, the 1-photon-loop, 1-graviton-loop, and counterterm contributions to the electron--graviton vertex are given by the following Feynman diagrams:
\begin{align}
 &-\frac{i}{2}\sqrt{\frac{\kappa c}{\hbar}}\Gamma_\mathrm{1L,photon}^\mathrm{\mu\nu}\nonumber\\
 &=\adjustbox{valign=c}{\includegraphics{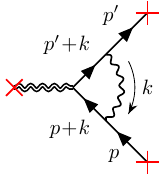}}
\;+\;
\adjustbox{valign=c}{\includegraphics{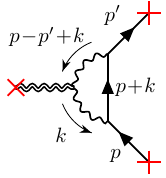}}\nonumber\\
&\;+\;
\adjustbox{valign=c}{\includegraphics{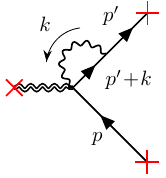}}
\;+\;
\adjustbox{valign=c}{\includegraphics{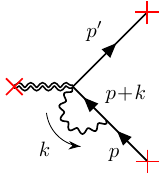}},
\label{eq:vertex2photon}
\end{align}
\begin{align}
 &-\frac{i}{2}\sqrt{\frac{\kappa c}{\hbar}}\Gamma_\mathrm{1L,graviton}^\mathrm{\mu\nu}=\adjustbox{valign=c}{\includegraphics{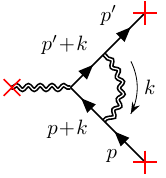}},
\label{eq:vertex2graviton}
\end{align}
\begin{align}
 &-\frac{i}{2}\sqrt{\frac{\kappa c}{\hbar}}\Gamma_\mathrm{1L,CT}^\mathrm{\mu\nu}=\adjustbox{valign=c}{\includegraphics{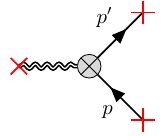}}\nonumber\\
&=-\dfrac{i}{2}\sqrt{\dfrac{\kappa c}{\hbar}}P^{\mu\nu,\rho\sigma}[\delta Z_\mathrm{g\psi}^{(1)}\boldsymbol{\gamma}_{\mathrm{F}\rho}(p+p')_\sigma\nonumber\\
 &\hspace{0.4cm}-(\delta Z_\mathrm{g\psi}^{(1)}+\delta Z_{\mathrm{g}m}^{(1)})m_\mathrm{e}c\eta_{\rho\sigma}\mathbf{I}_4]\nonumber\\
 &=\dfrac{i}{4}\sqrt{\dfrac{\kappa c}{\hbar}}\{\delta Z_\mathrm{g\psi}^{(1)}\eta^{\mu\nu}(\slashed{p}+\slashed{p}')
 -2(\delta Z_\mathrm{g\psi}^{(1)}+\delta Z_{\mathrm{g}m}^{(1)})\eta^{\mu\nu}m_\mathrm{e}c\mathbf{I}_4\nonumber\\
 &\hspace{0.4cm}-\delta Z_\mathrm{g\psi}^{(1)}[\boldsymbol{\gamma}_\mathrm{F}^\mu(p+p')^\nu+\boldsymbol{\gamma}_\mathrm{F}^\nu(p+p')^\mu]\}.
\label{eq:vertex2CT}
\end{align}

The total 1-loop contribution to the electron--graviton vertex is given by
\begin{equation}
 \Gamma_\mathrm{1L}^{\mu\nu}=\Gamma_\mathrm{1L,photon}^\mathrm{\mu\nu}+\Gamma_\mathrm{1L,graviton}^\mathrm{\mu\nu}+\Gamma_\mathrm{1L,CT}^\mathrm{\mu\nu}.
\end{equation}
The counterterm can also be split into photon and graviton parts $\Gamma_\mathrm{1L,CT,photon}^\mathrm{\mu\nu}$ and $\Gamma_\mathrm{1L,CT,graviton}^\mathrm{\mu\nu}$ based on the different contributions to the renormalization factors obtained below.

In analogy with the on-shell renormalization condition of the electron--photon vertex in equation~\eqref{eq:renorm_vertex_e-p}, we define the on-shell renormalization condition for the electron--graviton vertex as
\begin{equation}
 \bar{u}(p')\Gamma_\mathrm{1,1L}^{\mu\nu}u(p)\big|_{p=p'}=0.
 \label{eq:renorm_vertex_e-g}
\end{equation}
The vanishing of the quantum corrections of the electron--graviton vertex for $p=p'$ effectively means that the classical form of the Newtonian potential applies to the strength of the gravitational interaction at asymptotically long distances.

The 1-loop radiative correction to the electron--graviton vertex matrix element can be written in a Lorentz-invariant form as
\begin{align}
 &\bar{u}(p')\Gamma_\mathrm{1,1L}^{\mu\nu}u(p)\nonumber\\
&=\bar{u}(p')\Big\{F_\mathrm{1,1L}^\mathrm{(eg)}m_\mathrm{e}c\eta^{\mu\nu}\mathbf{I}_4\!+\!F_\mathrm{2,1L}^\mathrm{(eg)}\frac{(p\!+\!p')^\mu(p\!+\!p')^\nu}{2m_\mathrm{e}c}\mathbf{I}_4\nonumber\\
&\hspace{0.4cm}+\!F_\mathrm{3,1L}^\mathrm{(eg)}\frac{iq_\rho}{\hbar m_\mathrm{e}c}[(p+p')^\mu\hat{S}_\mathrm{F}^{\nu\rho}\!+\!(p+p')^\nu\hat{S}_\mathrm{F}^{\mu\rho}]\nonumber\\
&\hspace{0.4cm}+\!F_\mathrm{4,1L}^\mathrm{(eg)}q^\mu q^\nu\mathbf{I}_4\Big\}u(p).
\label{eq:vertexstructure2}
\end{align}
Here $F_\mathrm{1,1L}^\mathrm{(eg)}$, $F_\mathrm{2,1L}^\mathrm{(eg)}$, $F_\mathrm{3,1L}^\mathrm{(eg)}$, and $F_\mathrm{4,1L}^\mathrm{(eg)}$ are the 1-loop radiative corrections to the electron-graviton vertex form factors. Through the Gordon identity \cite{Peskin2018}, there is some freedom in the representation of the terms of equation~\eqref{eq:vertexstructure2}.

At $q=p'-p=0$, the last terms of equation~\eqref{eq:vertexstructure2}, associated with $F_\mathrm{3,1L}^\mathrm{(eg)}$ and $F_\mathrm{4,1L}^\mathrm{(eg)}$, are zero. Therefore, the renormalization condition in equation~\eqref{eq:renorm_vertex_e-g} implies that $F_\mathrm{1,1L}^\mathrm{(eg)}$ and $F_\mathrm{2,1L}^\mathrm{(eg)}$ must be zero at the renormalization point. Since the 1-photon-loop and 1-graviton-loop radiative corrections and the associated counterterm parts are proportional to different physical constants, the quantities $F_\mathrm{1,1L}^\mathrm{(eg)}$ and $F_\mathrm{2,1L}^\mathrm{(eg)}$ can also be split into the 1-photon-loop and 1-graviton-loop associated parts, which are separately zero for $q^2=0$ as
\begin{equation}
 F_\mathrm{1,1L,photon}^\mathrm{(eg)}=F_\mathrm{1,1L,photon}^\mathrm{(eg)}+F_\mathrm{1,1L,CT,photon}^\mathrm{(eg)}=0,
 \label{eq:Frenorm_eg1}
\end{equation}
\begin{equation}
 F_\mathrm{2,1L,photon}^\mathrm{(eg)}=F_\mathrm{2,1L,photon}^\mathrm{(eg)}+F_\mathrm{2,1L,CT,photon}^\mathrm{(eg)}=0,
 \label{eq:Frenorm_eg2}
\end{equation}
\begin{equation}
 F_\mathrm{1,1L,graviton}^\mathrm{(eg)}=F_\mathrm{1,1L,graviton}^\mathrm{(eg)}+F_\mathrm{1,1L,CT,graviton}^\mathrm{(eg)}=0,
 \label{eq:Frenorm_eg3}
\end{equation}
\begin{equation}
 F_\mathrm{2,1L,graviton}^\mathrm{(eg)}=F_\mathrm{2,1L,graviton}^\mathrm{(eg)}+F_\mathrm{2,1L,CT,graviton}^\mathrm{(eg)}=0.
 \label{eq:Frenorm_eg4}
\end{equation}

The 1-photon-loop contributions to the form factor $F_\mathrm{1,1L}^\mathrm{(eg)}$ following from each diagram in equation~\eqref{eq:vertex2photon} when compared with the structure of the vertex matrix element in equation~\eqref{eq:vertexstructure2} are given by
\begin{equation}
 F_\mathrm{1,1L,photon}^\mathrm{(eg,diag1)}=\frac{\alpha_\mathrm{e}}{12\pi}\Big[\frac{1}{\epsilon_\mathrm{UV}}-\frac{22}{3}+\log\!\Big(\frac{4\pi\mu^2e^{-\gamma}}{m_\mathrm{e}^2c^2}\Big)\Big],
 \label{eq:F1egdiag1}
\end{equation}
\begin{equation}
 F_\mathrm{1,1L,photon}^\mathrm{(eg,diag2)}=-\frac{\alpha_\mathrm{e}}{3\pi}\Big[\frac{1}{\epsilon_\mathrm{UV}}-\frac{1}{3}+\log\!\Big(\frac{4\pi\mu^2e^{-\gamma}}{m_\mathrm{e}^2c^2}\Big)\Big],
\end{equation}
\begin{align}
 &F_\mathrm{1,1L,photon}^\mathrm{(eg,diag3)}=F_\mathrm{1,1L,photon}^\mathrm{(eg,diag4)}\nonumber\\
 &=\frac{\alpha_\mathrm{e}}{2\pi}\Big[\frac{1}{\epsilon_\mathrm{UV}}+\frac{3}{2}+\log\!\Big(\frac{4\pi\mu^2e^{-\gamma}}{m_\mathrm{e}^2c^2}\Big)\Big].
\end{align}
The corresponding contributions to the form factor $F_\mathrm{2,1L}^\mathrm{(eg)}$, following from each diagram in equation~\eqref{eq:vertex2photon} and the structure of the vertex matrix element in equation~\eqref{eq:vertexstructure2}, are given by
\begin{equation}
 F_\mathrm{2,1L,photon}^\mathrm{(eg,diag1)}=\frac{\alpha_\mathrm{e}}{12\pi}\Big[\frac{1}{\epsilon_\mathrm{UV}}+\frac{6}{\epsilon_\mathrm{IR}}+\frac{56}{3}+7\log\!\Big(\frac{4\pi\mu^2e^{-\gamma}}{m_\mathrm{e}^2c^2}\Big)\Big],
\end{equation}
\begin{equation}
 F_\mathrm{2,1L,photon}^\mathrm{(eg,diag2)}=\frac{2\alpha_\mathrm{e}}{3\pi}\Big[\frac{1}{\epsilon_\mathrm{UV}}+\frac{17}{12}+\log\!\Big(\frac{4\pi\mu^2e^{-\gamma}}{m_\mathrm{e}^2c^2}\Big)\Big],
\end{equation}
\begin{align}
 &F_\mathrm{2,1L,photon}^\mathrm{(eg,diag3)}=F_\mathrm{2,1L,photon}^\mathrm{(eg,diag4)}\nonumber\\
 &=-\frac{\alpha_\mathrm{e}}{4\pi}\Big[\frac{1}{\epsilon_\mathrm{UV}}+3+\log\!\Big(\frac{4\pi\mu^2e^{-\gamma}}{m_\mathrm{e}^2c^2}\Big)\Big].
 \label{eq:F2egdiag4}
\end{align}

Based on equations \eqref{eq:F1egdiag1}--\eqref{eq:F2egdiag4} and the form-factor renormalization conditions in equations \eqref{eq:Frenorm_eg1} and \eqref{eq:Frenorm_eg2}, the form-factor counterterm parts $F_\mathrm{1,1L,CT,photon}^\mathrm{(eg)}$ and $F_\mathrm{2,1L,CT,photon}^\mathrm{(eg)}$ are determined to be
\begin{equation}
 F_\mathrm{1,1L,CT,photon}^\mathrm{(eg)}=-\frac{3\alpha_\mathrm{e}}{4\pi}\Big[\frac{1}{\epsilon_\mathrm{UV}}+\frac{4}{3}+\log\!\Big(\frac{4\pi\mu^2e^{-\gamma}}{m_\mathrm{e}^2c^2}\Big)\Big],
\end{equation}
\begin{equation}
 F_\mathrm{2,1L,CT,photon}^\mathrm{(eg)}=-\frac{\alpha_\mathrm{e}}{4\pi}\Big[\frac{1}{\epsilon_\mathrm{UV}}+\frac{2}{\epsilon_\mathrm{IR}}+4+3\log\!\Big(\frac{4\pi\mu^2e^{-\gamma}}{m_\mathrm{e}^2c^2}\Big)\Big].
\end{equation}
Using the counterterm in equation~\eqref{eq:vertex2CT} and the electron--graviton vertex matrix element in equation~\eqref{eq:vertexstructure2}, we then obtain the 1-photon-loop contributions to the renormalization factors $Z_{\mathrm{g}m}$ and $Z_\mathrm{g\psi}$ as
\begin{equation}
 \delta Z_{\mathrm{g}m,\mathrm{photon}}^{(1)}=-\frac{3\alpha_\mathrm{e}}{4\pi}\Big[\frac{1}{\epsilon_\mathrm{UV}}+\frac{4}{3}+\log\!\Big(\frac{4\pi\mu^2e^{-\gamma}}{m_\mathrm{e}^2c^2}\Big)\Big],
 \label{eq:Z_gm_photon}
\end{equation}
\begin{equation}
 \delta Z_\mathrm{g\psi,photon}^{(1)}=-\frac{\alpha_\mathrm{e}}{4\pi}\Big[\frac{1}{\epsilon_\mathrm{UV}}+\frac{2}{\epsilon_\mathrm{IR}}+4+3\log\!\Big(\frac{4\pi\mu^2e^{-\gamma}}{m_\mathrm{e}^2c^2}\Big)\Big].
 \label{eq:Z_gpsi_photon}
\end{equation}
The comparison of these equations with the 1-photon-loop contributions to the renormalization factors $Z_m$ and $Z_\psi$ in equations \eqref{eq:Z_m_photon} and \eqref{eq:Z_psi_photon}, we observe that the contributions are identical as $\delta Z_{\mathrm{g}m,\mathrm{photon}}^{(1)}=\delta Z_{m,\mathrm{photon}}^{(1)}$ and $\delta Z_\mathrm{g\psi,photon}^{(1)}=\delta Z_\mathrm{\psi,photon}^{(1)}$. This means that electromagnetic interaction makes no difference between the inertial and gravitational masses. Below, we show that a similar equivalence is not satisfied for 1-graviton-loop contributions.

The 1-graviton-loop contribution to the form factor $F_\mathrm{1,1L}^\mathrm{(eg)}$ following from the Feynman diagram in equation~\eqref{eq:vertex2graviton} when compared with the structure of the vertex matrix element in equation~\eqref{eq:vertexstructure2} is given by
\begin{equation}
 F_\mathrm{1,1L,graviton}^\mathrm{(eg)}=-\frac{5\kappa cp^2}{192\pi^2\hbar}\Big[\frac{1}{\epsilon_\mathrm{UV}}-\frac{31}{30}+\log\!\Big(\frac{4\pi\mu^2e^{-\gamma}}{m_\mathrm{e}^2c^2}\Big)\Big].
\end{equation}
The corresponding contribution to the form factor $F_\mathrm{2,1L}^\mathrm{(eg)}$ following from equation~\eqref{eq:vertex2graviton} and the structure of the vertex matrix element in equation~\eqref{eq:vertexstructure2} is given by
\begin{align}
 &F_\mathrm{2,1L,graviton}^\mathrm{(eg)}\nonumber\\
 &=-\frac{3\kappa cp^2}{576\pi^2\hbar}\Big[\frac{11}{\epsilon_\mathrm{UV}}+\frac{12}{\epsilon_\mathrm{IR}}+\frac{172}{3}+23\log\!\Big(\frac{4\pi\mu^2e^{-\gamma}}{m_\mathrm{e}^2c^2}\Big)\Big].
\end{align}

Based on the renormalization conditions in equations \eqref{eq:Frenorm_eg3} and \eqref{eq:Frenorm_eg4}, the 1-graviton-loop form-factor contributions $F_\mathrm{1,1L,CT,graviton}^\mathrm{(eg)}$ and $F_\mathrm{2,1L,CT,graviton}^\mathrm{(eg)}$ are given by
\begin{equation}
 F_\mathrm{1,1L,CT,graviton}^\mathrm{(eg)}=\frac{5\kappa cp^2}{192\pi^2\hbar}\Big[\frac{1}{\epsilon_\mathrm{UV}}-\frac{31}{30}+\log\!\Big(\frac{4\pi\mu^2e^{-\gamma}}{m_\mathrm{e}^2c^2}\Big)\Big],
\end{equation}
\begin{align}
 &F_\mathrm{2,1L,CT,graviton}^\mathrm{(eg)}\nonumber\\
 &=\frac{3\kappa cp^2}{576\pi^2\hbar}\Big[\frac{11}{\epsilon_\mathrm{UV}}+\frac{12}{\epsilon_\mathrm{IR}}+\frac{172}{3}+23\log\!\Big(\frac{4\pi\mu^2e^{-\gamma}}{m_\mathrm{e}^2c^2}\Big)\Big].
 \label{eq:F2eggraviton}
\end{align}
Using the counterterm in equation~\eqref{eq:vertex2CT} and the electron--graviton vertex matrix element in equation~\eqref{eq:vertexstructure2}, we then obtain the 1-graviton-loop contributions to the renormalization factors $Z_{\mathrm{g}m}$ and $Z_\mathrm{g\psi}$ as
\begin{equation}
 \delta Z_{\mathrm{g}m,\mathrm{graviton}}^{(1)}=\frac{5\kappa cp^2}{192\pi^2\hbar}\Big[\frac{1}{\epsilon_\mathrm{UV}}-\frac{31}{30}+\log\!\Big(\frac{4\pi\mu^2e^{-\gamma}}{m_\mathrm{e}^2c^2}\Big)\Big],
 \label{eq:Z_gm_graviton}
\end{equation}
\begin{align}
 &\delta Z_\mathrm{g\psi,graviton}^{(1)}\nonumber\\
 &=\frac{\kappa cp^2}{192\pi^2\hbar}\Big[\frac{11}{\epsilon_\mathrm{UV}}+\frac{12}{\epsilon_\mathrm{IR}}+\frac{172}{3}+23\log\!\Big(\frac{4\pi\mu^2e^{-\gamma}}{m_\mathrm{e}^2c^2}\Big)\Big].
 \label{eq:Z_gpsi_graviton}
\end{align}
In contrast to the case of the 1-photon-loop contributions in equations \eqref{eq:Z_gm_photon} and \eqref{eq:Z_gpsi_photon}, the 1-graviton-loop contributions to $Z_{\mathrm{g}m}$ and $Z_\mathrm{g\psi}$ in equations \eqref{eq:Z_gm_graviton} and \eqref{eq:Z_gpsi_graviton} are not equivalent to the 1-graviton-loop contributions to the renormalization factors $Z_m$ and $Z_\psi$ in equations \eqref{eq:Z_m_graviton} and \eqref{eq:Z_psi_graviton}. This shows that the different renormalization factors of the inertial and gravitational masses are necessary for the renormalizability of unified gravity.

\subsection{Renormalization of the photon--graviton vertex}

Next, we study the renormalization of the photon--graviton vertex. In unified gravity, the renormalization of this vertex is related to the renormalization factor $Z_{\mathrm{g}A}$ in the Lagrangian density terms of equation~\eqref{eq:LUGMrenormalized}. The 1-loop contribution to the renormalization factor $Z_{\mathrm{g}A}$ can already be determined based on the equality $Z_{\mathrm{g}A}=Z_AZ_\mathrm{g\psi}/Z_\psi$, following from equation~\eqref{eq:Zequalities1}, and the 1-loop contributions to the renormalization factors $Z_A$, $Z_\mathrm{g\psi}$, and $Z_\psi$, determined in the previous sections. Using equations \eqref{eq:Z_psi_photon}, \eqref{eq:Z_psi_graviton}, \eqref{eq:ZAe}, \eqref{eq:ZAg}, \eqref{eq:Z_gpsi_photon}, and \eqref{eq:Z_gpsi_graviton} we then obtain
\begin{align}
 \delta Z_{\mathrm{g}A,\mathrm{electron}}^{(1)} &=\delta Z_{A,\mathrm{electron}}^{(1)}\nonumber\\
 &=-\frac{\alpha_\mathrm{e}}{3\pi}\Big[\frac{1}{\epsilon_\mathrm{UV}}+\log\!\Big(\frac{4\pi\mu^2e^{-\gamma}}{m_\mathrm{e}^2c^2}\Big)\Big],
 \label{eq:ZgAe}
\end{align}
\begin{align}
 \delta Z_{\mathrm{g}A,\mathrm{graviton}}^{(1)} &=\delta Z_{A,\mathrm{graviton}}^{(1)}+\delta Z_\mathrm{g\psi,graviton}^{(1)}-\delta Z_\mathrm{\psi,graviton}^{(1)} &\nonumber\\
 &=\frac{\kappa c p^2}{8\pi^2\hbar}\Big[\frac{1}{\epsilon_\mathrm{UV}}+\frac{43}{12}+\log\!\Big(\frac{4\pi\mu^2e^{-\gamma}}{(m_\mathrm{e}^2c^2)^{4/3}}\Big)\Big].
 \label{eq:ZgAg}
\end{align}
Therefore, in the renormalization of the photon--graviton vertex below, we do not obtain any new relations for the renormalization factors but only verify that the renormalization condition of the photon--graviton vertex is satisfied identically.

We assume the same indices for the fields as in the photon--graviton vertex in table~\ref{tbl:Feynman} but, for compactness, do not show them in the diagrams below. Accordingly, the 1-electron-loop, 1-graviton-loop, and counterterm contributions to the photon--graviton vertex are given by
\begin{align}
 &-i\sqrt{\frac{\kappa c}{\hbar}}\Gamma_\mathrm{1L,electron}^\mathrm{\mu\nu,\sigma,\rho}\nonumber\\
 &=\adjustbox{valign=c, raise=0.4mm}{\includegraphics{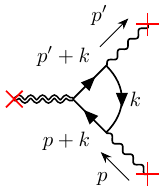}}
\;+\;
\adjustbox{valign=c, raise=0.3mm}{\includegraphics{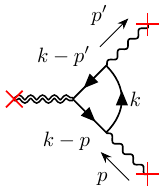}}\nonumber\\[-0.2cm]
&\;+\;
\adjustbox{valign=c, raise=0.8mm}{\includegraphics{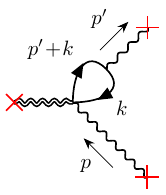}}
\;+\;
\adjustbox{valign=c, raise=-0.8mm}{\includegraphics{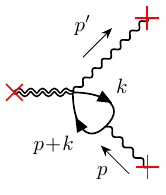}},
\label{eq:vertex4electron}
\end{align}
\begin{align}
 &-i\sqrt{\frac{\kappa c}{\hbar}}\Gamma_\mathrm{1L,graviton}^\mathrm{\mu\nu,\sigma,\rho}=\adjustbox{valign=c, raise=0.3mm}{\includegraphics{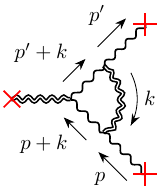}},
\label{eq:vertex4graviton}
\end{align}
\vspace{-0.2cm}
\begin{align}
 &-i\sqrt{\frac{\kappa c}{\hbar}}\Gamma_\mathrm{1L,CT}^\mathrm{\mu\nu,\sigma,\rho}=\adjustbox{valign=c, raise=0.3mm}{\includegraphics{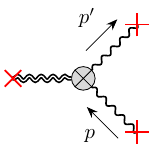}}\nonumber\\
&=-i\delta Z_{\mathrm{g}A}^{(1)}\sqrt{\dfrac{\kappa c}{\hbar}}P^{\mu\nu,\sigma\eta,\rho\lambda}p'_\eta p_\lambda.
\label{eq:vertex4CT}
\end{align}

The total 1-loop contribution to the photon--graviton vertex is given by
\begin{equation}
 \Gamma_\mathrm{1L}^{\mu\nu,\sigma,\rho}=\Gamma_\mathrm{1L,electron}^\mathrm{\mu\nu,\sigma,\rho}+\Gamma_\mathrm{1L,graviton}^\mathrm{\mu\nu,\sigma,\rho}+\Gamma_\mathrm{1L,CT}^\mathrm{\mu\nu,\sigma,\rho}.
\end{equation}
The counterterm can also be split into the electron and graviton parts $\Gamma_\mathrm{1L,CT,electron}^\mathrm{\mu\nu,\sigma,\rho}$ and $\Gamma_\mathrm{1L,CT,graviton}^\mathrm{\mu\nu,\sigma,\rho}$ based on substituting the renormalization factors of equations \eqref{eq:ZgAe} and \eqref{eq:ZgAg} into equation~\eqref{eq:vertex4CT}.

The on-shell renormalization condition of the photon--graviton vertex is imposed by requiring that $\Gamma_\mathrm{1L}^{\mu\nu,\sigma,\rho}$ and any higher-order-loop terms do not lead to corrections of the tree-level vertex for $p=p'$. This requirement is analogous to the renormalization conditions of the electron--photon and electron--graviton vertices above. For on-shell photons, we naturally have $p^2=0$. Therefore, the matrix element of $\Gamma_\mathrm{1L}^{\mu\nu,\sigma,\rho}$ must be zero for on-shell states with $p=p'$ as
\begin{equation}
 \epsilon_\sigma^*(p')\Gamma_\mathrm{1L}^{\mu\nu,\sigma,\rho}\epsilon_\rho(p)\big|_{p=p'}=0.
 \label{eq:renorm_vertex_p-g}
\end{equation}
The vanishing of the quantum corrections of the photon--graviton vertex for $p=p'$ effectively means that the photon--graviton interactions do not modify the classical forms of the Coulomb and Newtonian potentials, which apply to the strengths of the electromagnetic and gravitational interactions at asymptotically long distances.

The scalar product of $\Gamma_\mathrm{1L}^{\mu\nu,\sigma,\rho}$ with the photon polarization vectors in equation~\eqref{eq:renorm_vertex_p-g} effectively means that the terms of $\Gamma_\mathrm{1L}^{\mu\nu,\sigma,\rho}$ proportional to $p'^\sigma p^\rho$ do not contribute. This is due to the transversality of the photon momentum and polarization vectors, i.e., $p^\rho\epsilon_\rho(p)=0$. In more complex calculations of the S-matrix elements, the photon polarization vectors can be replaced by the photon propagators, which connect to fermion lines. In this case, through the summation over all places along the fermion line where the photon propagator could connect, one obtains that the terms proportional to $p^\rho$ or $p'^\sigma$ vanish. This is due to the Ward identity and is analogous to the discussion on the last term of the photon propagator below equation~\eqref{eq:photonpropagatorrenormalized}.

The 1-electron-loop contributions to the photon--graviton vertex resulting from each Feynman diagram in equation~\eqref{eq:vertex4electron} are given for $p=p'$ by
\vspace{-0.2cm}
\begin{align}
 &\Gamma_\mathrm{1L,electron}^\mathrm{(diag1)\mu\nu,\sigma,\rho}\big|_{p=p'}=\Gamma_\mathrm{1L,electron}^\mathrm{(diag2)\mu\nu,\sigma,\rho}\big|_{p=p'}\nonumber\\
 &=-\frac{\alpha_\mathrm{e}}{12\pi}\Big[\frac{1}{\epsilon_\mathrm{UV}}+\log\!\Big(\frac{4\pi\mu^2e^{-\gamma}}{m_\mathrm{e}^2c^2}\Big)\Big](4p^\mu p^\nu\eta^{\rho\sigma}\nonumber\\
 &\hspace{0.4cm}-p^\mu p^\rho\eta^{\nu\sigma}-p^\nu p^\rho\eta^{\mu\sigma}-p^\mu p^\sigma\eta^{\nu\rho}-p^\nu p^\sigma\eta^{\mu\rho})\nonumber\\
 &\hspace{0.4cm}+\frac{\alpha_\mathrm{e}}{15\pi m_\mathrm{e}^2c^2}p^\mu p^\nu p^\rho p^\sigma,
 \label{eq:egloop1}
\end{align}
\vspace{-0.5cm}
\begin{align}
 &\Gamma_\mathrm{1L,electron}^\mathrm{(diag3)\mu\nu,\sigma,\rho}\big|_{p=p'}\nonumber\\
 &=\frac{\alpha_\mathrm{e}}{6\pi}\Big[\frac{1}{\epsilon_\mathrm{UV}}+\log\!\Big(\frac{4\pi\mu^2e^{-\gamma}}{m_\mathrm{e}^2c^2}\Big)\Big]\nonumber\\
 &\hspace{0.4cm}\times p^\rho(p^\mu\eta^{\nu\sigma}+p^\nu\eta^{\mu\sigma}-p^\sigma\eta^{\mu\nu}),
\end{align}
\vspace{-0.6cm}
\begin{align}
 &\Gamma_\mathrm{1L,electron}^\mathrm{(diag4)\mu\nu,\sigma,\rho}\big|_{p=p'}\nonumber\\
 &=\frac{\alpha_\mathrm{e}}{6\pi}\Big[\frac{1}{\epsilon_\mathrm{UV}}+\log\!\Big(\frac{4\pi\mu^2e^{-\gamma}}{m_\mathrm{e}^2c^2}\Big)\Big]\nonumber\\
 &\hspace{0.4cm}\times p^\sigma(p^\mu\eta^{\nu\rho}+p^\nu\eta^{\mu\rho}-p^\rho\eta^{\mu\nu}).
 \label{eq:egloop4}
\end{align}

The summation of the contributions of all diagrams, given in equations \eqref{eq:egloop1}--\eqref{eq:egloop4}, results in the total 1-electron-loop contribution to the photon--graviton vertex without the counterterm as
\vspace{-0.2cm}
\begin{align}
 &\Gamma_\mathrm{1L,electron}^{\mu\nu,\sigma,\rho}\big|_{p=p'}\nonumber\\
 &=\frac{\alpha_\mathrm{e}}{3\pi}\Big[\frac{1}{\epsilon_\mathrm{UV}}+\log\!\Big(\frac{4\pi\mu^2e^{-\gamma}}{m_\mathrm{e}^2c^2}\Big)\Big]P^{\mu\nu,\sigma\eta,\rho\lambda}p'_\eta p_\lambda\nonumber\\
 &\hspace{0.4cm}+\frac{\alpha_\mathrm{e}}{15\pi m_\mathrm{e}^2c^2}p^\mu p^\nu p^\rho p^\sigma.
 \label{eq:egloops}
\end{align}

\vspace{-0.2cm}\noindent
The coefficient of the first term of equation~\eqref{eq:egloops} is found to be equal in magnitude and opposite in sign to the 1-electron-loop contribution to the renormalization factor $Z_{\mathrm{g}A}$, given in equation~\eqref{eq:ZgAe}. Therefore, this term becomes cancelled by the electron-loop part of the counterterm in equation~\eqref{eq:vertex4CT}. The contribution of the second term of equation~\eqref{eq:egloops} vanishes in the S-matrix element calculations according to the discussion below equation~\eqref{eq:renorm_vertex_p-g}. Therefore, this term can be dropped out, and we conclude that the renormalization condition in equation~\eqref{eq:renorm_vertex_p-g} is satisfied.

Correspondingly, the 1-graviton-loop contribution to the photon--graviton vertex resulting from the Feynman diagram in equation~\eqref{eq:vertex4graviton} is given for $p=p'$ by
\begin{align}
 &\Gamma_\mathrm{1L,graviton}^{\mu\nu,\sigma,\rho}\big|_{p=p'}\nonumber\\
 &=-\frac{\kappa c}{8\pi^2\hbar}\Big[\frac{1}{\epsilon_\mathrm{UV}}\!+\!\frac{11}{6}\!+\!\log\!\Big(\!-\!\frac{4\pi\mu^2e^{-\gamma}}{p^2}\Big)\Big]p^\mu p^\nu p^\rho p^\sigma.
 \label{eq:eggraviton}
\end{align}
Here we have used the on-shell condition of photons, $p^2=0$, apart from the logarithm, where $p^2$ appears in the denominator. By the same argument as discussed in the case of the electron-loop contribution below equations \eqref{eq:renorm_vertex_e-g} and \eqref{eq:egloops}, the terms proportional to $p^\rho$ or $p'^\sigma$ do not contribute to the S-matrix elements. Therefore, we conclude that the renormalization condition in equation~\eqref{eq:renorm_vertex_p-g} is satisfied, and the 1-graviton loop does not contribute to the renormalization of the photon--graviton vertex. The counterterm in equation~\eqref{eq:vertex4CT} is in agreement with this result since the 1-graviton-loop contribution to $Z_{\mathrm{g}A}$ is zero at the physically meaningful on-shell renormalization point as seen from equation~\eqref{eq:ZgAg} with $p^2=0$.

\subsection{Renormalization of the electron--photon--graviton vertex}

Finally, we study the renormalization of the electron--photon--graviton vertex. This vertex is related to the renormalization factor $Z_{\mathrm{g}\psi}$ in the Lagrangian density terms of equation~\eqref{eq:LUGMrenormalized}. Since the 1-loop contributions to this renormalization factor is already determined in equations \eqref{eq:Z_gpsi_photon} and \eqref{eq:Z_gpsi_graviton}, in this section we do not obtain any new relations for the renormalization factors but only verify that the renormalization condition of the electron--photon--graviton vertex is satisfied identically.

We assume the same indices for the fields as in the electron--photon--graviton vertex in table~\ref{tbl:Feynman} but, for compactness, do not show them in the diagrams below. Accordingly, the 1-photon-loop, 1-graviton-loop, and counterterm contributions to the electron--photon--graviton vertex are given by the following Feynman diagrams:
\begin{widetext}
\vspace{-0.8cm}
\begin{align}
 &\dfrac{iq_\mathrm{e}}{\hbar}\sqrt{\dfrac{\kappa}{\varepsilon_0}}\Gamma_\mathrm{1L,photon}^\mathrm{\mu\nu,\rho}\nonumber\\
 &=\adjustbox{valign=c}{\includegraphics{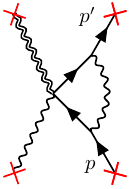}}
+
\adjustbox{valign=c}{\includegraphics{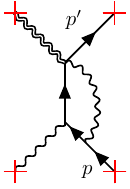}}
+
\adjustbox{valign=c}{\includegraphics{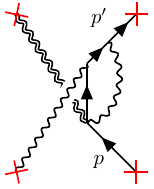}}
+
\adjustbox{valign=c}{\includegraphics{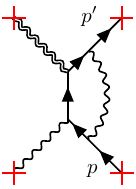}}
+
\adjustbox{valign=c}{\includegraphics{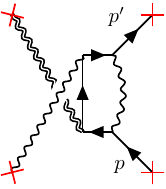}}
+
\adjustbox{valign=c}{\includegraphics{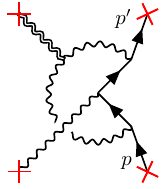}},
\label{eq:vertex_epg_photon}
\end{align}
\begin{align}
 &\dfrac{iq_\mathrm{e}}{\hbar}\sqrt{\dfrac{\kappa}{\varepsilon_0}}\Gamma_\mathrm{1L,graviton}^\mathrm{\mu\nu,\rho}\nonumber\\
 &=\adjustbox{valign=c}{\includegraphics{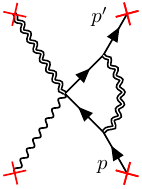}}
+
\adjustbox{valign=c}{\includegraphics{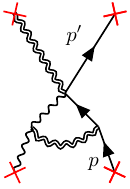}}
+
\adjustbox{valign=c}{\includegraphics{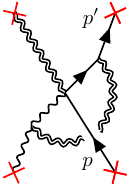}}
+
\adjustbox{valign=c}{\includegraphics{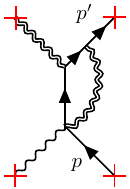}}
+
\adjustbox{valign=c}{\includegraphics{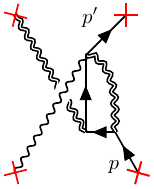}}
+
\adjustbox{valign=c}{\includegraphics{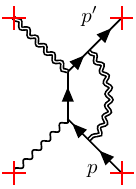}}\nonumber\\
&\;+
\adjustbox{valign=c}{\includegraphics{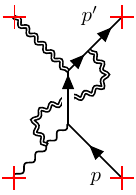}}
+
\adjustbox{valign=c}{\includegraphics{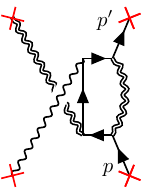}}
+
\adjustbox{valign=c}{\includegraphics{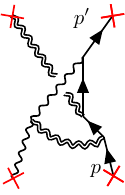}}
+
\adjustbox{valign=c}{\includegraphics{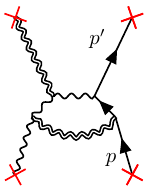}}
+
\adjustbox{valign=c}{\includegraphics{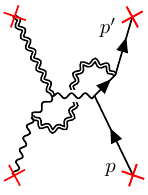}}
+
\adjustbox{valign=c}{\includegraphics{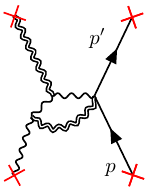}},
\label{eq:vertex_epg_graviton}
\end{align}
\begin{align}
 &\dfrac{iq_\mathrm{e}}{\hbar}\sqrt{\dfrac{\kappa}{\varepsilon_0}}\Gamma_\mathrm{1L,2P\&3P}^\mathrm{\mu\nu,\rho}=
\adjustbox{valign=c}{\includegraphics{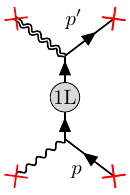}}
+
\adjustbox{valign=c}{\includegraphics{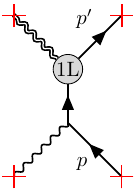}}
+
\adjustbox{valign=c}{\includegraphics{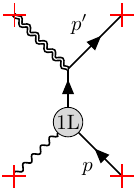}}
+
\adjustbox{valign=c}{\includegraphics{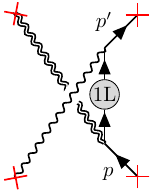}}
+
\adjustbox{valign=c}{\includegraphics{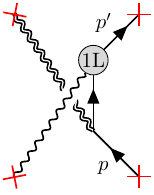}}\nonumber\\
&\;+
\adjustbox{valign=c}{\includegraphics{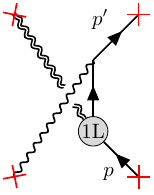}}
+
\adjustbox{valign=c}{\includegraphics{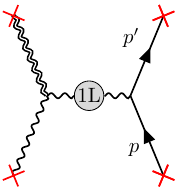}}
+
\adjustbox{valign=c}{\includegraphics{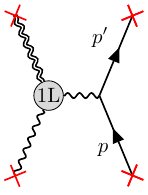}}
+
\adjustbox{valign=c}{\includegraphics{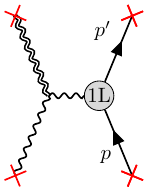}},
\label{eq:vertex4p2p3}
\end{align}
\vspace{-0.2cm}
\end{widetext}

\begin{align}
 &\dfrac{iq_\mathrm{e}}{\hbar}\sqrt{\dfrac{\kappa}{\varepsilon_0}}\Gamma_\mathrm{1L,CT}^\mathrm{\mu\nu,\rho}=\adjustbox{valign=c}{\includegraphics{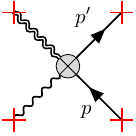}}\nonumber\\
&=(Z_{\mathrm{g}\psi}-1)\dfrac{iq_\mathrm{e}}{\hbar}\sqrt{\dfrac{\kappa}{\varepsilon_0}}P^{\mu\nu,\rho\sigma}\boldsymbol{\gamma}_{\mathrm{F}\sigma}.
\label{eq:vertex_epg_CT}
\end{align}

Equation \eqref{eq:vertex4p2p3} represents the Feynman diagrams, which are already renormalized by the counterterms of the two-point and three-point functions discussed in the previous sections. Accordingly, each 1L-blob in the Feynman diagrams of equation~\eqref{eq:vertex4p2p3} contains the 1-loop diagrams and counterterms associated with the pertinent propagator or vertex.

The total 1-loop contribution to the electron--photon--graviton vertex corresponding to the Feynman diagrams in equations \eqref{eq:vertex_epg_photon}, \eqref{eq:vertex_epg_graviton} and \eqref{eq:vertex_epg_CT} is given by
\begin{equation}
 \Gamma_\mathrm{1L}^{\mu}=\Gamma_\mathrm{1L,photon}^{\mu\nu,\rho}+\Gamma_\mathrm{1L,graviton}^\mathrm{\mu\nu,\rho}+\Gamma_\mathrm{1L,CT}^{\mu\nu,\rho}.
\end{equation}
The counterterm can also be split into photon and graviton parts $\Gamma_\mathrm{1L,CT,photon}^{\mu\nu,\rho}$ and $\Gamma_\mathrm{1L,CT,graviton}^{\mu\nu,\rho}$ based on the different contributions to the renormalization factors obtained below.

In analogy with the on-shell renormalization conditions of the electron--photon vertex in equation~\eqref{eq:renorm_vertex_e-p}, the electron--graviton vertex in equation~\eqref{eq:renorm_vertex_e-g}, and the photon--graviton vertex in equation~\eqref{eq:renorm_vertex_p-g}, we define the on-shell renormalization condition for the electron--photon--graviton vertex. Since it is a four-point vertex, following previous literature \cite{Schwartz2014}, the renormalization condition is defined in terms of the Mandelstam variables. Accordingly, we write the renormalization condition as
\begin{equation}
 \bar{u}(p')\Gamma_\mathrm{1,1L}^{\mu\nu,\rho}u(p)\epsilon_{\mu\nu}(q')\epsilon_\rho(q)\Big|_{\begin{array}{l}\scriptstyle s=u=m_\mathrm{e}^2c^2\\[-6pt]\scriptstyle t=0\end{array}}=0.
 \label{eq:renorm_vertex_e-p-g}
\end{equation}
Here the photon and graviton four-momenta $q$ and $q'$ satisfy $q'-q=p'-p$.
For the Mandelstam variables, we use the values $s=u=m_\mathrm{e}^2c^2$ and $t=0$.

The Mandelstam variable $t=0$ implies $p=p'$ and $q=q'$. We consider the rest frame of the electron where $p=p'=(m_\mathrm{e}c,0,0,0)$. The Mandelstam variables $s=u=m_\mathrm{e}^2c^2$ then imply $q=q'=(0,0,0,0)$. Through the use of the Gordon identity, all terms of $\bar{u}(p')\Gamma_\mathrm{1,1L}^{\mu\nu,\rho}u(p)$ resulting from the 1-photon and 1-graviton loop diagrams in equations \eqref{eq:vertex_epg_photon} and \eqref{eq:vertex_epg_graviton} and from the counterterm in equation~\eqref{eq:vertex_epg_CT} are found to be either zero or proportional to $p^\mu$, $p^\nu$, or $p^\rho$ at the renormalization point. In the rest frame of the electron, the only nonzero component of the electron four-momentum is the time component. In contrast, since photons and gravitons are massless, the photon polarization four-vector $\epsilon_\rho(0)$ and the graviton polarization tensor $\epsilon_{\mu\nu}(0)$ are defined to have only spatial components. Therefore, the scalar product of the electron four-momentum and the photon polarization four-vector is zero independently of the spatial orientation of the photon polarization four-vector as $p^\rho\epsilon_\rho(0)=0$. The same applies to the contraction of the electron four-momentum with either of the two indices of the graviton polarization tensor as $p^\mu\epsilon_{\mu\nu}(0)=0$ and $p^\nu\epsilon_{\mu\nu}(0)=0$. Thus, the 1-photon and 1-graviton loop diagrams in equations \eqref{eq:vertex_epg_photon} and \eqref{eq:vertex_epg_graviton} do not contribute to the S-matrix element calculations, and we conclude that the renormalization condition in equation~\eqref{eq:renorm_vertex_e-p-g} is satisfied.

\section{\label{sec:radiative}Radiative corrections}

In this section, we study the radiative corrections of unified gravity in the UGM formulation. In quantum field theories, radiative corrections refer to the modifications in particle interactions and the related physical quantities due to the effects of virtual particles \cite{Schwartz2014,Peskin2018}. The virtual-particle interactions are described by the Feynman diagrams as discussed in the previous sections. The radiative corrections follow from the loop-order Feynman diagrams. The radiative corrections play a crucial role in obtaining accurate quantum-field-theoretical predictions for scattering processes, which can be verified experimentally. For example, the prediction of the Standard Model for the electron magnetic moment has been experimentally verified to an astonishing accuracy of one part in a trillion \cite{Fan2023}. In the sections below, we consider how unified gravity leads to the radiative corrections to the Coulomb and Newtonian potentials and to the electron magnetic moment.

\subsection{\label{sec:radiativeCoulomb}Radiative corrections to the Coulomb potential}

As an example of the radiative corrections to the Coulomb potential, we calculate the 1-electron-loop and 1-graviton-loop corrections resulting from the vacuum polarization diagrams and their counterterm in equations \eqref{eq:photon1Lelectron}--\eqref{eq:photon1LCT}. Detailed study of the contributions resulting from the box and crossed box diagrams, the triangular diagrams, the circular diagram, and the vertex corrections, similar to those studied in \cite{Butt2006}, is left to a separate work.

In the nonrelativistic limit, for the four-momentum of the exchanged photon, we have $p^2\rightarrow-|\mathbf{p}|^2$, where $\mathbf{p}$ is the three-dimensional momentum vector. Thus, the 1-electron-loop and 1-graviton-loop renormalized scalar amplitude factors of the photon self-energy in equations \eqref{eq:Pi_1L_electron} and \eqref{eq:Pi_1L_graviton} become
\begin{align}
 &\Pi_\mathrm{1L,electron}^\mathrm{(r)}(p^2)\approx
 \Pi_\mathrm{1L,electron}^\mathrm{(r)}(-|\mathbf{p}|^2)\nonumber\\
 &=-\frac{\alpha_\mathrm{e}}{3\pi}\Big\{\frac{5}{3}-\frac{4m_\mathrm{e}^2c^2}{|\mathbf{p}|^2}+\Big(1-\frac{2m_\mathrm{e}^2c^2}{|\mathbf{p}|^2}\Big)\sqrt{1\!+\!\frac{4m_\mathrm{e}^2c^2}{|\mathbf{p}|^2}}\nonumber\\
&\hspace{0.4cm}\times\log\!\Big[1\!-\!\frac{|\mathbf{p}|^2}{2m_\mathrm{e}^2c^2}\Big(\sqrt{1\!+\!\frac{4m_\mathrm{e}^2c^2}{|\mathbf{p}|^2}}\!-\!1\Big)\Big]\Big\},
\label{eq:Pi_1L_electron_p}
\end{align}
\begin{align}
 \Pi_\mathrm{1L,graviton}^\mathrm{(r)}(p^2) &\approx
 \Pi_\mathrm{1L,graviton}^\mathrm{(r)}(-|\mathbf{p}|^2)\nonumber\\
 &=-\frac{\kappa c|\mathbf{p}|^2}{24\pi^2\hbar}\log(|\mathbf{p}|^2).
 \label{eq:Pi_1L_graviton_p}
\end{align}

The Fourier transform of the Coulomb potential obtains correction terms, which are additions to the tree-level Fourier transform of the Coulomb potential in equation~(S27) of the supplementary material. Using equations \eqref{eq:Pi_1L_electron_p} and \eqref{eq:Pi_1L_graviton_p}, the renormalized 1-electron-loop and 1-graviton-loop correction terms are given by
\begin{equation}
 \widetilde{V}_\mathrm{e,1L,electron}(\mathbf{p})=\frac{\hbar^2e^2}{\varepsilon_0(|\mathbf{p}|^2-i\epsilon)}\Pi_\mathrm{1L,electron}^\mathrm{(r)}(-|\mathbf{p}|^2),
 \label{eq:Ve_Fourier1}
\end{equation}
\begin{equation}
 \widetilde{V}_\mathrm{e,1L,graviton}(\mathbf{p})=\frac{\hbar^2e^2}{\varepsilon_0(|\mathbf{p}|^2-i\epsilon)}\Pi_\mathrm{1L,graviton}^\mathrm{(r)}(-|\mathbf{p}|^2).
 \label{eq:Ve_Fourier2}
\end{equation}

\subsubsection{Renormalized 1-electron-loop contribution}

First, we consider the renormalized 1-electron-loop contribution to the radiative correction of the Coulomb potential. This radiative correction is known from QED \cite{Peskin2018,Landau1982} and is presented here for completeness and as a background for the 1-graviton-loop contribution calculated below and for the radiative correction of the Newtonian potential studied in section~\ref{sec:radiativeNewton}.

Using the renormalized 1-electron-loop contribution to the Fourier transform of the Coulomb potential correction in equation~\eqref{eq:Ve_Fourier1}, and performing the inverse Fourier transform, we obtain the renormalized 1-electron-loop contribution to the Coulomb potential correction as
\begin{align}
 &V_\mathrm{e,1L,electron}(\mathbf{r})\nonumber\\
 &\!=\!\int\widetilde{V}_\mathrm{e,1L,electron}(\mathbf{p})e^{i\mathbf{p}\cdot\mathbf{r}/\hbar}\frac{d^3p}{(2\pi\hbar)^3}\nonumber\\
 &\!=\!\frac{\hbar^2e^2}{(2\pi\hbar)^3\varepsilon_0}\int_0^\infty\!\int_0^\pi\int_0^{2\pi}\!\frac{e^{ip_\mathrm{r}r\cos\theta_\mathrm{r}/\hbar}}{(p_\mathrm{r}^2-i\epsilon)}\Pi_\mathrm{1L,electron}^\mathrm{(r)}(-p_\mathrm{r}^2)\nonumber\\
 &\hspace{0.4cm}\times p_\mathrm{r}^2\sin\theta_\mathrm{r}d\phi_\mathrm{r}d\theta_\mathrm{r}dp_\mathrm{r}\nonumber\\
 &\!=\!\frac{\hbar^2e^2}{(2\pi\hbar)^3\varepsilon_0}\!\frac{2\pi\hbar}{r}\int_0^\infty\!\Big|_0^\pi\frac{ip_\mathrm{r}e^{ip_\mathrm{r}r\cos\theta_\mathrm{r}/\hbar}}{p_\mathrm{r}^2-i\epsilon}\Pi_\mathrm{1L,electron}^\mathrm{(r)}(-p_\mathrm{r}^2)dp_\mathrm{r}\nonumber\\
 &\!=\!\frac{e^2}{4\pi^2\varepsilon_0r}\!\!\int_0^\infty\!\frac{ip_\mathrm{r}(e^{-ip_\mathrm{r}r/\hbar}\!-\!e^{ip_\mathrm{r}r/\hbar})}{p_\mathrm{r}^2-i\epsilon}\Pi_\mathrm{1L,electron}^\mathrm{(r)}(-p_\mathrm{r}^2)dp_\mathrm{r}\nonumber\\
 &\!=\!\frac{e^2}{8\pi^2\varepsilon_0r}\!\!\int_{-\infty}^\infty\!\!\!\frac{ip_\mathrm{r}(e^{-ip_\mathrm{r}r/\hbar}\!\!-\!e^{ip_\mathrm{r}r/\hbar})}{p_\mathrm{r}^2-i\epsilon}\Pi_\mathrm{1L,electron}^\mathrm{(r)}(-p_\mathrm{r}^2)dp_\mathrm{r}.
 \label{eq:Ve_1L_electron_calculation1}
\end{align}
In the second equality of equation~\eqref{eq:Ve_1L_electron_calculation1}, we have expressed the momentum-space integral in spherical coordinates $(p_\mathrm{r},\theta_\mathrm{r},\phi_\mathrm{r})$. In the third and fourth equalities, we have carried out the integration with respect to the angular variables $\theta_\mathrm{r}$ and $\phi_\mathrm{r}$. In the last equality of equation~\eqref{eq:Ve_1L_electron_calculation1}, we have applied the symmetry of the integrand with respect to $p_\mathrm{r}$ to change the lower bound of the integral to the negative infinity.

As a complex function, the imaginary part of $\Pi_\mathrm{1L,electron}^\mathrm{(r)}(-p_\mathrm{r}^2)$ in the integrand of equation~\eqref{eq:Ve_1L_electron_calculation1} has branch cuts at the positive and negative imaginary axes starting at $p_\mathrm{r}=\pm 2im_\mathrm{e}c$ and extending to infinities. This must be accounted for when applying the residue theorem to the integral of equation~\eqref{eq:Ve_1L_electron_calculation1}. Writing $p_\mathrm{r}=\pm\epsilon+ip_\mathrm{ri}$, where $p_\mathrm{ri}=\mathrm{Im}(p_\mathrm{r})$, we obtain
\begin{equation}
 \mathrm{Im}[\Pi_\mathrm{1L,electron}^\mathrm{(r)}(-p_\mathrm{r}^2)]=\pm\frac{\alpha_\mathrm{e}}{3}\sqrt{1-\frac{4m_\mathrm{e}^2c^2}{p_\mathrm{ri}^2}}\Big(1+\frac{2m_\mathrm{e}^2c^2}{p_\mathrm{ri}^2}\Big).
 \label{eq:branchcuts1}
\end{equation}
The positive and negative signs of equation~\eqref{eq:branchcuts1} correspond to $\mathrm{Re}(p_\mathrm{r})>0$ and $\mathrm{Re}(p_\mathrm{r})<0$, respectively.

Applying the residue theorem and the branch cuts of the integrand to calculate the integral of equation~\eqref{eq:Ve_1L_electron_calculation1}, we obtain 
\begin{align}
 &V_\mathrm{e,1L,electron}(\mathbf{r})\nonumber\\
 &\!=\!\frac{e^2}{8\pi^2\varepsilon_0r}\Big\{\!\!-\!2\pi i\!\!\!\underset{p_\mathrm{r}=-\sqrt{i\epsilon}}{\mathrm{Res}}\Big[ip_\mathrm{r}\frac{e^{-ip_\mathrm{r}r/\hbar}}{p_\mathrm{r}^2-i\epsilon}\Pi_\mathrm{1L,electron}^\mathrm{(r)}(-p_\mathrm{r}^2)\Big]\nonumber\\
 &\hspace{0.25cm}-\!2\pi i\!\!\underset{p_\mathrm{r}=\sqrt{i\epsilon}}{\mathrm{Res}}\Big[ip_\mathrm{r}\frac{e^{ip_\mathrm{r}r/\hbar}}{p_\mathrm{r}^2-i\epsilon}\Pi_\mathrm{1L,electron}^\mathrm{(r)}(-p_\mathrm{r}^2)\Big]\nonumber\\
 &\hspace{0.25cm}-2\int_{\epsilon-i\infty}^{\epsilon-2im_\mathrm{e}c}\frac{ip_\mathrm{r}e^{-ip_\mathrm{r}r/\hbar}}{(p_\mathrm{r}^2-i\epsilon)}i\mathrm{Im}[\Pi_\mathrm{1L,electron}^\mathrm{(r)}(-p_\mathrm{r}^2)]dp_\mathrm{r}\nonumber\\
 &\hspace{0.25cm}+2\int_{-\epsilon+2im_\mathrm{e}c}^{-\epsilon+i\infty}\frac{ip_\mathrm{r}e^{ip_\mathrm{r}r/\hbar}}{(p_\mathrm{r}^2-i\epsilon)}i\mathrm{Im}[\Pi_\mathrm{1L,electron}^\mathrm{(r)}(-p_\mathrm{r}^2)]dp_\mathrm{r}\Big\}\nonumber\\
 &\!=\!\frac{e^2}{8\pi^2\varepsilon_0r}\Big\{-2\pi i\Big[\frac{1}{2}e^{ir\sqrt{i\epsilon}/\hbar}\Pi_\mathrm{1L,electron}^\mathrm{(r)}(-i\epsilon)\Big]\nonumber\\
 &\hspace{0.25cm}-2\pi i\Big[\frac{1}{2}e^{ir\sqrt{i\epsilon}/\hbar}\Pi_\mathrm{1L,electron}^\mathrm{(r)}(-i\epsilon)\Big]\nonumber\\
 &\hspace{0.25cm}+\!\frac{2\alpha_\mathrm{e}}{3}\!\!\int_{\epsilon-i\infty}^{\epsilon-2im_\mathrm{e}c}\!\frac{p_\mathrm{r}e^{-ip_\mathrm{r}r/\hbar}}{(p_\mathrm{r}^2-i\epsilon)}\sqrt{1\!-\!\frac{4m_\mathrm{e}^2c^2}{p_\mathrm{ri}^2}}\Big(1\!+\!\frac{2m_\mathrm{e}^2c^2}{p_\mathrm{ri}^2}\Big)dp_\mathrm{r}\nonumber\\
 &\hspace{0.25cm}+\!\frac{2\alpha_\mathrm{e}}{3}\!\!\int_{-\epsilon+2im_\mathrm{e}c}^{-\epsilon+i\infty}\!\frac{p_\mathrm{r}e^{ip_\mathrm{r}r/\hbar}}{(p_\mathrm{r}^2-i\epsilon)}\sqrt{1\!-\!\frac{4m_\mathrm{e}^2c^2}{p_\mathrm{ri}^2}}\Big(1\!+\!\frac{2m_\mathrm{e}^2c^2}{p_\mathrm{ri}^2}\Big)dp_\mathrm{r}\Big\}.
 \label{eq:Ve_1L_electron_calculation2_1}
\end{align}
In the second equality of equation~\eqref{eq:Ve_1L_electron_calculation2_1}, we have calculated the residues and applied equation~\eqref{eq:branchcuts1}. In the limit of $\epsilon$ approaching zero as $\epsilon\rightarrow0$, we obtain
\begin{align}
 &V_\mathrm{e,1L,electron}(\mathbf{r})=\frac{e^2}{8\pi^2\varepsilon_0r}\nonumber\\
 &\hspace{0.25cm}\times\!\Big[\frac{2\alpha_\mathrm{e}}{3}\!\!\int_{-\infty}^{-2m_\mathrm{e}c}\!\frac{e^{p_\mathrm{ri}r/\hbar}}{p_\mathrm{ri}}\sqrt{1\!-\!\frac{4m_\mathrm{e}^2c^2}{p_\mathrm{ri}^2}}\Big(1\!+\!\frac{2m_\mathrm{e}^2c^2}{p_\mathrm{ri}^2}\Big)dp_\mathrm{ri}\nonumber\\
 &\hspace{0.25cm}+\!\frac{2\alpha_\mathrm{e}}{3}\!\!\int_{2m_\mathrm{e}c}^{\infty}\!\frac{e^{-p_\mathrm{ri}r/\hbar}}{p_\mathrm{ri}}\sqrt{1\!-\!\frac{4m_\mathrm{e}^2c^2}{p_\mathrm{ri}^2}}\Big(1\!+\!\frac{2m_\mathrm{e}^2c^2}{p_\mathrm{ri}^2}\Big)dp_\mathrm{ri}\Big]\nonumber\\
 &\!=\!\frac{2\hbar c\alpha_\mathrm{e}^2}{3\pi r}\!\!\int_{2m_\mathrm{e}c}^{\infty}\frac{e^{-p_\mathrm{ri}r/\hbar}}{p_\mathrm{ri}}\sqrt{1-\frac{4m_\mathrm{e}^2c^2}{p_\mathrm{ri}^2}}\Big(1+\frac{2m_\mathrm{e}^2c^2}{p_\mathrm{ri}^2}\Big)dp_\mathrm{ri}\nonumber\\
 &\!=\!\frac{2\hbar c\alpha_\mathrm{e}^2}{3\pi r}\!\!\int_{2m_\mathrm{e}c}^{\infty}\frac{e^{-p_\mathrm{ri}r/\hbar}}{p_\mathrm{ri}^2}\sqrt{(p_\mathrm{ri}-2m_\mathrm{e}c)(p_\mathrm{ri}+2m_\mathrm{e}c)}\nonumber\\
 &\hspace{0.25cm}\times\Big(1+\frac{2m_\mathrm{e}^2c^2}{p_\mathrm{ri}^2}\Big)dp_\mathrm{ri}.
 \label{eq:Ve_1L_electron_calculation2_2}
\end{align}
While taking the limit $\epsilon\rightarrow0$, we have used $p_\mathrm{r}=\epsilon+ip_\mathrm{ri}$ and $p_\mathrm{r}=-\epsilon+ip_\mathrm{ri}$ to change the integration variables of the first and second integrals, respectively. In the last two equalitites of equation~\eqref{eq:Ve_1L_electron_calculation2_2}, we have made a change of variables, $p_\mathrm{ri}\rightarrow -p_\mathrm{ri}$, to the first integral and used some algebra to obtain the last form.

Next, we evaluate the integral in the last form of equation~\eqref{eq:Ve_1L_electron_calculation2_2} in the long distance limit of $r\gg1/(m_\mathrm{e}c)$. In this limit, the integral is dominated by the region where $p_\mathrm{ri}\approx2m_\mathrm{e}c$. Approximating the integrand in this region and changing the variable to $p_\mathrm{t}=p_\mathrm{ri}-2m_\mathrm{e}c$, we obtain
\begin{align}
 &V_\mathrm{e,1L,electron}(\mathbf{r})\nonumber\\
 &\approx\frac{\hbar c\alpha_\mathrm{e}^2}{\pi r}\int_0^{\infty}\frac{e^{-(p_\mathrm{t}+2m_\mathrm{e}c)r/\hbar}}{4m_\mathrm{e}^2c^2}\sqrt{4m_\mathrm{e}cp_\mathrm{t}}dp_\mathrm{t}\nonumber\\
 &=\frac{\hbar c\alpha_\mathrm{e}^2e^{-2m_\mathrm{e}cr/\hbar}}{2\pi r(m_\mathrm{e}c)^{3/2}}\int_0^{\infty}e^{-p_\mathrm{t}r/\hbar}\sqrt{p_\mathrm{t}}dp_\mathrm{t}\nonumber\\
 &=\frac{\hbar c\alpha_\mathrm{e}^2e^{-2m_\mathrm{e}cr/\hbar}}{2\pi r(m_\mathrm{e}c)^{3/2}}\frac{\sqrt{\pi}}{2(r/\hbar)^{3/2}}\nonumber\\
 &=\frac{\hbar c\alpha_\mathrm{e}^2e^{-2m_\mathrm{e}c|\mathbf{r}|/\hbar}}{4\sqrt{\pi}|\mathbf{r}|(m_\mathrm{e}c|\mathbf{r}|/\hbar)^{3/2}}.
 \label{eq:Ve_1L_electron_calculation3}
\end{align}
This radiative correction of the Coulomb potential is known from QED, where the classical Coulomb potential corrected with this term is known as the Uehling potential \cite{Peskin2018,Landau1982}.

\subsubsection{Renormalized 1-graviton-loop contribution}

Next, we calculate the 1-graviton-loop correction to the Coulomb potential. Using the renormalized 1-graviton-loop contribution to the Fourier transform of the Coulomb potential correction in equation~\eqref{eq:Ve_Fourier2} together with equation~\eqref{eq:Pi_1L_graviton_p}, and performing the inverse Fourier transform, we obtain the renormalized 1-graviton-loop contribution to the Coulomb potential correction as
\begin{align}
 &V_\mathrm{e,1L,graviton}(\mathbf{r})\nonumber\\
 &=\int\widetilde{V}_\mathrm{e,1L,graviton}(\mathbf{p})e^{i\mathbf{p}\cdot\mathbf{r}/\hbar}\frac{d^3p}{(2\pi\hbar)^3}\nonumber\\
 &=-\frac{\alpha_\mathrm{e}\kappa c^2}{48\pi^4\hbar}\int\frac{|\mathbf{p}|^2\log(|\mathbf{p}|^2)}{|\mathbf{p}|^2-i\epsilon}e^{i\mathbf{p}\cdot\mathbf{r}/\hbar}d^3p\nonumber\\
 &=\frac{\alpha_\mathrm{e}\kappa\hbar^2 c^2}{12\pi^2|\mathbf{r}|^3}e^{i|\mathbf{r}|\sqrt{i\epsilon}/\hbar}.
 \label{eq:Ve_1L_graviton_calculation1}
\end{align}
In the third equality, we have applied the second Fourier transform integral given in section~7 of the supplementary material.

In the limit of $\epsilon$ approaching zero as $\epsilon\rightarrow0$, the renormalized 1-graviton-loop contribution of the Coulomb potential correction in equation~\eqref{eq:Ve_1L_graviton_calculation1} becomes
\begin{equation}
 V_\mathrm{e,1L,graviton}(\mathbf{r})
 =\frac{\kappa\hbar^2c^2\alpha_\mathrm{e}}{12\pi^2|\mathbf{r}|^3}
 =\frac{2G\hbar^2\alpha_\mathrm{e}}{3\pi c^2|\mathbf{r}|^3}.
\end{equation}
This result corresponds to the conventional effective-field-theory-based gravity quantization result for the vacuum polarization term of the Coulomb potential correction, which can be found from some works in previous literature \cite{Butt2006}.

\subsubsection{Total vacuum-polarization-corrected Coulomb potential}

The sum of the classical Coulomb potential and the 1-electron-loop and 1-graviton-loop vacuum polarization corrections to the Coulomb potential is given by
\begin{align}
 V_\mathrm{e}(\mathbf{r}) &=V_\mathrm{e}(\mathbf{r})+V_\mathrm{e,1L,electron}(\mathbf{r})+V_\mathrm{e,1L,graviton}(\mathbf{r})\nonumber\\
 &=\frac{\hbar c\alpha_\mathrm{e}}{|\mathbf{r}|}
 +\frac{\hbar c\alpha_\mathrm{e}^2e^{-2m_\mathrm{e}c|\mathbf{r}|/\hbar}}{4\sqrt{\pi}|\mathbf{r}|(m_\mathrm{e}c|\mathbf{r}|/\hbar)^{3/2}}
 +\frac{2G\hbar^2\alpha_\mathrm{e}}{3\pi c^2|\mathbf{r}|^3}\nonumber\\
 &=\frac{\hbar c\alpha_\mathrm{e}}{|\mathbf{r}|}\Big[1
 +\frac{\alpha_\mathrm{e}e^{-2m_\mathrm{e}c|\mathbf{r}|/\hbar}}{4\sqrt{\pi}(m_\mathrm{e}c|\mathbf{r}|/\hbar)^{3/2}}
 +\frac{2G\hbar}{3\pi c^3|\mathbf{r}|^2}\Big].
\end{align}
The Uehling potential of QED corresponds to the first two terms of this equation \cite{Schwartz2014,Peskin2018,Landau1982}, and the last term is the modification of the QED result by the gravitational interaction.

\subsection{\label{sec:radiativeNewton}Radiative corrections to the Newtonian potential}

Next, we study the radiative corrections to the Newtonian potential. As an example, we calculate the 1-electron-loop and 1-photon-loop corrections resulting from the vacuum polarization diagrams and their counterterm in equations \eqref{eq:graviton1Lelectron}--\eqref{eq:graviton1LCT}. Detailed study of the contributions resulting from the other relevant Feynman diagrams, partly similar to those studied in \cite{Donoghue1994a}, is left to a separate work.

In the nonrelativistic limit, for the four-momentum of the exchanged graviton, we have $p^2\rightarrow-|\mathbf{p}|^2$, where $\mathbf{p}$ is the three-dimensional momentum vector. Thus, the 1-electron-loop and 1-photon-loop renormalized scalar amplitude factors of the graviton self-energy in equations \eqref{eq:Xi_1L_electron} and \eqref{eq:Xi_1L_photon} become
\begin{align}
 &\Xi_\mathrm{1L,electron}^\mathrm{(r)}(p^2)\approx
 \Xi_\mathrm{1L,electron}^\mathrm{(r)}(-|\mathbf{p}|^2)\nonumber\\
 &=\frac{\kappa c|\mathbf{p}|^2}{32\pi^2\hbar}\Big\{\frac{8m_\mathrm{e}^2c^2}{3|\mathbf{p}|^2}\Big(1-\frac{6m_\mathrm{e}^2c^2}{|\mathbf{p}|^2}\Big)-\log(m_\mathrm{e}^2c^2)\nonumber\\
 &\hspace{0.4cm}+\Big(1-\frac{2m_\mathrm{e}^2c^2}{|\mathbf{p}|^2}\Big)\Big(1+\frac{4m_\mathrm{e}^2c^2}{|\mathbf{p}|^2}\Big)^{3/2}\nonumber\\
 &\hspace{0.4cm}\times\log\Big[1-\frac{|\mathbf{p}|^2}{2m_\mathrm{e}^2c^2}\Big(\sqrt{1+\frac{4m_\mathrm{e}^2c^2}{|\mathbf{p}|^2}}-1\Big)\Big]\Big\},
 \label{eq:Xi_1L_electron2}
\end{align}
\begin{align}
 \Xi_\mathrm{1L,photon}^\mathrm{(r)}(p^2) &\approx
 \Xi_\mathrm{1L,photon}^\mathrm{(r)}(-|\mathbf{p}|^2)\nonumber\\
 &=-\frac{\kappa c|\mathbf{p}|^2}{16\pi^2\hbar}\log(|\mathbf{p}|^2).
 \label{eq:Xi_1L_photon2}
\end{align}

The Fourier transform of the Newtonian potential obtains corrections, which are additions to the tree-level Fourier transform of the Newtonian potential in equation~\eqref{eq:Vg_Fourier}. Using equations \eqref{eq:Xi_1L_electron2} and \eqref{eq:Xi_1L_photon2}, the renormalized 1-electron-loop and 1-photon-loop correction terms are given by
\begin{equation}
 \tilde{V}_\mathrm{g,1L,electron}(\mathbf{p})
 =-\frac{\kappa\hbar^2m_\mathrm{e}^2c^4}{2(|\mathbf{p}|^2-i\epsilon)}\Xi_\mathrm{1L,electron}^\mathrm{(r)}(-|\mathbf{p}|^2),
 \label{eq:Vg_Fourier1}
\end{equation}
\begin{equation}
 \tilde{V}_\mathrm{g,1L,photon}(\mathbf{p})
 =-\frac{\kappa\hbar^2m_\mathrm{e}^2c^4}{2(|\mathbf{p}|^2-i\epsilon)}\Xi_\mathrm{1L,photon}^\mathrm{(r)}(-|\mathbf{p}|^2).
 \label{eq:Vg_Fourier2}
\end{equation}

\subsubsection{Renormalized 1-electron-loop contribution}

Here we study the renormalized 1-electron-loop contribution to the radiative correction of the Newtonian potential. Using the renormalized 1-electron-loop contribution to the Fourier transform of the Newtonian potential correction in equation~\eqref{eq:Vg_Fourier1}, and performing the inverse Fourier transform, we obtain the renormalized 1-electron-loop contribution to the Newtonian potential correction as
\begin{align}
 &V_\mathrm{g,1L,electron}(\mathbf{r})\nonumber\\
 &\!=\!\int\widetilde{V}_\mathrm{g,1L,electron}(\mathbf{p})e^{i\mathbf{p}\cdot\mathbf{r}/\hbar}\frac{d^3p}{(2\pi\hbar)^3}\nonumber\\
 &\!=\!-\frac{\kappa m_\mathrm{e}^2c^4}{8\pi^2r}\!\!\int_{-\infty}^\infty\!\!\!\frac{ip_\mathrm{r}(e^{-ip_\mathrm{r}r/\hbar}\!\!-\!e^{ip_\mathrm{r}r/\hbar})}{p_\mathrm{r}^2-i\epsilon}\Xi_\mathrm{1L,electron}^\mathrm{(r)}(-p_\mathrm{r}^2)dp_\mathrm{r}.
 \label{eq:Vg_1L_electron_calculation1}
\end{align}
The intermediate steps in obtaining the last form of equation~\eqref{eq:Vg_1L_electron_calculation1} are analogous to those in equation~\eqref{eq:Ve_1L_electron_calculation1}.

As a complex function, the imaginary part of $\Xi_\mathrm{1L,electron}^\mathrm{(r)}(-p_\mathrm{r}^2)$ in the integrand of equation~\eqref{eq:Vg_1L_electron_calculation1} has branch cuts at the positive and negative imaginary axes starting at $p_\mathrm{r}=\pm 2im_\mathrm{e}c$ and extending to infinities. Writing $p_\mathrm{r}=\pm\epsilon+ip_\mathrm{ri}$, where $p_\mathrm{ri}=\mathrm{Im}(p_\mathrm{r})$, we obtain
\begin{align}
 &\mathrm{Im}[\Xi_\mathrm{1L,electron}^\mathrm{(r)}(-p_\mathrm{r}^2)]\nonumber\\
 &=\pm\frac{\kappa cp_\mathrm{ri}^2}{32\pi\hbar}\Big(1+\frac{2m_\mathrm{e}^2c^2}{p_\mathrm{ri}^2}\Big)\Big(1-\frac{4m_\mathrm{e}^2c^2}{p_\mathrm{ri}^2}\Big)^{3/2}.
 \label{eq:branchcuts2}
\end{align}
The positive and negative signs of equation~\eqref{eq:branchcuts2} correspond to $\mathrm{Re}(p_\mathrm{r})>0$ and $\mathrm{Re}(p_\mathrm{r})<0$, respectively.

Applying the residue theorem and the branch cuts of the integrand to calculate the integral of equation~\eqref{eq:Vg_1L_electron_calculation1}, we obtain 
\begin{align}
 &V_\mathrm{g,1L,electron}(\mathbf{r})\nonumber\\
 &\!=\!-\frac{\kappa m_\mathrm{e}^2c^4}{8\pi^2r}\Big\{\!\!-\!2\pi i\!\!\!\underset{p_\mathrm{r}=-\sqrt{i\epsilon}}{\mathrm{Res}}\Big[ip_\mathrm{r}\frac{e^{-ip_\mathrm{r}r/\hbar}}{p_\mathrm{r}^2-i\epsilon}\Xi_\mathrm{1L,electron}^\mathrm{(r)}(-p_\mathrm{r}^2)\Big]\nonumber\\
 &\hspace{0.25cm}-\!2\pi i\!\!\underset{p_\mathrm{r}=\sqrt{i\epsilon}}{\mathrm{Res}}\Big[ip_\mathrm{r}\frac{e^{ip_\mathrm{r}r/\hbar}}{p_\mathrm{r}^2-i\epsilon}\Xi_\mathrm{1L,electron}^\mathrm{(r)}(-p_\mathrm{r}^2)\Big]\nonumber\\
 &\hspace{0.25cm}-2\int_{\epsilon-i\infty}^{\epsilon-2im_\mathrm{e}c}\frac{ip_\mathrm{r}e^{-ip_\mathrm{r}r/\hbar}}{(p_\mathrm{r}^2-i\epsilon)}i\mathrm{Im}[\Xi_\mathrm{1L,electron}^\mathrm{(r)}(-p_\mathrm{r}^2)]dp_\mathrm{r}\nonumber\\
 &\hspace{0.25cm}+2\int_{-\epsilon+2im_\mathrm{e}c}^{-\epsilon+i\infty}\frac{ip_\mathrm{r}e^{ip_\mathrm{r}r/\hbar}}{(p_\mathrm{r}^2-i\epsilon)}i\mathrm{Im}[\Xi_\mathrm{1L,electron}^\mathrm{(r)}(-p_\mathrm{r}^2)]dp_\mathrm{r}\Big\}\nonumber\\
 &\!=\!-\frac{\kappa m_\mathrm{e}^2c^4}{8\pi^2r}\Big\{-2\pi i\Big[\frac{1}{2}e^{ir\sqrt{i\epsilon}/\hbar}\Xi_\mathrm{1L,electron}^\mathrm{(r)}(-i\epsilon)\Big]\nonumber\\
 &\hspace{0.25cm}-2\pi i\Big[\frac{1}{2}e^{ir\sqrt{i\epsilon}/\hbar}\Xi_\mathrm{1L,electron}^\mathrm{(r)}(-i\epsilon)\Big]\nonumber\\
 &\hspace{0.25cm}+\!\frac{\kappa c}{16\pi\hbar}\!\!\int_{\epsilon-i\infty}^{\epsilon-2im_\mathrm{e}c}\!\frac{p_\mathrm{r}^3e^{-ip_\mathrm{r}r/\hbar}}{(p_\mathrm{r}^2-i\epsilon)}\nonumber\\
 &\hspace{0.25cm}\times\Big(1+\frac{2m_\mathrm{e}^2c^2}{p_\mathrm{ri}^2}\Big)\Big(1-\frac{4m_\mathrm{e}^2c^2}{p_\mathrm{ri}^2}\Big)^{3/2}dp_\mathrm{r}\nonumber\\
 &\hspace{0.25cm}+\!\frac{\kappa c}{16\pi\hbar}\!\!\int_{-\epsilon+2im_\mathrm{e}c}^{-\epsilon+i\infty}\!\frac{p_\mathrm{r}^3e^{ip_\mathrm{r}r/\hbar}}{(p_\mathrm{r}^2-i\epsilon)}\nonumber\\
 &\hspace{0.25cm}\times\Big(1+\frac{2m_\mathrm{e}^2c^2}{p_\mathrm{ri}^2}\Big)\Big(1-\frac{4m_\mathrm{e}^2c^2}{p_\mathrm{ri}^2}\Big)^{3/2}dp_\mathrm{r}\Big\}.
 \label{eq:Vg_1L_electron_calculation2_1}
\end{align}
In the second equality of equation~\eqref{eq:Vg_1L_electron_calculation2_1}, we have calculated the residues and applied equation~\eqref{eq:branchcuts2}. In the limit of $\epsilon$ approaching zero as $\epsilon\rightarrow0$, we obtain
\begin{align}
 &V_\mathrm{g,1L,electron}(\mathbf{r})=-\frac{\kappa m_\mathrm{e}^2c^4}{8\pi^2r}\!\Big[\frac{\kappa c}{16\pi\hbar}\!\!\int_{-\infty}^{-2m_\mathrm{e}c}\!p_\mathrm{ri}e^{p_\mathrm{ri}r/\hbar}\nonumber\\
 &\hspace{0.25cm}\times\Big(1+\frac{2m_\mathrm{e}^2c^2}{p_\mathrm{ri}^2}\Big)\Big(1-\frac{4m_\mathrm{e}^2c^2}{p_\mathrm{ri}^2}\Big)^{3/2}dp_\mathrm{ri}\nonumber\\
 &\hspace{0.25cm}+\!\frac{\kappa c}{16\pi\hbar}\!\!\int_{2m_\mathrm{e}c}^{\infty}\!p_\mathrm{ri}e^{-p_\mathrm{ri}r/\hbar}\nonumber\\
 &\hspace{0.25cm}\times\Big(1+\frac{2m_\mathrm{e}^2c^2}{p_\mathrm{ri}^2}\Big)\Big(1-\frac{4m_\mathrm{e}^2c^2}{p_\mathrm{ri}^2}\Big)^{3/2}dp_\mathrm{ri}\Big]\nonumber\\
 &\!=\!-\frac{\kappa^2 m_\mathrm{e}^2c^5}{64\pi^3\hbar r}\!\!\int_{2m_\mathrm{e}c}^{\infty}\!p_\mathrm{ri}e^{-p_\mathrm{ri}r/\hbar}\nonumber\\
 &\hspace{0.25cm}\times\Big(1+\frac{2m_\mathrm{e}^2c^2}{p_\mathrm{ri}^2}\Big)\Big(1-\frac{4m_\mathrm{e}^2c^2}{p_\mathrm{ri}^2}\Big)^{3/2}dp_\mathrm{ri}\nonumber\\
 &\!=\!-\frac{\kappa^2 m_\mathrm{e}^2c^5}{64\pi^3\hbar r}\!\!\int_{2m_\mathrm{e}c}^{\infty}\frac{e^{-p_\mathrm{ri}r/\hbar}}{p_\mathrm{ri}^2}[(p_\mathrm{ri}-2m_\mathrm{e}c)(p_\mathrm{ri}+2m_\mathrm{e}c)]^{3/2}\nonumber\\
 &\hspace{0.25cm}\times\Big(1+\frac{2m_\mathrm{e}^2c^2}{p_\mathrm{ri}^2}\Big)dp_\mathrm{ri}.
 \label{eq:Vg_1L_electron_calculation2_2}
\end{align}
While taking the limit $\epsilon\rightarrow0$, we have used $p_\mathrm{r}=\epsilon+ip_\mathrm{ri}$ and $p_\mathrm{r}=-\epsilon+ip_\mathrm{ri}$ to change the integration variables of the first and second integrals, respectively. In the last two equalitites of equation~\eqref{eq:Vg_1L_electron_calculation2_2}, we have made a change of variables, $p_\mathrm{ri}\rightarrow -p_\mathrm{ri}$, to the first integral and used some algebra to obtain the last form.

Next, we evaluate the integral in the last form of equation~\eqref{eq:Vg_1L_electron_calculation2_2} in the long distance limit of $r\gg1/(m_\mathrm{e}c)$. In this limit, the integral is dominated by the region where $p_\mathrm{ri}\approx2m_\mathrm{e}c$. Approximating the integrand in this region and changing the variable to $p_\mathrm{t}=p_\mathrm{ri}-2m_\mathrm{e}c$, we obtain
\begin{align}
 &V_\mathrm{g,1L,electron}(\mathbf{r})\nonumber\\
 &\approx-\frac{3\kappa^2 m_\mathrm{e}^2c^5}{128\pi^3\hbar r}\int_0^{\infty}\frac{e^{-(p_\mathrm{t}+2m_\mathrm{e}c)r/\hbar}}{4m_\mathrm{e}^2c^2}(4m_\mathrm{e}cp_\mathrm{t})^{3/2}dp_\mathrm{t}\nonumber\\
 &=-\frac{3\kappa^2 m_\mathrm{e}^2c^5e^{-2m_\mathrm{e}cr/\hbar}}{64\pi^3\hbar\sqrt{m_\mathrm{e}c}\,r}\int_0^{\infty}e^{-p_\mathrm{t}r/\hbar}p_\mathrm{t}^{3/2}dp_\mathrm{t}\nonumber\\
 &=-\frac{3\kappa^2 m_\mathrm{e}^2c^5e^{-2m_\mathrm{e}cr/\hbar}}{64\pi^3\hbar\sqrt{m_\mathrm{e}c}\,r}\frac{3\sqrt{\pi}}{4(r/\hbar)^{5/2}}\nonumber\\
 &=-\frac{9\kappa^2c^3(m_\mathrm{e}c\hbar)^{3/2}e^{-2m_\mathrm{e}c|\mathbf{r}|/\hbar}}{256\pi^2\sqrt{\pi}\,|\mathbf{r}|^{7/2}}\nonumber\\
 &=-\frac{9G^2(m_\mathrm{e}c\hbar)^{3/2}e^{-2m_\mathrm{e}c|\mathbf{r}|/\hbar}}{4\sqrt{\pi}c^5|\mathbf{r}|^{7/2}}.
 \label{eq:Vg_1L_electron_calculation3}
\end{align}
This is the 1-electron-loop contribution to the radiative correction of the Newtonian potential in unified gravity. The exponential factor of equation~\eqref{eq:Vg_1L_electron_calculation3} is equivalent to that in the QED contribution to the radiative correction of the Coulomb potential in equation~\eqref{eq:Ve_1L_electron_calculation3} but, otherwise, these corrections depend on different powers of $|\mathbf{r}|$. The authors are not aware of the appearance of the correction term in equation~\eqref{eq:Vg_1L_electron_calculation3} in previous literature. Previously, arguments have been presented in favor of omitting all the diagrams with internal lines of massive particles in the calculations of quantum gravity corrections \cite{Donoghue1994a,Netto2022}. However, these arguments have been constructed for scalar particles, and thus, they do not apply to electron loops.

\subsubsection{Renormalized 1-photon-loop contribution}

Next, we calculate the 1-photon-loop correction to the Newtonian potential. Using the renormalized 1-photon-loop contribution to the Fourier transform of the Newtonian potential correction in equation~\eqref{eq:Vg_Fourier2} together with equation~\eqref{eq:Xi_1L_photon2}, and performing the inverse Fourier transform, we obtain the renormalized 1-photon-loop contribution to the Newtonian potential correction as
\begin{align}
 &V_\mathrm{g,1L,photon}(\mathbf{r})\nonumber\\
 &=\int\widetilde{V}_\mathrm{g,1L,photon}(\mathbf{p})e^{i\mathbf{p}\cdot\mathbf{r}/\hbar}\frac{d^3p}{(2\pi\hbar)^3}\nonumber\\
 &=\frac{\kappa^2m_\mathrm{e}^2c^5}{256\pi^5\hbar^2}\int\frac{|\mathbf{p}|^2\log(|\mathbf{p}|^2)}{|\mathbf{p}|^2-i\epsilon}e^{i\mathbf{p}\cdot\mathbf{r}/\hbar}d^3p\nonumber\\
 &=-\frac{\kappa^2\hbar m_\mathrm{e}^2c^5}{64\pi^3|\mathbf{r}|^3}e^{i|\mathbf{r}|\sqrt{i\epsilon}/\hbar}.
 \label{eq:Vg_1L_photon_calculation1}
\end{align}
In the third equality, we have applied the second Fourier transform integral given in section~7 of the supplementary material.

In the limit of $\epsilon$ approaching zero as $\epsilon\rightarrow0$, the renormalized 1-photon-loop contribution of the Newtonian potential correction in equation~\eqref{eq:Vg_1L_photon_calculation1} becomes
\begin{equation}
 V_\mathrm{g,1L,photon}(\mathbf{r})
 =-\frac{\kappa^2\hbar m_\mathrm{e}^2c^5}{64\pi^3|\mathbf{r}|^3}
 =-\frac{G^2\hbar m_\mathrm{e}^2}{\pi c^3|\mathbf{r}|^3}.
\end{equation}
Apart from the prefactor, the form of this term is identical to the form of the quantum correction term of the Newtonian potential obtained using the conventional effective field theory in previous literature, where other Feynman diagrams were also accounted for \cite{Donoghue1994a}.

\subsubsection{Total vacuum-polarization-corrected Newtonian potential}

The sum of the classical Newtonian potential and the 1-electron-loop and 1-photon-loop vacuum polarization corrections to the Newtonian potential is given by
\begin{align}
 V_\mathrm{g}(\mathbf{r}) &=V_\mathrm{g}(\mathbf{r})+V_\mathrm{g,1L,photon}(\mathbf{r})+V_\mathrm{g,1L,electron}(\mathbf{r})\nonumber\\
 &=-\frac{Gm_\mathrm{e}^2}{|\mathbf{r}|}
 -\frac{G^2\hbar m_\mathrm{e}^2}{\pi c^3|\mathbf{r}|^3}
 -\frac{9G^2(m_\mathrm{e}c\hbar)^{3/2}e^{-2m_\mathrm{e}c|\mathbf{r}|/\hbar}}{4\sqrt{\pi}c^5|\mathbf{r}|^{7/2}}\nonumber\\
 &=-\frac{Gm_\mathrm{e}^2}{|\mathbf{r}|}\Big(1
 +\frac{G\hbar}{\pi c^3|\mathbf{r}|^2}
 +\frac{9G\hbar e^{-2m_\mathrm{e}c|\mathbf{r}|/\hbar}}{4c^3\sqrt{\pi m_\mathrm{e}c/\hbar}|\mathbf{r}|^{5/2}}\Big).
\end{align}

\subsection{\label{sec:radiativemoment}Radiative corrections to the electron magnetic moment}

The form factor $F_2^\mathrm{e\gamma}$ in the electron--photon vertex matrix element in equation~\eqref{eq:vertexstructure} is called the anomalous magnetic moment since the total magnetic moment of an electron is given by $g_\mathrm{s}=2+2F_2^\mathrm{e\gamma}$ \cite{Schwartz2014}. Order by order in perturbation theory, we obtain more and more accurate approximation for $F_2^\mathrm{e\gamma}$. Experiments have verfied the Standard Model prediction for the electron magnetic moment to an astonishing accuracy of one part in a trillion \cite{Fan2023}. More experimental accuracy is, however, needed to probe the corrections associated with quantum gravity.

The 1-photon-loop radiative correction of the form factor $F_2^\mathrm{e\gamma}$ following from equation~\eqref{eq:vertex1photon} is the famous result first calculated by Schwinger \cite{Schwinger1948}, given by
\begin{equation}
 F_\mathrm{2,1L,photon}^\mathrm{e\gamma}=\frac{\alpha_\mathrm{e}}{2\pi}.
\end{equation}

Correspondingly, the 1-graviton-loop radiative corrections of the form factor $F_2^\mathrm{e\gamma}$ following from equation~\eqref{eq:vertex1graviton} is given by
\begin{equation}
 F_\mathrm{2,1L,graviton}^\mathrm{e\gamma}=\frac{7\alpha_\mathrm{g}}{4\pi}.
 \label{eq:F2g}
\end{equation} 
The 1-graviton-loop contributions to equation~\eqref{eq:F2g} following from each Feynman diagram in equation~\eqref{eq:vertex1graviton} are given by
\begin{equation}
 F_\mathrm{2,1L,graviton}^\mathrm{(e\gamma,diag1)}=-\frac{\alpha_\mathrm{g}}{6\pi}\Big[\frac{1}{\epsilon_\mathrm{UV}}+\frac{61}{6}+\log\!\Big(\frac{4\pi\mu^2e^{-\gamma}}{m_\mathrm{e}^2c^2}\Big)\Big],
 \label{eq:F2g1}
\end{equation}
\begin{align}
 &F_\mathrm{2,1L,graviton}^\mathrm{(e\gamma,diag2)}
 =F_\mathrm{2,1L,graviton}^\mathrm{(e\gamma,diag3)}\nonumber\\
 &=\frac{\alpha_\mathrm{g}}{\pi}\Big[\frac{1}{\epsilon_\mathrm{UV}}+\frac{7}{2}+\log\!\Big(\frac{4\pi\mu^2e^{-\gamma}}{m_\mathrm{e}^2c^2}\Big)\Big],
\end{align}
\begin{align}
 &F_\mathrm{2,1L,graviton}^\mathrm{(e\gamma,diag4)}
 =F_\mathrm{2,1L,graviton}^\mathrm{(e\gamma,diag5)}\nonumber\\
 &=-\frac{33\alpha_\mathrm{g}}{36\pi}\Big[\frac{1}{\epsilon_\mathrm{UV}}+\frac{64}{33}+\log\!\Big(\frac{4\pi\mu^2e^{-\gamma}}{m_\mathrm{e}^2c^2}\Big)\Big],
 \label{eq:F2g4}
\end{align}
\begin{equation}
 F_\mathrm{2,1L,graviton}^\mathrm{(e\gamma,diag6)}=0.
\end{equation}
Equation \eqref{eq:F2g} agrees with the result first calculated by Berends and Gastmans \cite{Berends1975}. Since the counterterm does not contribute to equation~\eqref{eq:F2g}, this result was possible to calculate without the renormalizable theory of quantum gravity presented in this work. The exact cancellation of the UV divergences and the scale and Euler--Mascheroni constants of dimensional regularization in the terms of equations \eqref{eq:F2g1}--\eqref{eq:F2g4} when substituted into equation~\eqref{eq:F2g} is necessary for physical observables to be finite and independent of the regularization parameters. In previous literature, another method for the calculation of radiative corrections of the electron magnetic moment has been based on dimensional reduction of supergravity \cite{delAguila1984,Grisaru1985,Georges1985,Ottoni2006}. The results of this approach differ from our results.

\section{\label{sec:teleparallel}Obtaining TEGR from unified gravity}

In this section, we show that by an appropriate choice of the tetrad and the metric tensor, we obtain the theory of teleparallel equivalent of general relativity in the Weitzenböck gauge (TEGRW) \cite{Bahamonde2023a,Aldrovandi2012,Heisenberg2019,Jimenez2019,Maluf2013,Cabral2020,Blagojevic2013} directly from unified gravity presented above. The geometric conditions of this section can be called the Weitzenböck gauge fixing. The Weitzenböck gauge fixing does not fix the gauge unambiguously since there are remaining redundant degrees of freedom associated with the class of reference frames where the teleparallel spin connection vanishes \cite{Bahamonde2023a,Aldrovandi2012}.

\subsection{\label{sec:teleparalleltetrad}Geometric conditions of TEGRW: tetrad and metric tensor}

The form of the gauge-invariant Lagrangian density of unified gravity in equation~\eqref{eq:LL} explicitly depends on the gauge field $H_{a\nu}$ in analogy to how the Lagrangian density explicitly depends on the electromagnetic gauge field $A_\nu$ when the electromagnetic-gauge-covariant derivatives have been written out using equation~\eqref{eq:Da}. In contrast to the gauge theories of the Standard Model, in the case of the gauge theory of gravity, it is possible to absorb the explicit dependence of the Lagrangian density on the gauge field into the definition of the tetrad. This possibility is enabled by the equivalence principle in equation~\eqref{eq:equivalenceprinciple}. Using the equivalence principle of scale, $g'_\mathrm{g}=g_\mathrm{g}$, with the reduced form of the Lagrangian density of unified gravity in equation~\eqref{eq:LL}, we observe that the explicit dependence on $H_{a\nu}$ can be absorbed in the generalized tetrad $\oset{\bullet}{e}_{a\nu}$, defined as
\begin{equation}
 \oset{\bullet}{e}_{a\nu}=\partial_\nu x_a+H_{a\nu}.
 \label{eq:tetrad}
\end{equation}
Equations \eqref{eq:geom} and \eqref{eq:tetrad} and the resulting definition of the metric tensor below are called in this work \emph{the geometric conditions of TEGRW}. The first term of $\oset{\bullet}{e}_{a\nu}$ in equation~\eqref{eq:tetrad} is related to inertial effects, and the second term describes gravitational effects. The division of the tetrad into $\partial_\nu x_a$ and $H_{a\nu}$ in equation~\eqref{eq:tetrad} reminds to how one divides the tetrad into trivial and nontrivial parts in TEGRW in previous literature \cite{Aldrovandi2012}.

Even if the expression of the space-time-dependent tetrad in equation~\eqref{eq:tetrad} is qualitatively of the same form as that in the conventional representation of TEGRW \cite{Bahamonde2023a,Aldrovandi2012}, in the present theory, it has been introduced in a fundamentally different way. In the conventional gauge theory approach to TEGRW, the gauge field is the translation gauge field, which is added to make the expression of the tetrad invariant in translations of the tangent-space coordinates $x_a$. In contrast, in the present theory, equation~\eqref{eq:tetrad} is introduced as a geometric condition that hides the explicit local gauge invariance of the Lagrangian density with respect to the gauge symmetry transformations of $X_a$ and $H_{a\nu}$ in equations \eqref{eq:xatransformation} and \eqref{eq:Hanutransformation}.

Having defined the space-time-dependent tetrad in equation~\eqref{eq:tetrad}, we then write the general space-time-dependent metric tensor in the conventional way in terms of the tetrad as
\begin{equation}
 g_{\mu\nu}=\eta^{ab}\oset{\bullet}{e}_{a\mu}\oset{\bullet}{e}_{b\nu}.
 \label{eq:metric}
\end{equation}
As a consistency check, when the gauge field $H_{a\nu}$ is zero, the definition of the tetrad in equation~\eqref{eq:tetrad} and the metric tensor in equation~\eqref{eq:metric} straightforwardly simplify to the well-known representations of the tetrad and the metric tensor in flat space-time in the chosen coordinates.

\subsection{\label{sec:LTEGR}Geometric Lagrangian density of TEGRW}
Next, we present the \emph{geometric} Lagrangian density of TEGRW obtained from unified gravity. We call the Lagrangian density geometric after the geometric condition of TEGRW in equation~\eqref{eq:tetrad} has been applied to the reduced form of the locally gauge-invariant Lagrangian density of unified gravity in equation~\eqref{eq:LL}. We also present different forms of this geometric Lagrangian density. Using the definition of the tetrad in the Weitzenböck gauge fixing in equation~\eqref{eq:tetrad} and setting $m'_\mathrm{e}=m_\mathrm{e}$ according to the equivalence principle in equation~\eqref{eq:equivalenceprinciple}, the Lagrangian density of unified gravity in equation~\eqref{eq:LL} becomes the Lagrangian density of TEGRW, given by
\begin{align}
 \mathcal{L}_\mathrm{TEGRW} &=\Big\{\!\oset{\bullet}{e}_{a\nu}\!\Big[\frac{i\hbar c}{4}\bar{\psi}_8(\cev{D}\boldsymbol{\gamma}_\mathrm{B}^5\boldsymbol{\gamma}_\mathrm{B}^\nu\mathbf{t}^a\boldsymbol{\gamma}_\mathrm{F}-\bar{\boldsymbol{\gamma}}_\mathrm{F}\boldsymbol{\gamma}_\mathrm{B}^5\boldsymbol{\gamma}_\mathrm{B}^\nu\mathbf{t}^a\vec{D})\psi_8\nonumber\\
 &\hspace{0.4cm}-\frac{m_\mathrm{e}c^2}{2}\bar{\psi}_8\mathbf{t}^{a}\boldsymbol{\gamma}_\mathrm{B}^\nu\boldsymbol{\gamma}_\mathrm{B}^5\psi_8
 -\bar{\Psi}\mathbf{t}^{a}\boldsymbol{\gamma}_\mathrm{B}^\nu\boldsymbol{\gamma}_\mathrm{B}^5\Psi\Big]\nonumber\\
 &\hspace{0.4cm}-m_\mathrm{e}c^2\bar{\psi}_8\psi_8
 +\bar{\Psi}\Psi
 +\frac{1}{4\kappa}H_{a\mu\nu}S^{a\mu\nu}\Big\}\sqrt{-g}.
 \label{eq:LTEGR}
\end{align}
Using the identities $\oset{\bullet}{e}_{a\nu}\boldsymbol{\gamma}_\mathrm{B}^5\boldsymbol{\gamma}_\mathrm{B}^\nu\mathbf{t}^a\vec{D}=2\vec{D}$,
$\oset{\bullet}{e}_{a\nu}\boldsymbol{\gamma}_\mathrm{B}^5\boldsymbol{\gamma}_\mathrm{B}^\nu\mathbf{t}^a\boldsymbol{\gamma}_\mathrm{F}=2\boldsymbol{\gamma}_\mathrm{F}$, $\oset{\bullet}{e}_{a\nu}\bar{\psi}_8\boldsymbol{\gamma}_\mathrm{B}^5\boldsymbol{\gamma}_\mathrm{B}^\nu\bar{\mathbf{t}}^a\psi_8=-4\bar{\psi}\psi$, $\bar{\psi}_8\psi_8=-\bar{\psi}\psi$, $\oset{\bullet}{e}_{a\nu}\boldsymbol{\gamma}_\mathrm{B}^5\boldsymbol{\gamma}_\mathrm{B}^\nu\bar{\mathbf{t}}^a\Psi=\mathbf{0}$, $\bar{\Psi}\Psi=\bar{\Theta}\cev{\partial}_\rho\boldsymbol{\gamma}_\mathrm{B}^\rho(\mathbf{I}_8+\boldsymbol{\mathfrak{e}}_0\bar{\boldsymbol{\mathfrak{e}}}_0)^2\boldsymbol{\gamma}_\mathrm{B}^\sigma\vec{\partial}_\sigma\Theta$, and $H_{a\mu\nu}S^{a\mu\nu}=H_{\rho\mu\nu}S^{\rho\mu\nu}$, and writing out the electromagnetic-gauge-covariant derivatives using equation~\eqref{eq:eightspinorderivatives}, the Lagrangian density in equation~\eqref{eq:LTEGR} becomes
\begin{align}
 \mathcal{L}_\mathrm{TEGRW} &\!=\!\Big[\frac{i\hbar c}{2}\bar{\psi}(\bar{\boldsymbol{\gamma}}_\mathrm{F}\vec{\partial}-\cev{\partial}\boldsymbol{\gamma}_\mathrm{F})\psi-m_\mathrm{e}c^2\bar{\psi}\psi
 +\bar{\Phi}\Theta+\bar{\Theta}\Phi\nonumber\\
 &\hspace{0.4cm}+\bar{\Theta}\cev{\partial}_\rho\boldsymbol{\gamma}_\mathrm{B}^\rho(\mathbf{I}_8+\boldsymbol{\mathfrak{e}}_0\bar{\boldsymbol{\mathfrak{e}}}_0)^2\boldsymbol{\gamma}_\mathrm{B}^\sigma\vec{\partial}_\sigma\Theta\nonumber\\
 &\hspace{0.4cm}+\frac{1}{4\kappa}H_{\rho\mu\nu}S^{\rho\mu\nu}\Big]\sqrt{-g}.
 \label{eq:LTEGRMaxwell}
\end{align}
Here the electromagnetic interaction Lagrangian density $\bar{\Phi}\Theta+\bar{\Theta}\Phi$ is written using the charge-current spinor $\Phi$, given in equation~\eqref{eq:chargecurrentspinor}. The form of the Lagrangian density of TEGRW in equation~\eqref{eq:LTEGRMaxwell} is used in the derivation of Maxwell's equations of TEGRW in the eight-spinor formalism, discussed in section~\ref{sec:Maxwell}.

Using $\bar{\boldsymbol{\gamma}}_\mathrm{F}\vec{\partial}=\boldsymbol{\gamma}_\mathrm{F}^\nu\vec{\partial}_\nu$, $\cev{\partial}\boldsymbol{\gamma}_\mathrm{F}=\cev{\partial}_\nu\boldsymbol{\gamma}_\mathrm{F}^\nu$, $\bar{\Phi}\Theta+\bar{\Theta}\Phi=-J_\mathrm{e}^\nu A_\nu=-q_\mathrm{e}c\bar{\psi}\boldsymbol{\gamma}_\mathrm{F}^\nu\psi A_\nu$, and rewriting the electromagnetic Lagrangian density in terms of the electromagnetic field strength tensor $F^{\mu\nu}$ using equation~\eqref{eq:scalar}, the Lagrangian density of TEGRW in equation~\eqref{eq:LTEGRMaxwell} becomes
\begin{align}
 \mathcal{L}_\mathrm{TEGRW} &\!=\!\Big[\frac{i\hbar c}{2}\bar{\psi}(\boldsymbol{\gamma}_\mathrm{F}^\nu\vec{\partial}_\nu-\cev{\partial}_\nu\boldsymbol{\gamma}_\mathrm{F}^\nu)\psi-m_\mathrm{e}c^2\bar{\psi}\psi
 -J_\mathrm{e}^\nu A_\nu\nonumber\\
 &\hspace{0.4cm}-\frac{1}{4\mu_0}F_{\mu\nu}F^{\mu\nu}
 +\frac{1}{4\kappa}H_{\rho\mu\nu}S^{\rho\mu\nu}\Big]\sqrt{-g}.
 \label{eq:LTEGRDirac}
\end{align}
This form of the Lagrangian density of TEGRW is used in the derivations of the Maxwell's equations in TEGRW in the four-vector and tensor formalism, discussed in section~\ref{sec:Maxwell}, and Dirac equation of TEGRW, discussed in section~\ref{sec:Dirac}. Apart from the last term of the Lagrangian density in equation~\eqref{eq:LTEGRDirac} describing the gravity gauge field, equation~\eqref{eq:LTEGRDirac} is directly seen to be equal to the Lagrangian density of QED, discussed in section~2.8 of the supplementary material.

Writing the Lagrangian density in terms of the tetrad and the inverse metric, and using again the electromagnetic-gauge-covariant derivatives instead of the explicit electromagnetic interaction term $-J_\mathrm{e}^\nu A_\nu$, we obtain
\begin{align}
 &\mathcal{L}_\mathrm{TEGRW}\nonumber\\
 &=\!\Big\{\frac{\!i\hbar c}{4}\bar{\psi}[\oset{\bullet}{e}_b^{\;\,\nu}(\boldsymbol{\gamma}_\mathrm{F}^b\vec{D}_\nu\!+\!\boldsymbol{\gamma}_{\mathrm{F}\nu}\vec{D}^b)\!-\!(\cev{D}_\nu\boldsymbol{\gamma}_\mathrm{F}^b\!+\!\cev{D}^b\boldsymbol{\gamma}_{\mathrm{F}\nu})\oset{\bullet}{e}_b^{\;\,\nu}]\psi\nonumber\\
 &\hspace{0.4cm}-\!m_\mathrm{e}c^2\bar{\psi}\psi
 \!-\!\frac{1}{4\mu_0}g^{\mu\rho}g^{\nu\sigma}F_{\mu\nu}F_{\rho\sigma}
 \!+\!\frac{1}{4\kappa}H_{\rho\mu\nu}S^{\rho\mu\nu}\Big\}\sqrt{-g}.
 \label{eq:LTEGR3}
\end{align}
This form of the Lagrangian density of TEGRW is used as the starting point of the derivation of the field equation of gravity in TEGRW, discussed in section~\ref{sec:TEGRgravity}. The Lagrangian density in equation~\eqref{eq:LTEGR3} is found \emph{to be equivalent} to the Lagrangian density of TEGRW in previous literature \cite{Aldrovandi2012}. Detailed comparison of equation~\eqref{eq:LTEGR3} to the Lagrangian density of TEGRW in previous literature, including also references to pertinent equations of a well-known textbook of TEGR, is presented in section~6 of the supplementary material. The relation to the Lagrangian density of general relativity is also discussed in section~6 of the supplementary material. Further comparison of unified gravity with previous theories of gravity can be found in section~\ref{sec:discussion_grav}.

\subsection{\label{sec:dynamicsTEGR}Dynamical equations of TEGRW}
The dynamics of all fields appearing in the Lagrangian density in equation~\eqref{eq:L} are described by the well-known Euler--Lagrange equations. In the sections below, we derive the dynamical equations for the gravity gauge field, the electromagnetic gauge field, and the Dirac field in TEGRW.

\subsubsection{\label{sec:TEGRgravity}Field equation of gravity in TEGRW}
Next, we present the dynamical equation of the gravity gauge field, which we call the field equation of gravity in TEGRW. Starting from the Euler--Lagrange equation for $H_{a\nu}$, after some algebra, presented in section~5.1 of the supplementary material, we obtain the field equation of gravity in TEGRW, given by
\begin{equation}
 \widetilde{\nabla}_\rho S^{a\nu\rho} =\oset{\circ}{\nabla}_\rho S^{a\nu\rho}
 =\kappa T_\mathrm{TEGRW}^{a\nu}.
 \label{eq:tensorgravity}
\end{equation}
The source term $T_\mathrm{TEGRW}^{a\nu}=\oset{\bullet}{e}_{\;\,\mu}^aT_\mathrm{TEGRW}^{\mu\nu}$ in equation~\eqref{eq:tensorgravity} is the total SEM tensor of the Dirac, electromagnetic, and gravitational fields in TEGRW, given by
\begin{align}
 T_\mathrm{TEGRW}^{\mu\nu} &=T_\mathrm{D}^{\mu\nu}+T_\mathrm{D,diff}^{\mu\nu}+T_\mathrm{em}^{\mu\nu}+T_\mathrm{g}^{\mu\nu},\nonumber\\
 T_\mathrm{D}^{\mu\nu} &=\frac{i\hbar c}{4}\bar{\psi}(\boldsymbol{\gamma}_\mathrm{F}^\mu\vec{D}^\nu
 +\boldsymbol{\gamma}_\mathrm{F}^\nu\vec{D}^\mu
 -\cev{D}^\nu\boldsymbol{\gamma}_\mathrm{F}^\mu
 -\cev{D}^\mu\boldsymbol{\gamma}_\mathrm{F}^\nu)\psi\nonumber\\
 &\hspace{0.4cm}-\frac{1}{2}g^{\mu\nu}\Big[\frac{i\hbar c}{2}\bar{\psi}(\boldsymbol{\gamma}_\mathrm{F}^\rho\vec{D}_\rho-\cev{D}_\rho\boldsymbol{\gamma}_\mathrm{F}^\rho)\psi-m_\mathrm{e}c^2\bar{\psi}\psi\Big],\nonumber\\
 T_\mathrm{D,diff}^{\mu\nu} &=-\frac{1}{2}g^{\mu\nu}\Big[\frac{i\hbar c}{2}\bar{\psi}(\boldsymbol{\gamma}_\mathrm{F}^\rho\vec{D}_\rho-\cev{D}_\rho\boldsymbol{\gamma}_\mathrm{F}^\rho)\psi-m_\mathrm{e}c^2\bar{\psi}\psi\Big],\nonumber\\
 T_\mathrm{em}^{\mu\nu} &=\frac{1}{\mu_0}\Big(F_{\;\;\rho}^{\mu}F^{\rho\nu}+\frac{1}{4}g^{\mu\nu}F_{\rho\sigma}F^{\rho\sigma}\Big),\nonumber\\
 T_\mathrm{g}^{\mu\nu} &=\frac{1}{\kappa}\Big(H^{\;\,\;\;\mu}_{\sigma\rho}S^{\sigma\rho\nu}-\frac{1}{4}g^{\mu\nu}H_{\rho\sigma\lambda}S^{\rho\sigma\lambda}\Big).
 \label{eq:semtensors}
\end{align}
The terms $T_\mathrm{D}^{\mu\nu}$ and $T_\mathrm{em}^{\mu\nu}$ of equation~\eqref{eq:semtensors} are the SEM tensors of the Dirac and electromagnetic fields, equal to those in equation~\eqref{eq:semtensors0}. The term $T_\mathrm{D,diff}^{\mu\nu}$ of equation~\eqref{eq:semtensors} is an additional Dirac field SEM tensor contribution that is obtained from the Euler--Lagrange equation of gravity in TEGRW. As shown in section~5.4 of the supplementary material, this term is, however, zero when the Dirac equation of TEGRW is satisfied, i.e., for on-shell states. The term $T_\mathrm{g}^{\mu\nu}$ of equation~\eqref{eq:semtensors} is the SEM tensor of the gravitational gauge field, which is also known from previous literature on TEGR \cite{Aldrovandi2012,Bahamonde2023a,Andrade2000}. Similarly to the SEM tensor of the electromagnetic field, the SEM tensor of the gravitational field is traceless. Thus, the associated force carriers, the gravitons, are massless. The trace of the total SEM tensor in equation~\eqref{eq:semtensors}, given by ${T_\mathrm{TEGRW}}_{\;\,\nu}^\nu=m_\mathrm{e}c^2\bar{\psi}\psi$, is then equal to the trace of the SEM tensor of QED, given in equation~\eqref{eq:contraction}.

By operating both sides of equation~\eqref{eq:tensorgravity} by $\widetilde{\nabla}_\nu$, the left-hand side becomes $\widetilde{\nabla}_\nu\widetilde{\nabla}_\rho S^{a\nu\rho}=\frac{1}{\sqrt{-g}}\partial_\nu\partial_\rho(\sqrt{-g}\,S^{a\nu\rho})=0$ since $\partial_\nu\partial_\rho$ is symmetric and $S^{a\nu\rho}$ is antisymmetric with respect to the indices $\nu$ and $\rho$. Therefore, the total SEM tensor $T_\mathrm{TEGRW}^{a\nu}$ is conserved as
\begin{equation}
 \widetilde{\nabla}_\nu T_\mathrm{TEGRW}^{a\nu}=\oset{\circ}{\nabla}_\nu T_\mathrm{TEGRW}^{a\nu}=0.
 \label{eq:Tanuconservation}
\end{equation}
This conservation law is necessary for the consistency of the theory \cite{Bahamonde2023a}.

As shown in section~5.1 of the supplementary material, the field equation of TEGRW in equation~\eqref{eq:tensorgravity} can also be written in a form showing only space-time indices as
\begin{equation}
 \widetilde{\nabla}_\rho S^{\mu\nu\rho}=\kappa\mathfrak{T}_\mathrm{TEGRW}^{\mu\nu}.
 \label{eq:tensorgravity2}
\end{equation}
On the right-hand side of equation~\eqref{eq:tensorgravity2}, the SEM pseudotensor $\mathfrak{T}_\mathrm{TEGRW}^{\mu\nu}$ is given by
\begin{equation}
 \mathfrak{T}_\mathrm{TEGRW}^{\mu\nu}=T_\mathrm{TEGRW}^{\mu\nu}+\frac{1}{\kappa}S^{\sigma\rho\nu}\oset{\bullet}{\Gamma}^{\setexp{-1}{\mu}}_{\;\,\sigma\rho}.
 \label{eq:SEMpseudotensor}
\end{equation}
By operating both sides of equation~\eqref{eq:tensorgravity2} by $\widetilde{\nabla}_\nu$, the left-hand side becomes $\widetilde{\nabla}_\nu\widetilde{\nabla}_\rho S^{\mu\nu\rho}=\frac{1}{\sqrt{-g}}\partial_\nu\partial_\rho(\sqrt{-g}\,S^{\mu\nu\rho})=0$ since $\partial_\nu\partial_\rho$ is symmetric and $S^{\mu\nu\rho}$ is antisymmetric with respect to the indices $\nu$ and $\rho$. Therefore, the total SEM pseudotensor $\mathfrak{T}^{\mu\nu}$ is conserved as
\begin{equation}
 \widetilde{\nabla}_\nu\mathfrak{T}_\mathrm{TEGRW}^{\mu\nu}=0.
 \label{eq:pseudoconservation}
\end{equation}
Due to the pseudotensor nature of $\mathfrak{T}_\mathrm{TEGRW}^{\mu\nu}$, the conservation law of $\mathfrak{T}_\mathrm{TEGRW}^{\mu\nu}$ in equation~\eqref{eq:pseudoconservation} cannot be written in terms of the Levi-Civita coordinate-covariant derivative in the same way as the conservation laws of $T_\mathrm{m}^{\mu\nu}$ and $T_\mathrm{TEGRW}^{a\nu}$ in equations \eqref{eq:Tmconservation} and \eqref{eq:Tanuconservation}.

\subsubsection{\label{sec:Maxwell}Maxwell's equations in TEGRW}
Next, we present the dynamical equation of the electromagnetic field in TEGRW. In the eight-spinor formalism, the dynamical equation of the electromagnetic field is obtained from the Euler--Lagrange equation for $\Theta$ or $\bar{\Theta}$ \cite{Partanen2024a}. After some algebra, presented in section~5.2 of the supplementary material, we obtain the spinorial Maxwell equation, given by
\begin{equation}
 \widetilde{\nabla}_\rho(\boldsymbol{\gamma}_\mathrm{B}^\rho\Psi)
 =\boldsymbol{\gamma}_\mathrm{B}^\rho(\vec{\partial}_\rho+\oset{\bullet}{e}_{\;\,\rho}^a\widetilde{\nabla}_\sigma \oset{\bullet}{e}_a^{\;\,\sigma})\Psi
 =-\Phi.
 \label{eq:MaxwellTEGR}
\end{equation}
Equation \eqref{eq:MaxwellTEGR} corresponds to the spinorial Maxwell equation, introduced in \cite{Partanen2024a}, generalized to include coupling to gravity. The coupling to gravity comes through $\widetilde{\nabla}_\sigma \oset{\bullet}{e}_a^{\;\,\sigma}$. In flat space-time, we have $\widetilde{\nabla}_\sigma \oset{\bullet}{e}_a^{\;\,\sigma}=0$. In Cartesian coordinates, this relation is trivial, but it also holds in curvilinear coordinate systems, such as in spherical coordinates. Therefore, the spinorial Maxwell equation in general coordinates in flat space-time can be written as $\boldsymbol{\gamma}_\mathrm{B}^\rho\vec{\partial}_\rho\Psi=-\Phi$. The only change to the Cartesian Minkowski space-time form studied in \cite{Partanen2024a} is the use of general coordinates.

The derivation of the spinorial Maxwell equation in equation~\eqref{eq:MaxwellTEGR} is equivalent but different from the conventional electromagnetic four-potential-based derivation of Maxwell's equations. Starting from the Euler--Lagrange equation for $A_\nu$ and using also the Bianchi identity, after some algebra, presented in section~5.2 of the supplementary material, we obtain the conventional form of Maxwell's equations in general relativity as a set of two equations, given by \cite{Misner1973}
\begin{equation}
 \widetilde{\nabla}_\rho F^{\rho\nu}=\oset{\circ}{\nabla}_\rho F^{\rho\nu}=\mu_0J_\mathrm{e}^\nu,
 \label{eq:MaxwellTEGR_conventional}
\end{equation}
\begin{equation}
 \oset{\circ}{\nabla}_\rho F_{\mu\nu}+\oset{\circ}{\nabla}_\mu F_{\nu\rho}+\oset{\circ}{\nabla}_\nu F_{\rho\mu}=0.
 \label{eq:Bianchi1}
\end{equation}
As shown in \cite{Partanen2024a}, the spinorial Maxwell equation, i.e., equation~\eqref{eq:MaxwellTEGR}, is equivalent to the full set of conventional Maxwell's equations. In contrast, in the four-vector and tensor formalism, one obtains only Gauss's law for electricity and the Amp\'ere--Maxwell law, i.e., equation~\eqref{eq:MaxwellTEGR_conventional}, directly from the Euler--Lagrange equations. Gauss's law for magnetism and Faraday's law, i.e., equation~\eqref{eq:Bianchi1}, are obtained by using the Bianchi identity of the electromagnetic field tensor \cite{Misner1973}.

By operating both sides of equation~\eqref{eq:MaxwellTEGR_conventional} by $\widetilde{\nabla}_\nu$, the left-hand side becomes $\widetilde{\nabla}_\nu\widetilde{\nabla}_\rho F^{\rho\nu}=\frac{1}{\sqrt{-g}}\partial_\nu\partial_\rho(\sqrt{-g}\,F^{\rho\nu})=0$ since $\partial_\nu\partial_\rho$ is symmetric and $F^{\rho\nu}$ is antisymmetric with respect to the indices $\nu$ and $\rho$. The right-hand side becomes $\mu_0\widetilde{\nabla}_\nu J_\mathrm{e}^\nu=\mu_0\oset{\circ}{\nabla}_\nu J_\mathrm{e}^\nu$. Therefore, the electromagnetic four-current density is conserved as
\begin{equation}
 \widetilde{\nabla}_\nu J_\mathrm{e}^\nu=\oset{\circ}{\nabla}_\nu J_\mathrm{e}^\nu=0.
 \label{eq:Jconservation}
\end{equation}
Another derivation of the conservation law in equation~\eqref{eq:Jconservation}, based on the U(1) symmetry of QED, is given in section~2.4 of the supplementary material. In the case of electromagnetism in TEGRW, the two derivations lead to the same conservation law of the electric four-current density $J_\mathrm{e}^\nu$, whereas, in the case of gravity in TEGRW, the derivations lead to conservation laws of different objects: the form $T_\mathrm{m}^{\mu\nu}$ of the SEM tensor of the Dirac and electromagnetic fields in equation~\eqref{eq:Tmconservation} and the form $T_\mathrm{TEGRW}^{a\nu}$ of the total SEM tensor including the gravitational field in equation~\eqref{eq:Tanuconservation}. Furthermore, in the case of gravity in TEGRW, we have the conservation law of the SEM pseudotensor $\mathfrak{T}_\mathrm{TEGRW}^{\mu\nu}$, given in equation~\eqref{eq:pseudoconservation}.

\subsubsection{\label{sec:Dirac}Dirac equation in TEGRW}
Next, we present the dynamical equation of the Dirac field in TEGRW. Starting from the Euler--Lagrange equation for $\psi$ or $\bar{\psi}$, after some algebra, presented in section~5.3 of the supplementary material, we obtain the dynamical equation of the Dirac field, given by
\begin{equation}
 i\hbar c\boldsymbol{\gamma}_\mathrm{F}^\rho\Big(\vec{D}_\rho+\frac{1}{2}\oset{\bullet}{e}_{\;\,\rho}^a\widetilde{\nabla}_\sigma \oset{\bullet}{e}_a^{\;\,\sigma}\Big)\psi
 -m_\mathrm{e}c^2\psi=0.
 \label{eq:DiracTEGR}
\end{equation}
Equation \eqref{eq:DiracTEGR} is the Dirac equation in TEGRW. We note that, in flat space-time, we have $\widetilde{\nabla}_\sigma \oset{\bullet}{e}_a^{\;\,\sigma}=0$. Therefore, the Dirac equation in general coordinates in flat space-time simplifies to its conventional form, given by $i\hbar c\boldsymbol{\gamma}_\mathrm{F}^\rho\vec{D}_\rho\psi-m_\mathrm{e}c^2\psi=0$, as expected. Comparison of equations \eqref{eq:MaxwellTEGR} and \eqref{eq:DiracTEGR} shows that the quantity that multiplies the term $\oset{\bullet}{e}_{\;\,\rho}^a\widetilde{\nabla}_\sigma \oset{\bullet}{e}_a^{\;\,\sigma}$ is the spin since it is $S=1$ for vector bosons and $S=\frac{1}{2}$ for Dirac fermions.

The Dirac equation of TEGRW in equation~\eqref{eq:DiracTEGR} can be compared with previous formulations of the Dirac equation in TEGRW. We write the term $\oset{\bullet}{e}_{\;\,\rho}^a\widetilde{\nabla}_\sigma \oset{\bullet}{e}_a^{\;\,\sigma}$ in equation~\eqref{eq:DiracTEGR} as $\oset{\bullet}{e}_{\;\,\rho}^a\widetilde{\nabla}_\sigma \oset{\bullet}{e}_a^{\;\,\sigma}=\oset{\bullet}{e}_{\;\,\rho}^a\partial_\sigma \oset{\bullet}{e}_a^{\;\,\sigma}+\oset{\circ}{\Gamma}_{\;\,\rho\sigma}^{\setexp{-1}{\sigma}}=-\oset{\bullet}{K}_{\;\,\rho\sigma}^{\setexp{-1}{\sigma}}=-\oset{\bullet}{T}_{\;\,\sigma\rho}^{\setexp{-1}{\sigma}}=-\mathcal{V}_\rho$. In the first equality, we have applied partial differentiation. In the second equality, we have applied equation~\eqref{eq:Christoffel2}. In the third equality, we have applied the definition of the contortion tensor in equation~\eqref{eq:contortion0} and the antisymmetry of the torsion tensor in its last two indices. In the last equality, we have defined $\mathcal{V}_\rho=\oset{\bullet}{T}_{\;\,\sigma\rho}^{\setexp{-1}{\sigma}}$, which is the vector part of the torsion decomposition \cite{Aldrovandi2012}. Thus, the Dirac equation in equation~\eqref{eq:DiracTEGR} becomes $i\hbar c\boldsymbol{\gamma}_\mathrm{F}^\rho(\vec{D}_\rho-\frac{1}{2}\mathcal{V}_\rho)\psi-m_\mathrm{e}c^2\psi=0$. In the absence of the electromagnetic coupling, we set $\vec{D}_\rho=\vec{\partial}_\rho$, in which case, the Dirac equation above corresponds the Dirac equation in equation (12.36) of \cite{Aldrovandi2012} for zero spin connection.

\section{\label{sec:StandardModel}Unified gravity and the full Standard Model}

In this section, our goal is to explain how unified gravity is extended to cover all quantum fields of the full Standard Model. We start with a review of the representations of quantum fields of different particle types. For the quantum fields of the Standard Model, we present two different but mathematically equivalent representations. The first one of these representations is the standard representation, and the second one is the eight-spinor representation, which was originally introduced in the case of the electromagnetic field in \cite{Partanen2024a}. Finally, we include the fields of all particle types in the Lagrangian density, and thus, unify gravity with the Lagrangian density of the Standard Model.

\subsection{Fermions}

\subsubsection{Standard representation}

The Standard Model fermions have spin $S=\frac{1}{2}$. They are divided into quarks and leptons. The quark fields are described by the conventional Dirac spinors $q^i$ for SU(3) color charge $i\in\{r,g,b\}$ and flavor $q\in\{u,d,c,s,t,b\}$. For a given flavor, we write \cite{Peskin2018,Schwartz2014}
\begin{equation}
 q=
 \left[\begin{array}{c}
  q^r\\
  q^g\\
  q^b
 \end{array}\right].
 \label{eq:q}
\end{equation}
Using this notation, the color charge index of quark fields is not shown in the equations below.

There are three generations of SU(2) doublet pairs of left-handed quarks and leptons, indexed by $i\in\{1,2,3\}$, as \cite{Peskin2018,Schwartz2014}
\begin{align}
 Q_\mathrm{L}^i &\in\left\{
 \left[\begin{array}{c}
   u_\mathrm{L}\\
   d_\mathrm{L}
 \end{array}\right],
 \left[\begin{array}{c}
   c_\mathrm{L}\\
   s_\mathrm{L}
 \end{array}\right],
 \left[\begin{array}{c}
   t_\mathrm{L}\\
   b_\mathrm{L}
 \end{array}\right]\right\},\nonumber\\
 L_\mathrm{L}^i &\in\left\{
 \left[\begin{array}{c}
   \nu_{e\mathrm{L}}\\
   e_{\mathrm{L}}
 \end{array}\right],
 \left[\begin{array}{c}
   \nu_{\mu\mathrm{L}}\\
   \mu_{\mathrm{L}}
 \end{array}\right],
 \left[\begin{array}{c}
   \nu_{\tau\mathrm{L}}\\
   \tau_{\mathrm{L}}
 \end{array}\right]\right\},
 \label{eq:QL}
\end{align}
and there are three generations of SU(2) singlet right-handed quarks and leptons, indexed by $i\in\{1,2,3\}$, as \cite{Peskin2018,Schwartz2014}
\begin{equation}
 u_\mathrm{R}^i\in\{u_\mathrm{R},c_\mathrm{R},t_\mathrm{R}\},\hspace{0.5cm}
 d_\mathrm{R}^i\in\{d_\mathrm{R},s_\mathrm{R},b_\mathrm{R}\},
\end{equation}
\begin{equation}
 e_\mathrm{R}^i\in\{e_\mathrm{R},\mu_\mathrm{R},\tau_\mathrm{R}\},\hspace{0.5cm}
 \nu_\mathrm{R}^i\in\{\nu_{e\mathrm{R}},\nu_{\mu\mathrm{R}},\nu_{\tau\mathrm{R}}\}.
\end{equation}
For simplicity, we introduce a notation combining the quark and lepton fields, indexed by $j\in\{1,2,...,6\}$, as
\begin{equation}
 \psi_j^i\in\{Q_\mathrm{L}^i,u_\mathrm{R}^i,d_\mathrm{R}^i,L_\mathrm{L}^i,\nu_\mathrm{R}^i,e_\mathrm{R}^i\}.
 \label{eq:psi}
\end{equation}

\subsubsection{Eight-spinor representation}

The eight-spinor representation of fermions is obtained from the standard representation above in a simple way generalizing equation~\eqref{eq:Diraceightspinor} as
\begin{equation}
 \psi_{8j}^i=\psi_j^i\boldsymbol{\mathfrak{e}}_0=[\mathbf{0},\mathbf{0},\mathbf{0},\mathbf{0},\psi_j^i,\mathbf{0},\mathbf{0},\mathbf{0}]^T.
 \label{eq:DiraceightspinorSM}
\end{equation}
Again, the transpose in equation~\eqref{eq:DiraceightspinorSM} only operates on the eight-spinor degree of freedom. For a discussion of selected aspects of Dirac eight-spinors, see section~\ref{sec:eightspinors}.

\subsection{\label{sec:vectorbosons}Vector bosons}

The vector bosons of the Standard Model, having spin $S=1$, are the force carriers of the fundamental interactions for elementary fermions. They are introduced as gauge fields through the gauge-covariant derivative acting on fermion fields. Therefore, vector bosons are often called gauge bosons. In the absence of vector bosons, the Lagrangian density of the Standard Model satisfies the symmetries U(1), SU(2), and SU(3) globally, i.e., for constant, space-time-independent, symmetry transformation parameters. These global symmetries are then promoted to local symmetries, with space-time-dependent symmetry transformation parameters, by introducing the gauge-covariant derivative and utilizing gauge theory. To describe dynamics of the gauge fields, the kinetic field strength terms of each gauge field are also added in the Lagrangian density.

\subsubsection{Standard representation}

The gauge-covariant derivative, through which the vector bosons of the Standard Model have been introduced, is given by \cite{Peskin2018,Schwartz2014}
\begin{align}
 \vec{\mathbf{D}}_\nu &=\vec{\partial}_\nu-i\frac{g_\mathrm{s}}{\sqrt{\hbar c}}G_{l\nu}\frac{\boldsymbol{\lambda}^l}{2}
 -i\frac{g_\mathrm{ew}}{\sqrt{\hbar c}}W_{i\nu}\frac{\boldsymbol{\sigma}_\mathrm{F}^i}{2}-i\frac{g'_\mathrm{ew}}{\sqrt{\hbar c}}B_\nu\frac{Y_\mathrm{w}}{2},\nonumber\\
 \cev{\mathbf{D}}_\nu &=\cev{\partial}_\nu+i\frac{g_\mathrm{s}}{\sqrt{\hbar c}}G_{l\nu}\frac{\boldsymbol{\lambda}^l}{2}
 +i\frac{g_\mathrm{ew}}{\sqrt{\hbar c}}W_{i\nu}\frac{\boldsymbol{\sigma}_\mathrm{F}^i}{2}+i\frac{g'_\mathrm{ew}}{\sqrt{\hbar c}}B_\nu\frac{Y_\mathrm{w}}{2}.
\end{align}
Here $G_{la}$, with $l\in\{1,2,...,8\}$, is the SU(3) gauge field of eight gluons, and $g_\mathrm{s}$ is the associated coupling constant of strong interaction. The SU(3) generators are the Gell-Mann matrices $\boldsymbol{\lambda}^l/2$, which act on the SU(3) color charge degrees of freedom in equation~\eqref{eq:q}. The quantity $W_{i\nu}$, with $i\in\{1,2,3\}$, is the SU(2) gauge field associated with weak isospin and coupling constant $g_\mathrm{ew}$. The canonically normalized SU(2) generators $\boldsymbol{\sigma}_\mathrm{F}^i/2$ are given by the Pauli matrices, which act on the SU(2) doublet degrees of freedom in equation~\eqref{eq:QL} and in the Higgs field described below. The quantity $B_\nu$ is the U(1)$_\mathrm{Y}$ gauge field associated with weak hypercharge $Y_\mathrm{w}$ and coupling constant $g'_\mathrm{ew}$. In the matrix form, the gauge fields $\mathbf{G}_\nu$ and $\mathbf{W}_\nu$ of the strong and weak interactions are written as \cite{Peskin2018,Schwartz2014}
\begin{equation}
 \mathbf{G}_\nu=G_{l\nu}\frac{\boldsymbol{\lambda}^l}{2},\hspace{0.5cm}
 \mathbf{W}_\nu=W_{i\nu}\frac{\boldsymbol{\sigma}_\mathrm{F}^i}{2}
\end{equation}
The commutator of the gauge-covariant derivative defines the corresponding field strength tensors $\mathbf{G}_{\mu\nu}$ and $\mathbf{W}_{\mu\nu}$, for which we obtain \cite{Peskin2018,Schwartz2014}
\begin{align}
 &\mathbf{G}_{\mu\nu}=\partial_\mu\mathbf{G}_\nu-\partial_\nu\mathbf{G}_\mu-ig_\mathrm{s}[\mathbf{G}_\mu,\mathbf{G}_\nu]=G_{l\mu\nu}\frac{\boldsymbol{\lambda}^l}{2},\nonumber\\
 &\hspace{0.2cm}G_{l\mu\nu}=\partial_\mu G_{l\nu}-\partial_\nu G_{l\mu}+g_\mathrm{s}(f_\mathrm{s})_l^{\;\,mn}G_{m\mu}G_{n\nu},
 \label{eq:Gdefinition}
\end{align}
\begin{align}
 &\mathbf{W}_{\mu\nu}=\partial_\mu\mathbf{W}_\nu-\partial_\nu\mathbf{W}_\mu-ig_\mathrm{ew}[\mathbf{W}_\mu,\mathbf{W}_\nu]=W_{i\mu\nu}\frac{\boldsymbol{\sigma}_\mathrm{F}^i}{2},\nonumber\\
 &\hspace{0.2cm}W_{i\mu\nu}=\partial_\mu W_{i\nu}-\partial_\nu W_{i\mu}+g_\mathrm{ew}(f_\mathrm{w})_i^{\;\,jk}W_{j\mu}W_{k\nu}.
 \label{eq:Wdefinition}
\end{align}
Here $(f_\mathrm{s})_l^{\;\,mn}=-\frac{i}{4}\mathrm{Tr}(\boldsymbol{\lambda}^l[\boldsymbol{\lambda}^m,\boldsymbol{\lambda}^n])$, with $l,m,n\in\{1,2,...,8\}$, are the totally antisymmetric structure constants of the strong interaction, and $(f_\mathrm{w})_i^{\;\,jk}=-\frac{i}{4}\mathrm{Tr}(\boldsymbol{\sigma}_\mathrm{F}^i[\boldsymbol{\sigma}_\mathrm{F}^j,\boldsymbol{\sigma}_\mathrm{F}^k])=\varepsilon^{ijk}$, with $i,j,k\in\{1,2,3\}$, are the corresponding structure constants of the weak interaction.

The diagonal Lie algebra metric is determined to lower and raise the Lie algebra indices. For strong interaction, the Lie algebra metric is $(\eta_\mathrm{s})^{mn}=\delta^{mn}$. Correspondingly, for weak interaction, we have $(\eta_\mathrm{w})^{jk}=\delta^{jk}$. Therefore, the Lie algebra indices of the strong and weak interactions can be written lowered or raised without paying attention to it.

The electroweak unification is based on the symmetry breaking of SU(2)$\otimes$U(1)$_\mathrm{Y}\rightarrow$U(1)$_\mathrm{em}$ \cite{Schwartz2014,Glashow1961,Weinberg1967,Salam1968}. The high-energy $\mathrm{U(1)}_\mathrm{Y}$ is not to be confused with the low-energy U(1)$_\mathrm{em}$. The U(1)$_\mathrm{em}$ symmetry of QED is generated by a linear combination the weak hypercharge and the third SU(2) generator of the weak isospin, i.e., $Q=\frac{1}{2}(\boldsymbol{\sigma}_\mathrm{F}^3+Y_\mathrm{w}\mathbf{I}_2)$. The neutral W$^0$ boson of the three vector bosons of the weak isospin (W$^+$, W$^-$, W$^0$) mixes with the weak hypercharge gauge boson (B). This results in the observed Z$^0$ intermediate vector boson and the photon. Thus, there are three intermediate vector bosons (W$^+$, W$^-$, Z$^0$) and one photon.

\subsubsection{Eight-spinor representation}

In analogy to how the electromagnetic spinor in equation~\eqref{eq:electromagneticspinor} is formed from the Cartesian Minkowski coordinate form of the electromagnetic field strength tensor $F_{ab}$ in equation~\eqref{eq:Ftensor} \cite{Partanen2024a}, we form eight gluon spinors $\mathcal{G}_l$ from the gluon field strength tensor $G_{lab}$, one for each $l$, three intermediate vector boson spinors $\mathcal{W}_i$ from the field strength tensor $W_{iab}$, one for each $i$, and one weak hypercharge field spinor $\mathcal{B}$ from the field strength tensor $B_{ab}$ as
\begin{equation}
 \mathcal{G}_l=[0,G_{l0x},G_{l0y},G_{l0z},0,iG_{lzy},iG_{lxz},iG_{lyx}]^T,
\end{equation}
\begin{equation}
 \mathcal{W}_i=[0,W_{i0x},W_{i0y},W_{i0z},0,iW_{izy},iW_{ixz},iW_{iyx}]^T,
\end{equation}
\begin{equation}
 \mathcal{B}=[0,B_{0x},B_{0y},B_{0z},0,iB_{zy},iB_{xz},iB_{yx}]^T.
\end{equation}
For simplicity, we introduce a notation combining the different eight-spinor gauge field strengths, indexed by $i\in\{1,2,...,12\}$, as
\begin{equation}
 \Psi_i\in\{\mathcal{G}_1,\mathcal{G}_2,...,\mathcal{G}_8,\mathcal{W}_1,\mathcal{W}_2,\mathcal{W}_3,\mathcal{B}\}.
 \label{eq:Psi}
\end{equation}

In analogy to the electromagnetic potential spinor in equation~\eqref{eq:potentialspinor} \cite{Partanen2024a}, we form the gluon potential spinors $G_l$, the intermediate vector boson potential spinors $W_i$, and the weak hypercharge potential spinor $B$. These spinors are four-vector-type eight-spinors \cite{Partanen2024a}, formed from the components of $G_l^{\;\,a}$, $W_i^{\;\,a}$, and $B^a$ as
\begin{align}
 G_l &=[0,G_l^{\;\,x},G_l^{\;\,y},G_l^{\;\,z},G_l^{\;\,0},0,0,0]^T\nonumber\\
 &=[0,-G_{lx},-G_{ly},-G_{lz},G_{l0},0,0,0]^T,
\end{align}
\begin{align}
 W_i &=[0,W_i^{\;\,x},W_i^{\;\,y},W_i^{\;\,z},W_i^{\;\,0},0,0,0]^T\nonumber\\
 &=[0,-W_{ix},-W_{iy},-W_{iz},W_{i0},0,0,0]^T,
\end{align}
\begin{align}
 B &=[0,B^x,B^y,B^z,B^0,0,0,0]^T,\nonumber\\
 &=[0,-B_x,-B_y,-B_z,B_0,0,0,0]^T.
\end{align}

Using the eight-spinor gauge potentials above, we define the eight-spinor gauge-covariant derivative operators as
\begin{align}
 \vec{\mathbf{D}} &=\vec{\partial}-i\frac{g_\mathrm{s}}{\sqrt{\hbar c}}G_l\frac{\boldsymbol{\lambda}^l}{2}-i\frac{g_\mathrm{ew}}{\sqrt{\hbar c}}W_i\frac{\boldsymbol{\sigma}_\mathrm{F}^i}{2}-i\frac{g'_\mathrm{ew}}{\sqrt{\hbar c}}B\frac{Y_\mathrm{w}}{2}\nonumber\\
 &=[\mathbf{0},\vec{\mathbf{D}}_x,\vec{\mathbf{D}}_y,\vec{\mathbf{D}}_z,-\vec{\mathbf{D}}_0,\mathbf{0},\mathbf{0},\mathbf{0}]^T,\nonumber\\
 \cev{\mathbf{D}} &=\cev{\partial}+i\frac{g_\mathrm{s}}{\sqrt{\hbar c}}G_l\frac{\boldsymbol{\lambda}^l}{2}+i\frac{g_\mathrm{ew}}{\sqrt{\hbar c}}W_i\frac{\boldsymbol{\sigma}_\mathrm{F}^i}{2}+i\frac{g'_\mathrm{ew}}{\sqrt{\hbar c}}B\frac{Y_\mathrm{w}}{2}\nonumber\\
 &=[\mathbf{0},\cev{\mathbf{D}}_x,\cev{\mathbf{D}}_y,\cev{\mathbf{D}}_z,\cev{\mathbf{D}}_0,\mathbf{0},\mathbf{0},\mathbf{0}].
 \label{eq:chromodynamicD}
\end{align}
The transpose in the last expression of $\vec{\mathbf{D}}$ only makes this operator an eight-component column vector and it does not apply to the component matrix operators $\vec{\mathbf{D}}_a$.

\subsection{Scalar boson}

\subsubsection{Standard representation}

The scalar boson of the Standard Model is known as the Higgs boson with spin $S=0$. It is represented by an SU(2) doublet of complex-valued scalar fields as \cite{Weinberg1996}
\begin{equation}
 \varphi=
 \left[\begin{array}{c}
 \varphi^+\\
 \varphi^0
 \end{array}\right].
\end{equation}
Here the superscripts + and 0 indicate the electric charge of the components. Through the Higgs mechanism \cite{Higgs1964,Englert1964,Higgs1966}, the nonzero vacuum expectation value of the Higgs field results in the electroweak symmetry breaking and the generation of masses for the intermediate vector bosons and the fermions of the Standard Model \cite{Schwartz2014}. Enabling the generation of masses while preserving the gauge symmetry of the electroweak theory \cite{Glashow1961,Weinberg1967,Salam1968}, the Higgs mechanism became an essential part of the Standard Model.

\subsubsection{Eight-spinor representation}

The eight-spinor representation of the Higgs field is obtained from the standard representation above in analogy to the eight-spinor representation of the fermionic fields in equation~\eqref{eq:DiraceightspinorSM}, as
\begin{equation}
 \varphi_8=\varphi\boldsymbol{\mathfrak{e}}_0=[\mathbf{0},\mathbf{0},\mathbf{0},\mathbf{0},\varphi,\mathbf{0},\mathbf{0},\mathbf{0}]^T.
 \label{eq:HiggseightspinorSM}
\end{equation}
Again, the transpose in equation~\eqref{eq:HiggseightspinorSM} only operates on the eight-spinor degree of freedom. The adjoint spinor is given by $\bar{\varphi}_8=\varphi_8^\dag\boldsymbol{\gamma}_\mathrm{B}^0=\varphi^\dag\bar{\boldsymbol{\mathfrak{e}}}_0=[\mathbf{0},\mathbf{0},\mathbf{0},\mathbf{0},-\varphi^\dag,\mathbf{0},\mathbf{0},\mathbf{0}]$.

\subsection{Tensor boson}

The Standard Model does not contain the tensor boson of spin $S=2$, the graviton, since it does not describe gravity. In the present gauge theory of unified gravity, the tensor boson associated with the four U(1) symmetry transformations of the components of the space-time dimension field in equation~\eqref{eq:psitransformation} is described by the tensor gauge field $H_{a\nu}$ and the related field strength tensor $H_{a\mu\nu}$, introduced in equations \eqref{eq:covariantderivative} and \eqref{eq:Hdefinition}.

\subsection{Lagrangian density of the Standard Model extended to include gravity}

We can now generalize the Lagrangian density of unified gravity, which involved the Dirac electron--positron field, electromagnetic field, and the gravitational field, to include all quantum fields of the Standard Model. Here we explicitly write the extension of the locally gauge-invariant Lagrangian density in equation~\eqref{eq:L}. The pertinent gauge-fixed Lagrangian density would follow from the gauge-invariant Lagrangian density by the Faddeev--Popov gauge fixing method in analogy to the discussion in section~\ref{sec:gaugefixing}. An essential difference is that the interactions of ghost fields take a more essential role in the Yang-Mills gauge theories of the Standard Model as is well known \cite{Peskin2018,Schwartz2014}.

The complete generalized gauge-invariant Lagrangian density of the Standard Model including gravity can be considered as the master Lagrangian of the Universe since it contains all known fundamental interactions of nature. It is written as
\begin{align}
 \mathcal{L} &=\mathcal{L}_{S=0}+\mathcal{L}_{S=\frac{1}{2}}+\mathcal{L}_{S=1}+\mathcal{L}_{S=2}+\mathcal{L}_\mathrm{pot}+\mathcal{L}_\mathrm{Yukawa}.
 \label{eq:LSM}
\end{align}
Here $\mathcal{L}_{S=0}$, $\mathcal{L}_{S=\frac{1}{2}}$, $\mathcal{L}_{S=1}$, and $\mathcal{L}_{S=2}$ are the generalized derivative terms of the spin $S=0$ Higgs boson field, spin $S=\frac{1}{2}$ fermion fields, spin $S=1$ vector boson fields, and the spin $S=2$ tensor boson gravitational field, respectively. The generalized derivative terms are given by
\begin{align}
 \mathcal{L}_{S=0} &=\hbar c(i\bar{\varphi}_8\cev{\mathbf{D}}\bar{\mathbf{I}}_\mathrm{g}\boldsymbol{\gamma}_\mathrm{B}^5\boldsymbol{\gamma}_\mathrm{B}^\nu\vec{\boldsymbol{\mathcal{D}}}_\nu\mathbf{I}_\mathrm{g}\vec{\mathbf{D}}\varphi_8
 -\bar{\varphi}_8\cev{\mathbf{D}}\vec{\mathbf{D}}\varphi_8)\sqrt{-g},\nonumber\\
 \mathcal{L}_{S=\frac{1}{2}}
 &=\sum_{i,j}\frac{\hbar c}{4}\bar{\psi}_{8j}^i(\bar{\boldsymbol{\gamma}}_\mathrm{F}\bar{\mathbf{I}}_\mathrm{g}\boldsymbol{\gamma}_\mathrm{B}^5\boldsymbol{\gamma}_\mathrm{B}^\nu\vec{\boldsymbol{\mathcal{D}}}_\nu\mathbf{I}_\mathrm{g}\vec{\mathbf{D}}\nonumber\\
 &\hspace{0.5cm}-\cev{\mathbf{D}}\bar{\mathbf{I}}_\mathrm{g}\boldsymbol{\gamma}_\mathrm{B}^5\boldsymbol{\gamma}_\mathrm{B}^\nu\vec{\boldsymbol{\mathcal{D}}}_\nu\mathbf{I}_\mathrm{g}\boldsymbol{\gamma}_\mathrm{F})\psi_{8j}^i\sqrt{-g},\nonumber\\
 \mathcal{L}_{S=1} &=\sum_i(i\bar{\Psi}_i\mathbf{I}_\mathrm{g}^\dag\boldsymbol{\gamma}_\mathrm{B}^5\boldsymbol{\gamma}_\mathrm{B}^\nu\lowerbar{\vec{\boldsymbol{\mathcal{D}}}}_\nu^\lowerdag\bar{\mathbf{I}}_\mathrm{g}^\dag \Psi_i
 +\bar{\Psi}_i\Psi_i)\sqrt{-g},\nonumber\\
 \mathcal{L}_{S=2} &=\frac{1}{4\kappa}H_{\rho\mu\nu}S^{\rho\mu\nu}\sqrt{-g}.
 \label{eq:Lkin}
\end{align}
The term $\mathcal{L}_\mathrm{pot}$ of equation~\eqref{eq:LSM} is the Higgs field potential term, and $\mathcal{L}_\mathrm{Yukawa}$ is the Yukawa coupling term. These terms are responsible for the generation of masses of the intermediate vector bosons and fermions. They are given by
\begin{align}
 \mathcal{L}_\mathrm{pot} &=\hbar c(i\bar{\varphi}_8\mathbf{I}_\mathrm{g}^\dag\boldsymbol{\gamma}_\mathrm{B}^5\boldsymbol{\gamma}_\mathrm{B}^\nu\lowerbar{\vec{\boldsymbol{\mathcal{D}}}}_\nu^\lowerdag\bar{\mathbf{I}}_\mathrm{g}^\dag\varphi_8\!+\!3\bar{\varphi}_8\varphi_8)\nonumber\\
 &\hspace{0.4cm}\times(\lambda_\mathrm{H}\bar{\varphi}_8\varphi_8\!+\!\mu_\mathrm{H}^2)\sqrt{-g},
\end{align}
\begin{align}
 \mathcal{L}_\mathrm{Yukawa}
 &=\sum_{i,j}\Big[-\frac{i}{2}(Y'_\mathrm{u})_{ij}(\bar{Q}_\mathrm{L8})_i\tilde\varphi\mathbf{I}_\mathrm{g}^\dag\boldsymbol{\gamma}_\mathrm{B}^5\boldsymbol{\gamma}_\mathrm{B}^\nu\lowerbar{\vec{\boldsymbol{\mathcal{D}}}}_\nu^\lowerdag\bar{\mathbf{I}}_\mathrm{g}^\dag(u_\mathrm{R8})_j\nonumber\\
 &\hspace{0.4cm}-\frac{i}{2}(Y'_\mathrm{d})_{ij}(\bar{Q}_\mathrm{L8})_i\varphi\mathbf{I}_\mathrm{g}^\dag\boldsymbol{\gamma}_\mathrm{B}^5\boldsymbol{\gamma}_\mathrm{B}^\nu\lowerbar{\vec{\boldsymbol{\mathcal{D}}}}_\nu^\lowerdag\bar{\mathbf{I}}_\mathrm{g}^\dag(d_\mathrm{R8})_j\nonumber\\
 &\hspace{0.4cm}-\frac{i}{2}(Y'_\mathrm{e})_{ij}(\bar{L}_\mathrm{L8})_i\varphi\mathbf{I}_\mathrm{g}^\dag\boldsymbol{\gamma}_\mathrm{B}^5\boldsymbol{\gamma}_\mathrm{B}^\nu\lowerbar{\vec{\boldsymbol{\mathcal{D}}}}_\nu^\lowerdag\bar{\mathbf{I}}_\mathrm{g}^\dag(e_\mathrm{R8})_j\nonumber\\
 &\hspace{0.4cm}+(2Y'_\mathrm{u}-Y_\mathrm{u})_{ij}(\bar{Q}_\mathrm{L8})_i\tilde\varphi(u_\mathrm{R8})_j\nonumber\\
 &\hspace{0.4cm}+(2Y'_\mathrm{d}-Y_\mathrm{d})_{ij}(\bar{Q}_\mathrm{L8})_i\varphi(d_\mathrm{R8})_j\nonumber\\
 &\hspace{0.4cm}+(2Y'_\mathrm{e}-Y_\mathrm{e})_{ij}(\bar{L}_\mathrm{L8})_i\varphi(e_\mathrm{R8})_j
 +h.c.\Big]\sqrt{-g}.
\end{align}
Here $\tilde\varphi=i\boldsymbol{\sigma}_\mathrm{F}^2\varphi^*$ is the charge conjugate state of the Higgs doublet, $\mu_\mathrm{H}$ and $\lambda_\mathrm{H}$ are parameters of the Higgs potential, $Y_\mathrm{u}$, $Y_\mathrm{d}$, and $Y_\mathrm{e}$ are $3\times3$ Yukawa coupling matrices corresponding to the inertial masses, $Y'_\mathrm{u}$, $Y'_\mathrm{d}$, and $Y'_\mathrm{e}$ are Yukawa coupling matrices corresponding to the gravitational masses, and $h.c.$ denotes the Hermitian conjugate of the preceding terms. The gravitational gauge field $H_{a\nu}$ enters the theory trough the gravitational-gauge covariant derivative $\vec{\boldsymbol{\mathcal{D}}}_\nu$, defined in equation~\eqref{eq:covariantderivative}. This derivative preserves its form when all interactions of the Standard Model are included in the theory. Thus the Lagrangian densities in equations \eqref{eq:L} and \eqref{eq:LSM} are invariant in the same local U(1) symmetry transformation of equation~\eqref{eq:psitransformation}.

The relation between the Lagrangian density of the gauge theory of gravity for QED in equation~\eqref{eq:L} and the complete Lagrangian density of the Standard Model including gravity in equation~\eqref{eq:LSM} is the following: The first term of equation~\eqref{eq:L}, the derivative term of the Dirac\vspace{-0.07cm} electron--positron field, is described through $\mathcal{L}_{S=\frac{1}{2}}$, where the right- and left-handed electron--positron fields $e_\mathrm{R}$ and $e_\mathrm{L}$ in $\psi_j^i$, defined in equation~\eqref{eq:psi}, give the contribution of the Dirac electron--positron field spinor $\psi$ of equation~\eqref{eq:L}. The second and third terms of equation~\eqref{eq:L}, the mass terms of the Dirac field, are described through the Yukawa coupling $\mathcal{L}_\mathrm{Yukawa}$. The fourth and fifth terms of equation~\eqref{eq:L}, the derivative terms of the electromagnetic field, are described through $\mathcal{L}_{S=1}$, where the linear combination of $\mathcal{W}_3$ and $\mathcal{B}$ terms of $\Psi_i$, defined in equation~\eqref{eq:Psi}, forms the electromagnetic field spinor $\Psi$ of equation~\eqref{eq:L}. The interaction between the Dirac and electromagnetic fields is described through the gauge-covariant derivative. The sixth term of equation~\eqref{eq:L}, the derivative term of the gravitational gauge field, is equal to $\mathcal{L}_\mathrm{g,kin}=\mathcal{L}_{S=2}$, given in equation~\eqref{eq:Lg}.

\begin{table*}
\footnotesize
\setlength{\tabcolsep}{6.9pt}
\renewcommand{\arraystretch}{1.7}
\caption{\label{tbl:comparison}
Comparison of unified gravity with QED and QCD as examples of Abelian and Yang--Mills gauge theories of the Standard Model and with the conventional TEGRW, which is the established gauge theory of space-time translations. For unified gravity, we use the scaled representation given in section~\ref{sec:scaled}. For QED, we use the scaled representation discussed in section~2.9 of the supplementary material. For QCD, we use its conventional representation. For TEGRW, we use the scaled representation given in section~3.5 of the supplementary material. All representations are given in the SI units. The gauge-covariant derivative row shows only the relevant terms of the full gauge-covariant derivative that is common to all gauge theories. The Lagrangian density row shows the Lagrangian density of the gauge field strength.}
\begin{tabular}{p{2.7cm}p{3.2cm}p{2.8cm}p{3.3cm}p{3.5cm}}
   \hline\hline
   Theory & Unified gravity & QED & QCD & Conventional TEGRW \\[4pt]
   \hline
Gauge symmetry & $4\times$U(1) of $\mathbf{I}_\mathrm{g}$ & Fermionic U(1) & Fermionic SU(3) & Tangent-space translations \\[3pt]
   \hline
\vspace{-0.63cm}\begin{flushleft}Dimension of symmetry\end{flushleft}\vspace{-0.63cm} & 4 & 1 & 8 & $\infty$ \\[3pt]
   \hline
\vspace{-0.63cm}\begin{flushleft}Compactness of symmetry\end{flushleft}\vspace{-0.63cm} & Compact & Compact & Compact & Noncompact \\[3pt]
   \hline
\vspace{-0.63cm}\begin{flushleft}Symmetry   transformation\end{flushleft}\vspace{-0.63cm} & $\bigotimes_a e^{i\phi_{(a)}\mathbf{t}^{(a)}}$ of $\mathbf{I}_\mathrm{g}$ & $e^{i\theta Q}$ of fermions & $e^{i\theta_l\boldsymbol{\lambda}^l/2}$ of quark triplets & \vspace{-0.63cm}\begin{flushleft}$e^{\xi^a\vec{\partial}_a}$ of tangent-space coordinates\end{flushleft}\vspace{-0.63cm}  \\[3pt]
   \hline
\vspace{-0.63cm}\begin{flushleft}Symmetry   generators\end{flushleft}\vspace{-0.63cm} & $\mathbf{t}^a$ (4 generators) & $Q$ (1 generator) & $\boldsymbol{\lambda}^l/2$ (8 generators) & \vspace{-0.63cm}\begin{flushleft}$\vec{\partial}_a$ (4 continuum generators)\end{flushleft}\vspace{-0.63cm} \\[3pt]
   \hline
Symmetry generator dimension & Dimensionless & Dimensionless & Dimensionless & Mass dimension $1$ \\[3pt]
   \hline
Coupling constant & $E_\mathrm{g}'=E_\mathrm{g}\sqrt{\frac{\kappa}{\hbar c}}$ & $e'=e\sqrt{\frac{\mu_0c}{\hbar}}$ & $g_\mathrm{s}$ & $k=\sqrt{\kappa\hbar c}$ \\[3pt]
   \hline
\vspace{-0.63cm}\begin{flushleft}Coupling constant dimension\end{flushleft}\vspace{-0.63cm} & Dimensionless & Dimensionless & Dimensionless & Mass dimension $-1$ \\[3pt]
   \hline
Gauge field & $H'_{a\nu}$ (gravity gauge field) & $A'_\nu$ (four-potential) & $G_{l\nu}$ (8 gluon gauge fields) & \vspace{-0.63cm}\begin{flushleft}${B'}^a_{\;\,\nu}$ (translation gauge field)\end{flushleft}\vspace{-0.63cm} \\[3pt]
   \hline
Gauge-covariant derivative & $\vec{\partial}_\nu-i\frac{E_\mathrm{g}'}{\sqrt{\hbar c}}H'_{a\nu}\mathbf{t}^a$ & $\vec{\partial}_\nu+i\frac{e'}{\sqrt{\hbar c}}A'_\nu Q$ & $\vec{\partial}_\nu-i\frac{g_\mathrm{s}}{\sqrt{\hbar c}}G_{l\nu}\frac{\boldsymbol{\lambda}^l}{2}$ & $\vec{\partial}_\nu+i\frac{k}{\sqrt{\hbar c}}{B'}_{\;\,\nu}^a\vec{\partial}_a$ \\[3pt]
   \hline
Gauge field strength & $H'_{a\mu\nu}=\partial_\mu H'_{a\nu}-\partial_\nu H'_{a\mu}$ & $F'_{\mu\nu}=\partial_\mu A'_\nu-\partial_\nu A'_\mu$ & $G_{l\mu\nu}=\partial_\mu G_{l\nu}-\partial_\nu G_{l\mu}+g_\mathrm{s}(f_\mathrm{s})_l^{mn}G_{m\mu}G_{n\nu}$ & ${\osetsmall{\bullet}{T}}{}^{\prime a}_{\;\;\,\mu\nu}=\partial_\mu {B'}_{\;\,\nu}^a-\partial_\nu {B'}_{\;\,\mu}^a$ \\[3pt]
   \hline
Lagrangian density & $\frac{1}{8}H'_{a\mu\nu}{\widetilde{H}}'^a_{\;\;\,\sigma\lambda}\varepsilon^{\mu\nu\sigma\lambda}\sqrt{-g}$ & $-\frac{1}{8}F'_{\mu\nu}\widetilde{F}'_{\sigma\lambda}\varepsilon^{\mu\nu\sigma\lambda}\sqrt{-g}$ & $-\frac{1}{8}G_{l\mu\nu}\widetilde{G}_{\;\sigma\lambda}^l\varepsilon^{\mu\nu\sigma\lambda}\sqrt{-g}$ & $\frac{1}{8}{\osetsmall{\bullet}{T}}{}'_{\!a\mu\nu}{\widetilde{\osetsmall{\bullet}{T}}}{}^{\prime a}_{\;\;\,\sigma\lambda}\varepsilon^{\mu\nu\sigma\lambda}\sqrt{-g}$ \\[3pt]
   \hline\hline
 \end{tabular}
\end{table*}

\section{\label{sec:discussion}Discussion and comparison with previous theories}

\subsection{\label{sec:comparisonSM}Comparison with the gauge theories of the Standard Model}

Unified gravity is compared with QED and QCD as examples of Abelian and Yang--Mills gauge theories of the Standard Model in table~\ref{tbl:comparison}. The use of unitary and special unitary gauge symmetries is the clearest similarity between unified gravity and the gauge theories of the Standard Model. All these theories are based on compact, finite-dimensional symmetry groups. However, there are also certain differences. The main difference is the action of the gauge symmetry transformation on the space-time dimension field in unified gravity. In contrast, the gauge symmetry transformations of QED and QCD act on fermionic fields and their triplets, respectively. Thus, the gauge symmetry of unified gravity belongs to a different hierarchy in comparison with the Standard Model symmetries related to fermionic and Higgs fields. A related difference is the soldered character of unified gravity, in which a tetrad maps the indices of the four U(1) gauge symmetry generators of gravity to space-time indices. This is expected to be necessary for the description of the tensor gauge field describing gravitational interaction.

As seen in table~\ref{tbl:comparison}, the expressions of the gauge-covariant derivatives in terms of the dimensionless coupling constants, the gauge fields, and the dimensionless symmetry generators in unified gravity are very similar to those of QED and QCD. The expression of the gauge field strength tensor in unified gravity is similar to that in QED due to the Abelian gauge theory nature of these theories. The gauge field strength tensor expression of QCD differs from those of unified gravity and QED by the Lie algebra commutator term originating from the Yang-Mills gauge theory nature of QCD. The expressions of the gauge field Lagrangian densities of the three theories in table~\ref{tbl:comparison} are similar apart from the different sign of the gravity gauge field Lagrangian density in unified gravity. This sign was found to be necessary for obtaining complete agreement between unified gravity and TEGR.

\subsection{\label{sec:discussion_grav}Comparison with the conventional TEGRW}

In table~\ref{tbl:comparison}, unified gravity is also compared with the conventional translation gauge theory formulation of TEGRW. For selected technical aspects in the conventional TEGRW, see section~3 of the supplementary material. The main difference between unified gravity and the conventional TEGRW is the gauge symmetry. Unified gravity is based on four U(1) symmetries, whose gauge symmetry groups are compact and whose gauge symmetry transformations operate on the four components of the space-time dimension field. In contrast, the conventional gauge theory formulation of TEGRW is based on the translation group, which is noncompact. The continuum of tangent-space coordinates in TEGRW acts in the same role as the discrete-valued index of finite number of field or field-multiplet components in unified gravity and in the gauge theories of the Standard Model. The gauge symmetry transformations of TEGRW correspond to translations of the tangent-space coordinates of all fields in the Lagrangian density. Based on the points above, the foundations of the symmetry of unified gravity seem very different from those of the translation gauge symmetry of the conventional TEGRW.

However, as discussed in section~\ref{sec:symmetry}, the gauge symmetries of unified gravity can be described as the translations of the phase factors of $\mathbf{I}_\mathrm{g}$. Since $\mathbf{I}_\mathrm{g}$ is only a single field in the Lagrangian density, its gauge symmetry transformations are fundamentally different from how the translation gauge symmetry is applied in TEGRW. Nevertheless, as discussed in section~\ref{sec:teleparallel}, we observe that there is no difference \emph{at the level of classical field equations} between unified gravity and the conventional TEGRW \cite{Bahamonde2023a,Aldrovandi2012,Krssak2019,Jimenez2020a}. Furthermore, even if teleparallel gravity is based on torsion and general relativity is based on curvature, it is well known that their action integrals differ only by a boundary term, and therefore, the theories have no differences at the level of classical field equations. Therefore, the predictions of unified gravity for classical fields also agree with those of general relativity.

The gauge symmetry generators in the conventional TEGRW are the four differential operators $\vec{\partial}_a$, which can be considered as dimensional continuum generators due to their dependence on the space-time coordinates. In contrast, in unified gravity, the gauge symmetry generators are the four dimensionless $8\times 8$ kernel matrices $\mathbf{t}^a$. This property makes unified gravity to resemble the gauge theories of the Standard Model more than the conventional TEGRW. However, in certain aspects, the translation gauge field of the conventional TEGRW resembles the gravity gauge field of unified gravity. For example, the expressions of the gauge field strength tensors and the gauge field Lagrangian densities are very similar between unified gravity and the conventional TEGRW. This is due to the Abelian gauge theory nature of these theories.

As seen in table~\ref{tbl:comparison}, the gauge-covariant derivative of the scaled representation of the conventional TEGRW has a coupling constant of mass dimension $-1$. This contrasts to the cases of dimensionless coupling constants of unified gravity and the gauge theories of the Standard Model. A coupling constant of negative mass dimension is typically interpreted to indicate nonrenormalizability of the gravitational interaction \cite{Zee2010,Schwartz2014,Maggiore2005}. In this perspective, the dimensionless coupling in unified gravity suggests that unified gravity is renormalizable in the same sense as the gauge theories of the Standard Model. Accordingly, it is expected that the 1-loop renormalizability studied in section~\ref{sec:renormalization} extends to all loop orders.

\subsection{Comparison with Fermi's theory of weak interaction and the electroweak unification}

Based on the conventional theory of gravity, it can be argued that the theory of gravity is closer in similarities to Fermi's theory of the weak interaction than to QED and QCD. Both the weak interaction and the conventional general relativity or TEGRW have dimensionful coupling constants and this is what makes these theories nonrenormalizable. Fermi's original four-fermion description of the weak interaction works well at low energies but leads to infinities if it is applied at higher energies. This issue arises because the theory lacks the self-consistency provided by the gauge symmetry present in QED and QCD. Unlike these renormalizable theories, Fermi's theory does not possess a natural mechanism for cancelling the infinities that arise in higher-order loop calculations, making it nonrenormalizable. The solution to this problem came with the development of the electroweak theory \cite{Glashow1961,Weinberg1967,Salam1968}, where the weak interaction was unified with electromagnetism by utilizing the $\mathrm{SU(2)}\otimes\mathrm{U(1)_\mathrm{Y}}$ gauge symmetry group.

Based on the renormalization of the weak interaction, discussed above, one might think that gravity should be renormalized through a similar process of unification. However, in the present theory of unified gravity, the process is entirely different. The four U(1) symmetries of gravity are brought to the theory by introducing the concept of the space-time dimension field, which does not exist in the original Standard Model. Therefore, the four U(1) symmetries of gravity are not associated with the phase rotation symmetries of the quantum fields of the Standard Model. Instead, these symmetries are associated with the phase rotations of the four components of the space-time dimension field. These phase rotations effectively describe space-time translations. One can conclude that the introduction of the space-time dimension field and its gauge symmetries is all that is needed to provide the self-consistency for the theory enabling successful renormalization.

\subsection{Potential in providing the ultimate quantum field theory of gravity}

Unified gravity provides a completely new approach to the gauge theory of gravity. It inherits the mathematical elegance of the Standard Model, and the following aspects indicate its high potential of being the basis for the ultimate quantum field theory of gravity:

\subsubsection{Minimal addition to the Standard Model}

The Standard Model has been extremely successful in its predictions on particle physics phenomena. Therefore, one of the guiding principles of unified gravity is that there is no reason to modify the description of the three fundamental interactions of the Standard Model. In unified gravity, the description of gravity is enabled by introducing an additional structure in the Lagrangian density of the Standard Model in such a way that the theory remains unchanged at first. This structure is the spacetime dimension field. The spacetime dimension field possesses global symmetries, which are made local, i.e., spacetime dependent, by introducing a tensor gauge field in analogy with the introduction of the vector gauge fields in the Standard Model. The tensor gauge field is the only actual addition to the Standard Model in unified gravity.

\subsubsection{Symmetries similar to those of the Standard Model}

As discussed in section~\ref{sec:comparisonSM}, the use of finite-dimensional unitary symmetries is the prominent similarity between unified gravity and the gauge theories of the Standard Model. Through these symmetries and the dimensionless coupling constant, unified gravity circumvents the difficulties of the conventional effective field theory of gravity \cite{Schwartz2014,Bambi2023,Casadio2022,Donoghue1994b}, which is inherently nonrenormalizable \cite{Zee2010,Maggiore2005}. The investigation of the renormalization of unified gravity in section~\ref{sec:renormalization} strongly indicates that unified gravity provides a clear path to renormalizable quantum gravity and provides a solid foundation for the theory of all fundamental interactions of nature.

\subsubsection{Absence of free parameters}

One of the most prominent features of unified gravity is that it is capable of being a valid candidate as the ultimate theory of quantum gravity without introducing a single free parameter that should be fixed by experiments. In contrast, for example, string theory involves a wide range of free parameters that have not yet been experimentally measured. These parameters include the string coupling constant, compactification parameters of extra dimensions, gauge group couplings, moduli fields, and parameters related to supersymmetry \cite{Green1987,Becker2007,Dine2007}. Therefore, being expressed in terms of known physical constants, all results of unified gravity are quantitative and can be directly compared with the results of possible future laboratory experiments or astronomical observations \cite{Addazi2022}.

\subsubsection{SEM tensor source term of gravity}

Using the concept of the spacetime dimension field and the kernel matrices of the eight-spinor theory, it became possible to write the Lagrangian density of the theory so that the symmetric stress-energy-momentum tensor source term of gravity and its conservation law can be derived in a new way: Following Noether's theorem and by performing differentiation of the Lagrangian density with respect to the parameters of the symmetry transformation of the spacetime dimension field, the stress-energy-momentum tensor follows completely analogously to the four-vector source terms of the three fundamental interactions of the Standard Model. This provides direct evidence that the gauge symmetry of the spacetime dimension field is key to improved understanding of gravitational interaction within the framework of quantum field theory. Before this derivation, the symmetric stress-energy-momentum tensor has been possible to derive only by utilizing the conventional spacetime symmetries and the Belinfante--Rosenfeld symmetrization \cite{Belinfante1940,Rosenfeld1940} or the related variation of the Lagrangian density with respect to the classical metric tensor \cite{Landau1989}.

\subsubsection{Classical limit: TEGR}

The spacetime metric tensor enters in unified gravity through geometric conditions. Using a space-time-dependent geometric condition, unified gravity was shown to reproduce TEGRW. Therefore, unified gravity is consistent with all classical predictions of general relativity, such as the precession of the perihelion of Mercury, the bending of light by the Sun, and the gravitational redshift of light \cite{Misner1973}. Accordingly, it is also consistent with the recent measurements on the waveforms of gravitational waves \cite{Abbott2016a,Abbott2016b,Abbott2017}, on the shadow of black holes \cite{Akiyama2019}, and on the motion of antimatter \cite{Anderson2023}. Any theory that can be considered a valid candidate for a theory of quantum gravity must explain these results.

\subsubsection{Exact description of gravity in the Minkowski metric}

In UGM, one uses the geometric condition, in which the spacetime metric tensor is independent of the gravity gauge field. Consequently, one can investigate unified gravity in the Minkowski metric in an exact way without dealing with an infinite number of terms in the Lagrangian density. This differs from the use of metric in general relativity, where the metric depends on the gravitational field by definition \cite{Einstein1916,Misner1973,Landau1989,Dirac1975}. In general relativity, the effective quantization requires expansion of the metric about the flat or smooth background  \cite{Gupta1952,Voronov1973,vanNieuwenhuizen1973,Donoghue1994b,Choi1995,Olyaei2018}. The expansion of the metric about the flat or smooth background introduces an infinite number of terms in the Lagrangian density of the conventional effective field theory of gravity. It is this expansion that makes the renormalization of the conventional theory impossible by a finite number of counterterms. Thanks to its geometric condition, UGM avoids this problem.

\subsubsection{BRST invariance}

The Lagrangian density of unified gravity was shown to satisfy the BRST invariance. The BRST invariance provides a rigorous framework for dealing with the redundancies in gauge theories, allowing for consistent quantization so that only gauge-invariant observables contribute to physical predictions. It also ensures that quantum field theories respect the symmetries of the classical theory while they extend to the description of quantum effects. The BRST invariance of unified gravity strongly suggests that unified gravity is a renormalizable gauge theory like the gauge theories of the Standard Model. The BRST invariance also makes unified gravity simpler than the conventional theories of gravity, where the BRST invariance must be replaced by a more general Batalin--Vilkovisky formalism \cite{Batalin1981, Batalin1983,Weinberg1996,Costello2011,Henneaux1992,Gomis1995}.

\subsection{Remaining challenges in quantum gravity}

\subsubsection{Lack of experimental data on quantum gravity}

The development of the Standard Model has been deeply intertwined with experimental discoveries, which have played a crucial role in shaping and validating the theory \cite{Schwartz2014}. Initially, the Standard Model emerged as a theoretical framework to describe the fundamental particles and their interactions. However, it was through a series of groundbreaking experiments that its predictions were tested, and the Standard Model itself obtained its present form. In this respect, the lack of experimental data on quantum gravity, due to the weakness of the gravitational interaction, has so far been a notable challenge for the development of a well-functioning theory of quantum gravity. Experimental advances can, however, take place in the following years \cite{Addazi2022,Brown2023,Nezami2023,Oppenheim2023c}. Any experiments must be planned carefully to clearly distinguish between the classical and quantum effects \cite{Carney2024}. For example, the gravitational Aharonov--Bohm effect \cite{Audretsch1983,Hohensee2012} has already been measured \cite{Overstreet2022}, but it can be explained semiclassically using the classical gravitational potential without requiring the full quantization of the gravitational field.

\subsubsection{Possibility of divergences in high-order loop diagrams}

In the conventional effective field theory approach to quantum gravity \cite{Schwartz2014,Bambi2023,Casadio2022,Donoghue1994b}, the loop diagrams are problematic because they lead to divergences that cannot be renormalized in the usual sense \cite{Deser1974a,Deser1974b,Deser1974c,Hooft1974}. In the case of unified gravity, one can speculate with the possibility that some high-order loop diagrams could not be renormalized. Even though the complete proof of the renormalizability of unified gravity to all loop orders remains a topic for future work, we have strong arguments against this scenario. One of these arguments is the fact that the known gauge theories of the Standard Model are based on similar unitary or special unitary groups and also have dimensionless coupling constants, and they have turned out to be renormalizable \cite{Schwartz2014}. Another, closely related, argument is the BRST invariance.

\begin{table*}
\setlength{\tabcolsep}{4.5pt}
\renewcommand{\arraystretch}{2.0}
\caption{\label{tbl:steps}
Summary of the key steps in the emergence of unified gravity}
\begin{tabular}{p{17.5cm}}
   \hline\hline
   (i) The generating Lagrangian density of gravity is written using the space-time dimension field to enable additional compact, finite-dimensional symmetries similar to those of the Standard Model. The form of the space-time dimension field is strongly restricted by the requirement that the generating Lagrangian density of gravity reduces to the Lagrangian density of the Standard Model. Without the space-time dimension field, the Lagrangian density satisfies only the well-known symmetries of the Standard Model and the external space-time symmetries, which have a noncompact, infinite-dimensional gauge group.\\
   \hline
   (ii) The theory is written using the inertial and gravitational masses and the scale and coupling constants. Einstein's equivalence principle is formulated by requiring that the renormalized values of the inertial and gravitational masses are equal. Equivalence is also required for the scale and coupling constants of unified gravity. The equivalence principle and the scale invariance guarantee that the theory does not introduce any free parameters beyond the physical constants determined in previous experiments.\\
   \hline
   (iii) The gauge symmetry is applied. The variation of the action integral of the theory with respect to the parameters of the four U(1) gauge symmetry transformations of the space-time dimension field leads to the conservation law of the SEM tensor. The generating Lagrangian density is made locally gauge-invariant by introducing a tensor gauge field through the gauge-covariant derivative. Apart from the soldered character, the gauge theory of unified gravity follows from its symmetries analogously in comparison with the gauge theories of the Standard Model.\\
   \hline
   (iv) The space-time metric tensor enters unified gravity through geometric conditions, which can either depend on or be independent of the gravity gauge field. The gravity-gauge-field-independent geometric condition of UGM allows one to study unified gravity in the Minkowski metric in an exact way. This differs from the use of metric in general relativity, where the metric depends on the gravitational field by definition. Alternatively, the gravity-gauge-field-dependent geometric condition, called the Weitzenböck gauge fixing approach, allows one to derive TEGRW from unified gravity. Thus, unified gravity is in perfect agreement with the known nonlinear field equations of general relativity.\\
   \hline
   (v) The Faddeev--Popov gauge fixing approach in the path integral formulation of UGM leads to the locally gauge-fixed Lagrangian density of UGM. This Lagrangian density satisfies the global BRST invariance in analogy with the gauge theories of the Standard Model. This contrasts to conventional theories of gravity, where the BRST symmetry must be replaced by a more general Batalin--Vilkovisky formalism. The BRST symmetry is satisfied at each loop order, which supports the renormalizability of the theory by forbidding gauge-violating counterterms.\\
   \hline
   (vi) The Feynman rules for unified gravity are derived based on the Lagrangian density, including the counterterms needed in the renormalization. Selected examples of the application of the Feynman rules at tree level indicate that the theory is physically meaningful. The nonrelativistic limit of the gravitational scattering of electrons leads to Newton's law of universal gravitation. In this limit, the only difference of the gravitational scattering in comparison with the electromagnetic scattering is that the electromagnetic fine-structure constant is replaced by the gravitational fine-structure constant.\\
   \hline
   (vii) The renormalizability of unified gravity is first proven at 1-loop order using the conventional on-shell renormalization scheme and dimensional regularization. The values of all renormalization factors are determined. Selected radiative corrections are calculated as examples of the use of the renormalized theory. The general proof of the renormalizability of unified gravity at any loop order and in the nonperturbative regime is left as a topic of further works.\\
   \hline\hline
 \end{tabular}
\vspace{-0.2cm}
\end{table*}

\subsubsection{Nonperturbative regime of the theory}

One challenge for unified gravity is provided by its nonperturbative regime at high energies. Previous quantum field theories, such as QCD at low energies \cite{Schwartz2014}, have shown that the nonperturbative regime is theoretically challenging to approach. This is primarily because, in the nonperturbative regime, the coupling constant is large, and perturbative methods fail \cite{Schwartz2014}. Therefore, alternative approaches, such as lattice gauge theory simulations \cite{Gattringer2010,Hattori2023} or functional methods \cite{Wen2023}, are required. These methods are computationally very intensive. Even after complete unification of all fundamental interactions of nature, a comprehensive understanding of the nonperturbative regimes of quantum field theories may still remain one of the most challenging and important goals in theoretical physics.

\subsubsection{Eventual fundamental limitations of unified gravity}

Unified gravity is a powerful and mathematically transparent framework with the potential of being an ultimate quantum theory of all fundamental forces of nature. In the end, a physical theory must be grounded on experimental verification. Unified gravity does not contain a single free parameter that has not been measured in previous experiments. Since unified gravity contains the Standard Model, it is equally predictive in related phenomena. The classical limit of unified gravity is equivalent to general relativity, and thus, consistent with the observations on gravitational interaction. However, the predictability of unified gravity in the explanation of quantum gravity phenomena is yet to be proven by future experiments \cite{Addazi2022}.

\section{\label{sec:conclusion}Conclusion}

We have investigated the possibility of formulating a gauge theory of gravity using compact, finite-dimensional symmetry groups instead of the noncompact, infinite-dimensional translation gauge group of conventional theories of gravity. The resulting gauge theory, unified gravity, was made possible \emph{without a single free parameter} by introducing the concept of the space-time dimension field and utilizing the recent eight-spinor formulation of QED \cite{Partanen2024a} extended to cover the full Standard Model. Four U(1) symmetries of the components of the space-time dimension field lead to unified gravity in a way that resembles the gauge theories of the Standard Model. Thus, our theory differs from conventional theories of gravity, which are typically based on external translation symmetry of the Lagrangian density. Compactness of the gauge group of unified gravity represents a fundamental change in the understanding of the structure of space-time and the emergence of gravity. The key steps in the emergence of unified gravity are summarized in table~\ref{tbl:steps}.

Within unified gravity, the entire dynamics of the known particles and fields, including gravity, can be described by a single master Lagrangian of the Universe through compact, finite-dimensional unitary symmetries and the resulting dynamical equations in a unified way. Therefore, our theory brings the gauge theory of gravity closer to the gauge theories of the Standard Model as compared with the conventional gauge theories of gravity. Several aspects of unified gravity, gauge theories of the Standard Model, and the conventional translation gauge theory of TEGRW were compared in detail. We have also discussed the potential of unified gravity in providing the ultimate quantum field theory of gravity as well as the remaining challenges that persist in understanding quantum gravity. After extending the proof of renormalizability of the theory to all loop orders and obtaining further understanding of the nonperturbative regime of the theory, physicists may finally have the long-sought tool for the investigation of intense gravitational fields in black holes and at the possible beginning of time. Full understanding of the implications of unified gravity on the field theories will be obtained only after extensive further work.

\vspace{0.3cm}\noindent\textbf{\begin{center}\small DATA AVAILABILITY STATEMENT\end{center}}
\vspace{0.0cm}
\phantomsection
\addcontentsline{toc}{section}{\hspace{0.6cm}Data availability statement}

All data that support the findings of this study are included within the article (and any supplementary files).

\vspace{0.5cm}\noindent\textbf{\begin{center}\small ACKNOWLEDGMENTS\end{center}}
\vspace{0.3cm}
\phantomsection
\addcontentsline{toc}{section}{\hspace{0.6cm}Acknowledgments}
 This work has been funded by the Research Council of Finland under Contract No.~349971. Wolfram Mathematica has been extensively used to verify the equations of the present work. We thank Will Barker and Grigory Volovik for useful discussions. Will Barker pointed out \cite{Barker2023} that the theory following from the Lagrangian density term $\mathcal{L}_\mathrm{g,kin}$ of the first version of our manuscript was not directly comparable to the known theories of gravity, which led us to the redefinition of this term by using degrees of freedom enabled by the theory. We also thank anonymous referees for asking relevant questions, answering to which has substantially improved our work.

\onecolumngrid
\clearpage
\begin{figure}
 \centering
 \hspace*{-1.95cm}
 \includegraphics[trim={0 0 0 1.8cm},clip]{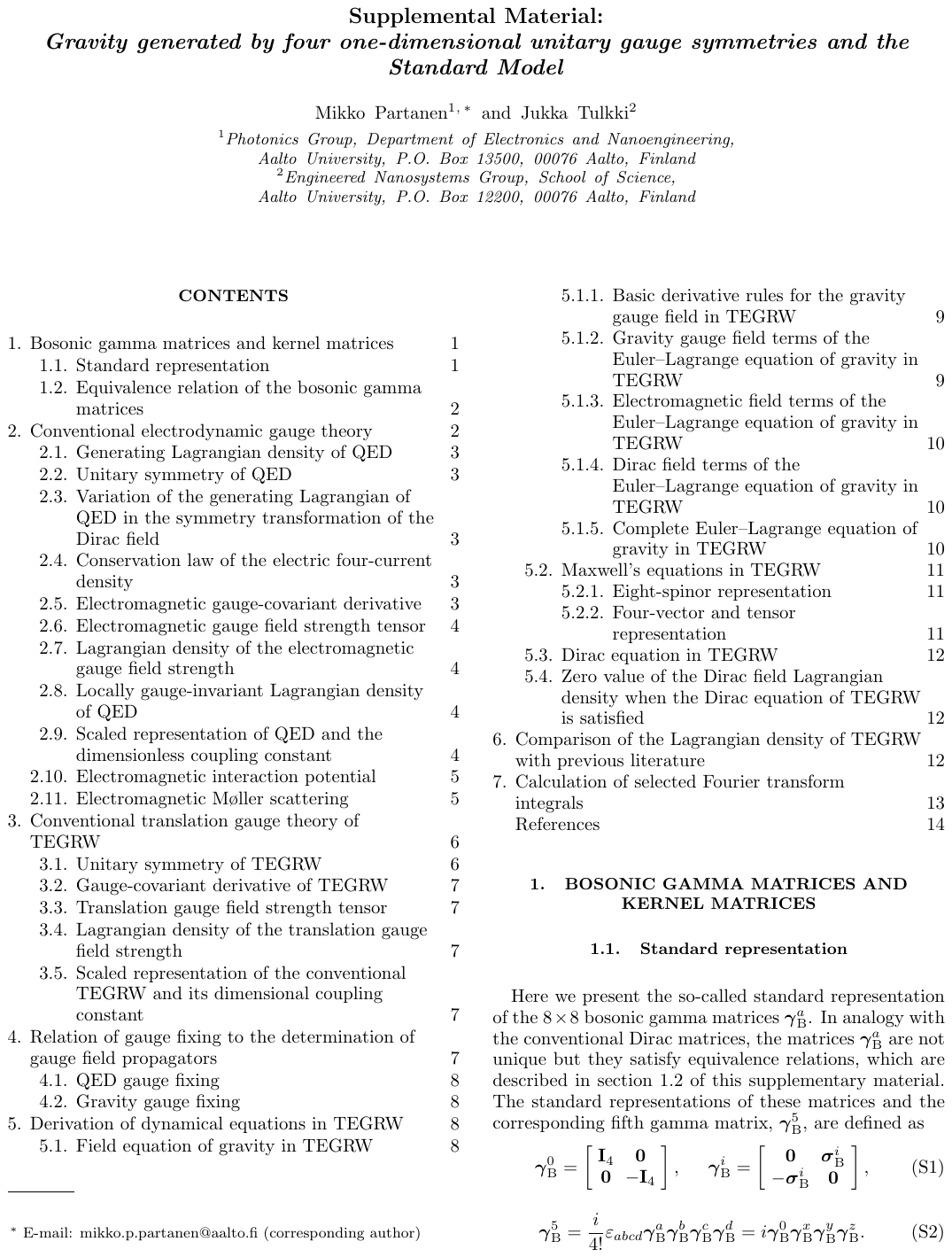}
\end{figure}
\clearpage
\setcounter{page}{1}
\begin{figure}
 \centering
 \hspace*{-1.95cm}
 \includegraphics[trim={0 0 0 1.8cm},clip]{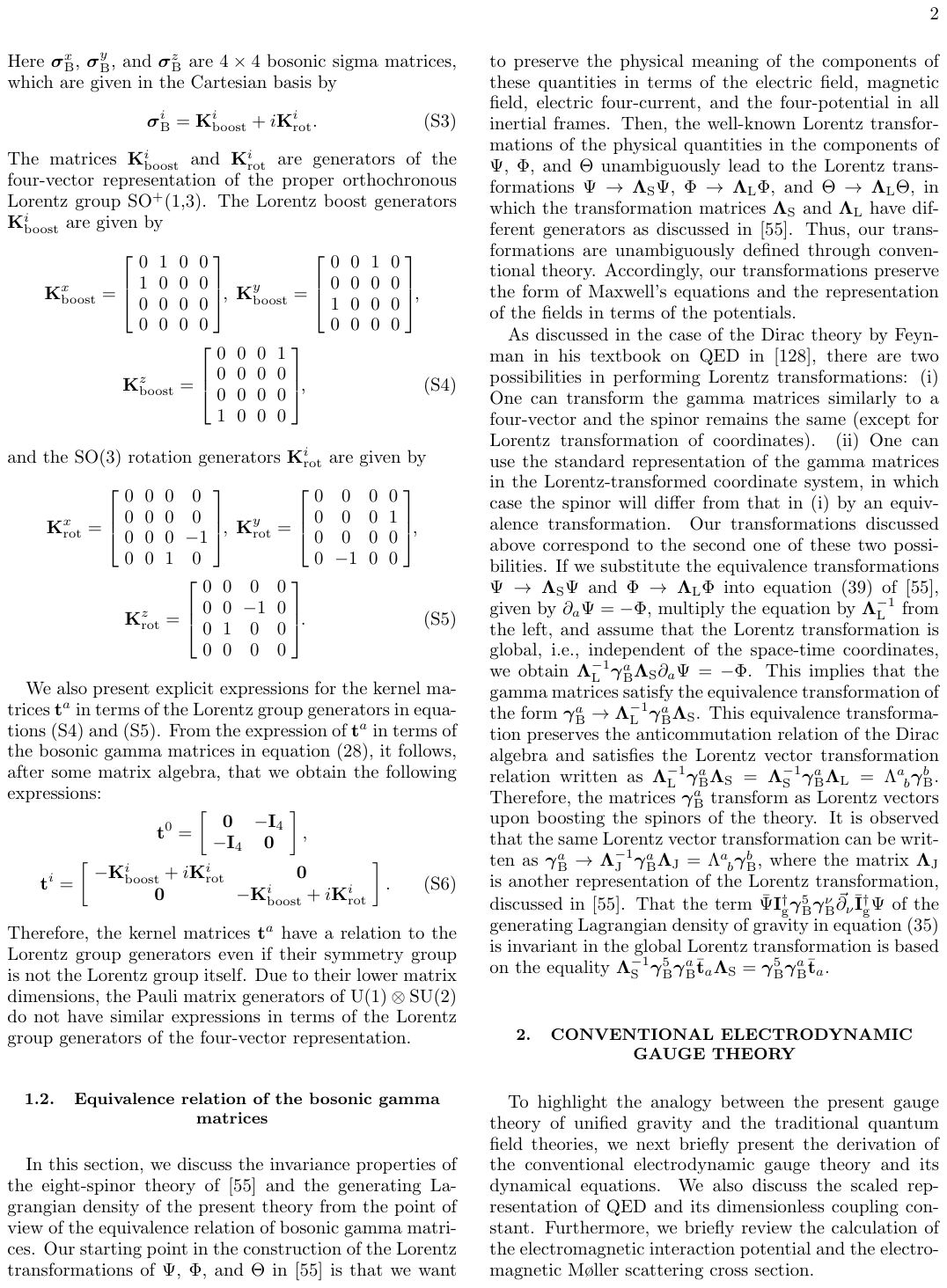}
\end{figure}
\clearpage
\begin{figure}
 \centering
 \hspace*{-1.95cm}
 \includegraphics[trim={0 0 0 1.8cm},clip]{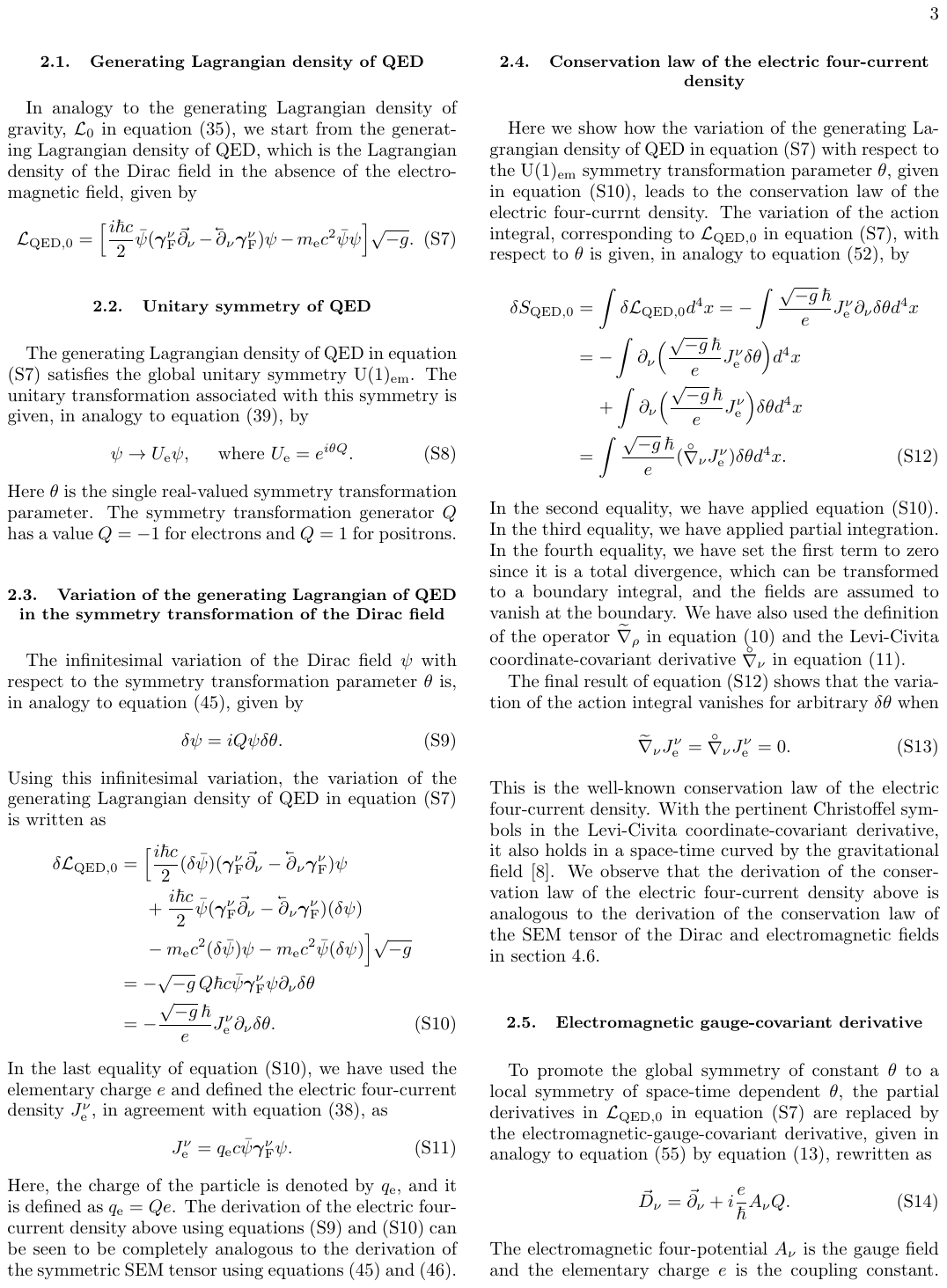}
\end{figure}
\clearpage
\begin{figure}
 \centering
 \hspace*{-1.95cm}
 \includegraphics[trim={0 0 0 1.8cm},clip]{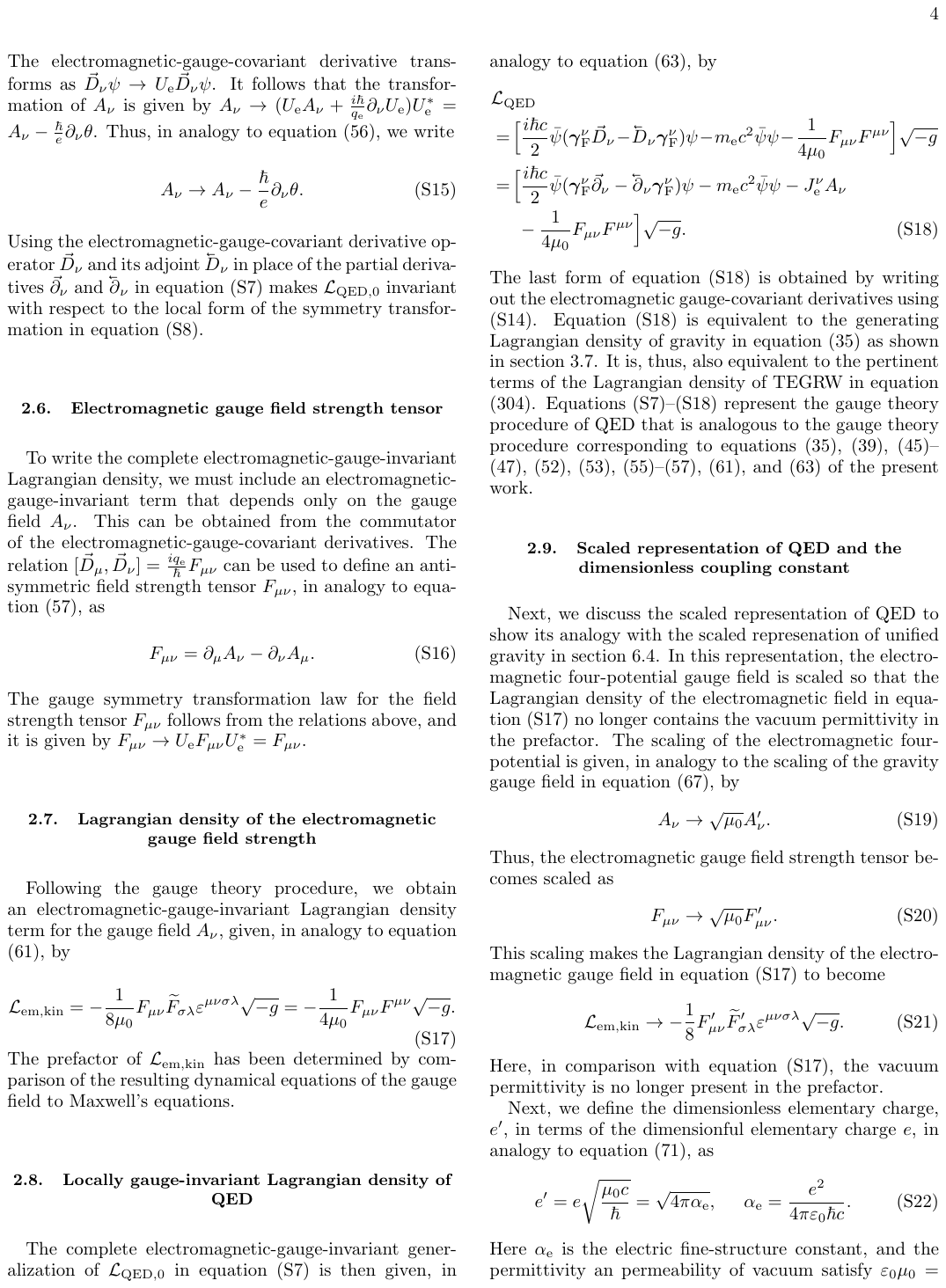}
\end{figure}
\clearpage
\begin{figure}
 \centering
 \hspace*{-1.95cm}
 \includegraphics[trim={0 0 0 1.8cm},clip]{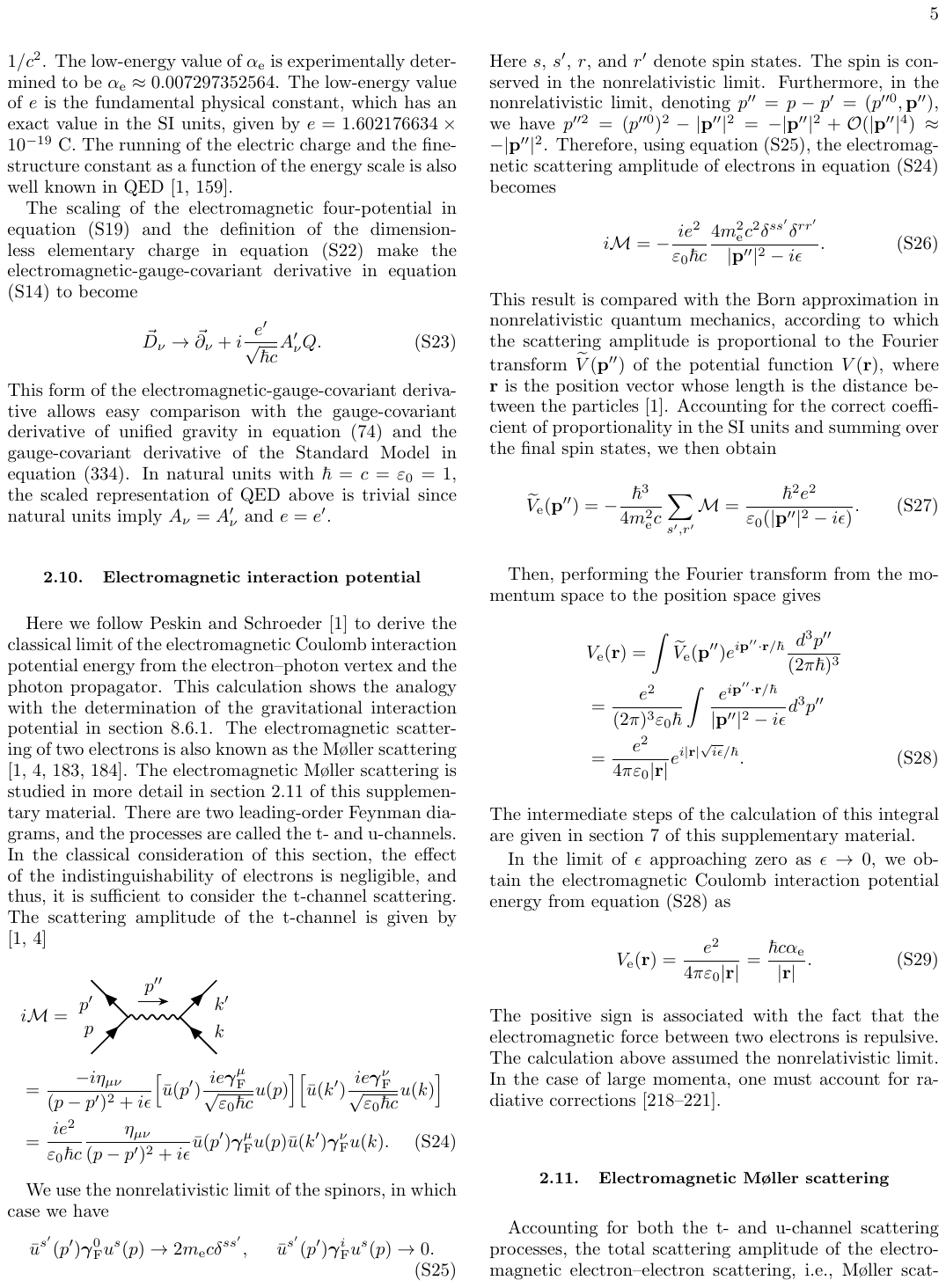}
\end{figure}
\clearpage
\begin{figure}
 \centering
 \hspace*{-1.95cm}
 \includegraphics[trim={0 0 0 1.8cm},clip]{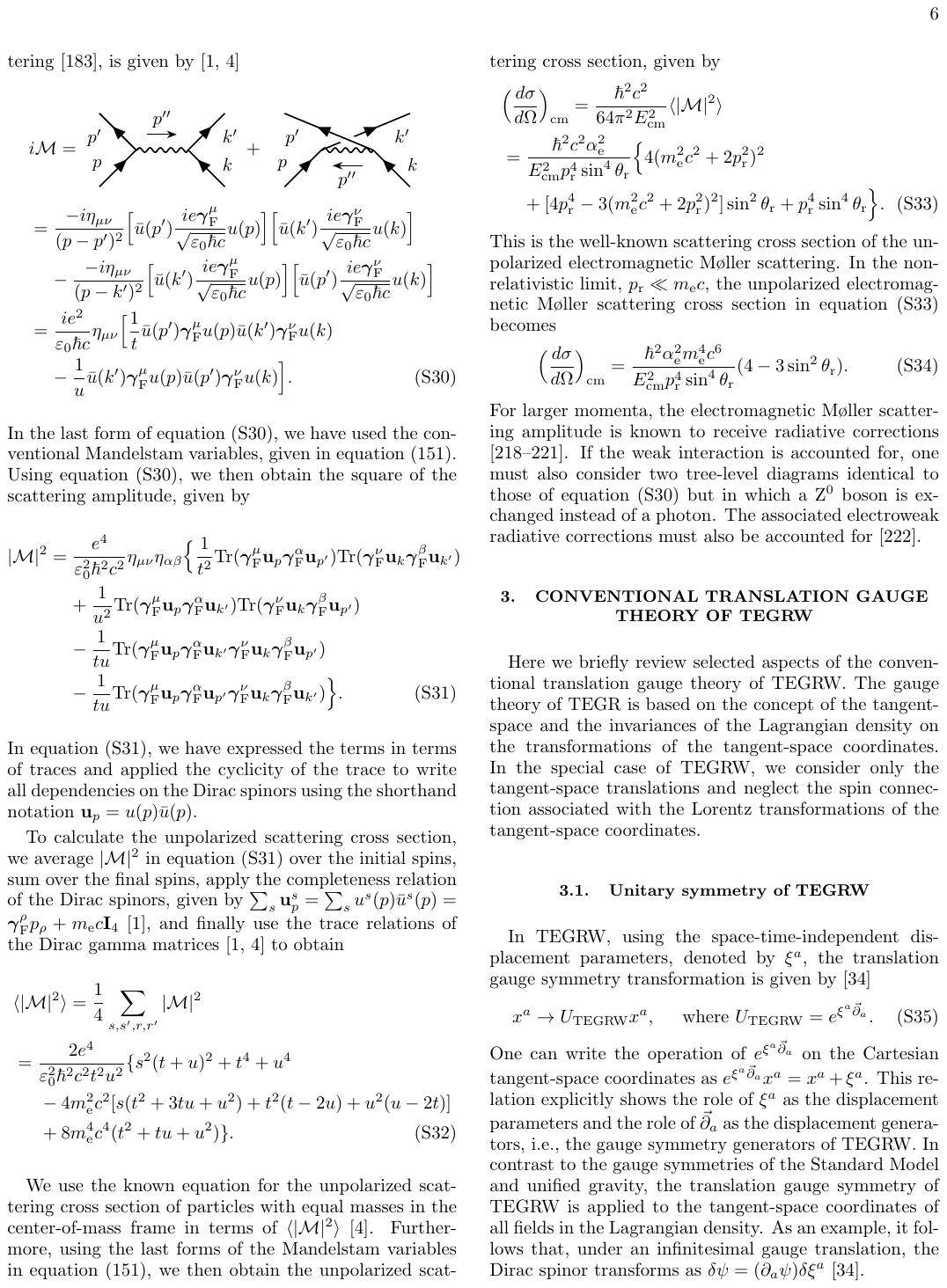}
\end{figure}
\clearpage
\begin{figure}
 \centering
 \hspace*{-1.95cm}
 \includegraphics[trim={0 0 0 1.8cm},clip]{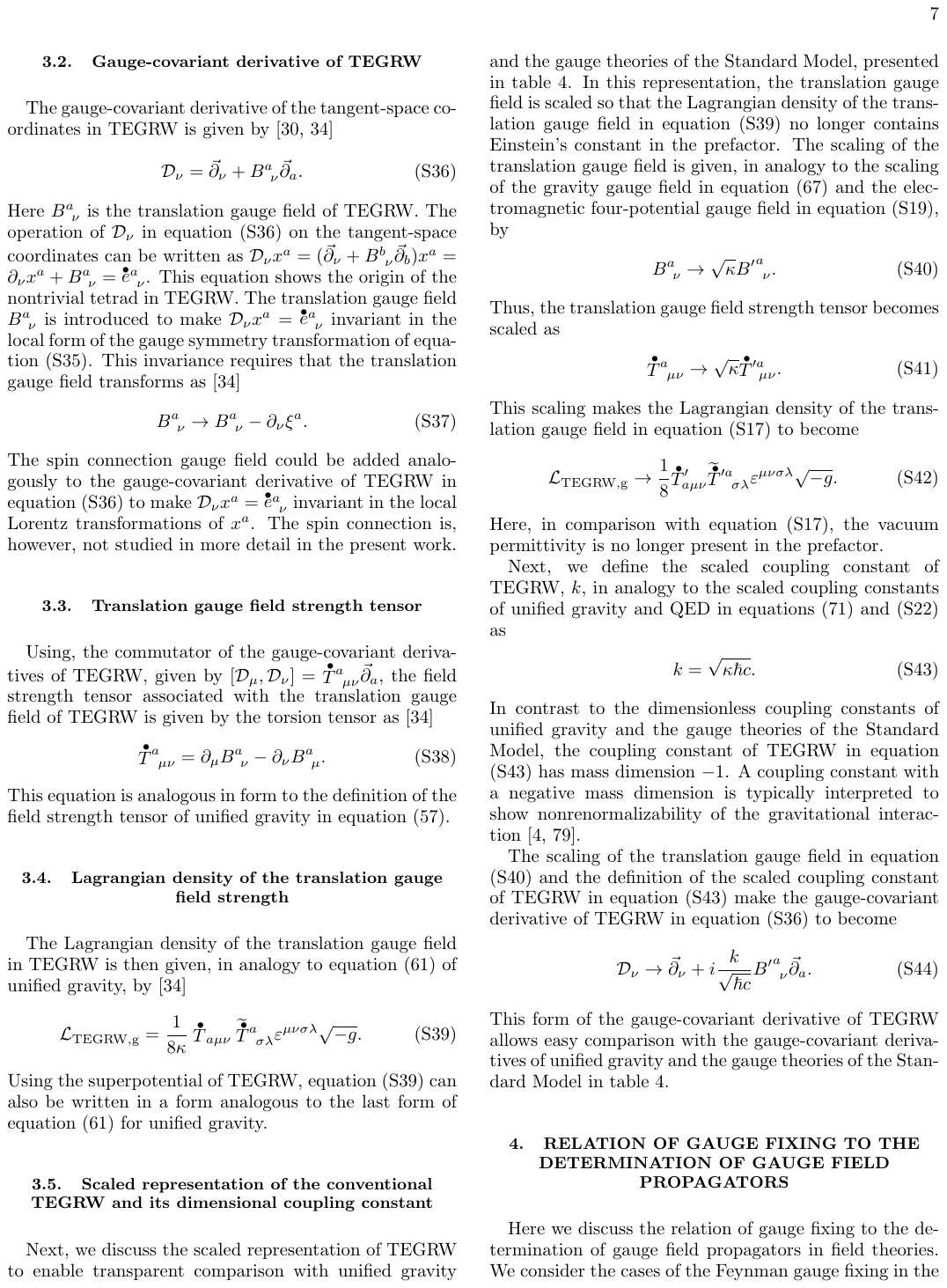}
\end{figure}
\clearpage
\begin{figure}
 \centering
 \hspace*{-1.95cm}
 \includegraphics[trim={0 0 0 1.8cm},clip]{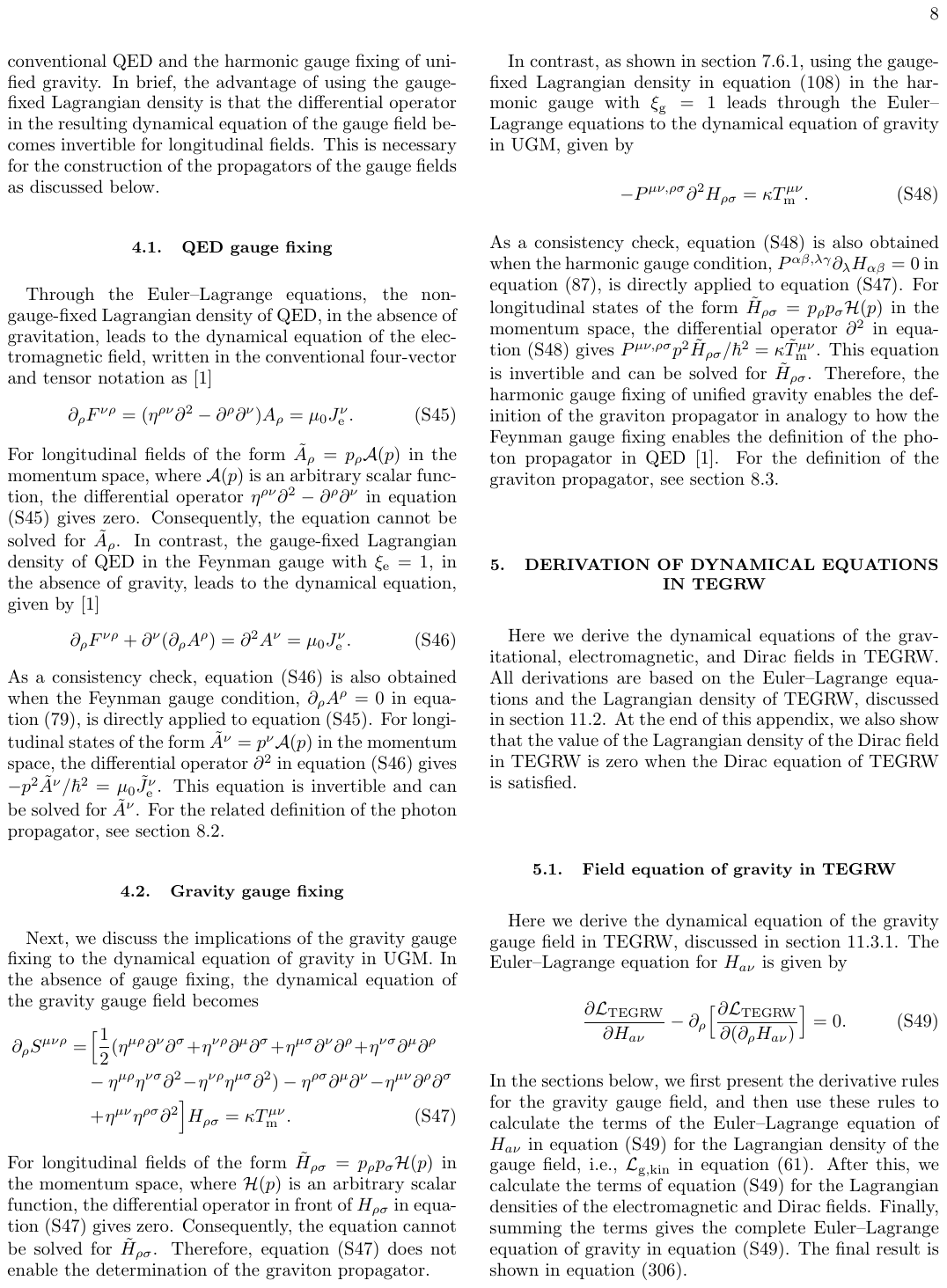}
\end{figure}
\clearpage
\begin{figure}
 \centering
 \hspace*{-1.95cm}
 \includegraphics[trim={0 0 0 1.8cm},clip]{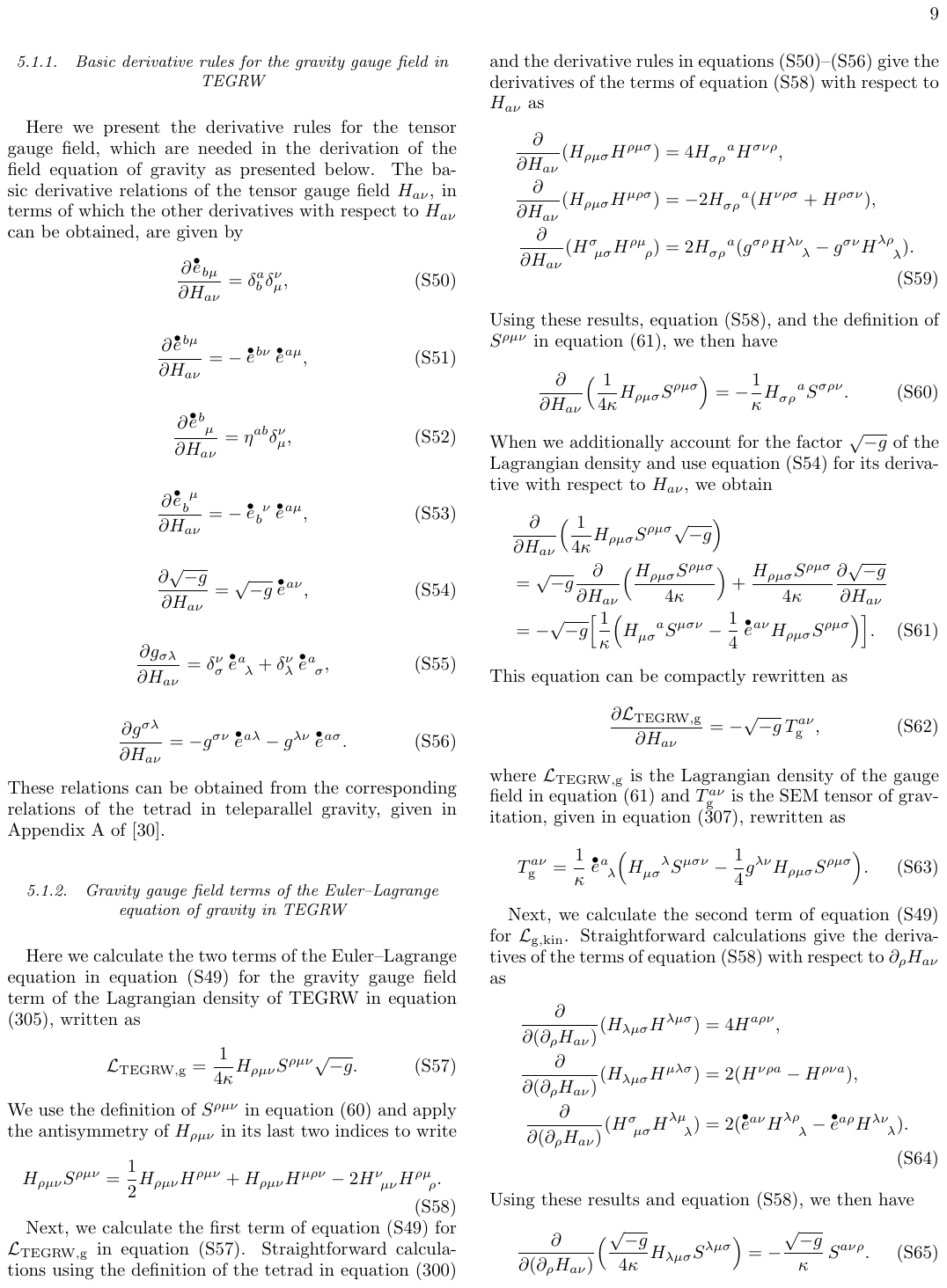}
\end{figure}
\clearpage
\begin{figure}
 \centering
 \hspace*{-1.95cm}
 \includegraphics[trim={0 0 0 1.8cm},clip]{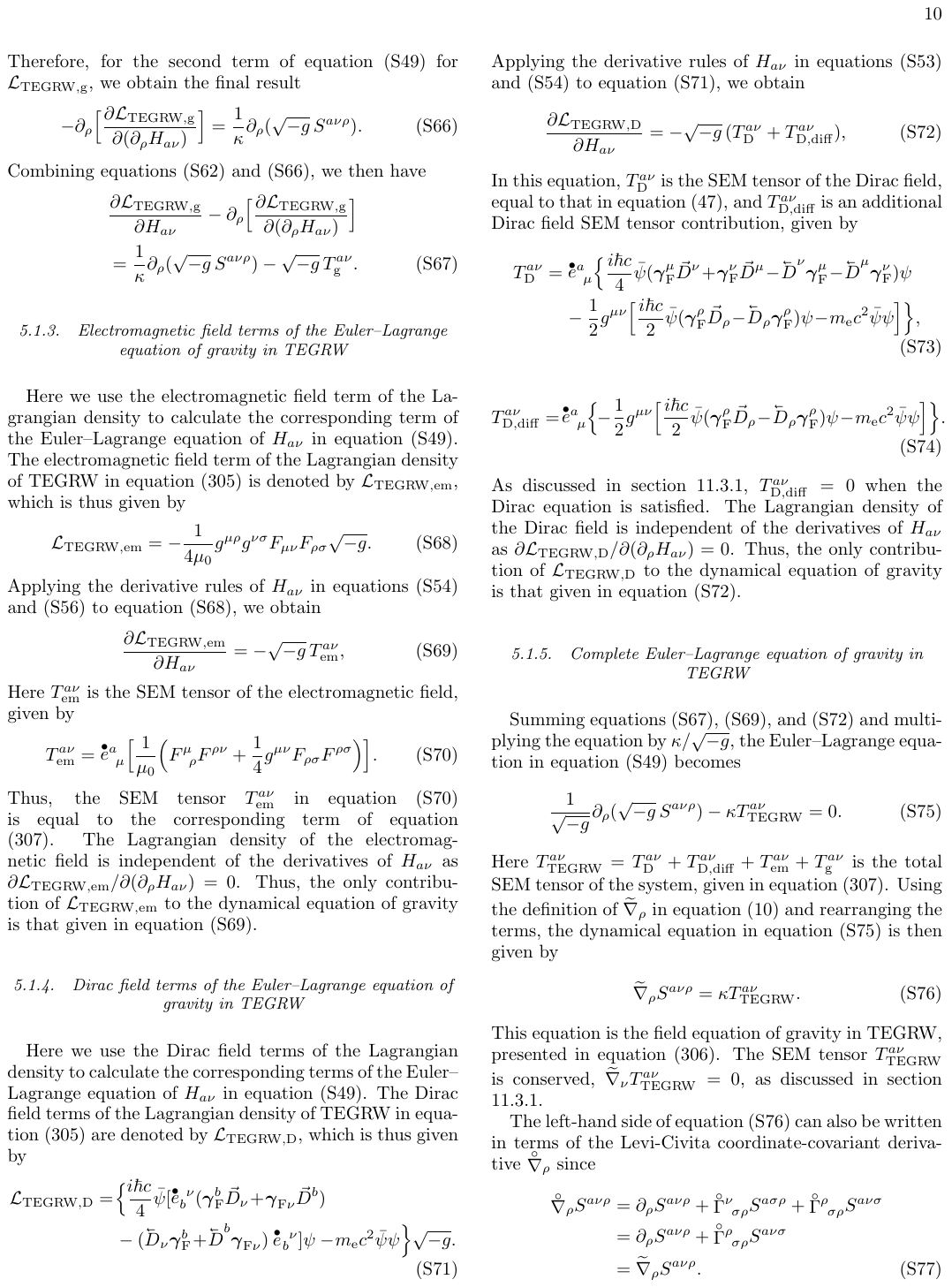}
\end{figure}
\clearpage
\begin{figure}
 \centering
 \hspace*{-1.95cm}
 \includegraphics[trim={0 0 0 1.8cm},clip]{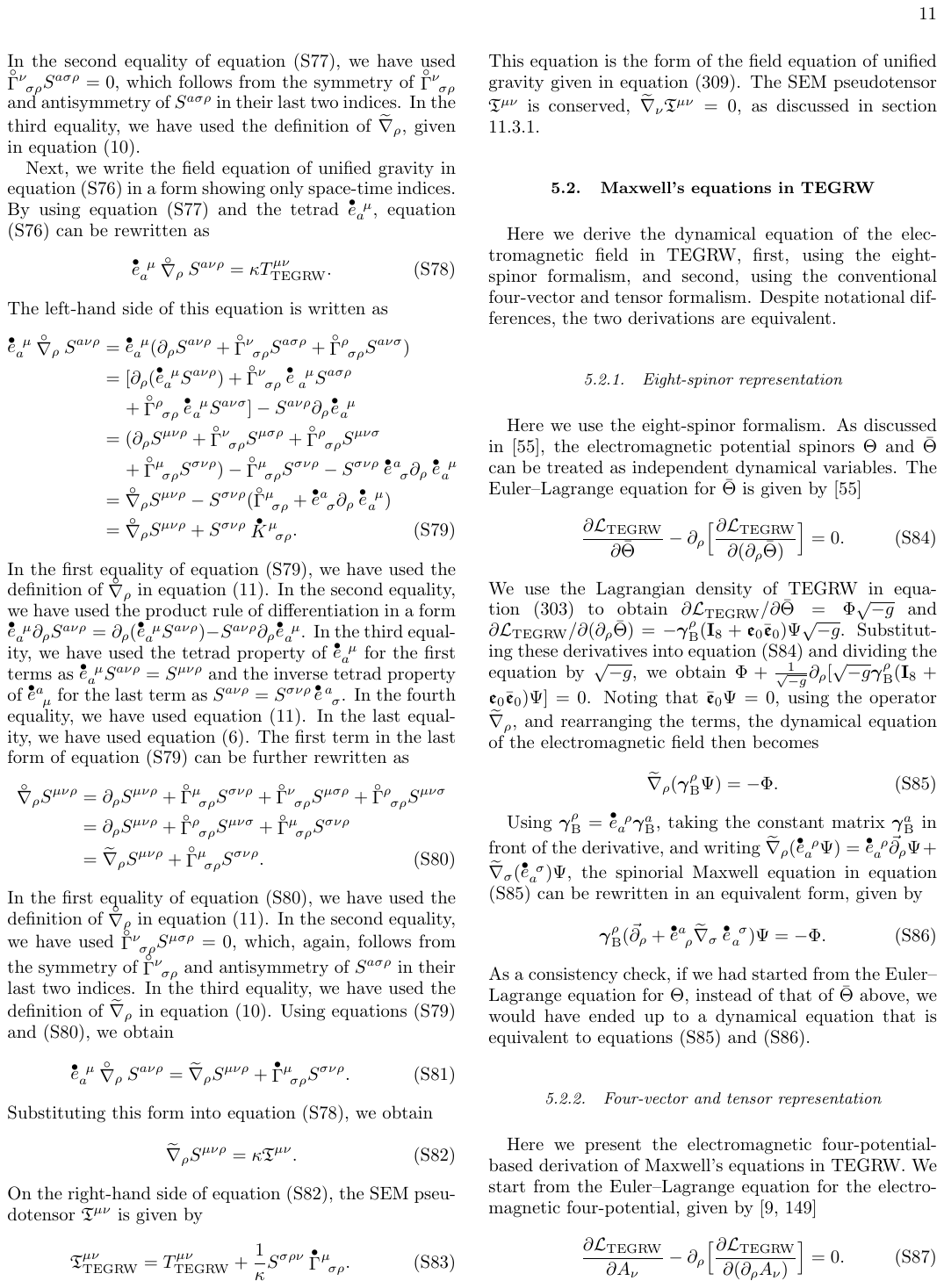}
\end{figure}
\clearpage
\begin{figure}
 \centering
 \hspace*{-1.95cm}
 \includegraphics[trim={0 0 0 1.8cm},clip]{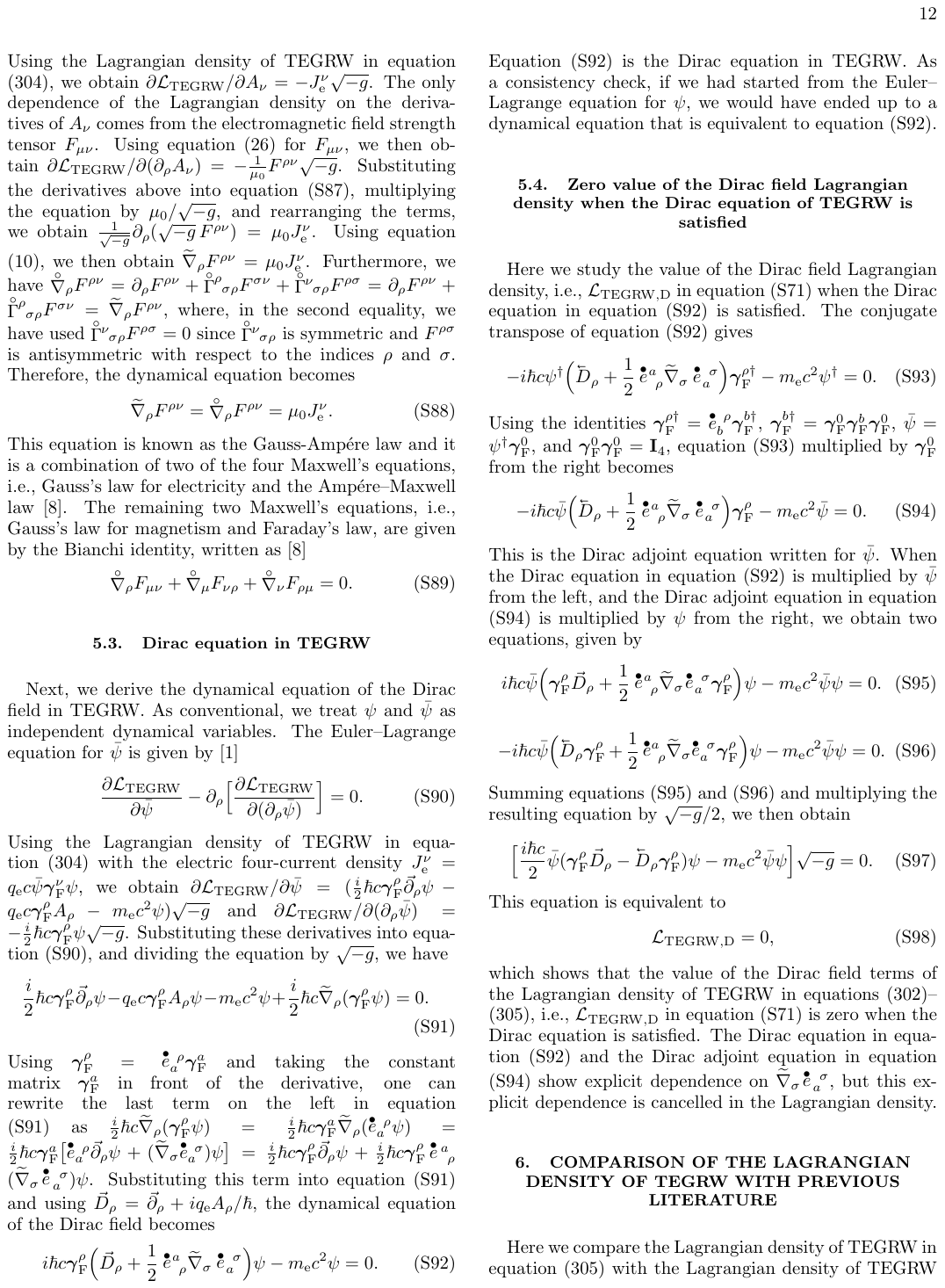}
\end{figure}
\clearpage
\begin{figure}
 \centering
 \hspace*{-1.95cm}
 \includegraphics[trim={0 0 0 1.8cm},clip]{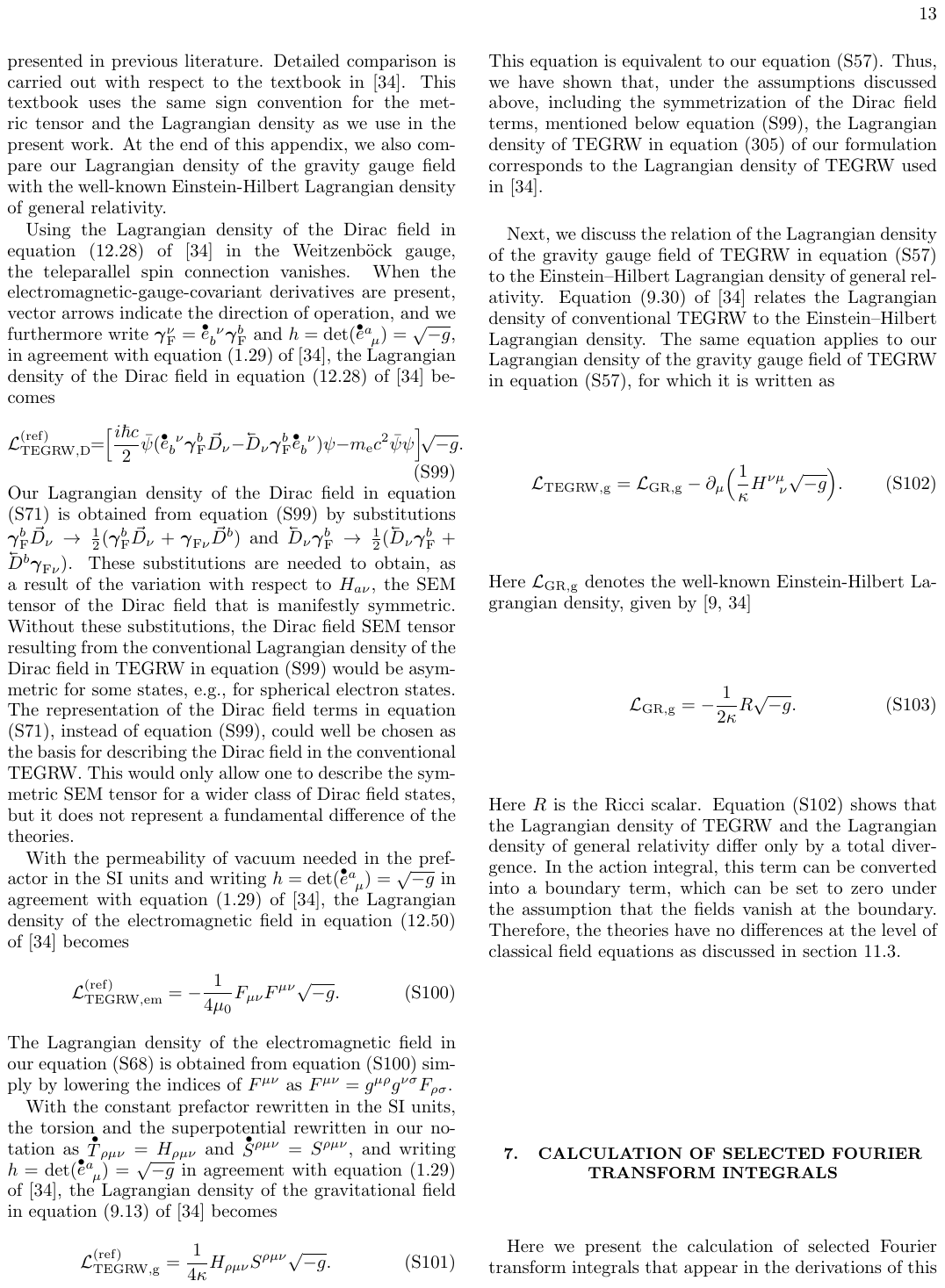}
\end{figure}
\clearpage
\begin{figure}
 \centering
 \hspace*{-1.95cm}
 \includegraphics[trim={0 0 0 1.8cm},clip]{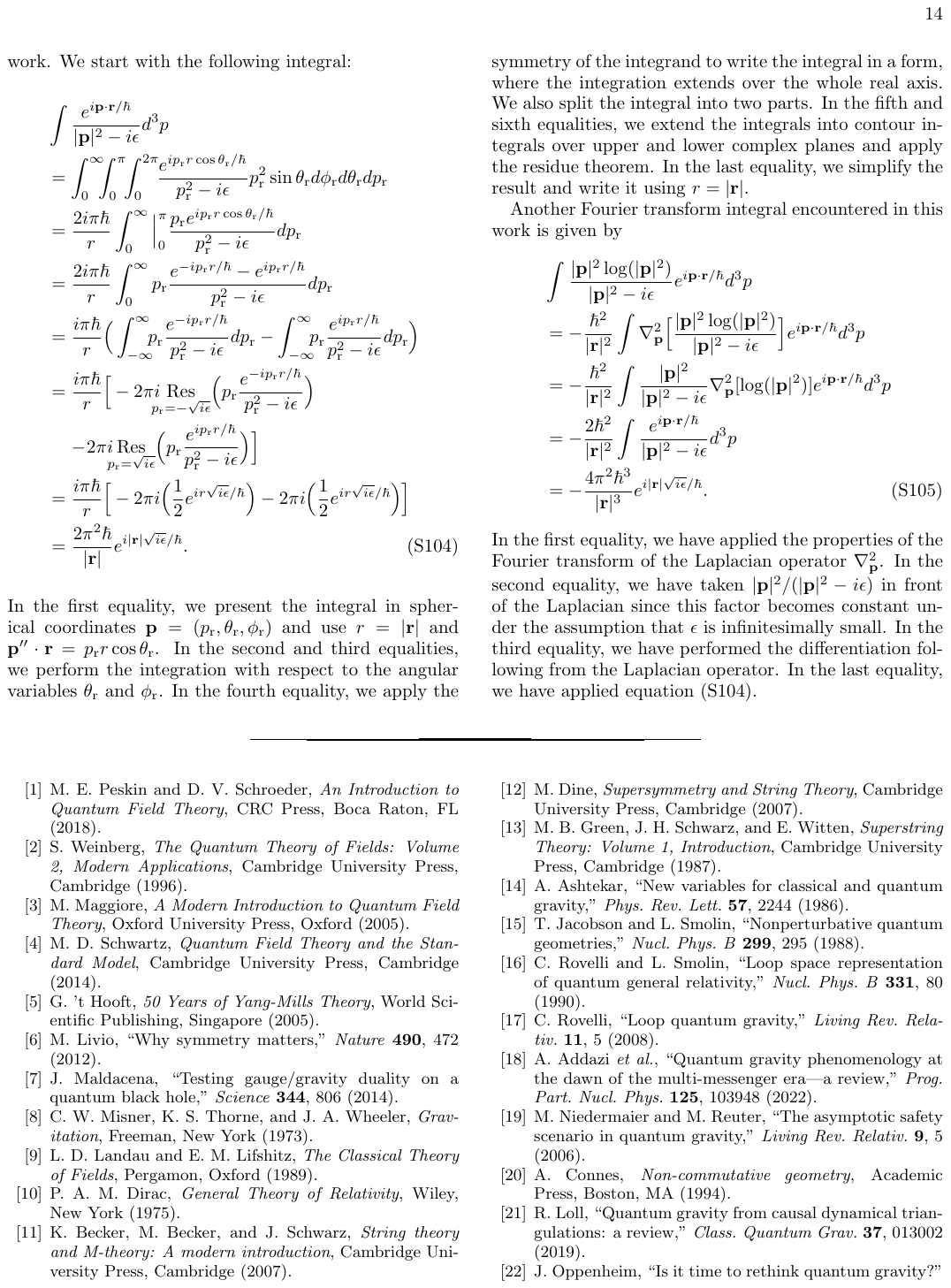}
\end{figure}
\clearpage
\begin{figure}
 \centering
 \hspace*{-1.95cm}
 \includegraphics[trim={0 0 0 1.8cm},clip]{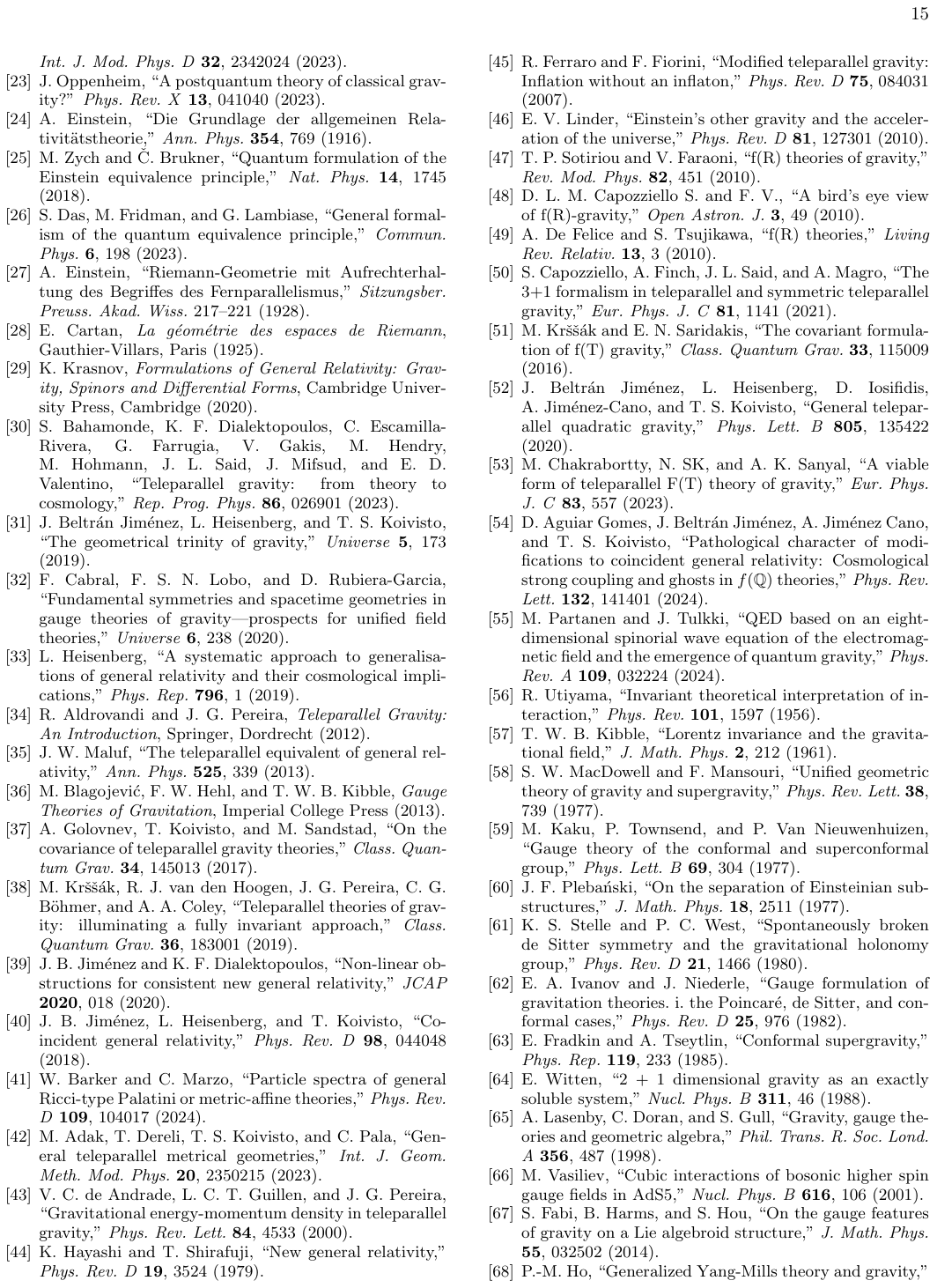}
\end{figure}
\clearpage
\begin{figure}
 \centering
 \hspace*{-1.95cm}
 \includegraphics[trim={0 0 0 1.8cm},clip]{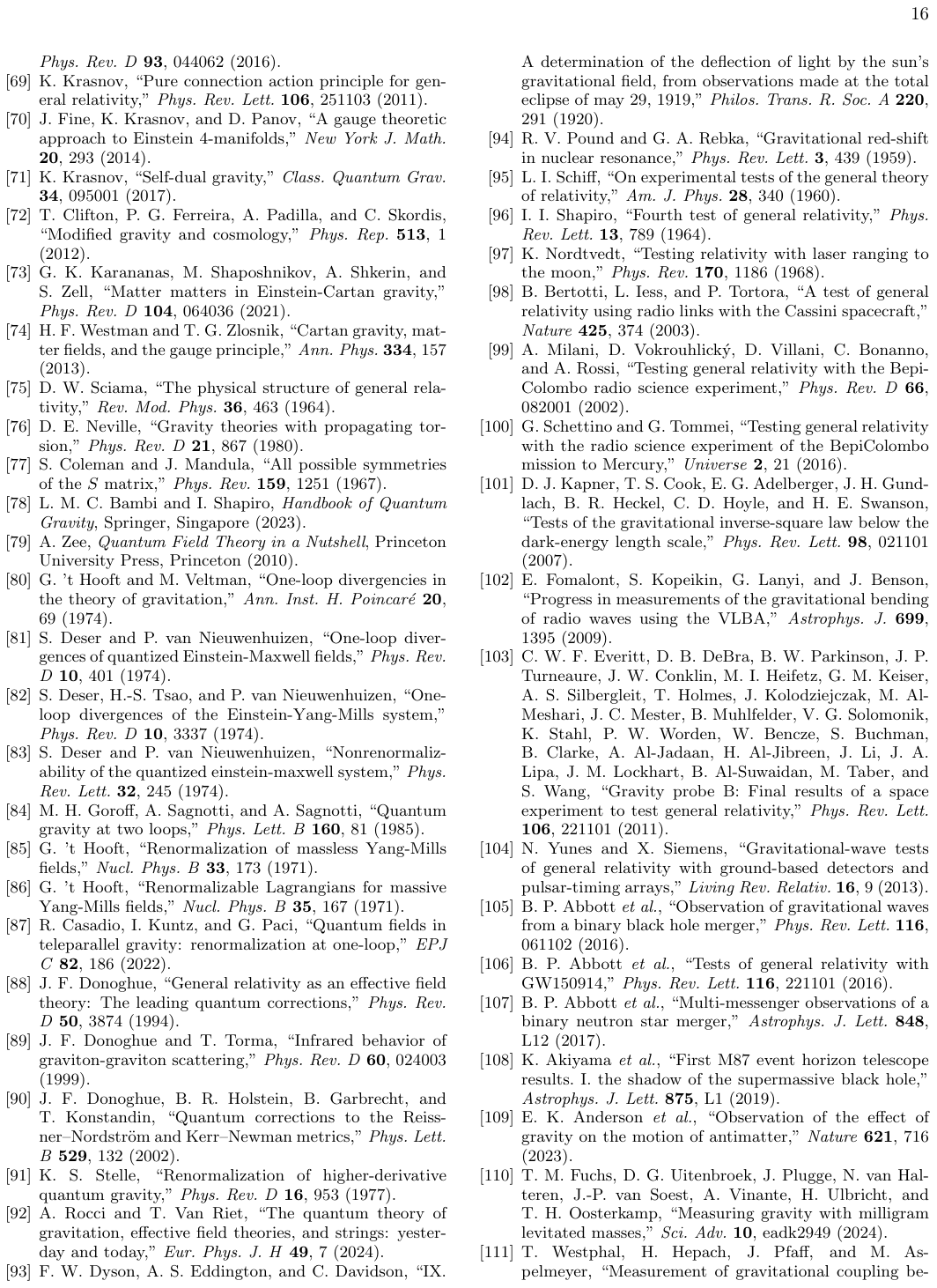}
\end{figure}
\clearpage
\begin{figure}
 \centering
 \hspace*{-1.95cm}
 \includegraphics[trim={0 0 0 1.8cm},clip]{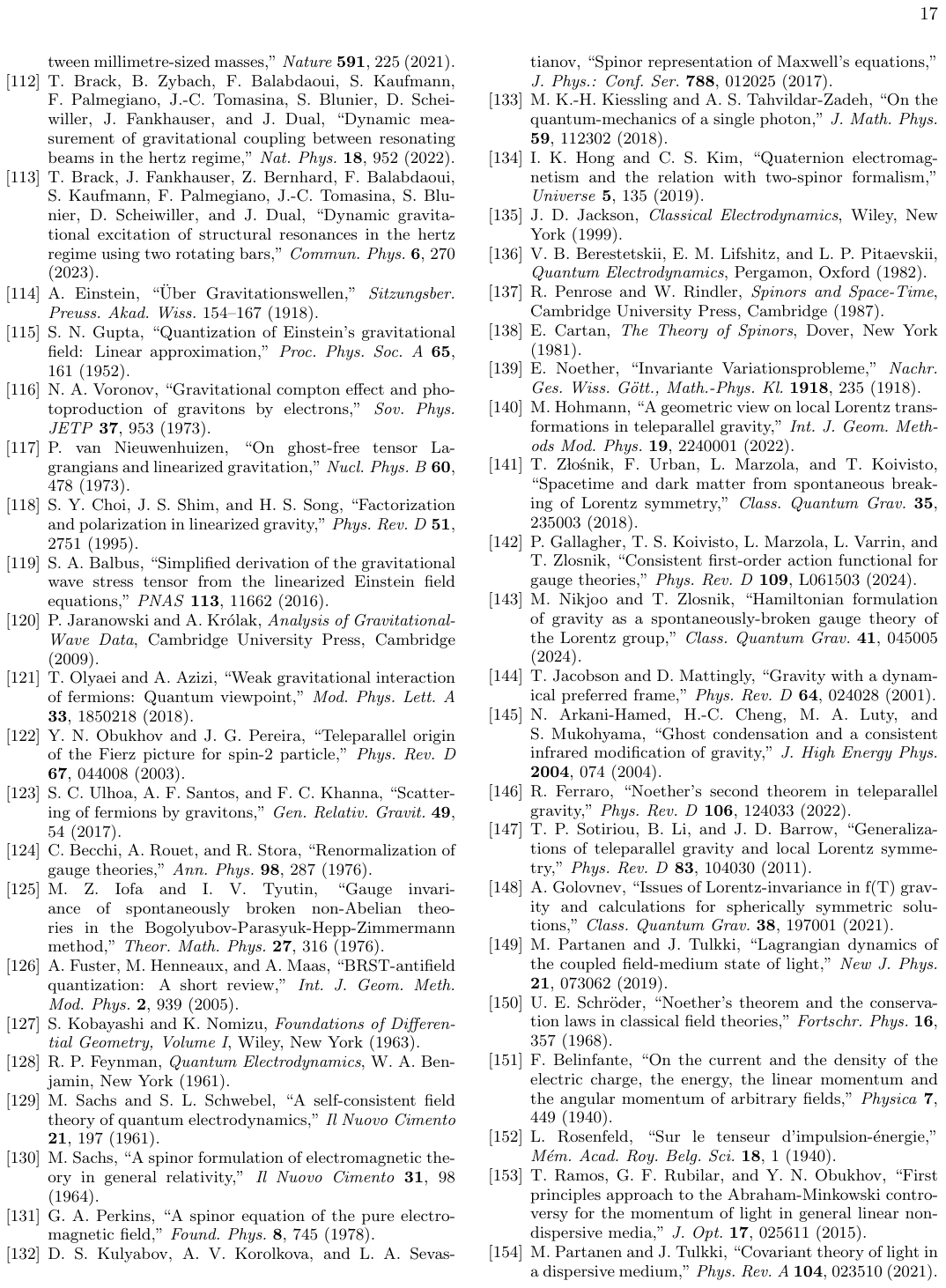}
\end{figure}
\clearpage
\begin{figure}
 \centering
 \hspace*{-1.95cm}
 \includegraphics[trim={0 0 0 1.8cm},clip]{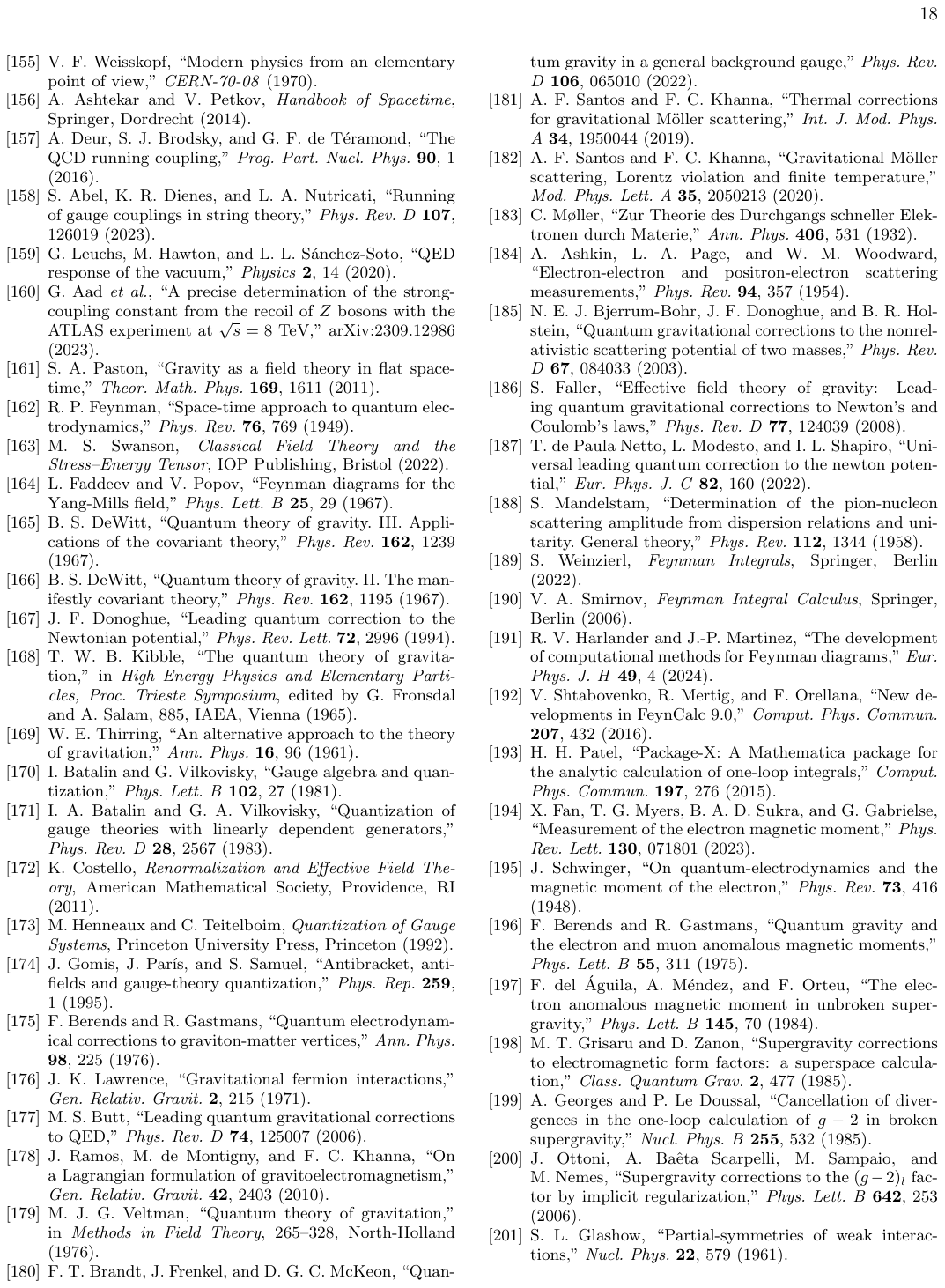}
\end{figure}
\clearpage
\begin{figure}
 \centering
 \hspace*{-1.95cm}
 \includegraphics[trim={0 0 0 1.8cm},clip]{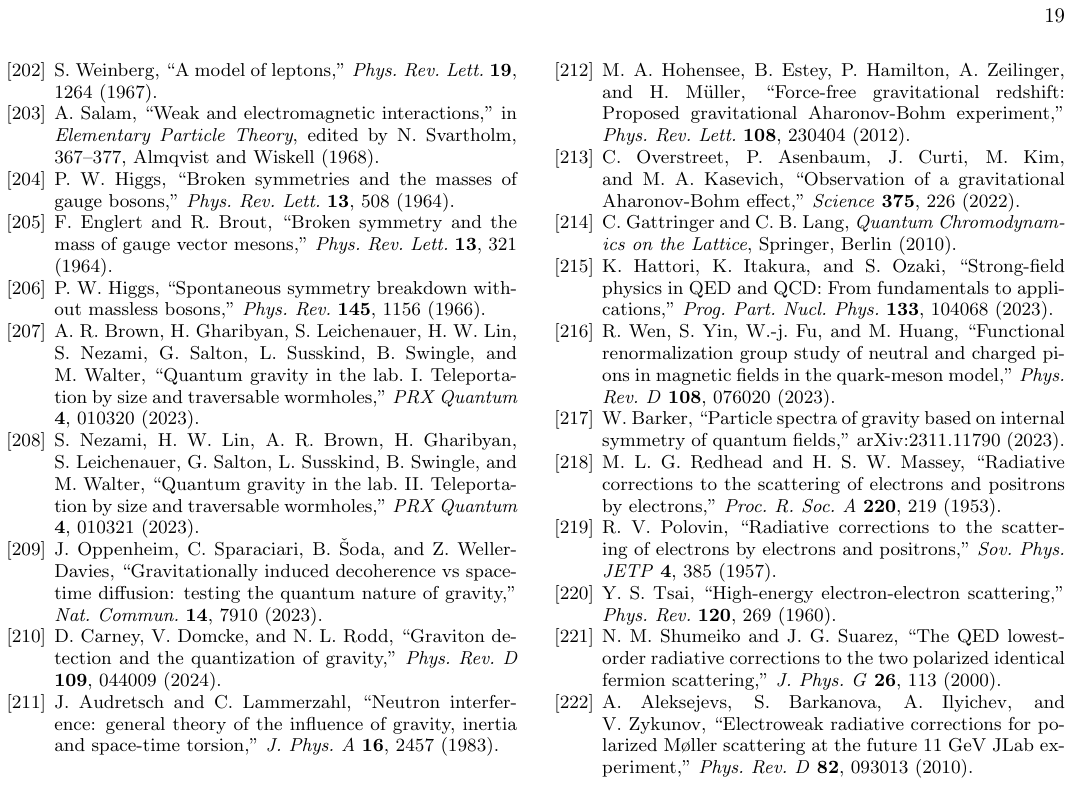}
\end{figure}


\clearpage





\begin{thebibliography}{100}
\newcommand{\enquote}[1]{``#1''}

\bibitem{Peskin2018}
M.~E. Peskin and D.~V. Schroeder, \emph{An Introduction to Quantum Field
  Theory}, CRC Press, Boca Raton, FL (2018).

\bibitem{Weinberg1996}
S.~Weinberg, \emph{The Quantum Theory of Fields: Volume 2, Modern
  Applications}, Cambridge University Press, Cambridge (1996).

\bibitem{Maggiore2005}
M.~Maggiore, \emph{A Modern Introduction to Quantum Field Theory}, Oxford
  University Press, Oxford (2005).

\bibitem{Schwartz2014}
M.~D. Schwartz, \emph{Quantum Field Theory and the Standard Model}, Cambridge
  University Press, Cambridge (2014).

\bibitem{Hooft2005}
G.~'t~Hooft, \emph{50 Years of Yang-Mills Theory}, World Scientific Publishing,
  Singapore (2005).

\bibitem{Livio2012}
M.~Livio, \enquote{Why symmetry matters,} \emph{Nature} \textbf{490}, 472
  (2012).

\bibitem{Maldacena2014}
J.~Maldacena, \enquote{Testing gauge/gravity duality on a quantum black hole,}
  \emph{Science} \textbf{344}, 806 (2014).

\bibitem{Misner1973}
C.~W. Misner, K.~S. Thorne, and J.~A. Wheeler, \emph{Gravitation}, Freeman, New
  York (1973).

\bibitem{Landau1989}
L.~D. Landau and E.~M. Lifshitz, \emph{The Classical Theory of Fields},
  Pergamon, Oxford (1989).

\bibitem{Dirac1975}
P.~A.~M. Dirac, \emph{General Theory of Relativity}, Wiley, New York (1975).

\bibitem{Becker2007}
K.~Becker, M.~Becker, and J.~Schwarz, \emph{String theory and {M}-theory: A
  modern introduction}, Cambridge University Press, Cambridge (2007).

\bibitem{Dine2007}
M.~Dine, \emph{Supersymmetry and String Theory}, Cambridge University Press,
  Cambridge (2007).

\bibitem{Green1987}
M.~B. Green, J.~H. Schwarz, and E.~Witten, \emph{Superstring Theory: Volume 1,
  Introduction}, Cambridge University Press, Cambridge (1987).

\bibitem{Ashtekar1986}
A.~Ashtekar, \enquote{New variables for classical and quantum gravity,}
  \emph{Phys. Rev. Lett.} \textbf{57}, 2244 (1986).

\bibitem{Jacobson1988}
T.~Jacobson and L.~Smolin, \enquote{Nonperturbative quantum geometries,}
  \emph{Nucl. Phys. B} \textbf{299}, 295 (1988).

\bibitem{Rovelli1990}
C.~Rovelli and L.~Smolin, \enquote{Loop space representation of quantum general
  relativity,} \emph{Nucl. Phys. B} \textbf{331}, 80 (1990).

\bibitem{Rovelli2008}
C.~Rovelli, \enquote{Loop quantum gravity,} \emph{Living Rev. Relativ.}
  \textbf{11}, 5 (2008).

\bibitem{Addazi2022}
{A. Addazi \emph{et al.}}, \enquote{Quantum gravity phenomenology at the dawn
  of the multi-messenger era—a review,} \emph{Prog. Part. Nucl. Phys.}
  \textbf{125}, 103948 (2022).

\bibitem{Niedermaier2006}
M.~Niedermaier and M.~Reuter, \enquote{The asymptotic safety scenario in
  quantum gravity,} \emph{Living Rev. Relativ.} \textbf{9}, 5 (2006).

\bibitem{Connes1994}
A.~Connes, \emph{Non-commutative geometry}, Academic Press, Boston, MA (1994).

\bibitem{Loll2020}
R.~Loll, \enquote{Quantum gravity from causal dynamical triangulations: a
  review,} \emph{Class. Quantum Grav.} \textbf{37}, 013002 (2019).

\bibitem{Oppenheim2023a}
J.~Oppenheim, \enquote{Is it time to rethink quantum gravity?} \emph{Int. J.
  Mod. Phys. D} \textbf{32}, 2342024 (2023).

\bibitem{Oppenheim2023b}
J.~Oppenheim, \enquote{A postquantum theory of classical gravity?} \emph{Phys.
  Rev. X} \textbf{13}, 041040 (2023).

\bibitem{Einstein1916}
A.~Einstein, \enquote{Die {G}rundlage der allgemeinen {R}elativitätstheorie,}
  \emph{Ann. Phys.} \textbf{354}, 769 (1916).

\bibitem{Zych2018}
M.~Zych and {\v{C}}.~Brukner, \enquote{Quantum formulation of the {E}instein
  equivalence principle,} \emph{Nat. Phys.} \textbf{14}, 1745 (2018).

\bibitem{Das2023}
S.~Das, M.~Fridman, and G.~Lambiase, \enquote{General formalism of the quantum
  equivalence principle,} \emph{Commun. Phys.} \textbf{6}, 198 (2023).

\bibitem{Einstein1928}
A.~Einstein, \enquote{Riemann-{G}eometrie mit {A}ufrechterhaltung des
  {B}egriffes des {F}ernparallelismus,} \emph{Sitzungsber. Preuss. Akad. Wiss.}
  217--221 (1928).

\bibitem{Cartan1925}
E.~Cartan, \emph{La g\'eom\'etrie des espaces de Riemann}, Gauthier-Villars,
  Paris (1925).

\bibitem{Krasnov2020}
K.~Krasnov, \emph{Formulations of General Relativity: Gravity, Spinors and
  Differential Forms}, Cambridge University Press, Cambridge (2020).

\bibitem{Bahamonde2023a}
S.~Bahamonde, K.~F. Dialektopoulos, C.~Escamilla-Rivera, G.~Farrugia, V.~Gakis,
  M.~Hendry, M.~Hohmann, J.~L. Said, J.~Mifsud, and E.~D. Valentino,
  \enquote{Teleparallel gravity: from theory to cosmology,} \emph{Rep. Prog.
  Phys.} \textbf{86}, 026901 (2023).

\bibitem{Jimenez2019}
J.~Beltrán~Jiménez, L.~Heisenberg, and T.~S. Koivisto, \enquote{The
  geometrical trinity of gravity,} \emph{Universe} \textbf{5}, 173 (2019).

\bibitem{Cabral2020}
F.~Cabral, F.~S.~N. Lobo, and D.~Rubiera-Garcia, \enquote{Fundamental
  symmetries and spacetime geometries in gauge theories of gravity—prospects
  for unified field theories,} \emph{Universe} \textbf{6}, 238 (2020).

\bibitem{Heisenberg2019}
L.~Heisenberg, \enquote{A systematic approach to generalisations of general
  relativity and their cosmological implications,} \emph{Phys. Rep.}
  \textbf{796}, 1 (2019).

\bibitem{Aldrovandi2012}
R.~Aldrovandi and J.~G. Pereira, \emph{Teleparallel Gravity: An Introduction},
  Springer, Dordrecht (2012).

\bibitem{Maluf2013}
J.~W. Maluf, \enquote{The teleparallel equivalent of general relativity,}
  \emph{Ann. Phys.} \textbf{525}, 339 (2013).

\bibitem{Blagojevic2013}
M.~Blagojević, F.~W. Hehl, and T.~W.~B. Kibble, \emph{Gauge Theories of
  Gravitation}, Imperial College Press (2013).

\bibitem{Golovnev2017}
A.~Golovnev, T.~Koivisto, and M.~Sandstad, \enquote{On the covariance of
  teleparallel gravity theories,} \emph{Class. Quantum Grav.} \textbf{34},
  145013 (2017).

\bibitem{Krssak2019}
M.~Krššák, R.~J. van~den Hoogen, J.~G. Pereira, C.~G. Böhmer, and A.~A.
  Coley, \enquote{Teleparallel theories of gravity: illuminating a fully
  invariant approach,} \emph{Class. Quantum Grav.} \textbf{36}, 183001 (2019).

\bibitem{Jimenez2020a}
J.~B. Jiménez and K.~F. Dialektopoulos, \enquote{Non-linear obstructions for
  consistent new general relativity,} \emph{JCAP} \textbf{2020}, 018 (2020).

\bibitem{Jimenez2018}
J.~B. Jim\'enez, L.~Heisenberg, and T.~Koivisto, \enquote{Coincident general
  relativity,} \emph{Phys. Rev. D} \textbf{98}, 044048 (2018).

\bibitem{Barker2024a}
W.~Barker and C.~Marzo, \enquote{Particle spectra of general {R}icci-type
  {P}alatini or metric-affine theories,} \emph{Phys. Rev. D} \textbf{109},
  104017 (2024).

\bibitem{Adak2023}
M.~Adak, T.~Dereli, T.~S. Koivisto, and C.~Pala, \enquote{General teleparallel
  metrical geometries,} \emph{Int. J. Geom. Meth. Mod. Phys.} \textbf{20},
  2350215 (2023).

\bibitem{Andrade2000}
V.~C. de~Andrade, L.~C.~T. Guillen, and J.~G. Pereira, \enquote{Gravitational
  energy-momentum density in teleparallel gravity,} \emph{Phys. Rev. Lett.}
  \textbf{84}, 4533 (2000).

\bibitem{Hayashi1979}
K.~Hayashi and T.~Shirafuji, \enquote{New general relativity,} \emph{Phys. Rev.
  D} \textbf{19}, 3524 (1979).

\bibitem{Ferraro2007}
R.~Ferraro and F.~Fiorini, \enquote{Modified teleparallel gravity: Inflation
  without an inflaton,} \emph{Phys. Rev. D} \textbf{75}, 084031 (2007).

\bibitem{Linder2010}
E.~V. Linder, \enquote{Einstein's other gravity and the acceleration of the
  universe,} \emph{Phys. Rev. D} \textbf{81}, 127301 (2010).

\bibitem{Sotiriou2010}
T.~P. Sotiriou and V.~Faraoni, \enquote{{f(R)} theories of gravity,} \emph{Rev.
  Mod. Phys.} \textbf{82}, 451 (2010).

\bibitem{Capozziello2010}
D.~L.~M. Capozziello~S. and F.~V., \enquote{A bird’s eye view of
  {f(R)}-gravity,} \emph{Open Astron. J.} \textbf{3}, 49 (2010).

\bibitem{DeFelice2010}
A.~De~Felice and S.~Tsujikawa, \enquote{{f(R)} theories,} \emph{Living Rev.
  Relativ.} \textbf{13}, 3 (2010).

\bibitem{Capozziello2021}
S.~Capozziello, A.~Finch, J.~L. Said, and A.~Magro, \enquote{The 3+1 formalism
  in teleparallel and symmetric teleparallel gravity,} \emph{Eur. Phys. J. C}
  \textbf{81}, 1141 (2021).

\bibitem{Krssak2016}
M.~Krššák and E.~N. Saridakis, \enquote{The covariant formulation of {f(T)}
  gravity,} \emph{Class. Quantum Grav.} \textbf{33}, 115009 (2016).

\bibitem{Jimenez2020b}
J.~{Beltrán Jiménez}, L.~Heisenberg, D.~Iosifidis, A.~Jiménez-Cano, and
  T.~S. Koivisto, \enquote{General teleparallel quadratic gravity,} \emph{Phys.
  Lett. B} \textbf{805}, 135422 (2020).

\bibitem{Chakrabortty2023}
M.~Chakrabortty, N.~SK, and A.~K. Sanyal, \enquote{A viable form of
  teleparallel {F(T)} theory of gravity,} \emph{Eur. Phys. J. C} \textbf{83},
  557 (2023).

\bibitem{Gomes2024}
D.~Aguiar~Gomes, J.~Beltr\'an~Jim\'enez, A.~Jim\'enez~Cano, and T.~S. Koivisto,
  \enquote{Pathological character of modifications to coincident general
  relativity: Cosmological strong coupling and ghosts in $f(\mathbb{Q})$
  theories,} \emph{Phys. Rev. Lett.} \textbf{132}, 141401 (2024).

\bibitem{Partanen2024a}
M.~Partanen and J.~Tulkki, \enquote{{QED} based on an eight-dimensional
  spinorial wave equation of the electromagnetic field and the emergence of
  quantum gravity,} \emph{Phys. Rev. A} \textbf{109}, 032224 (2024).

\bibitem{Utiyama1956}
R.~Utiyama, \enquote{Invariant theoretical interpretation of interaction,}
  \emph{Phys. Rev.} \textbf{101}, 1597 (1956).

\bibitem{Kibble1961}
T.~W.~B. Kibble, \enquote{Lorentz invariance and the gravitational field,}
  \emph{J. Math. Phys.} \textbf{2}, 212 (1961).

\bibitem{MacDowell1977}
S.~W. MacDowell and F.~Mansouri, \enquote{Unified geometric theory of gravity
  and supergravity,} \emph{Phys. Rev. Lett.} \textbf{38}, 739 (1977).

\bibitem{Kaku1977}
M.~Kaku, P.~Townsend, and P.~{Van Nieuwenhuizen}, \enquote{Gauge theory of the
  conformal and superconformal group,} \emph{Phys. Lett. B} \textbf{69}, 304
  (1977).

\bibitem{Plebanski1977}
J.~F. Plebański, \enquote{On the separation of {E}insteinian substructures,}
  \emph{J. Math. Phys.} \textbf{18}, 2511 (1977).

\bibitem{Stelle1980}
K.~S. Stelle and P.~C. West, \enquote{Spontaneously broken de {S}itter symmetry
  and the gravitational holonomy group,} \emph{Phys. Rev. D} \textbf{21}, 1466
  (1980).

\bibitem{Ivanov1982}
E.~A. Ivanov and J.~Niederle, \enquote{Gauge formulation of gravitation
  theories. i. the {P}oincar\'e, de {S}itter, and conformal cases,} \emph{Phys.
  Rev. D} \textbf{25}, 976 (1982).

\bibitem{Fradkin1985}
E.~Fradkin and A.~Tseytlin, \enquote{Conformal supergravity,} \emph{Phys. Rep.}
  \textbf{119}, 233 (1985).

\bibitem{Witten1988}
E.~Witten, \enquote{2 + 1 dimensional gravity as an exactly soluble system,}
  \emph{Nucl. Phys. B} \textbf{311}, 46 (1988).

\bibitem{Lasenby1998}
A.~Lasenby, C.~Doran, and S.~Gull, \enquote{Gravity, gauge theories and
  geometric algebra,} \emph{Phil. Trans. R. Soc. Lond. A} \textbf{356}, 487
  (1998).

\bibitem{Vasiliev2001}
M.~Vasiliev, \enquote{Cubic interactions of bosonic higher spin gauge fields in
  {A}d{S}5,} \emph{Nucl. Phys. B} \textbf{616}, 106 (2001).

\bibitem{Fabi2014}
S.~Fabi, B.~Harms, and S.~Hou, \enquote{On the gauge features of gravity on a
  {L}ie algebroid structure,} \emph{J. Math. Phys.} \textbf{55}, 032502 (2014).

\bibitem{Ho2016}
P.-M. Ho, \enquote{Generalized {Y}ang-{M}ills theory and gravity,} \emph{Phys.
  Rev. D} \textbf{93}, 044062 (2016).

\bibitem{Krasnov2011}
K.~Krasnov, \enquote{Pure connection action principle for general relativity,}
  \emph{Phys. Rev. Lett.} \textbf{106}, 251103 (2011).

\bibitem{Fine2014}
J.~Fine, K.~Krasnov, and D.~Panov, \enquote{A gauge theoretic approach to
  {E}instein 4-manifolds,} \emph{New York J. Math.} \textbf{20}, 293 (2014).

\bibitem{Krasnov2017}
K.~Krasnov, \enquote{Self-dual gravity,} \emph{Class. Quantum Grav.}
  \textbf{34}, 095001 (2017).

\bibitem{Clifton2012}
T.~Clifton, P.~G. Ferreira, A.~Padilla, and C.~Skordis, \enquote{Modified
  gravity and cosmology,} \emph{Phys. Rep.} \textbf{513}, 1 (2012).

\bibitem{Karananas2021}
G.~K. Karananas, M.~Shaposhnikov, A.~Shkerin, and S.~Zell, \enquote{Matter
  matters in {E}instein-{C}artan gravity,} \emph{Phys. Rev. D} \textbf{104},
  064036 (2021).

\bibitem{Westman2013}
H.~F. Westman and T.~G. Zlosnik, \enquote{Cartan gravity, matter fields, and
  the gauge principle,} \emph{Ann. Phys.} \textbf{334}, 157 (2013).

\bibitem{Sciama1964}
D.~W. Sciama, \enquote{The physical structure of general relativity,}
  \emph{Rev. Mod. Phys.} \textbf{36}, 463 (1964).

\bibitem{Neville1980}
D.~E. Neville, \enquote{Gravity theories with propagating torsion,} \emph{Phys.
  Rev. D} \textbf{21}, 867 (1980).

\bibitem{Coleman1967}
S.~Coleman and J.~Mandula, \enquote{All possible symmetries of the {$S$}
  matrix,} \emph{Phys. Rev.} \textbf{159}, 1251 (1967).

\bibitem{Bambi2023}
L.~M. C.~Bambi and I.~Shapiro, \emph{Handbook of Quantum Gravity}, Springer,
  Singapore (2023).

\bibitem{Zee2010}
A.~Zee, \emph{Quantum Field Theory in a Nutshell}, Princeton University Press,
  Princeton (2010).

\bibitem{Hooft1974}
G.~'t~Hooft and M.~Veltman, \enquote{One-loop divergencies in the theory of
  gravitation,} \emph{Ann. Inst. H. Poincaré} \textbf{20}, 69 (1974).

\bibitem{Deser1974a}
S.~Deser and P.~van Nieuwenhuizen, \enquote{One-loop divergences of quantized
  {E}instein-{M}axwell fields,} \emph{Phys. Rev. D} \textbf{10}, 401 (1974).

\bibitem{Deser1974b}
S.~Deser, H.-S. Tsao, and P.~van Nieuwenhuizen, \enquote{One-loop divergences
  of the {E}instein-{Y}ang-{M}ills system,} \emph{Phys. Rev. D} \textbf{10},
  3337 (1974).

\bibitem{Deser1974c}
S.~Deser and P.~van Nieuwenhuizen, \enquote{Nonrenormalizability of the
  quantized einstein-maxwell system,} \emph{Phys. Rev. Lett.} \textbf{32}, 245
  (1974).

\bibitem{Goroff1985}
M.~H. Goroff, A.~Sagnotti, and A.~Sagnotti, \enquote{Quantum gravity at two
  loops,} \emph{Phys. Lett. B} \textbf{160}, 81 (1985).

\bibitem{Hooft1971a}
G.~'t~Hooft, \enquote{Renormalization of massless {Y}ang-{M}ills fields,}
  \emph{Nucl. Phys. B} \textbf{33}, 173 (1971).

\bibitem{Hooft1971b}
G.~'t~Hooft, \enquote{Renormalizable {L}agrangians for massive {Y}ang-{M}ills
  fields,} \emph{Nucl. Phys. B} \textbf{35}, 167 (1971).

\bibitem{Casadio2022}
R.~Casadio, I.~Kuntz, and G.~Paci, \enquote{Quantum fields in teleparallel
  gravity: renormalization at one-loop,} \emph{EPJ C} \textbf{82}, 186 (2022).

\bibitem{Donoghue1994b}
J.~F. Donoghue, \enquote{General relativity as an effective field theory: The
  leading quantum corrections,} \emph{Phys. Rev. D} \textbf{50}, 3874 (1994).

\bibitem{Donoghue1999}
J.~F. Donoghue and T.~Torma, \enquote{Infrared behavior of graviton-graviton
  scattering,} \emph{Phys. Rev. D} \textbf{60}, 024003 (1999).

\bibitem{Donoghue2002}
J.~F. Donoghue, B.~R. Holstein, B.~Garbrecht, and T.~Konstandin,
  \enquote{Quantum corrections to the {R}eissner–{N}ordström and
  {K}err–{N}ewman metrics,} \emph{Phys. Lett. B} \textbf{529}, 132 (2002).

\bibitem{Stelle1977}
K.~S. Stelle, \enquote{Renormalization of higher-derivative quantum gravity,}
  \emph{Phys. Rev. D} \textbf{16}, 953 (1977).

\bibitem{Rocci2024}
A.~Rocci and T.~Van~Riet, \enquote{The quantum theory of gravitation, effective
  field theories, and strings: yesterday and today,} \emph{Eur. Phys. J. H}
  \textbf{49}, 7 (2024).

\bibitem{Watson1920}
F.~W. Dyson, A.~S. Eddington, and C.~Davidson, \enquote{{IX}. {A} determination
  of the deflection of light by the sun's gravitational field, from
  observations made at the total eclipse of may 29, 1919,} \emph{Philos. Trans.
  R. Soc. A} \textbf{220}, 291 (1920).

\bibitem{Pound1959}
R.~V. Pound and G.~A. Rebka, \enquote{Gravitational red-shift in nuclear
  resonance,} \emph{Phys. Rev. Lett.} \textbf{3}, 439 (1959).

\bibitem{Schiff1960}
L.~I. Schiff, \enquote{On experimental tests of the general theory of
  relativity,} \emph{Am. J. Phys.} \textbf{28}, 340 (1960).

\bibitem{Shapiro1964}
I.~I. Shapiro, \enquote{Fourth test of general relativity,} \emph{Phys. Rev.
  Lett.} \textbf{13}, 789 (1964).

\bibitem{Nordtvedt1968}
K.~Nordtvedt, \enquote{Testing relativity with laser ranging to the moon,}
  \emph{Phys. Rev.} \textbf{170}, 1186 (1968).

\bibitem{Bertotti2003}
B.~Bertotti, L.~Iess, and P.~Tortora, \enquote{A test of general relativity
  using radio links with the {C}assini spacecraft,} \emph{Nature} \textbf{425},
  374 (2003).

\bibitem{Milani2002}
A.~Milani, D.~Vokrouhlick\'y, D.~Villani, C.~Bonanno, and A.~Rossi,
  \enquote{Testing general relativity with the {B}epi{C}olombo radio science
  experiment,} \emph{Phys. Rev. D} \textbf{66}, 082001 (2002).

\bibitem{Schettino2016}
G.~Schettino and G.~Tommei, \enquote{Testing general relativity with the radio
  science experiment of the {B}epi{C}olombo mission to {M}ercury,}
  \emph{Universe} \textbf{2}, 21 (2016).

\bibitem{Kapner2007}
D.~J. Kapner, T.~S. Cook, E.~G. Adelberger, J.~H. Gundlach, B.~R. Heckel, C.~D.
  Hoyle, and H.~E. Swanson, \enquote{Tests of the gravitational inverse-square
  law below the dark-energy length scale,} \emph{Phys. Rev. Lett.} \textbf{98},
  021101 (2007).

\bibitem{Fomalont2009}
E.~Fomalont, S.~Kopeikin, G.~Lanyi, and J.~Benson, \enquote{Progress in
  measurements of the gravitational bending of radio waves using the {VLBA},}
  \emph{Astrophys. J.} \textbf{699}, 1395 (2009).

\bibitem{Everitt2011}
C.~W.~F. Everitt, D.~B. DeBra, B.~W. Parkinson, J.~P. Turneaure, J.~W. Conklin,
  M.~I. Heifetz, G.~M. Keiser, A.~S. Silbergleit, T.~Holmes, J.~Kolodziejczak,
  M.~Al-Meshari, J.~C. Mester, B.~Muhlfelder, V.~G. Solomonik, K.~Stahl, P.~W.
  Worden, W.~Bencze, S.~Buchman, B.~Clarke, A.~Al-Jadaan, H.~Al-Jibreen, J.~Li,
  J.~A. Lipa, J.~M. Lockhart, B.~Al-Suwaidan, M.~Taber, and S.~Wang,
  \enquote{Gravity probe {B}: Final results of a space experiment to test
  general relativity,} \emph{Phys. Rev. Lett.} \textbf{106}, 221101 (2011).

\bibitem{Yunes2013}
N.~Yunes and X.~Siemens, \enquote{Gravitational-wave tests of general
  relativity with ground-based detectors and pulsar-timing arrays,}
  \emph{Living Rev. Relativ.} \textbf{16}, 9 (2013).

\bibitem{Abbott2016a}
{B. P. Abbott \emph{et al.}}, \enquote{Observation of gravitational waves from
  a binary black hole merger,} \emph{Phys. Rev. Lett.} \textbf{116}, 061102
  (2016).

\bibitem{Abbott2016b}
{B. P. Abbott \emph{et al.}}, \enquote{Tests of general relativity with
  {GW150914},} \emph{Phys. Rev. Lett.} \textbf{116}, 221101 (2016).

\bibitem{Abbott2017}
{B. P. Abbott \emph{et al.}}, \enquote{Multi-messenger observations of a binary
  neutron star merger,} \emph{Astrophys. J. Lett.} \textbf{848}, L12 (2017).

\bibitem{Akiyama2019}
{K. Akiyama \emph{et al.}}, \enquote{First {M87} event horizon telescope
  results. {I}. the shadow of the supermassive black hole,} \emph{Astrophys. J.
  Lett.} \textbf{875}, L1 (2019).

\bibitem{Anderson2023}
{E. K. Anderson \emph{et al.}}, \enquote{Observation of the effect of gravity
  on the motion of antimatter,} \emph{Nature} \textbf{621}, 716 (2023).

\bibitem{Fuchs2024}
T.~M. Fuchs, D.~G. Uitenbroek, J.~Plugge, N.~van Halteren, J.-P. van Soest,
  A.~Vinante, H.~Ulbricht, and T.~H. Oosterkamp, \enquote{Measuring gravity
  with milligram levitated masses,} \emph{Sci. Adv.} \textbf{10}, eadk2949
  (2024).

\bibitem{Westphal2021}
T.~Westphal, H.~Hepach, J.~Pfaff, and M.~Aspelmeyer, \enquote{Measurement of
  gravitational coupling between millimetre-sized masses,} \emph{Nature}
  \textbf{591}, 225 (2021).

\bibitem{Brack2022}
T.~Brack, B.~Zybach, F.~Balabdaoui, S.~Kaufmann, F.~Palmegiano, J.-C. Tomasina,
  S.~Blunier, D.~Scheiwiller, J.~Fankhauser, and J.~Dual, \enquote{Dynamic
  measurement of gravitational coupling between resonating beams in the hertz
  regime,} \emph{Nat. Phys.} \textbf{18}, 952 (2022).

\bibitem{Brack2023}
T.~Brack, J.~Fankhauser, Z.~Bernhard, F.~Balabdaoui, S.~Kaufmann,
  F.~Palmegiano, J.-C. Tomasina, S.~Blunier, D.~Scheiwiller, and J.~Dual,
  \enquote{Dynamic gravitational excitation of structural resonances in the
  hertz regime using two rotating bars,} \emph{Commun. Phys.} \textbf{6}, 270
  (2023).

\bibitem{Einstein1918}
A.~Einstein, \enquote{{\"U}ber {G}ravitationswellen,} \emph{Sitzungsber.
  Preuss. Akad. Wiss.} 154--167 (1918).

\bibitem{Gupta1952}
S.~N. Gupta, \enquote{Quantization of {E}instein's gravitational field: Linear
  approximation,} \emph{Proc. Phys. Soc. A} \textbf{65}, 161 (1952).

\bibitem{Voronov1973}
N.~A. Voronov, \enquote{Gravitational compton effect and photoproduction of
  gravitons by electrons,} \emph{Sov. Phys. JETP} \textbf{37}, 953 (1973).

\bibitem{vanNieuwenhuizen1973}
P.~{van Nieuwenhuizen}, \enquote{On ghost-free tensor {L}agrangians and
  linearized gravitation,} \emph{Nucl. Phys. B} \textbf{60}, 478 (1973).

\bibitem{Choi1995}
S.~Y. Choi, J.~S. Shim, and H.~S. Song, \enquote{Factorization and polarization
  in linearized gravity,} \emph{Phys. Rev. D} \textbf{51}, 2751 (1995).

\bibitem{Balbus2016}
S.~A. Balbus, \enquote{Simplified derivation of the gravitational wave stress
  tensor from the linearized {E}instein field equations,} \emph{PNAS}
  \textbf{113}, 11662 (2016).

\bibitem{Jaranowski2009}
P.~Jaranowski and A.~Kr\'olak, \emph{Analysis of Gravitational-Wave Data},
  Cambridge University Press, Cambridge (2009).

\bibitem{Olyaei2018}
T.~Olyaei and A.~Azizi, \enquote{Weak gravitational interaction of fermions:
  Quantum viewpoint,} \emph{Mod. Phys. Lett. A} \textbf{33}, 1850218 (2018).

\bibitem{Obukhov2003b}
Y.~N. Obukhov and J.~G. Pereira, \enquote{Teleparallel origin of the {F}ierz
  picture for spin-2 particle,} \emph{Phys. Rev. D} \textbf{67}, 044008 (2003).

\bibitem{Ulhoa2017}
S.~C. Ulhoa, A.~F. Santos, and F.~C. Khanna, \enquote{Scattering of fermions by
  gravitons,} \emph{Gen. Relativ. Gravit.} \textbf{49}, 54 (2017).

\bibitem{Becchi1976}
C.~Becchi, A.~Rouet, and R.~Stora, \enquote{Renormalization of gauge theories,}
  \emph{Ann. Phys.} \textbf{98}, 287 (1976).

\bibitem{Iofa1976}
M.~Z. Iofa and I.~V. Tyutin, \enquote{Gauge invariance of spontaneously broken
  non-{A}belian theories in the {B}ogolyubov-{P}arasyuk-{H}epp-{Z}immermann
  method,} \emph{Theor. Math. Phys.} \textbf{27}, 316 (1976).

\bibitem{Fuster2005}
A.~Fuster, M.~Henneaux, and A.~Maas, \enquote{{BRST}-antifield quantization: A
  short review,} \emph{Int. J. Geom. Meth. Mod. Phys.} \textbf{2}, 939 (2005).

\bibitem{Kobayashi1963}
S.~Kobayashi and K.~Nomizu, \emph{Foundations of Differential Geometry, Volume
  I}, Wiley, New York (1963).

\bibitem{Feynman1961}
R.~P. Feynman, \emph{Quantum Electrodynamics}, W. A. Benjamin, New York (1961).

\bibitem{Sachs1961}
M.~Sachs and S.~L. Schwebel, \enquote{A self-consistent field theory of quantum
  electrodynamics,} \emph{Il Nuovo Cimento} \textbf{21}, 197 (1961).

\bibitem{Sachs1964}
M.~Sachs, \enquote{A spinor formulation of electromagnetic theory in general
  relativity,} \emph{Il Nuovo Cimento} \textbf{31}, 98 (1964).

\bibitem{Perkins1978}
G.~A. Perkins, \enquote{A spinor equation of the pure electromagnetic field,}
  \emph{Found. Phys.} \textbf{8}, 745 (1978).

\bibitem{Kulyabov2017}
D.~S. Kulyabov, A.~V. Korolkova, and L.~A. Sevastianov, \enquote{Spinor
  representation of {M}axwell’s equations,} \emph{J. Phys.: Conf. Ser.}
  \textbf{788}, 012025 (2017).

\bibitem{Kiessling2018}
M.~K.-H. Kiessling and A.~S. Tahvildar-Zadeh, \enquote{On the quantum-mechanics
  of a single photon,} \emph{J. Math. Phys.} \textbf{59}, 112302 (2018).

\bibitem{Hong2019}
I.~K. Hong and C.~S. Kim, \enquote{Quaternion electromagnetism and the relation
  with two-spinor formalism,} \emph{Universe} \textbf{5}, 135 (2019).

\bibitem{Jackson1999}
J.~D. Jackson, \emph{Classical Electrodynamics}, Wiley, New York (1999).

\bibitem{Landau1982}
V.~B. Berestetskii, E.~M. Lifshitz, and L.~P. Pitaevskii, \emph{Quantum
  Electrodynamics}, Pergamon, Oxford (1982).

\bibitem{Penrose1987}
R.~Penrose and W.~Rindler, \emph{Spinors and Space-Time}, Cambridge University
  Press, Cambridge (1987).

\bibitem{Cartan1981}
E.~Cartan, \emph{The Theory of Spinors}, Dover, New York (1981).

\bibitem{Noether1918}
E.~Noether, \enquote{Invariante {V}ariationsprobleme,} \emph{Nachr. Ges. Wiss.
  Gött., Math.-Phys. Kl.} \textbf{1918}, 235 (1918).

\bibitem{Hohmann2022}
M.~Hohmann, \enquote{A geometric view on local {L}orentz transformations in
  teleparallel gravity,} \emph{Int. J. Geom. Methods Mod. Phys.} \textbf{19},
  2240001 (2022).

\bibitem{Zlosnik2018}
T.~Złośnik, F.~Urban, L.~Marzola, and T.~Koivisto, \enquote{Spacetime and
  dark matter from spontaneous breaking of {L}orentz symmetry,} \emph{Class.
  Quantum Grav.} \textbf{35}, 235003 (2018).

\bibitem{Gallagher2024}
P.~Gallagher, T.~S. Koivisto, L.~Marzola, L.~Varrin, and T.~Zlosnik,
  \enquote{Consistent first-order action functional for gauge theories,}
  \emph{Phys. Rev. D} \textbf{109}, L061503 (2024).

\bibitem{Nikjoo2024}
M.~Nikjoo and T.~Zlosnik, \enquote{Hamiltonian formulation of gravity as a
  spontaneously-broken gauge theory of the {L}orentz group,} \emph{Class.
  Quantum Grav.} \textbf{41}, 045005 (2024).

\bibitem{Jacobson2001}
T.~Jacobson and D.~Mattingly, \enquote{Gravity with a dynamical preferred
  frame,} \emph{Phys. Rev. D} \textbf{64}, 024028 (2001).

\bibitem{ArkaniHamed2004}
N.~Arkani-Hamed, H.-C. Cheng, M.~A. Luty, and S.~Mukohyama, \enquote{Ghost
  condensation and a consistent infrared modification of gravity,} \emph{J.
  High Energy Phys.} \textbf{2004}, 074 (2004).

\bibitem{Ferraro2022}
R.~Ferraro, \enquote{Noether's second theorem in teleparallel gravity,}
  \emph{Phys. Rev. D} \textbf{106}, 124033 (2022).

\bibitem{Sotiriou2011}
T.~P. Sotiriou, B.~Li, and J.~D. Barrow, \enquote{Generalizations of
  teleparallel gravity and local {L}orentz symmetry,} \emph{Phys. Rev. D}
  \textbf{83}, 104030 (2011).

\bibitem{Golovnev2021}
A.~Golovnev, \enquote{Issues of {L}orentz-invariance in f({T}) gravity and
  calculations for spherically symmetric solutions,} \emph{Class. Quantum
  Grav.} \textbf{38}, 197001 (2021).

\bibitem{Partanen2019b}
M.~Partanen and J.~Tulkki, \enquote{Lagrangian dynamics of the coupled
  field-medium state of light,} \emph{New J. Phys.} \textbf{21}, 073062 (2019).

\bibitem{Schroder1968}
U.~E. Schr\"oder, \enquote{Noether's theorem and the conservation laws in
  classical field theories,} \emph{Fortschr. Phys.} \textbf{16}, 357 (1968).

\bibitem{Belinfante1940}
F.~Belinfante, \enquote{On the current and the density of the electric charge,
  the energy, the linear momentum and the angular momentum of arbitrary
  fields,} \emph{Physica} \textbf{7}, 449 (1940).

\bibitem{Rosenfeld1940}
L.~Rosenfeld, \enquote{Sur le tenseur d’impulsion-\'energie,} \emph{M\'em.
  Acad. Roy. Belg. Sci.} \textbf{18}, 1 (1940).

\bibitem{Ramos2015}
T.~Ramos, G.~F. Rubilar, and Y.~N. Obukhov, \enquote{First principles approach
  to the {A}braham-{M}inkowski controversy for the momentum of light in general
  linear non-dispersive media,} \emph{J. Opt.} \textbf{17}, 025611 (2015).

\bibitem{Partanen2021b}
M.~Partanen and J.~Tulkki, \enquote{Covariant theory of light in a dispersive
  medium,} \emph{Phys. Rev. A} \textbf{104}, 023510 (2021).

\bibitem{Weisskopf1970}
V.~F. Weisskopf, \enquote{Modern physics from an elementary point of view,}
  \emph{CERN-70-08}  (1970).

\bibitem{Ashtekar2014}
A.~Ashtekar and V.~Petkov, \emph{Handbook of Spacetime}, Springer, Dordrecht
  (2014).

\bibitem{Deur2016}
A.~Deur, S.~J. Brodsky, and G.~F. {de Téramond}, \enquote{The {QCD} running
  coupling,} \emph{Prog. Part. Nucl. Phys.} \textbf{90}, 1 (2016).

\bibitem{Abel2023}
S.~Abel, K.~R. Dienes, and L.~A. Nutricati, \enquote{Running of gauge couplings
  in string theory,} \emph{Phys. Rev. D} \textbf{107}, 126019 (2023).

\bibitem{Leuchs2020}
G.~Leuchs, M.~Hawton, and L.~L. Sánchez-Soto, \enquote{{QED} response of the
  vacuum,} \emph{Physics} \textbf{2}, 14 (2020).

\bibitem{Aad2023}
{G. Aad \emph{et al.}}, \enquote{A precise determination of the strong-coupling
  constant from the recoil of {$Z$} bosons with the {ATLAS} experiment at
  $\sqrt{s}=8$ {TeV},} \emph{\normalfont{arXiv:2309.12986}}  (2023).

\bibitem{Paston2011}
S.~A. Paston, \enquote{Gravity as a field theory in flat space-time,}
  \emph{Theor. Math. Phys.} \textbf{169}, 1611 (2011).

\bibitem{Feynman1949}
R.~P. Feynman, \enquote{Space-time approach to quantum electrodynamics,}
  \emph{Phys. Rev.} \textbf{76}, 769 (1949).

\bibitem{Swanson2022}
M.~S. Swanson, \emph{Classical Field Theory and the Stress–Energy Tensor},
  IOP Publishing, Bristol (2022).

\bibitem{Faddeev1967}
L.~Faddeev and V.~Popov, \enquote{Feynman diagrams for the {Y}ang-{M}ills
  field,} \emph{Phys. Lett. B} \textbf{25}, 29 (1967).

\bibitem{DeWitt1967c}
B.~S. DeWitt, \enquote{Quantum theory of gravity. {III}. {A}pplications of the
  covariant theory,} \emph{Phys. Rev.} \textbf{162}, 1239 (1967).

\bibitem{DeWitt1967b}
B.~S. DeWitt, \enquote{Quantum theory of gravity. {II}. {T}he manifestly
  covariant theory,} \emph{Phys. Rev.} \textbf{162}, 1195 (1967).

\bibitem{Donoghue1994a}
J.~F. Donoghue, \enquote{Leading quantum correction to the {N}ewtonian
  potential,} \emph{Phys. Rev. Lett.} \textbf{72}, 2996 (1994).

\bibitem{Kibble1965}
T.~W.~B. Kibble, \enquote{The quantum theory of gravitation,} in \emph{High
  Energy Physics and Elementary Particles, Proc. Trieste Symposium}, edited by
  G.~Fronsdal and A.~Salam, 885, IAEA, Vienna (1965).

\bibitem{Thirring1961}
W.~E. Thirring, \enquote{An alternative approach to the theory of gravitation,}
  \emph{Ann. Phys.} \textbf{16}, 96 (1961).

\bibitem{Batalin1981}
I.~Batalin and G.~Vilkovisky, \enquote{Gauge algebra and quantization,}
  \emph{Phys. Lett. B} \textbf{102}, 27 (1981).

\bibitem{Batalin1983}
I.~A. Batalin and G.~A. Vilkovisky, \enquote{Quantization of gauge theories
  with linearly dependent generators,} \emph{Phys. Rev. D} \textbf{28}, 2567
  (1983).

\bibitem{Costello2011}
K.~Costello, \emph{Renormalization and Effective Field Theory}, American
  Mathematical Society, Providence, RI (2011).

\bibitem{Henneaux1992}
M.~Henneaux and C.~Teitelboim, \emph{Quantization of Gauge Systems}, Princeton
  University Press, Princeton (1992).

\bibitem{Gomis1995}
J.~Gomis, J.~París, and S.~Samuel, \enquote{Antibracket, antifields and
  gauge-theory quantization,} \emph{Phys. Rep.} \textbf{259}, 1 (1995).

\bibitem{Berends1976}
F.~Berends and R.~Gastmans, \enquote{Quantum electrodynamical corrections to
  graviton-matter vertices,} \emph{Ann. Phys.} \textbf{98}, 225 (1976).

\bibitem{Lawrence1971}
J.~K. Lawrence, \enquote{Gravitational fermion interactions,} \emph{Gen.
  Relativ. Gravit.} \textbf{2}, 215 (1971).

\bibitem{Butt2006}
M.~S. Butt, \enquote{Leading quantum gravitational corrections to {QED},}
  \emph{Phys. Rev. D} \textbf{74}, 125007 (2006).

\bibitem{Ramos2010}
J.~Ramos, M.~de~Montigny, and F.~C. Khanna, \enquote{On a {L}agrangian
  formulation of gravitoelectromagnetism,} \emph{Gen. Relativ. Gravit.}
  \textbf{42}, 2403 (2010).

\bibitem{Veltman1976}
M.~J.~G. Veltman, \enquote{Quantum theory of gravitation,} in \emph{Methods in
  Field Theory}, 265--328, North-Holland (1976).

\bibitem{Brandt2022}
F.~T. Brandt, J.~Frenkel, and D.~G.~C. McKeon, \enquote{Quantum gravity in a
  general background gauge,} \emph{Phys. Rev. D} \textbf{106}, 065010 (2022).

\bibitem{Santos2019}
A.~F. Santos and F.~C. Khanna, \enquote{Thermal corrections for gravitational
  {M}öller scattering,} \emph{Int. J. Mod. Phys. A} \textbf{34}, 1950044
  (2019).

\bibitem{Santos2020}
A.~F. Santos and F.~C. Khanna, \enquote{Gravitational {M}öller scattering,
  {L}orentz violation and finite temperature,} \emph{Mod. Phys. Lett. A}
  \textbf{35}, 2050213 (2020).

\bibitem{Moller1932}
C.~Møller, \enquote{Zur {T}heorie des {D}urchgangs schneller {E}lektronen
  durch {M}aterie,} \emph{Ann. Phys.} \textbf{406}, 531 (1932).

\bibitem{Ashkin1954}
A.~Ashkin, L.~A. Page, and W.~M. Woodward, \enquote{Electron-electron and
  positron-electron scattering measurements,} \emph{Phys. Rev.} \textbf{94},
  357 (1954).

\bibitem{BjerrumBohr2003}
N.~E.~J. Bjerrum-Bohr, J.~F. Donoghue, and B.~R. Holstein, \enquote{Quantum
  gravitational corrections to the nonrelativistic scattering potential of two
  masses,} \emph{Phys. Rev. D} \textbf{67}, 084033 (2003).

\bibitem{Faller2008}
S.~Faller, \enquote{Effective field theory of gravity: Leading quantum
  gravitational corrections to {N}ewton's and {C}oulomb's laws,} \emph{Phys.
  Rev. D} \textbf{77}, 124039 (2008).

\bibitem{Netto2022}
T.~de~Paula Netto, L.~Modesto, and I.~L. Shapiro, \enquote{Universal leading
  quantum correction to the newton potential,} \emph{Eur. Phys. J. C}
  \textbf{82}, 160 (2022).

\bibitem{Mandelstam1958}
S.~Mandelstam, \enquote{Determination of the pion-nucleon scattering amplitude
  from dispersion relations and unitarity. {G}eneral theory,} \emph{Phys. Rev.}
  \textbf{112}, 1344 (1958).

\bibitem{Weinzierl2022}
S.~Weinzierl, \emph{Feynman Integrals}, Springer, Berlin (2022).

\bibitem{Smirnov2006}
V.~A. Smirnov, \emph{Feynman Integral Calculus}, Springer, Berlin (2006).

\bibitem{Harlander2024}
R.~V. Harlander and J.-P. Martinez, \enquote{The development of computational
  methods for {F}eynman diagrams,} \emph{Eur. Phys. J. H} \textbf{49}, 4
  (2024).

\bibitem{Shtabovenko2016}
V.~Shtabovenko, R.~Mertig, and F.~Orellana, \enquote{New developments in
  {F}eyn{C}alc 9.0,} \emph{Comput. Phys. Commun.} \textbf{207}, 432 (2016).

\bibitem{Patel2015}
H.~H. Patel, \enquote{Package-{X}: A {M}athematica package for the analytic
  calculation of one-loop integrals,} \emph{Comput. Phys. Commun.}
  \textbf{197}, 276 (2015).

\bibitem{Fan2023}
X.~Fan, T.~G. Myers, B.~A.~D. Sukra, and G.~Gabrielse, \enquote{Measurement of
  the electron magnetic moment,} \emph{Phys. Rev. Lett.} \textbf{130}, 071801
  (2023).

\bibitem{Schwinger1948}
J.~Schwinger, \enquote{On quantum-electrodynamics and the magnetic moment of
  the electron,} \emph{Phys. Rev.} \textbf{73}, 416 (1948).

\bibitem{Berends1975}
F.~Berends and R.~Gastmans, \enquote{Quantum gravity and the electron and muon
  anomalous magnetic moments,} \emph{Phys. Lett. B} \textbf{55}, 311 (1975).

\bibitem{delAguila1984}
F.~{del Águila}, A.~Méndez, and F.~Orteu, \enquote{The electron anomalous
  magnetic moment in unbroken supergravity,} \emph{Phys. Lett. B} \textbf{145},
  70 (1984).

\bibitem{Grisaru1985}
M.~T. Grisaru and D.~Zanon, \enquote{Supergravity corrections to
  electromagnetic form factors: a superspace calculation,} \emph{Class. Quantum
  Grav.} \textbf{2}, 477 (1985).

\bibitem{Georges1985}
A.~Georges and P.~{Le Doussal}, \enquote{Cancellation of divergences in the
  one-loop calculation of $g-2$ in broken supergravity,} \emph{Nucl. Phys. B}
  \textbf{255}, 532 (1985).

\bibitem{Ottoni2006}
J.~Ottoni, A.~{Ba$\hat{\text{e}}$ta Scarpelli}, M.~Sampaio, and M.~Nemes,
  \enquote{Supergravity corrections to the $(g-2)_l$ factor by implicit
  regularization,} \emph{Phys. Lett. B} \textbf{642}, 253 (2006).

\bibitem{Glashow1961}
S.~L. Glashow, \enquote{Partial-symmetries of weak interactions,} \emph{Nucl.
  Phys.} \textbf{22}, 579 (1961).

\bibitem{Weinberg1967}
S.~Weinberg, \enquote{A model of leptons,} \emph{Phys. Rev. Lett.} \textbf{19},
  1264 (1967).

\bibitem{Salam1968}
A.~Salam, \enquote{Weak and electromagnetic interactions,} in \emph{Elementary
  Particle Theory}, edited by N.~Svartholm, 367–377, Almqvist and Wiskell
  (1968).

\bibitem{Higgs1964}
P.~W. Higgs, \enquote{Broken symmetries and the masses of gauge bosons,}
  \emph{Phys. Rev. Lett.} \textbf{13}, 508 (1964).

\bibitem{Englert1964}
F.~Englert and R.~Brout, \enquote{Broken symmetry and the mass of gauge vector
  mesons,} \emph{Phys. Rev. Lett.} \textbf{13}, 321 (1964).

\bibitem{Higgs1966}
P.~W. Higgs, \enquote{Spontaneous symmetry breakdown without massless bosons,}
  \emph{Phys. Rev.} \textbf{145}, 1156 (1966).

\bibitem{Brown2023}
A.~R. Brown, H.~Gharibyan, S.~Leichenauer, H.~W. Lin, S.~Nezami, G.~Salton,
  L.~Susskind, B.~Swingle, and M.~Walter, \enquote{Quantum gravity in the lab.
  {I}. {T}eleportation by size and traversable wormholes,} \emph{PRX Quantum}
  \textbf{4}, 010320 (2023).

\bibitem{Nezami2023}
S.~Nezami, H.~W. Lin, A.~R. Brown, H.~Gharibyan, S.~Leichenauer, G.~Salton,
  L.~Susskind, B.~Swingle, and M.~Walter, \enquote{Quantum gravity in the lab.
  {II}. {T}eleportation by size and traversable wormholes,} \emph{PRX Quantum}
  \textbf{4}, 010321 (2023).

\bibitem{Oppenheim2023c}
J.~Oppenheim, C.~Sparaciari, B.~Šoda, and Z.~Weller-Davies,
  \enquote{Gravitationally induced decoherence vs space-time diffusion: testing
  the quantum nature of gravity,} \emph{Nat. Commun.} \textbf{14}, 7910 (2023).

\bibitem{Carney2024}
D.~Carney, V.~Domcke, and N.~L. Rodd, \enquote{Graviton detection and the
  quantization of gravity,} \emph{Phys. Rev. D} \textbf{109}, 044009 (2024).

\bibitem{Audretsch1983}
J.~Audretsch and C.~Lammerzahl, \enquote{Neutron interference: general theory
  of the influence of gravity, inertia and space-time torsion,} \emph{J. Phys.
  A} \textbf{16}, 2457 (1983).

\bibitem{Hohensee2012}
M.~A. Hohensee, B.~Estey, P.~Hamilton, A.~Zeilinger, and H.~M\"uller,
  \enquote{Force-free gravitational redshift: Proposed gravitational
  {A}haronov-{B}ohm experiment,} \emph{Phys. Rev. Lett.} \textbf{108}, 230404
  (2012).

\bibitem{Overstreet2022}
C.~Overstreet, P.~Asenbaum, J.~Curti, M.~Kim, and M.~A. Kasevich,
  \enquote{Observation of a gravitational {A}haronov-{B}ohm effect,}
  \emph{Science} \textbf{375}, 226 (2022).

\bibitem{Gattringer2010}
C.~Gattringer and C.~B. Lang, \emph{Quantum Chromodynamics on the Lattice},
  Springer, Berlin (2010).

\bibitem{Hattori2023}
K.~Hattori, K.~Itakura, and S.~Ozaki, \enquote{Strong-field physics in {QED}
  and {QCD}: From fundamentals to applications,} \emph{Prog. Part. Nucl. Phys.}
  \textbf{133}, 104068 (2023).

\bibitem{Wen2023}
R.~Wen, S.~Yin, W.-j. Fu, and M.~Huang, \enquote{Functional renormalization
  group study of neutral and charged pions in magnetic fields in the
  quark-meson model,} \emph{Phys. Rev. D} \textbf{108}, 076020 (2023).

\bibitem{Barker2023}
W.~Barker, \enquote{Particle spectra of gravity based on internal symmetry of
  quantum fields,} \emph{\normalfont{arXiv:2311.11790}}  (2023).

\end{thebibliography}
\end{document}